\documentclass[useAMS,usenatbib]{mn2e}

\newcommand{\bea}{\begin{eqnarray}}
\newcommand{\eea}{\end{eqnarray}}
\newcommand{\be}{\begin{equation}}
\newcommand{\ee}{\end{equation}}
\newcommand{\ben}{\begin{enumerate}}
\newcommand{\een}{\end{enumerate}}
\newcommand{\bi}{\begin{itemize}}
\newcommand{\ei}{\end{itemize}}
\newcommand{\bc}{\begin{center}}
\newcommand{\ec}{\end{center}}

\def\NOTE#1{{ }}

\def\abs#1{\left\vert #1 \right\vert}

\def\ave#1{\left\langle #1 \right\rangle}

\def\P{{\cal P}}

\def\eps{{\epsilon}}

\def\Real{{\rm I\mathchoice{\kern-0.70mm}{\kern-0.70mm}{\kern-0.65mm}%
  {\kern-0.50mm}R}}
\def\C{\rm C\kern-.42em\vrule width.03em height.58em depth-.02em
       \kern.4em}
\font \bolditalics = cmmib10
\def\bx#1{\leavevmode\thinspace\hbox{\vrule\vtop{\vbox{\hrule\kern1pt
        \hbox{\vphantom{\tt/}\thinspace{\bf#1}\thinspace}}
      \kern1pt\hrule}\vrule}\thinspace}

\def \vc #1{{\textfont1=\bolditalics \hbox{$\bf#1$}}}
{\catcode`\@=11
\gdef\SchlangeUnter#1#2{\lower2pt\vbox{\baselineskip 0pt \lineskip0pt
  \ialign{$\m@th#1\hfil##\hfil$\crcr#2\crcr\sim\crcr}}}
}

\def\deg      {{\ifmmode^\circ\else$^\circ$\fi}} 

\newcommand{\Elixir}{\texttt{Elixir}\space}

\newcommand{\MegaPrime}{\texttt{MegaPrime}\space}
\newcommand{\MegaCam}{\texttt{MegaCam}\space}
\newcommand{\VIMOS}{\texttt{VIMOS}\xspace}
\newcommand{\WFI}{\texttt{WFI}\xspace}

\newcommand{\SExtractor}{\texttt{SExtractor}\space}

\newcommand{\THELI}{\texttt{THELI}\space}

\usepackage{float,epsfig}
\usepackage{pstricks}
\usepackage{graphicx,color}
\usepackage{latexsym,amsfonts,amsmath,amssymb}
\usepackage{url}
\usepackage{natbib}
\usepackage{rotating}
\usepackage{xspace}

\title[Dark matter halo properties from GGL]{Dark matter halo properties from galaxy-galaxy lensing\thanks{Based on observations obtained with MegaPrime/MegaCam, a joint project of CFHT and CEA/DAPNIA, at the Canada-France-Hawaii Telescope (CFHT) which is operated by the National Research Council (NRC) of Canada, the Institut National des Sciences de l'Univers of the Centre National de la Recherche Scientifique (CNRS) of France, and the University of Hawaii. This work is based on data products produced at TERAPIX and the Canadian Astronomy Data Centre (CADC) as part of the Canada-France-Hawaii Telescope Legacy Survey, a collaborative project of NRC and CNRS.}}
\author[Brimioulle et al.]{F. Brimioulle$^{1,2}$\thanks{E-mail:
fabrice@usm.uni-muenchen.de}, S. Seitz$^{1,2}$, M. Lerchster$^{1,2}$, R. Bender$^{1,2}$, J. Snigula$^{1,2}$
\\
$^{1}$Universit\" ats-Sternwarte M\" unchen, Ludwig-Maximillians-Universit\" at M\" unchen, Scheinerstra\ss e 1, 81679 M\" unchen, Germany\\
$^{2}$Max-Planck-Institut f\" ur extraterrestrische Physik, Giessenbachstra\ss e, 85748 Garching, Germany}
\begin{document}



\maketitle

\label{firstpage}
\begin{abstract}
We present results for a galaxy-galaxy lensing study based on imaging data from the Canada-France-Hawaii Telescope Legacy Survey Wide. 
From a \mbox{12 million} object  multi-colour catalogue for 124 $\rm deg^2$ of photometric data in the $u^*g'r'i'z'$ filters, we compute photometric redshifts (with a scatter of  \mbox{$\sigma_{\Delta z/(1+z)} = 0.033$} and an outlier rate of $\eta=2.0$ per cent for $i' \le 22.5$) and extract galaxy shapes down to \mbox{$i'=24.0$}. 
We select a sample of lenses and sources with \mbox{$0.05 < z_{\rm d}\le 1$} and \mbox{$0.05 < z_{\rm s}\le 2$}. We fit three different galaxy halo \mbox{profiles to the lensing signal}, a singular isothermal sphere (SIS), a truncated isothermal sphere (BBS) and a \mbox{universal} density profile (NFW).
We derive velocity dispersions by fitting an SIS out to $100 \ h^{-1} \ \rm{kpc}$ to the excess surface mass density $\Delta \Sigma$ \mbox{and perform} maximum \mbox{likelihood analyses} out to a maximum scale of $2\ h^{-1}$ Mpc \mbox{to obtain} halo parameters and scaling relations.
We find luminosity scaling relations of \mbox{$\sigma_{\rm red} \propto L^{0.24\pm0.03}$} for the red lens sample, \mbox{$\sigma_{\rm blue} \propto L^{0.23\pm0.03}$} for blue lenses and \mbox{$\sigma \propto L^{0.29\pm0.02}$} for the combined lens sample with zeropoints of \mbox{$\sigma^{*}_{\rm red}=162\pm2\ \rm{km\ s^{-1}}$}, \mbox{$\sigma^{*}_{\rm blue}=115\pm3\ \rm{km\ s^{-1}}$} and \mbox{$\sigma^{*}=135\pm2\ \rm{km\ s^{-1}}$} at a chosen reference luminosity $L_{r'}^* = 1.6 \times 10^{10}\ h^{-2}\ L_{r',\odot}$. The steeper slope for the combined sample is due to the different zeropoints of the blue and red lenses and the fact that blue lenses dominate at low luminosities and red lenses at high luminosities. The \mbox{mean effective} redshifts for the lens samples are $\ave{z_{\rm red}}=0.28$ for red lenses, \mbox{$\ave{z_{\rm blue}}=0.35$ for} blue lenses and $\ave{z}=0.34$ for the combined lens sample. The BBS maximum likelihood analysis yields for the combined sample a velocity dispersion of \mbox{$\sigma^{*} = 131^{+2}_{-2}$ km\ $\rm{s^{-1}}$} and a truncation radius of \mbox{$s^{*} = 184^{+17}_{+14}  \ h^{-1}$ kpc, corresponding} to a total mass of \mbox{$M^{*}_{\rm{total,BBS}} = 2.32^{+0.28}_{-0.25} \times 10^{12} \ h^{-1} \ M_{\odot}$} and a mass-to-light (M/L) ratio of \mbox{$M^{*}_{\rm total,BBS}/L^{*}=178^{+22}_{-19} \ h \ M_{\odot}/L_{r',\odot}$} at $L_{r'}^{*}$. At a given \mbox{luminosity}, both velocity dispersion $\sigma$ and truncation radius $s$ are larger for red galaxies than for blue \mbox{galaxies}. For an NFW profile, we measure at $L_{r'}^{*}$ a virial radius of \mbox{$r^{*}_{200} = 133^{+3}_{-2} \ h^{-1}$ kpc} and a concentration  parameter of $c^{*} = 6.4^{+0.9}_{-0.7}$, implying a virial mass of \mbox{$M^{*}_{200} = 7.6^{+0.5}_{-0.3} \times 10^{11} \ h^{-1} \ M_{\odot}$}. At $L_{r'}^{*}$ for blue galaxies the concentration \mbox{parameter} is slightly higher than for red galaxies and $r_{200}$ is significantly lower. For the combined sample, if described as a single power law, the M/L-ratio scales as \mbox{$M_{\rm total,BBS}/L \propto L^{0.12^{+0.10}_{-0.11}}$}, the concentration parameter scales as $c \propto  L^{-0.07^{+0.11}_{-0.11}}$. Analysing the M/L-scaling for red and blue galaxies separately, we find that a broken power law (with a flat slope at high luminosities) provides a more appropriate description for the red and possibly also for the blue galaxies. We measure $M_{200}/M_{\rm star}$ for red galaxies over 2.5 decades in stellar mass. We find a minimum for this ratio at $M_{\rm star} \sim 3-4 \times 10^{10}\ h^{-2}\ M_{\odot}$ with a strong increase for lower stellar masses.
\end{abstract}
\begin{keywords}
gravitational lensing: weak -- galaxies: distances and redshifts -- galaxies: haloes -- galaxies: kinematics and dynamics -- dark matter
\end{keywords}
\section{Introduction}
Dark matter is the dominant mass component in the Universe, not only on scales of clusters or groups of galaxies but also on galaxy scales where dark matter haloes surround the luminous baryonic cores of galaxies. This brings up the question of how to measure and quantify the properties of dark matter galaxy haloes. On short scales the luminous matter can serve as a tracer to investigate the mass distribution with dynamical methods (e.g. measuring rotation curves of spiral galaxies or velocity dispersions of early type galaxies). On larger scales one can use satellite galaxies as tracers (see e.g. \citealt{prada03} or \citealt{conroy05}) or can study the gravitational lens effect that dark matter haloes impose on background galaxies: they introduce a small coherent distortion to the shapes of background galaxies with an amplitude depending on the angular diameter distances and relative positions of foreground and background galaxies and on the dark matter distribution of the galaxy haloes.
\\
As the gravitational impact of a single galaxy is too small to be detected individually, a large sample of foreground and background galaxies is required to measure this effect statistically. The first successful detection of galaxy-galaxy-lensing (GGL) was reported by \citet{bbs96}, although the analysis was limited by the small area covered by the observations, followed by \citet{dellantonio96} in the Hubble Deep Field (HDF). 
Later on GGL measurements from space with the Hubble Space Telescope (HST) have been done by \citet[HST Medium Deep Survey] {griffiths96} and \citet[HDF]{hudson98}. Further ground-based GGL analyses have been carried out by \citet{wilson01} with CFHT/UH8K or by \citet{hoekstra03} using data from the Canadian Network for Observational Cosmology Field Galaxy Redshift Survey (CNOC2) and by \citet{hoekstra04} investigating a significantly larger area of 45.5 $\rm{deg^2}$ Red-Sequence Cluster Survey (RCS) imaging data. While in both cases photometric redshift information was not available, \citet{kleinheinrich06} analysed Wide Field Imager (\WFI) data from the `Classifying Objects by Medium-Band Observations in 17 filters' (COMBO-17) survey, measuring the scaling 
for the Tully-Fisher relation \citep{tullyfisher} and investigating the halo properties assuming an NFW profile. Up to today the largest survey area used to quantify dark matter properties on galaxy scale is the Sloan Digital Sky Survey (SDSS, e.g. \citealt{fischer00}, \citealt{mckay01} or \citealt{seljak02}). \citet{sheldon04}, \citet{mandelbaum06b,mandelbaum08b} and \citet{schulz10} measured the excess surface density $\Delta\Sigma$ and mass profiles for various samples using spectroscopic redshift information for the lenses and a combination of photometric redshift and statistical redshift distributions for background objects. Finally, \citet{vanuitert11} combined high quality imaging data for the background shape estimation from the RCS2-$r'$-band data and the profound knowledge of the foreground redshifts available from the SDSS DR7 to quantify the dark matter halo properties as a function of luminosities, stellar and dynamical masses.

In this study we make use of 124 $\rm{deg}^2$ public CFHTLS-Wide imaging data, covering $u^{*}g'r'i'z'$ photometry, thus providing an excellent base for photometric redshift estimation and perform a GGL analysis.
We give an introduction to the theory of GGL and introduce our notation in Section~\ref{sec:theory}. In Sections \ref{sec:data} and \ref{sec:spectra} we discuss the imaging and the spectroscopic data used in this paper. The creation of colour catalogues and the computation of photometric redshifts are presented in Section~\ref{sec:photoz}. The background shape estimation is described in Section~\ref{sec:shape_shear}.  We present the properties of our sample catalogues in Section~\ref{sec:properties_catalogues}, define lens and background source samples  in Section~\ref{sec:lenssample} and discuss the lens mass errors introduced by possible errors in shape measurement and photometric redshift estimation in Section~\ref{sec:lensmassbias}. The weak lensing analysis, including measurement of the tangential shear $\gamma_{\rm t}$, measurement of the excess surface mass density $\Delta\Sigma$ and maximum likelihood analyses for various subsamples concerning spectral energy distributions (SEDs), luminosities and environments, is described in Section~\ref{sec:lensing}. In Section~\ref{sec:conclusions} we summarise and conclude our analysis. In the Appendix we measure the systematic errors to verify the integrity of our analysis and compare our lensing results to simulations of the 3D-line-of-sight(LOS)-projected weak lensing signal predicted by our results, i.e. using the measured best-fit parameters and luminosity scaling relations for the dark matter haloes.

Throughout this paper we adopt a cosmology with \mbox{$\Omega_{\rm M} = 0.27$}, \mbox{$ \Omega_{\rm \Lambda} = 0.73$} and a dimensionless Hubble parameter \mbox{$h=0.72$}, unless explicitly stated otherwise. All referred apparent magnitudes and colours are given in AB, all rest frame magnitudes are calculated in vega system, assuming a Hubble constant of \mbox{$H_{0}=72$ $\rm{km\ s^{-1}\ Mpc^{-1}}$}.

Our created catalogues (photometry, photometric redshifts and shears) are publicly available.\footnote{http://www.usm.uni-muenchen.de/people/stella/GGL/}
\section{Theory of galaxy-galaxy lensing}
\label{sec:theory}
Weak gravitational lensing describes the impact of line-of-sight structures on the shape, size and magnitude of background galaxies in the limit where no multiple images are produced. If the galaxy shape distortions are small, the shape distortions can be directly related to the density field producing them. The inversion of the shear-mass density relation into the mass density-shear relation is the Kaiser and Squires equation (\citealt{kaiser93}). It has paved the way for weak lensing analyses in general (for the equations we refer the reader to Section~\ref{sec:gammat,deltasigma}) and it also states the principle of weak lensing analyses using galaxy shapes: The mass density field can be inferred from galaxy shapes because they carry the imprint of the tidal field of the gravitational potential.
\\
Weak gravitational lensing analysis techniques can be divided into three branches: one can obtain mass maps for individual galaxy clusters, one can study the mean radial profile of an ensemble of clusters, groups and galaxies, and finally one can measure the statistics of dark matter fluctuations as a function of scale. For the first case one needs a high signal-to-noise lensing signal, i.e. a massive cluster and deep observations, providing a high number density of background galaxies. One can then turn a 2D gravitational shear map into a 2D mass map without requiring a reference to the light distribution of the cluster. For the second analysis method the signal of the individual structures can be lower. The stacking or averaging technique then yields the mean properties of the sample. The stacking requires a reference position, which is taken to be, e.g. the brightest cluster galaxy in case of cluster and group weak lensing and the centre of light for the galaxy of interest in case of GGL. The analysis therefore then results in a correlation between galaxy positions and the mass associated to them (galaxy mass correlation function). The third method (cosmic shear) analyses the correlations of gravitational shears along pairs of light rays traced by galaxies. The shear-shear correlation function has a very low amplitude, since it is the product of two small numbers. Therefore cosmic shear analyses require to analyse the largest amount of galaxy pairs to obtain a significant result. If one ignores the problem of measuring galaxy shapes, then the two remaining sources of bias in weak lensing measurements are due to incorrectly locating the galaxies along the line-of-sight and due to limited validity of the assumption that galaxies are randomly oriented on the sky.
\\
\\
In this Section we shortly summarise the theory of GGL in Section~\ref{sec:gammat,deltasigma} and introduce the tangential alignment $\gamma_{\rm t}$ and the excess surface density $\Delta\Sigma$. In  Section~\ref{sec:galaxy models} we describe the galaxy halo profiles investigated later on. Section~\ref{sec:scaling theory} introduces the scaling relations which we have assumed (and later on show to hold) when we analyse the profile halo parameters. In Section~\ref{sec:cluster halo} we investigate the influence of a group or cluster halo on the lensing signal of a galaxy. Finally in Section~\ref{sec:likelihood theory} we introduce the maximum likelihood analysis, which is a more sophisticated method to analyse the GGL signal, since it fully accounts for multiple deflections of light rays by multiple foreground galaxies.
\subsection{Tangential alignment and lens mass excess surface density}
\label{sec:gammat,deltasigma}
We briefly review some relations needed later on. We refer the reader to \citet{saasfee} for an extended overview. 
\\
In gravitational lensing the light deflection angle $\vc{{\alpha}}$  can be obtained from a deflection potential $\psi$ as  $\vc {\alpha}=\nabla \psi $, and the lens equation  reads $\vc \beta=\vc \theta-\vc\alpha = \vc\theta - \vc \nabla\psi$. The Jacobian describes the mapping of small light bundles \citep{saasfee,SSE94} and depends on the convergence
\begin{equation}
\kappa =\frac{1}{2} \nabla^2 \psi  = \frac{\Sigma}{\Sigma_{\rm c}}  
\label{eq:kappa}
\end{equation}
and the gravitational shear
\begin{equation}
\gamma = \gamma_1 + i \gamma_2 = \frac{1}{2} \left(\psi_{,11} -\psi_{,22}\right)  + i \psi_{,12}\ .
\label{eq:gamma}
\end{equation}
$\kappa $ is the ratio of the surface mass density $\Sigma$ of the lens and the critical density 
\begin{equation}
\Sigma_{\rm c}= \frac{c^2}{4 \pi G} \frac{D_{\rm{s}}}{D_{\rm{d}}\ D_{\rm{ds}}} \ ,
\end{equation}
depending on the angular diameter distances $D_{\rm d}$ and $D_{\rm s}$ from the observer to the lens and source and $D_{\rm ds}$ from the lens to the source. The gravitational shear $\gamma $ is a two component quantity describing the size and direction of the tidal forces distorting the shapes of light bundles.
Equations~(\ref{eq:kappa}) and (\ref{eq:gamma}) show that $\kappa$ and $\gamma$ are related to each other via the deflection potential. In fact \citet{kaiser95} has shown that
\begin{equation}
  \nabla\kappa = \left(\begin{array}{c}
    \gamma_{1,1}+\gamma_{2,2} \\
    \gamma_{2,1}-\gamma_{1,2} \\
  \end{array}\right) 
\label{eq:grad_kappa_gamma}
\end{equation}
holds and that one can obtain $\nabla\kappa$ from shape estimates in the very weak lensing limit, $\kappa\ll1$. For the more general case relations are given in \citet{kaiser95}, \citet{schneiderseitzC95}, \citet{seitzCschneider95} and are used e.g. in \citet{ksb95}, \citet{seitzschneider96} and \citet{seitzschneider01} to obtain $\kappa$-maps from local shape estimates. 
\\
We will now focus on the information that can be extracted from GGL, where the signal of many galaxies is stacked and only the mean tangential shear profile can be observed.
\\
\citet{saasfee} and \cite{kaiser95} have shown that for an arbitrary mass distribution the mean tangential shear $\ave{\gamma_{\rm t}}$ on a circle at radius $\theta$ can be written as a function of the mean surface mass density $\bar \kappa $ within that circle and the mean surface mass density $\ave{\kappa}$ on the edge of that circle: 
\begin{equation}
\ave{\gamma_{\rm t}}(\vc \theta)=\bar\kappa(\vc \theta)-\ave{\kappa} (\vc \theta)\ .
\label{eq:etan1}
\end{equation}
For circular mass distributions this relation was derived by \citet{bartelmann95} and \citet{kaiser96}.
\\
Equation~(\ref{eq:etan1}) can be rewritten as
\begin{equation}
 \Sigma_{\rm c} \ \ave{\gamma_{\rm t}}(R)=\bar\Sigma (R)-\ave{\Sigma}(R)\equiv \Delta \Sigma (R) \ ,
\label{eq:etan2}
\end{equation}
which is called the excess surface mass density $\Delta \Sigma$. This means that if for a lens system the source and lens distances and the mean tangential shear is known then $\Delta\Sigma(R)$ can be derived. \citet{mckay01} were the first who made use of this relation in the context of GGL. They directly measured $\Delta \Sigma$ for galaxies  as a function of the distance $R$ to the galaxy centres instead of just presenting the tangential gravitational shear $\ave{\gamma_{\rm t}}$ as a function of angular scale $\vc \theta$. Before that the mass density associated with galaxies was derived from the observations using parametric models for the gravitational shear and  statistical descriptions for the distribution of background and foreground objects \citep{bbs96,hoekstra03,hoekstra04,parker07}.
\subsection{Profiles for galaxy haloes}
\label{sec:galaxy models}
The measurements of $\Delta\Sigma(R)$ (equation~\ref{eq:etan2}) relative to the centres of galaxy light directly link galaxy evolution to structure formation. In the most simple view a galaxy can be pictured as being composed of an ensemble of stars in the centre surrounded by a dark matter halo which is described by an analytic profile. Since the dark matter halo is supposed to belong to the galaxy it has to be finite in extent and in mass.
\\
If galaxy haloes were all isolated one could just measure their profile directly, in particular how haloes end or are cut off in the outskirts. In a more complete picture one however has to take into account that galaxy haloes might be embedded as satellites in more massive galaxy, group or cluster haloes, or that their halo could contain satellite galaxies itself. Therefore, especially at larger distances, the individual profiles barely can be traced as we additionally measure the parent haloes in which the investigated galaxies are embedded in, or we measure the line-of-sight projection effects of other haloes within the same structure (e.g. in group and cluster cores). Furthermore, multiple gravitational deflections by different line-of-sight structures introduce further difficulties, which need to be taken into account in a quantitive analysis (see \citealt{brainerd10}).
\\
Although the situation seems complex it makes sense to nevertheless use simple halo profiles and constrain their parameters: On scales smaller than $~200\ h^{-1}$ kpc, the galaxy mass correlation function is dominated by the halo in which the galaxy resides (see Section~\ref{sec:cluster halo} and  \citealt{seljak05} or \citealt{mandelbaum05b}, \mbox{fig.~1}). This can be used to establish scaling relations and thus to model simultaneously galaxy haloes with various luminosity reasonably well. Then one can quantitatively describe all haloes that host a galaxy in their centre and vary their joint parameters such to match observations best. In this way only haloes which do not host a galaxy or which host a galaxy fainter than the detection limit get ignored. 
In the following we will model each galaxy halo as a function of the galaxy luminosity and scaling relations relative to a halo with a fiducial luminosity $L^*$. 
The motivation to analyse the galaxies as a function of luminosity and to assume and to investigate scaling relations are the following:
\ben
\item We do not only want to analyse `the' average dark matter halo (a result we could easily obtain by just stacking the signal with respect to all foreground galaxies).
\item We want to establish the link between dark matter halo properties and the optical properties of galaxies. The most basic galaxy property is the luminosity. In the redder bands it can also be taken as a proxy to stellar mass, which then allows to study how effectively baryons are turned into stars as a function of dark matter halo depth.
\item The galaxy luminosity is tightly related to the total halo potential at least for the central regions of a galaxy, partly because the stars are a major contributor to the potential in the centres of a galaxy, partly because of the luminous and dark matter conspiracy. This is causing the transition in the haloes from luminous to dark matter dominance to be such that spiral rotation curves are flat and ellipticals have a close to isothermal mass profile. From this picture we (in agreement with the Tully-Fisher and Faber-Jackson relation, \citealt{faberjackson} and \citealt{tullyfisher}) expect the halo parameters (in particular the halo `depth') to scale with luminosity.
\item If the halo parameters scale with luminosity we can assume scaling laws when analysing the GGL signal and obtain the halo parameters for a reference halo with high signal-to-noise by jointly analysing galaxies with various luminosities.
\item We can also investigate (by splitting into luminosity subsamples) whether the assumed scaling laws indeed hold and measure e.g. the coefficients for the power law scaling relations between luminosity and halo parameter.
\een
We will not distinguish between central galaxies and satellites at this point. This means that our results will describe the mean dark matter halo of a galaxy as a function of its luminosity. Then a galaxy which is in the centre of a halo is expected to have a larger mass than our mean, and a galaxy  which is a satellite is expected to have a lower mass (due to halo stripping) than this mean. The halo profiles investigated are the singular isothermal sphere (SIS), the truncated isothermals sphere (BBS, \citealt{bbs96}) and the universal dark matter profile of NFW (\citealt{nfw96}).
\subsubsection{Singular isothermal sphere (SIS)}
\label{sec:SIS_theory}
The singular isothermal sphere (SIS)
\begin{equation}
\rho_{\rm SIS}(R)=\frac{\sigma^2}{2\pi GR^2}
\label{eq:rho-SIS}
\end{equation}
has a projected surface density of 
\begin{equation}
\Sigma_{\rm SIS}(R)=\frac{\sigma^2}{ 2GR} \  ,
\label{eq:SIS1}
\end{equation}
with $\sigma$ being the line-of-sight velocity dispersion of test particles (stars) and $R$ the distance from the centre of mass in the plane of projection. For an SIS  the convergence and tangential shear are equal, 
\begin{equation}
\gamma_{\rm t, SIS}(\theta) =\kappa _{\rm SIS}(\theta) = {4 \pi} \left(\frac{\sigma}{c}\right)^2 \ \frac{D_{\rm ds}}{D_{\rm s}} \frac{1}{2\theta} \equiv \frac{\theta_{\rm E,SIS}}{2\theta} ,
\label{eq:SIS2}
\end{equation}
where we have used $R=\theta \ D_{\rm d}$. 
The line-of-sight projected mass within a sphere of radius  $R$ is given by
\begin{equation}
M_{\rm SIS}(<R) = \frac{2\sigma^2}{G} R \ ,
\end{equation}
which means that the total SIS mass diverges.
\subsubsection{Truncated isothermal sphere (BBS)}
\label{sec:BBS_theory}
\citet{bbs96} introduced the truncated isothermal sphere (BBS) with a volume density of 
\begin{equation}
\rho_{\rm BBS}(R)=\frac{\sigma^2}{2\pi GR^2} \frac{s^2}{R^2+s^2}\ ,
\end{equation}
which has one further free parameter compared to the SIS, the truncation radius $s$. For infinite truncation radii $s$ the density profile converges to that of an SIS. On short scales ($R\ll  s$) the difference to an SIS is negligible but at a radius of $R=s$  the volume density has dropped down to only half the value of an SIS with the same velocity dispersion. For larger distances $R\gg s$ the BBS volume density profile falls off $\propto R^{-4}$. 
The surface mass density equals
\begin{equation}
\Sigma_{\rm BBS}(R) = \frac{\sigma^2}{2 G R} \left(1 -\frac{R}{\sqrt{R^2+s^2}}\right) \ ,
\end{equation}
which can be rewritten with $\theta _{\rm s}= \theta/ s$ in terms of the dimensionless convergence $\kappa_{\rm BBS}$ as
\begin{equation}
\kappa_{\rm BBS}(\theta) = \frac{\theta_{\rm E, SIS}}{2 \theta} \left(1-\frac{\theta}{\sqrt{\theta^2 + \theta_{\rm s}^2 }}\right) \ .
\end{equation}
The gravitational shear equals
\begin{equation}
\gamma_{\rm BBS}(\theta) = \frac{\theta_{\rm E, SIS}}{2 \theta}
\left(\frac{\theta + 2\theta_{\rm s}}{\theta}
- \frac{\theta^2+2 \theta_{\rm s}^2}{\theta \sqrt{\theta^2 + \theta_{\rm s}^2 }}\right) \ .
\end{equation}
The line-of-sight projected mass within a sphere of radius  $R$ is given by
\begin{equation}
M_{\rm BBS}(<R) = \frac{2\sigma^2s}{G} {\rm arctan}\left(\frac{R}{s}\right) \ ,
\end{equation}
adding up to a total halo mass of
\begin{align}
M_{\rm total, BBS} = \frac{\pi \sigma^2 s}{G} = \ \ \ \ \ \ \ \ \ \ \ \ \ \ \ \ \ \ \ \ \ \ \nonumber \\
7.3 \times 10^{12} \ h^{-1} \ {\rm M_{\sun}} \left( \frac{\sigma}{1000 \ \rm{km\ s^{-1}}} \right)^2 \left( \frac{s}{1\ \rm{Mpc}} \right) \ .
\label{eq:massBBStotal}
\end{align}
\subsubsection{Universal density profile (NFW)}
\label{sec:NFW_theory}
The third profile considered is the universal density profile, also known as NFW profile, introduced by \citet{nfw96,nfw97} with a volume density of
\begin{equation}
\rho_{\rm NFW}(R)=\frac{\delta_{\rm c} \rho_{\rm c} }{(R/r_{\rm s})(1 + R/r_{\rm s})^2} \  ,
\end{equation}
where
\begin{equation}
\rho_{\rm c} = \frac{3 H(z)^2}{8 \pi G}
\label{eq:rho_crit}
\end{equation}
stands for the critical density of the Universe at redshift $z$.
The NFW profile has two parameters, the scale radius $r_s$, which can be expressed in terms of the radius where the density profile turns from $\rho_{\rm NFW}(R) \propto R^{-1}$ to \mbox{$\rho_{\rm NFW}(R) \propto R^{-3}$} and the density contrast 
\begin{equation}
\delta_{\rm c} = \frac{ 200}{3} \frac{c^3}{\rm{ln}(1+c) - c/(1+c)} \  .
\end{equation}
$c$ is called the concentration parameter and defined as
\begin{equation}
c = \frac{r_{\rm 200}}{r_{\rm s}} \  ,
\end{equation}
giving the ratio between the virial radius $r_{200}$ (within which the mass density equals $200 \rho_{\rm c}$) and the scale radius $r_{\rm s}$. 

The projected surface mass density of an NFW profile has been derived in \citet{bartelmann96}, the gravitational shear can be inferred from \citet{wright00} or can be calculated straightforwardly with equations 7-11 from \citet{bartelmann96}. Adopting a dimensionless radial distance $x=R/r_{\rm s} = \theta / \theta_{\rm r_s}$ we can write the convergence as 
\begin{align}
\kappa(x) =
\begin{cases}
\frac{2 r_s \delta_{\rm c} \rho_{\rm c}}{\Sigma_{\rm c} \left(x^2-1\right)} \left( 1 - \frac{2}{\sqrt{1-x}} \rm{artanh} \sqrt{\frac{1-x}{1+x}} \right) &  x <  1\ , \\
\frac{2 r_s \delta_{\rm c} \rho_{\rm c}}{3} & x = 1\ , \\
\frac{2 r_s \delta_{\rm c} \rho_{\rm c}}{\Sigma_{\rm c} \left(x^2-1\right)} \left( 1 - \frac{2}{\sqrt{1-x}} \rm{arctan} \sqrt{\frac{x-1}{x+1}} \right) &  x >1 \ .
\end{cases}
\end{align}
The gravitational shear is given by
\begin{align}
\gamma(x) = \begin{cases}
\frac{r_s\delta_{\rm c}\rho_{\rm c}}{\Sigma_{\rm c}} g_{<}(x) & x < 1\ ,\\
\frac{r_s\delta_{\rm c}\rho_{\rm c}}{\Sigma_{\rm c}} \left[ \frac{10}{3} + 4\ \rm{ln} \left(\frac{1}{2} \right) \right] & x = 1\ ,\\
\frac{r_s\delta_{\rm c}\rho_{\rm c}}{\Sigma_{\rm c}} g_{>}(x) & x < 1\ .
\end{cases}
\label{eq:gamma_nfw}
\end{align}
The functions $g_{<}(x)$ and $g_{>}(x)$ are independent from cosmology and do only depend on the dimensionless radial distance x:
\begin{multline}
g_{<}(x) = \frac{8\ {\rm{artanh}} \sqrt{(1-x)/(1+x)}}{x^2\sqrt{1-x^2}} + \frac{4}{x^2}\ \ln \left(\frac{x}{2}\right) \\
- \frac{2}{(x^2-1)} + \frac{4\ {\rm{artanh}} \sqrt{(1-x)/(1+x)}}{(x^2-1)(1-x^2)^{1/2}}
\end{multline}
\begin{multline}
g_{>}(x) = \frac{8\ {\rm{arctan}} \sqrt{(x-1)/(1+x)}}{x^2\sqrt{x^2-1}} + \frac{4}{x^2}\ \ln \left(\frac{x}{2}\right) \\
- \frac{2}{(x^2-1)} + \frac{4\ {\rm{arctan}} \sqrt{(x-1)/(1+x)}}{(x^2-1)^{3/2}} \ .
\end{multline}
The virial mass, the mass within the $r_{\rm 200 }$, is given by
\begin{equation}
M_{\rm 200} \equiv M _{\rm NFW}(<r_{\rm 200})= \frac{800\pi}{3}\rho_{\rm c} \ r_{\rm 200}^3 .
\label{eq:m200}
\end{equation}
The rotation velocity at the virial radius is 
\begin{equation}
v_{200} = \sqrt{\frac{GM_{200}}{r_{200}}} = \sqrt{\frac{800\pi\rho_{\rm c}G}{3}}\ r_{200}\ .
\label{eq:v200}
\end{equation}
If we fit BBS and NFW profiles to a mass distribution, we do not expect that the total BBS- and the $M_{200}$-NFW-masses are equal: the $M_{200}$-NFW-mass only gives that part of the halo mass which is already virialised, and does not specify the mass associated with the structure outside this radius. It however makes no sense to integrate this profile for much larger radii, since the total mass would diverge. \citet{baltz09} presented a modification of the dark matter profile that is NFW-like inside the virial radius and describes a cutoff relative to the original NFW profile for larger radii. The gravitational shear then falls off faster than for the NFW-profile (see their fig.~2). To measure these truncation details is not the goal of this work. We however point out that these truncated NFW haloes are also described approximately by a BBS halo. The total mass of the truncated NFW halo then is similar in interpretation to the total BBS mass, i.e. it equals the total mass associated with the halo. Therefore we expect that $M_{200}$-NFW-masses are lower than BBS masses obtained from our model fits.
\subsection{Scaling relations for the galaxy halo profiles}
\label{sec:scaling theory}
We assume the following scaling relations between halo parameters and galaxy luminosity:
\begin{equation}
 \Bigl(\frac{M}{M^{*}}\Bigr) = \Bigl(\frac{L_{}}{L^{*}_{}}\Bigr)^{\eta_{M}} \ ,
 \label{eq:scaling M}
\end{equation}
\begin{equation}
 \Bigl(\frac{M/L}{(M/L)^{*}}\Bigr) = \Bigl(\frac{L_{}}{L^{*}_{}}\Bigr)^{\eta_{M/L}} \ , \  \eta_{\rm M/L}=\eta_{\rm M}-1 \ .
 \label{eq:scaling M/L}
\end{equation}
In these two upper equations $M$ can be either the virial mass $M_{\rm 200}$ of the NFW or the total mass $M_{\rm total,BBS}$ of the BBS profile. The corresponding scaling indices $\eta_{M_{200}}$ and $\eta_{\rm M_{BBS}}$ can be different, in principle. The luminosity in this relation is not the bolometric one but that for a specific filter, which we choose to be the $r'$-band filter in the following.
\\
The halo velocity dispersion  vs. luminosity relation is parameterised by the 
Faber-Jackson relation (\citealt{faberjackson}) or (using $v_{\rm rot} \propto \sigma $) the Tully-Fisher relation (\citealt{tullyfisher}),
\begin{equation}
 \Bigl(\frac{\sigma}{\sigma^{*}}\Bigr) = \Bigl(\frac{L}{L^{*}}\Bigr)^{\eta_{\sigma}} \ .
 \label{eq:scaling sigma}
\end{equation}
The halo-size  vs. luminosity relation is described by
\begin{equation}
 \Bigl(\frac{s}{s^{*}}\Bigr) = \Bigl(\frac{L}{L^{*}}\Bigr)^{\eta_{s}} \ .
 \label{eq:scaling s}
\end{equation}
Inserting equations~(\ref{eq:scaling M}), (\ref{eq:scaling sigma}) and (\ref{eq:scaling s}) into equation~(\ref{eq:massBBStotal}) we obtain
\begin{equation}
\eta_{M_{\rm BBS }} = 2 \eta_{\sigma} + \eta_{s} \ .
\label{eq:eta_m}
\end{equation}
\\
Using the SDSS-$g'$-band for the mass-luminosity relation and analysing the GGL-signal in an early SDSS-data set \citet{guzik02} have found $\eta_M=1.2\pm0.2$, or $\eta_{M/L}=0.2$. This is also the scaling behavior found for the dynamical mass-to-light ratio for the centres of elliptical galaxies, the so-called fundamental plane (see e.g. \citealt{bbf92} or \citealt{saglia10}). The Tully-Fisher and Faber-Jackson power law indices that have been measured in various filters and galaxy samples (see e.g. \citealt{davies83} analysing faint early type galaxies and \citealt{matkovic07} analysing dwarf early type galaxies, both in $B$-band, or \citealt{nigoche-netro10} analysing SDSS early type galaxies in $g$- and $r$-band for the Faber-Jackson relation and analysing the Tully-Fisher relation for disc galaxies e.g. \citealt{bamford06}, \citealt{miller11} in $B$-band and \citealt{pizagno05}, \citealt{fernandez-lorenzo09} or \citealt{reyes11} in $I$-band). The `classical' Faber-Jackson index is 0.25  (\citealt{faberjackson}), but variations depending on the sample selection, e.g. luminosity interval and galaxy mixture (`pure ellipticals', S0s, spirals or a combination) have been reported (see e.g. \citealt{williams10} or \citealt{cappellari12}). When Equation~(\ref{eq:scaling sigma}) was assumed or attempted to be measured in the analysis of the GGL signal different assumptions were made and different results were obtained. \citet{seljak02}, for instance, followed the classical Faber-Jackson relation for elliptical galaxies, \citet{hoekstra04} assumed a scaling of $\sigma \propto L^{0.3}$ for a mixed SED sample. On the other hand \citet{kleinheinrich06} measured values of $\eta_{\sigma}$ between 0.3 and 0.4, depending on the maximum considered projected separations between source and lens.
\\
If one assumes  $\eta_M=1.2$ and $\eta_{\sigma}=0.3$ to hold then $\eta_s=0.6$ must hold as well. These three values provide kind of a generic model to analyse a combined galaxy sample and obtain the velocity dispersion and size of the `reference' halo with luminosity $L^*$. Our analysis will go beyond these assumptions and aims to measure these parameters for the scaling relations. In addition we'll measure the scaling relations and the values for the reference halo with luminosity $L^*$ for the red and blue galaxies separately. 
\\
We write the virial radius  vs. luminosity relation of the NFW profile as 
\begin{equation}
 \Bigl(\frac{r_{\rm 200}}{{r^{*}_{\rm 200}}}\Bigr) = \Bigl(\frac{L}{L^{*}}\Bigr)^{\eta_{r_{\rm 200}}} \ .
 \label{eq:scaling r200}
\end{equation}
This implies
\begin{equation}
\eta_{\rm M_{200}} =3 \eta_{\rm r_{200}} \ .
 \label{eq:M200vsr200}
\end{equation}
The zeropoint of this relations changes with redshift as the critical density $\rho_{\rm c}$ is redshift dependent as well (see equations~\ref{eq:rho_crit} and \ref{eq:m200}).
\\
For the concentration-mass-relation we take the relation of \citet{duffy08}, derived for simulated haloes made of dark matter only and which depends on the redshift of the halo, to hold,
\begin{equation}
c \propto M^{-0.084\pm0.006}_{\rm 200}\left(1+z\right)^{-0.47\pm0.04} \ .
\label{eq:duffy}
\end{equation}
Already e.g. \citet{bullock01} or \citet{shaw06} found a concentration-mass-relation, implying a slight decrease of the concentration parameter with increasing mass. In the past years also further mass-concentration relations have been discussed by e.g. \citet{bhattacharya11}, who found a relation with slightly higher amplitude, but very similar scaling behavior as the one of \citet{duffy08} or \citet{klypin11}, reporting an even higher amplitude and a shallower decrease with increasing mass. Recently \citet{prada12} presented a concentration-mass-relation, including a novel feature, as above a virial mass of approximately \mbox{$10^{15}\ h^{-1}\ M_{\odot}$} the concentration parameter begins to increase again instead of further decrease.
\\
Using the $M_{\rm 200}$-mass and luminosity relation in equation~(\ref{eq:scaling M}) we obtain the concentration-luminosity relation
\begin{equation}
 c \propto L^{\eta_c} \ ,
 \label{eq:scaling c}
\end{equation}
with
\begin{equation}
 \eta_{c} = \frac{-0.084\pm0.006}{\eta_{M_{\rm 200}}} \ .
 \label{eq:eta_c}
\end{equation}
\subsection{Influence of a cluster or group halo on $\Delta \Sigma$}
\label{sec:cluster halo}
\begin{figure*}
\centering
\includegraphics[width=17.0cm]{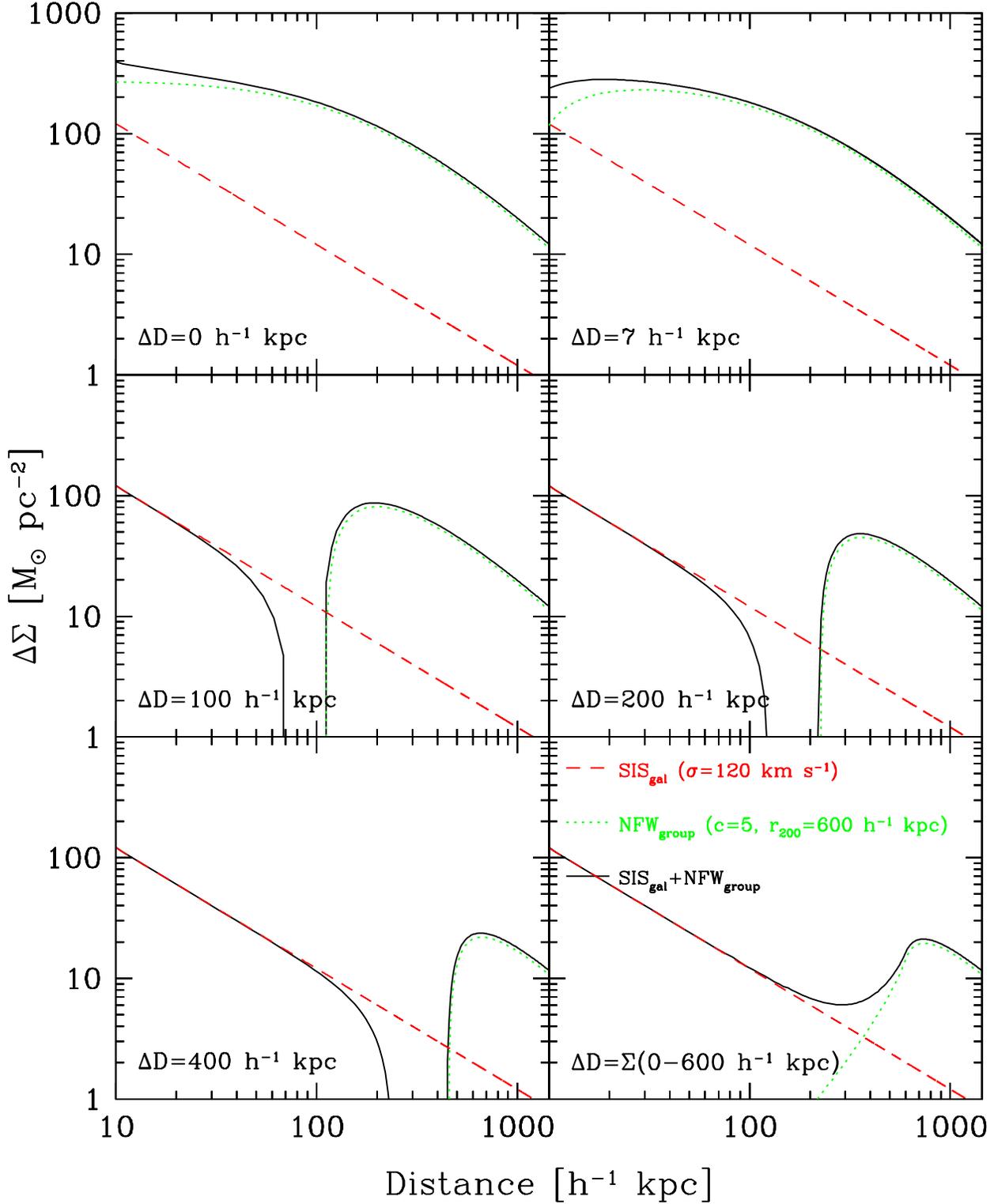}
\caption{Observable excess surface mass density $\Delta \Sigma$ for galaxies embedded in a group or cluster halo. For the galaxy halo we use an SIS with velocity dispersion $\sigma = 120$ km\ $\rm{s^{-1}}$, for the group an NFW profile with $c=5$ and $r_{200} = 600\ h^{-1}$ kpc. The red dashed lines show the excess surface densities $\Delta \Sigma$ of the galaxy SIS profile, the green dotted lines show that of the group NFW profile and the black solid lines show the sum of both.
The upper left panel shows the profile for a galaxy directly in the group centre, the upper right, middle and lower left panels show the profile of a galaxy at distances of $\Delta D = 7$, $100$, $200 $ and $ 400\ h^{-1}$kpc from the group centre. As can be seen the role of the group contribution to $\Delta \Sigma$ strongly decreases with increasing distance.
The lower right panel shows the mean excess surface mass density $\Delta \Sigma$ for an ensemble of galaxies distributed homogeneously within a disc of 600 $h^{-1}$ kpc radius. For the mean excess surface mass density the contribution of the group halo is negligible out to a scale of $200\ h^{-1}$ kpc. }
\label{fig:NFWSIS}
\end{figure*}

We investigate the influence of a parent cluster, group or massive galaxy halo on the observable excess surface mass density $\Delta \Sigma$ of a galaxy.
In this case the galaxy halo is a satellite in a more massive halo. This effect has been investigated and illustrated by  \citet{guzik02} or \citet{mandelbaum06b} in the past already. Since we need some numbers for later on, e.g. on which scales this effect is important, we give a relevant example here as well. 
\\
We assume that a galaxy halo described by an SIS with velocity dispersion $\sigma=120$ km\ $\rm{s^{-1}}$ is embedded in a group halo described by an NFW profile with concentration $c=5$ and $r_{200}=600\ h^{-1}$ kpc. We place this galaxy halo centrally in the group halo and with projected distances of 0, 7, 100, 200 and 400  $h^{-1}$ kpc. The total excess surface densities $\Delta \Sigma$ with respect to the galaxy halo centre are shown as a function of projected distances in Fig.~\ref{fig:NFWSIS} with black curves (upper two, middle two and lower left panel). The contribution of the galaxy halo and the group halo to the combined signals are shown in red (dashed) and green. If the galaxy halo is slightly offcentred from the group halo, a turnover of the combined excess surface density at small radii can be observed. The farther the galaxy from the group centre, the larger the scale at which the excess surface density of the combined signal equals the excess density of the galaxy halo. If one considers an ensemble of galaxies that is homogeneously spread over the group halo out to radii of  $600\ h^{-1}$ kpc (this was assumed to be the case in the lower right panel of Fig.~\ref{fig:NFWSIS}) then the group halo contribution averages out (i.e. if limited to small enough scales, then on average the group acts as a mass sheet which does not enter the mass excess density). We conclude that the galaxy environment has little impact on the measured excess surface mass density on scales up to $200\ h^{-1}$ kpc unless the galaxy haloes are limited to a very restricted area within their parent halo, i.e. if they would exist as central galaxies only (in this case we would measure the `group' excess surface density, potentially modulated by a miscentring term) or if they would preferentially live in the outskirts of group haloes. On scales  larger than the extent of the galaxy distribution the measured combined signal approximately equals that of the `parent halo'. I.e. the signature of satellite galaxies equals that of the isolated galaxies on small scales and on larger scales has a bump where the precise scale and amplitude depends on the (mean) properties of the central halo these satellites reside. 
\\
\subsection{Maximum likelihood analysis}
\label{sec:likelihood theory}
\citet{schneider_rix97} first applied the maximum likelihood method to GGL to infer halo parameters. Following their work we will compare the measured (background) galaxy ellipticities to the predictions of parameterised profiles. The profile parameter values are tied to that of a reference galaxy with luminosity $L^*$, the parameter values for other luminosities are determined by the scaling relations we have introduced in Section~\ref{sec:scaling theory}. The log-likelihood is given by
\begin{equation}
log \mathcal{L}= -\sum_{i,j}\Bigl(\frac{e_{i,j}-P^{\gamma}_{j} g_{i,j}^{model}}{\sigma_{\rm e_j}}\Bigr)^2
\end{equation}
where $e_{i,j}$ are the two components of the polarisations for the j-th galaxy and, $P^{\gamma}_{j}$ is the shear polarisability, $g_{i,j}$ are the analytic shear values for a given profile and $\sigma_{\rm e_j}$ quantifies the shape noise and measurement errors. The definition and a description of the polarisations is given in Section~\ref{sec:shape_shear} in more detail (see equations~\ref{eq:epsilon} to ~\ref{eq:WL_biascorr}).
\\
\\
The advantage of the maximum likelihood analysis is that 
`neighbouring' haloes in transverse separation (correlated structures like `parent' haloes or satellite galaxies) but also along the line-of-sight (multiple deflections, see \citealt{brainerd10} for a study of their relevance) are automatically accounted for correctly. 
\\
We use the observed quantities (locations, redshifts, SED types and luminosities of galaxies) and the best fit halo parameters obtained from the maximum likelihood analyses to perform 3-D-projected lensing signal simulations. In this way we create synthetic noise-free shear catalogues from which we then can predict the tangential shear and the excess surface mass density signal and compare them to the signal derived from our observations. 
%
\section{Data}
\label{sec:dataall}
In this Section we describe the data used for our weak lensing analysis. Section~\ref{sec:data} gives an overview over the CFHTLS-Wide imaging data. In Section~\ref{sec:spectra} we describe the spectroscopic data used for the photometric redshift calibration. The photometric catalogue creation and the photometric redshift estimation are presented in Section~\ref{sec:photoz}. We give the details on the galaxy shape estimation in Section~\ref{sec:shape_shear} and describe the properties of our lensing sample in Section~\ref{sec:properties_catalogues}. In Section~\ref{sec:lenssample} we define our weak lensing lens and source subsamples. In Section~\ref{sec:lensmassbias} we estimate the lens mass errors due to photometric redshift uncertainties and shape measurement errors and further refer to possible biases in Section~\ref{sec:IA} due to IA. In Section~\ref{sec:subsamples} we define lens subsamples with respect to galaxy luminosity, the galaxy SED type (`red' and `blue' galaxy types) and galaxy environment density.
\subsection{Imaging data}
\label{sec:data}
Our work uses publicly available CFHT Legacy Survey\footnote{http://www.cfht.hawaii.edu/Science/CFHTLS} Wide and Deep field data (CFHTLS-Wide and CFHTLS-Deep). The surveys are described in e.g. \citet{hoekstra06}, \citet{semboloni06}, \citet{coupon09} and \citet{astier06}. They map over 190  $\rm{deg^{2}}$ and 4  $\rm{deg^{2}}$, respectively, in the $u^*$, $g'$, $r'$, $i'$ and $z'$ broad band filters under superb seeing conditions of typically 0.8 arcsec with the \MegaPrime camera. \MegaPrime (see \citealt{boulade03}) is an optical multi-chip instrument with a 9 $\times$ 4 CCD array (2048 $\times$ 4096 pixels in each CCD; 0.186 arcsec pixel scale; FOV $\sim$ 1$^\circ$ $\times$ 1$^\circ$ total field of view).
The Wide and Deep survey fields are done in 4 patches each, W1 (D1), W2 (D2), W3 (D3) and W4 (D4). Coordinates and more details can be found at the CFHTLS-Wide webpage\footnote{http://terapix.iap.fr/cplt/oldSite/Descart/summarycfhtlswide.html}.
We restrict ourselves to  124  $\rm{deg^{2}}$ (49 in W1, 25 in W2, 30 in W3 and 20 in W4) of the Wide survey as the remaining ones only had incomplete photometric data for one or more bands at the time of our data reduced. We also consider the D1 and D3 Deep survey fields as they overlap with W1 and W3, respectively, allowing spectroscopic calibration for fainter objects than the Wide fields, and thus offering the possibility to estimate photometric redshift accuracy for the same filter set as for CFHTLS-Wide but over a broader magnitude range.
\\
We download the preprocessed single frame data from the the Canadian Astronomical Data Centre (CADC)\footnote{http://www1.cadc-ccda.hia-iha.nrc-cnrc.gc.ca/cadc/}. The preprocessing with the \Elixir (see \citealt{magnier04}) pipeline includes bias and dark subtraction, flatfielding, fringe correction in the $i'$- and $z'$-band data, as well as photometric calibration and a preliminary astrometric solution. We then use the \THELI-pipeline\footnote{http://www.astro.uni-bonn.de/\textasciitilde mischa/theli.html} (see also \citealt{erben05,erben09}) and derive a more accurate astrometric solution, remap and stack the single exposures. The work with the \THELI-pipeline closely follows the procedure of \citet{erben09}. They had used the United States Naval Observatory (USNO) catalogues as reference for astrometric calibrations. We found that an improved astrometric solution can be obtained when using SDSS-DR6-object catalogues \citep{adelman07} and Two Micron All-Sky Survey (2MASS) catalogues \citep{jarrett00} for those fields where SDSS data are not available. This improvement reduces shape artefacts in the remapping to the astrometric solution, and is relevant in the PSF anisotropy correction of the objects for the lensing analysis later on, while it is not required in photometry. We therefore generate a stack with improved astrometric solution for the $i'$-band data used for shape measurement and keep the USNO based astrometric solution for photometry of all five bands.
The data remain on one square degree tiles (with a pixel size of 0.186 arcsec), i.e. we do not coadd data in the small overlapping areas of adjacent pointings.  We instead analyse the data square degree by square degree later on. The depths of the stacked images are like those in \citet{erben09} (where 37  $\rm{deg^{2}}$ of CFHTLS-Wide were stacked), i.e. the limiting AB-magnitudes are about 25.3, 25.6, 24.5, 24.6 and 23.6 (5$\sigma$ detection within a 2 arcsec aperture for a point source) in the $u^*, g', r', i'$ and $z'$-bands. The i'-band PSF-width is close to 0.8 arcsec for all frames. 
The procedure for obtaining bad area masks is identical to that of \citet{erben09} and allows us to identify the area where good photometry and shape measurements can be obtained. After our data end-processing we thus have the stacked science data, their error frames (as weight images) and image masks.
%
\subsection{Spectroscopic data}
\label{sec:spectra}
We use public spectroscopic data from the Visible Multiobject Spectrograph (\VIMOS) VLT Deep Survey (VVDS)-Deep \citep{lefevre05}, VVDS-F22 \citep{lefevre04,lefevre05,garilli08} and the Deep Extragalactic Evolutionary Probe-2 (DEEP-2) programme \citep{davis03,davis07,vogt05,weiner05}. This provides us with 3\ 562, 7\ 986 and 3\ 746 high quality redshifts in the W1, W3 and W4 fields. From these spectroscopic redshifts 1\ 548 in the W1, 3\ 960 in the W3 and 3\ 573 in the W4 describe objects with $i' \le  22.5$. For the W2 we have also reduced spectroscopic \VIMOS data obtained with the low resolution LR-Blue and LR-Red grisms from the ESO Programme ID 082.A-0922(B). The data reduction was carried out as described on the zCOSMOS release webpage\footnote{http://irsa.ipac.caltech.edu/data/COSMOS/spectra/z-cosmos/Z-COSMOS\_INFO.html} and \citet{lilly07}. We obtain 960 high quality redshift estimates in the W2-field,  944 of those have  $i' \le  22.5$. In total we have 10\ 025 spectroscopic redshifts with $i' \le  22.5$ and additional 6\ 229 redshifts with $22.5 <  i' \le  24.0$.
%
\subsection{Catalogue creation and photometric redshift estimation}
\label{sec:photoz}
As next step we convolve for every square degree tile all data to the PSF which is the largest among all filters. These are often the $u^*$-band filter data which have a point spread function (PSF) FWHM between  $0.63$ to $1.22$ arcsec, with a median of $0.9$ arcsec. We convolve the (better) data with a Gaussian. Strictly speaking this procedure is not enough to map the data to the same PSF, because neither the `good' nor the `bad' data have a truly Gaussian PSF. For this reason convolving the good data with a Gaussian of $\sqrt{FWHM_{\rm bad}^2 - FWHM_{\rm good}^2}$ does not always give the best results for PSF photometry in practice. 
One should instead search for the kernel which maps the `good' PSF to the exact form of the `bad' PSF. This is a delicate procedure because the PSF-size and anisotropy also varies over the field (see also \citealt{darnell09} or \citealt{hildebrandt12}). Luckily the variation in PSF size over the field is  similar for the different filters at least for CFHTLS data. Therefore we follow a compromise: We test-convolve the good data with a sequence of Gaussians to match bad PSF data, and measure the colours of stars in different apertures in a range of 8 to 18 pixels or 1.5 to 3.3 arcsec diameter. The PSFs are considered to be well matched in practice when the colours of stars are independent of the aperture size, and the convolution width where this is achieved best is taken for the real convolution. This procedure does not make the two-dimensional PSF shapes of the good data identical to the bad data, but it ensures that the PSF is similar enough to allow matched aperture photometry.
After the PSF homogenisation we do PSF photometry within an aperture of 8 pixels (1.5 arcsec) diameter  with \SExtractor\footnote{http://www.astromatic.net/software/sextractor} (see \citealt{bertin96}) in dual-image mode, including the weight frames in the detection and extraction frame. The objects are detected in the unconvolved $i'$-band frames with a S/N threshold of 2$\sigma$ on at least four contiguous pixels. We make use of the \SExtractor option to convolve the data with a Gaussian before  detection on the unconvolved $i'$-band. The full width of half maximum of this pre-detection convolution is chosen to be 0.4 arcsec and suppresses correlated noise on a scale smaller than the PSF. Aperture fluxes and their errors are extracted on the PSF matched images. The total magnitudes in all filters are estimated using the difference between the fixed circular aperture magnitude ($\rm{mag_{aper}}$) and the automatic Kron-like aperture ($\rm{mag_{auto}}$) in the $i'$-band as a correction to the aperture magnitude in the band of interest $\rm mag_{auto_{\rm Filter}} \approx mag_{aper_{\rm Filter}} + \left(mag_{auto_{\rm i'}} - mag_{aper_{\rm i'}}\right)$.
\\
In these first photometric catalogues the colours of stars and galaxies still vary from field to field because of remaining zeropoint calibration errors and because of the galactic extinction. Since the CFHTLS-Wide fields are chosen to be off the galactic plane the extinction is rather small and does not vary a lot over one square degree tiles: the maximum and minimum extinction in all Wide fields is $0.03$ and $0.14$, and the difference between maximum and minimum extinction value per square degree can be up to  $0.03$ for high extinction fields and $0.01$ for fields with low extinction values. We apply one zeropoint and extinction correction value per square degree field by shifting the observed stellar colours to those predicted from the Pickles stellar library \citep{pickles98} for the given photometric system. In this way we obtain catalogues without remaining systematic field to field variations in the mean stellar and galaxy colours.
\\
We then derive first test photometric redshifts. The photometric redshift algorithm is that of \citet{bender01}. It was successfully applied in a variety of contexts (see \citealt{gabasch04a,gabasch06,gabasch08,feulner05,feulner06,drory01}, \citealt{brimioulle08}, \citealt{lerchster11}, \citealt{spinelli12} and recently \citealt{gruen13}). The SEDs used have been developed by \citet{bender01} for higher redshift and fainter HDF (see e.g. \citealt{arnouts99,cohen00,ferguson00}) and FDF (see e.g. \citealt{gabasch04a,gabasch04b,gabasch06,feulner05}) galaxies. Some of these SEDs actually were made to match galaxies at redshifts between 3 and 4, and between 4 and 5, respectively, which are a minority in the CFHTLS-Wide data. We therefore take over this SED sample but replace a few of them with 13 SEDs from \citet{ilbert06} based on models from \citet{cww80} to better match local, star-forming blue galaxies. We however do not further optimise the SED set for this work. We neither claim that we know the photometric throughput to ultimate precision (this is relevant mostly for the $u^*$-band), nor do we claim that our SED templates match the colours of CFHTLS galaxies perfectly. We therefore use spectroscopic redshifts in our fields and test which empirical zeropoint offsets have to be applied relative to the stellar template match in order to obtain a good overall photometric redshift performance. For each of the four CFHTLS-Wide fields we determine independently the necessary offsets from the average offsets for all particular subfields with spectroscopic data. We apply these offsets to the W1 to W4 fields without spectroscopic control sample. We have verified that this method yields reliable predictions by considering only a part of the spectroscopic subfields for the offset determination, and controlling the results with the remaining subfields. Also, these additional photometric offsets are rather small. The offsets applied are $0.01$ and $0.02$ for all filters and all four CFHTLS-Wide fields. The only exception is W2 and W4 where we apply an offset in the $u^*$ band of $-0.05$ and $-0.06$. 
\\
We then finally obtain the photometric redshifts for 124  $\rm{deg^{2}}$ of the CFHTLS-WIDE. We evaluate the photometric redshift accuracy using all available spectroscopic data (see Section~\ref{sec:spectra}). We only consider \VIMOS objects with a spectroscopic flag of 3, 4, 23 or 24 (confidence$\ge 95$ per cent) and DEEP2-objects with a spectroscopic flag of 4 (confidence=$100$ per cent). We use our bad area flag maps and compare photometric redshifts and spectroscopic data only in areas with good photometry. Also later on photometric redshift data are only used in good area regions. 
\begin{figure*}
\centering
\includegraphics[width=18.5cm]{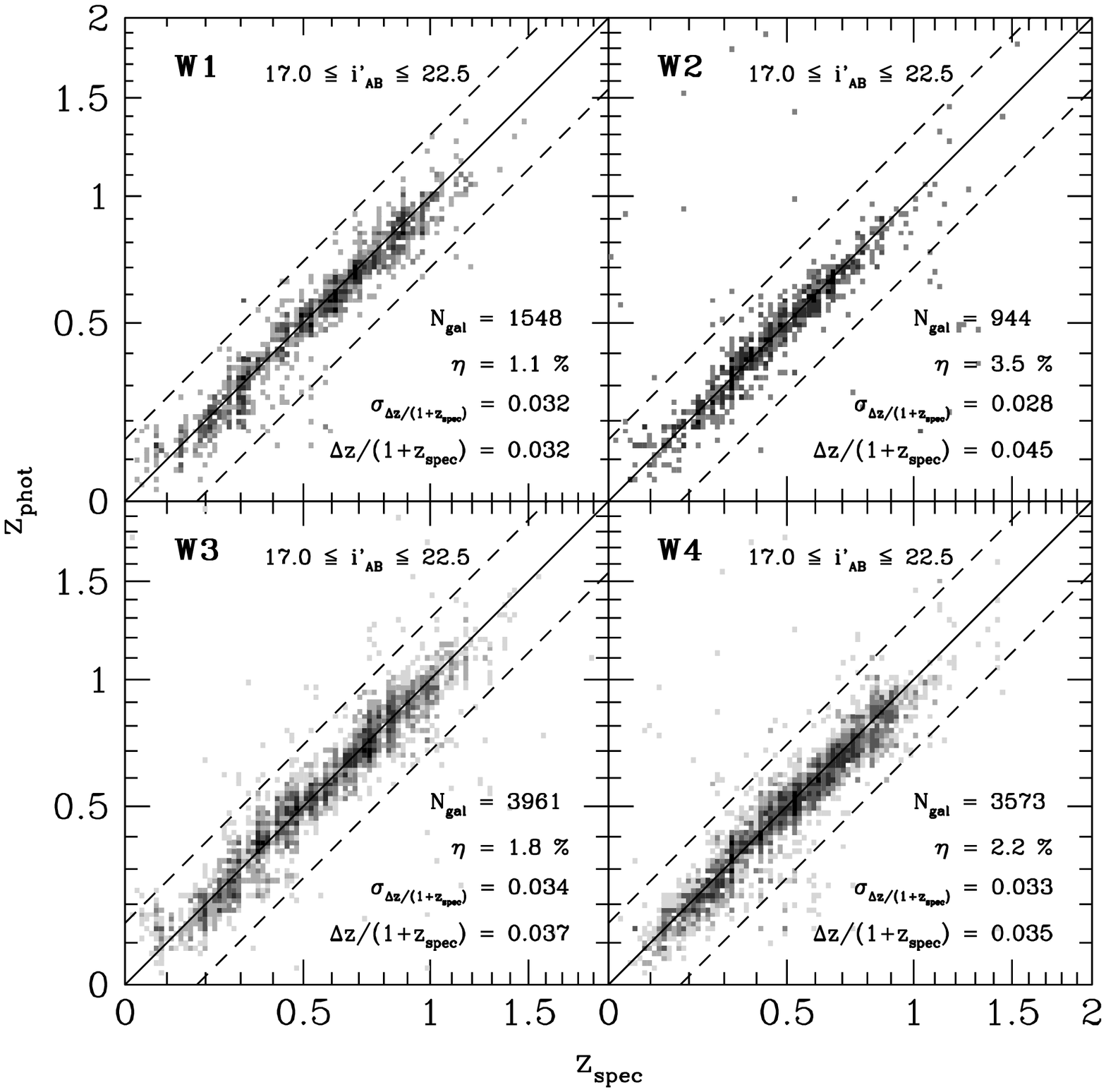}
\caption{Photometric redshift accuracy (density plot) for the CFHTLS W1 to W4 fields for a depth of 
$i'\le$22.5. Each panel shows the object number ${N_{\rm gal}}$, the outlier rate $\eta$, the photometric redshift scatter $\sigma_{\rm \Delta z /(1+z)}$ and the mean photometric redshift error $\Delta z/(1+z)$.}
\label{fig:WIDE.photoz_shallow}
\end{figure*}
\begin{figure*}
\centering
\includegraphics[width=18.5cm]{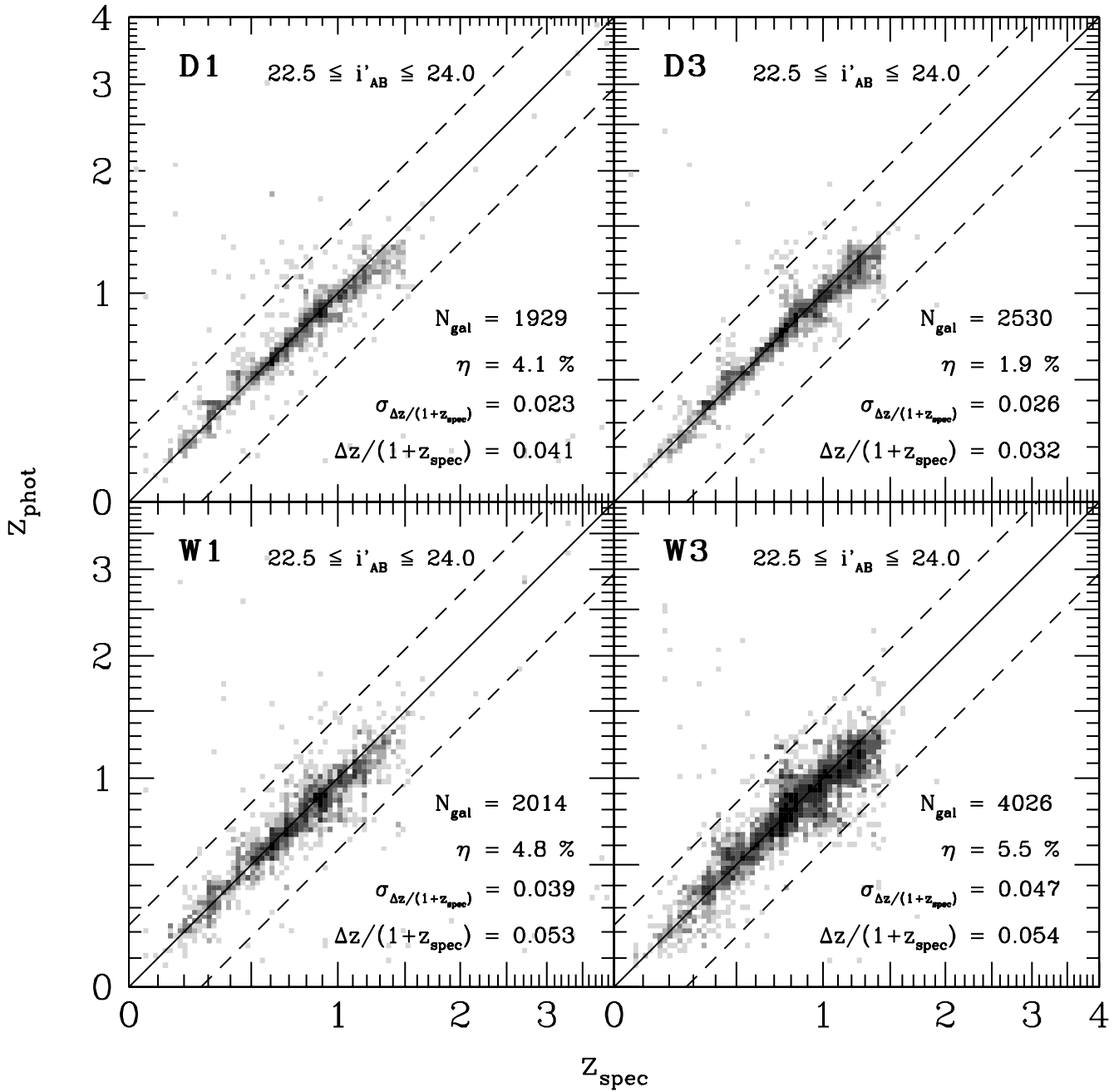}
\caption{Photometric redshift accuracy (density plot) for the CFHTLS D1- and D3-Deep and W1- and W3-Wide fields for a depth of  $22.5\le i' \le 24.0$. For these faint galaxies our photometric redshift are worse than they are for the bright sample in Fig.~\ref{fig:WIDE.photoz_shallow}. We have shown in the upper panel (by using CFHTLS-Deep D1 and D3 data) how good the photometric redshifts can become for the same spectroscopic control sample if the photometric errors could be drastically decreased. 
}
\label{fig:WIDE.photoz_deep}
\end{figure*}
\\
The quality of photometric redshifts is quantified with three numbers:
\ben
\item the outlier rate $\eta$, defined as the fraction of objects exceeding a rest frame error of $0.15$, i.e. 
\begin{equation}
\eta = {\rm fraction \  with } \left\lbrace\frac{\abs{z_{\rm spec}-z_{\rm phot}}}{1+z_{\rm spec}} \ge  0.15\right\rbrace \ ,
\end{equation}
\item the photometric redshift scatter, calculated from the width of the central part of the error distribution
\begin{equation}
\sigma_{\rm{\Delta z/(1+z)}}=
1.48 \times {\rm median \ of } \left\lbrace \frac{\abs{z_{\rm spec}-z_{\rm phot}}}{1+z_{\rm spec}} \right\rbrace_{\rm non-outliers} \ ,
\label{eq:sig}
\end{equation}
\\
\item the mean photometric redshift error 
\begin{equation}
\Delta z/(1+z) = 
\frac{1}{\rm{N_{\rm spec}}}  \sum_{i}^{N_{\rm spec}}
{\frac{z_{{\rm phot},i} - z_{{\rm spec},i}}{1+z_{\rm spec}}} \  . 
\label{eq:sigtotal}
\end{equation}
\een
In these equations $N_{\rm spec}$ is the spectroscopic sample size and $z_{\rm spec}$ and $z_{\rm phot}$ are the spectroscopic and photometric redshifts. 
\\
Fig.~\ref{fig:WIDE.photoz_shallow} compares photometric and spectroscopic redshifts for galaxies with $i'<22.5$ in all four Wide fields. For fainter magnitudes only W1 and W3 have spectroscopic redshifts. Fig.~\ref{fig:WIDE.photoz_shallow} shows that the photometric redshift performance is roughly equal in all four fields. The outlier rate is between 1 and 3.5 per cent and the root-mean-squared (rms) error of the distribution within the dashed lines (marking a rest frame redshift error of 0.15) equals $0.028$ to $0.034$. The photometric redshift quality for bright galaxies will mostly be relevant for the lenses under consideration. Most of the lensed objects will be between $22.5\le i'\le 24.0$ however. For these faint galaxies spectroscopic data are only available for the W1 and W3 fields (coming from the VVDS-Deep and DEEP2 data sets). 
We present the redshift accuracy for these galaxies in the two lower panels of Fig.~\ref{fig:WIDE.photoz_deep}. For these faint magnitudes we have outlier rates of about 5 per cent and an outlier-cleaned rms error of $0.04-0.05 \ (1+z)$. There are two potential reasons why the errors are increased relative to the bright sample: first of all, the photometric errors are increased; secondly, at fainter magnitudes the variety of SED types becomes larger and the  template set could be not flexible enough anymore or there are redshift degeneracies not present in bright samples because luminosity priors forbid break mismatches where the second, high redshift solution implies an extremely, unreasonably bright object. To investigate this point we also add our photometric redshift performance for the case of the much deeper D1 and D3 fields. These fields have integration times of 9 hours to 70 hours per filter compared to 20 minutes to 1.5 hours for the CFHTLS-WIDE. We see that the faint objects  ($i'=22.5-24$) perform better in the deep than in the shallow CFHTLS data sets. In fact their photometric redshift scatter in the Deep fields is not larger than the redshift scatter of the bright, $i'<22.5$ objects in the Deep fields, which means that our SEDs describe the bright and faint objects equally well, and the reduced performance for the $22.5<i'<24.0$-objects in the WIDE-data cannot be dominantly caused by an increased SED variety for the fainter objects.  We conclude that the photometric error (implying also insignificant flux estimates in some bands) is the major reason why the photometric redshifts for faint objects in CFHTLS-Wide deteriorate. 
\\
We also compare the accuracy of our redshifts with those of \citet{hildebrandt12}, who improved their CFHTLS data quality by executing a global homogenisation of the PSF over the field of view of each pointing. Fig.~\ref{fig:zpscatter}  shows the photometric redshift scatter $\sigma_{\Delta z/(1+z)}$, the outlier rate $\eta$ and the bias as a function of apparent magnitude $i'$ and photometric redshift $z _{\rm phot}$. Our results are in general similar. This implies that our pragmatic choice for PSF-`equalising' relative to a more rigorous PSF equalisation is not a limiting factor for the photometric accuracy that can be reached. The bias for our faintest luminosity bin is poorly estimated due to low number statistics.
\begin{figure*}
\includegraphics[width=18.5cm]{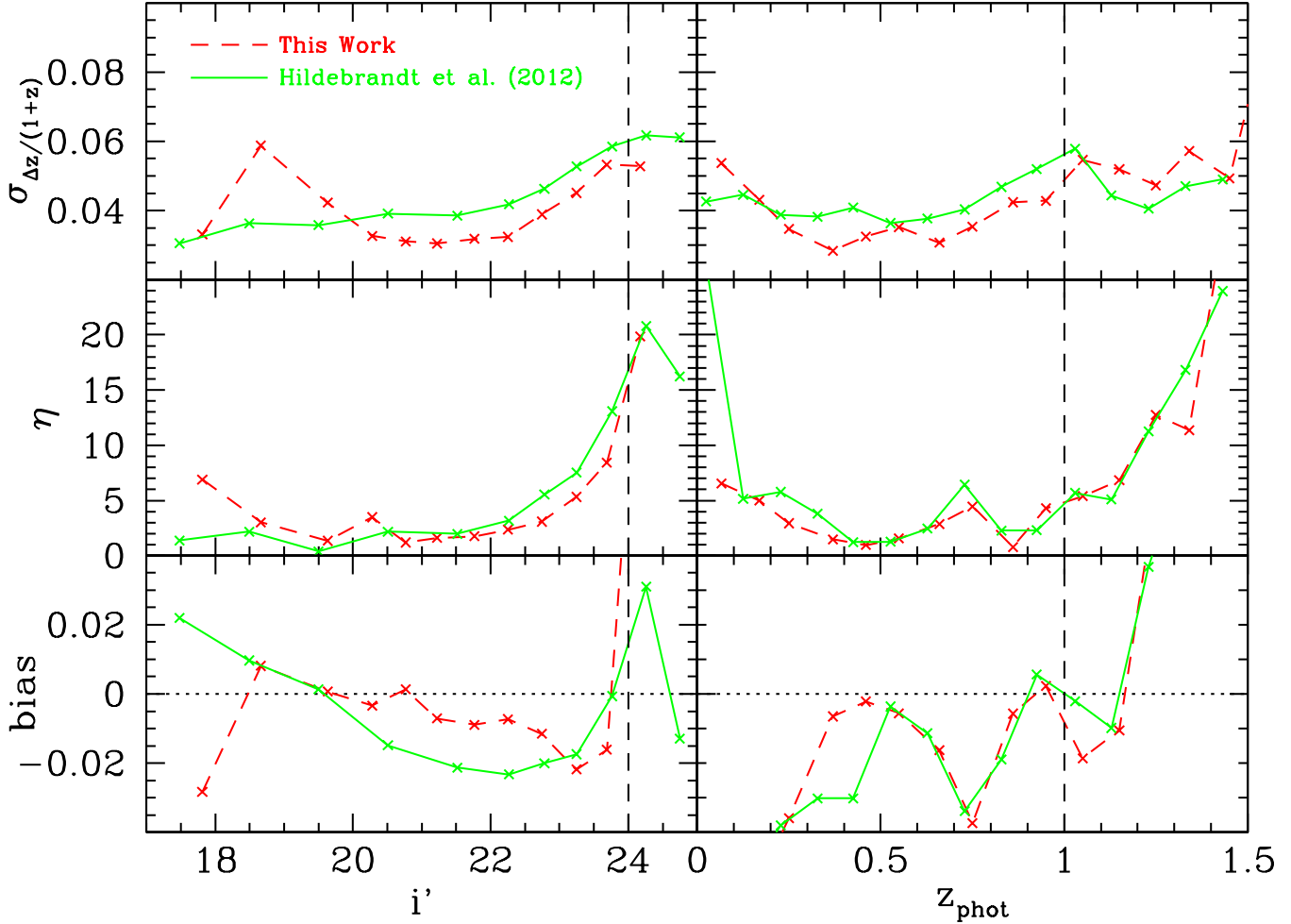}
\caption{Comparison of the photometric redshift accuracy. The red dashed lines show the accuracy of our photometric redshift estimation, the green solid lines show the results from \citet{hildebrandt12} on CFHTLS data. The plot shows the photometric redshift scatter $\sigma_{\Delta z/(1+z)}$ (upper panels), the outlier rate $\eta$ and the bias as a function of apparent magnitude $i'$ (left panels) and photometric redshift $z _{\rm phot}$ (right panels). The results are in general quite similar. There are no objects left in the source sample with magnitudes fainter than $i'$=24 (dashed line in the left panels). The bias for our faintest luminosity bin is poorly estimated due to a few outliers in a very small number sample. The dashed line at redshift z=1 in the right panels shows the redshift limit for our lens sample.}
\label{fig:zpscatter}
\end{figure*}
\\
We classify our catalogue objects with respect to size, photometric redshift uncertainties and saturation. Saturated objects are entirely excluded in the later analyses. Further objects with redshift uncertainties $\Delta z _{\rm phot} > 0.25\ (1+z _{\rm phot})$ and objects with sizes smaller than the PSF are excluded from the background sample.
\\
How photometric redshift errors may bias the lens surface mass density estimates is discussed in Section~\ref{sec:lensmassbias}.
\subsection{Shape and shear estimates}
\label{sec:shape_shear}
This section describes how we measure object shapes and obtain a gravitational shear estimate. We use the method of \citet[KSB]{ksb95}, extended by \citet[KSB+]{hoekstra98} and refer the reader to the detailed derivation in \citet{bartelmann_schneider01}. We follow the notation of \citet{bartelmann_schneider01} and summarise only the most important equations here.
\\
Object ellipticities are defined in terms of second moments of the surface brightness distribution  $I(\theta)$,
\begin{equation}
\epsilon = \epsilon_1 + i \epsilon_2 =  \frac{Q_{11}-Q_{22}-2iQ_{12}}{Q_{11}+Q_{22}} \   ,
\label{eq:epsilon}
\end{equation}
where
\begin{equation}
Q_{ij} = \int d^2\theta \ W_{r_{\rm g}}(|\theta|) \ \theta_i \ \theta_j \ I(\theta)
\   .
\end{equation}
Ellipticities (and gravitational shears) are  complex quantities describing the size and the major axis direction of the ellipticity (and size and direction of gravitational shear). The isotropic weight function $W_{r_{g}}$ is a Gaussian, with a width $r_{\rm g}$ adjusted to the object size. We choose $r_{\rm g}$ to be equal to the \SExtractor flux radius in our work later on.
\\
The apparent ellipticities of objects are altered by the reduced gravitational shear $g=\gamma /( 1-\kappa)$, by the PSF smearing (the isotropic part of the PSF, which makes galaxies larger) and by the anisotropic part of the PSF (which distorts objects). The relation of these quantities was derived by KSB in the linear regime of reduced gravitational shear, nearly Gaussian PSFs and small PSF anisotropies: 
\begin{equation}
\epsilon_{\rm obs} = \epsilon_{\rm s} + \frac{P^{\rm sm}}{P^{* \rm sm}}\epsilon^{*} 
+ P^{g} g \ ,
\label{eq:KSB}
\end{equation}
In this equation $\epsilon_{\rm obs}$ is the observed object ellipticity,  $\epsilon^{*}{}$ would be the observed ellipticity of a star at the same position, $\epsilon_{\rm s}$ is the ellipticity the object would have if there was only a PSF-smearing (circular PSF and no gravitational shear),  $g$ is the reduced gravitational shear and  $P^{g}$ is the pre-seeing polarisability introduced by \citet{luppino97},
\begin{equation}
P^{\rm g} = P^{\rm sh}-\frac{P^{sm}}{P^{sm*}} P^{\rm sh *} \  .
\label{eq:Pgam}
\end{equation}
The second order tensors $P^{sm}$ and  $P^{sh}$ describe the reaction of the surface brightness distribution of an object to smearing and shearing by the PSF and the gravitational field. They are calculated from the surface brightness distributions of the galaxies and stars, where quantities measured for stars carry an asterisk. The division by ${P^{sm*}}$ in equation~(\ref{eq:Pgam}) is strictly speaking a tensor inversion, but often replaced by dividing by half of the trace. 
In equations~(\ref{eq:KSB}) and (\ref{eq:Pgam}) the values for the stellar quantities at the locations of galaxies have to be used. They are estimated by using stars in the same field of view and fitting a two-dimensional polynomial to describe the spatial variation of these stellar quantities. 

We now define the corrected  ellipticities as
\begin{equation}
\epsilon_{\rm corr }= \left[\epsilon_{\rm obs}- \epsilon^{*} \frac{P^{\rm sm}}{P^{\rm sm*}}\right] \frac{1}{P^{\rm g}} \ .
\end{equation}
The assumption that sources are randomly oriented on the sky implies that their mean ellipticity and also their mean PSF-smeared ellipticity is zero, i.e. 
\begin{equation}
\ave{\vc \epsilon_{\rm s} } = 0 \  , 
\end{equation}
and thus
\begin{equation}
\langle e_{\rm corr} \rangle = g \equiv \frac{\gamma}{1-\kappa}
\label{eq:WL_true}
\end{equation}
holds.
In the  weak lensing case $\kappa \ll 1$ the equation simplifies to 
\begin{equation}
\langle e_{\rm corr} \rangle = \gamma \  .
\label{eq:WL_simple}
\end{equation}
We are using a KSB+-implementation adapted from the TS-pipeline \citep{schrabback07}, which was kindly provided by Thomas Erben and Tim Schrabback, mostly using code from \citet{erben01}, which itself is based on Nick Kaiser's original IMCAT tools\footnote{http://www.ifa.hawaii.edu/\textasciitilde kaiser/imcat/}. The TS- pipeline has been tested in the Shear TEsting Programme (STEP, see \citealt{step1}). The analysis of the first set of image simulations (STEP1) showed a significant bias \citep{step1} which could be mostly eliminated with a shear calibration factor $c_{cal} =1/ 0.91$, 
\begin{equation}
 \gamma  = c_{\rm cal} \langle \epsilon_{\rm corr} \rangle
\   . 
\label{eq:WL_biascorr}
\end{equation}
The analysis of STEP2 \citep{step2} seemed to indicate that this bias calibration is accurate to $\sim 3$ per cent. The GRavitational lEnsing Accuracy Testing 2008 (GREAT08) team (see \citealt{bridle10}) showed that the multiplicative shear bias can be still of the order of almost 5 per cent (at least for KSB in the Heymans implementation), and that this bias is present for the low signal to noise objects, \mbox{$\left(S/N\right)_{\rm GREAT08}\approx 10$}. However, the correct S/N for these GREAT08 objects with $\left(S/N\right)_{\rm GREAT08}=10$ is more like $\left(S/N\right)_{\rm GREAT08-true}=6$. We therefore conclude that even with the bias correction in  equation~(\ref{eq:WL_biascorr}) we still might have a bias of 5 per cent for objects with $S/N\approx 5$. The size of the bias can however be tested later on by comparing the shear signal of low and high signal to noise objects in the background of the same foreground structures. The  same test can be made to analyse whether the smallest objects in the shape catalogue have biases relative to the larger ones. According to the work of \citet{bridle10} we do not expect a bias if the ratio of galaxy and stellar FWHM exceeds 1.4, but we expect a bias of about 7 per cent in the case where this ratio is only 1.2.
\begin{figure*}
\centering
\includegraphics[width=8.8cm]{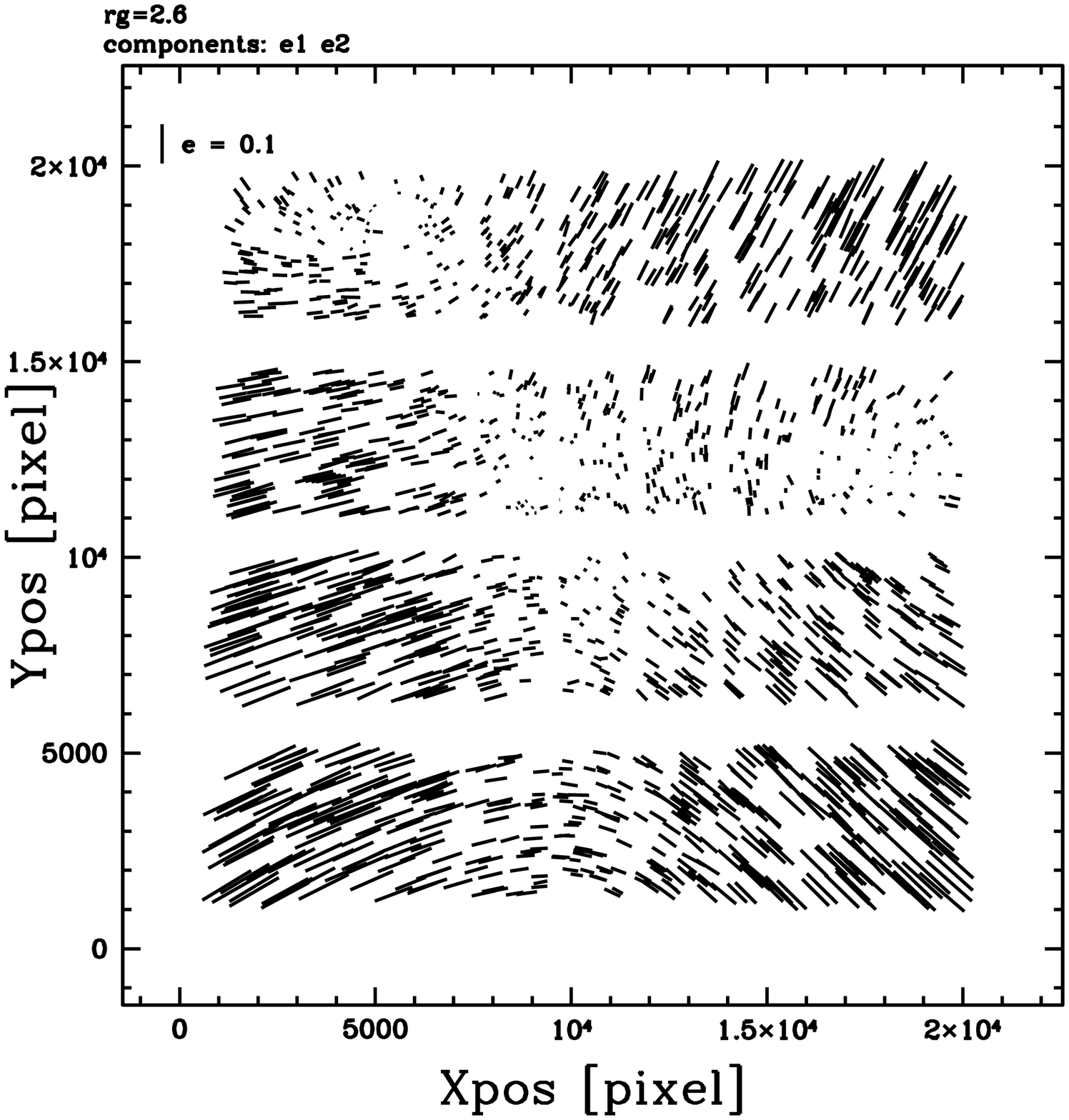}
\includegraphics[width=8.8cm]{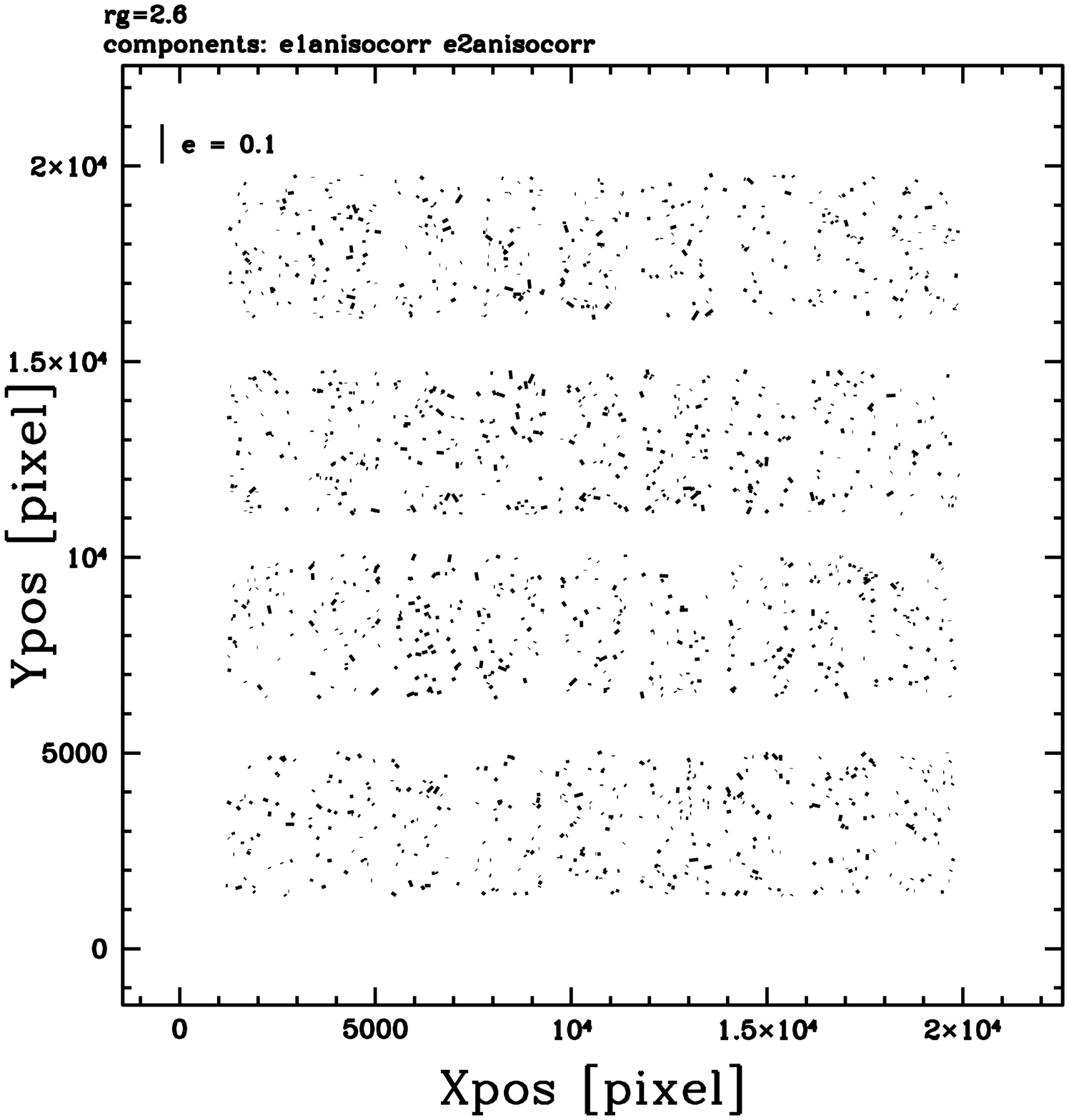}
\includegraphics[width=8.8cm]{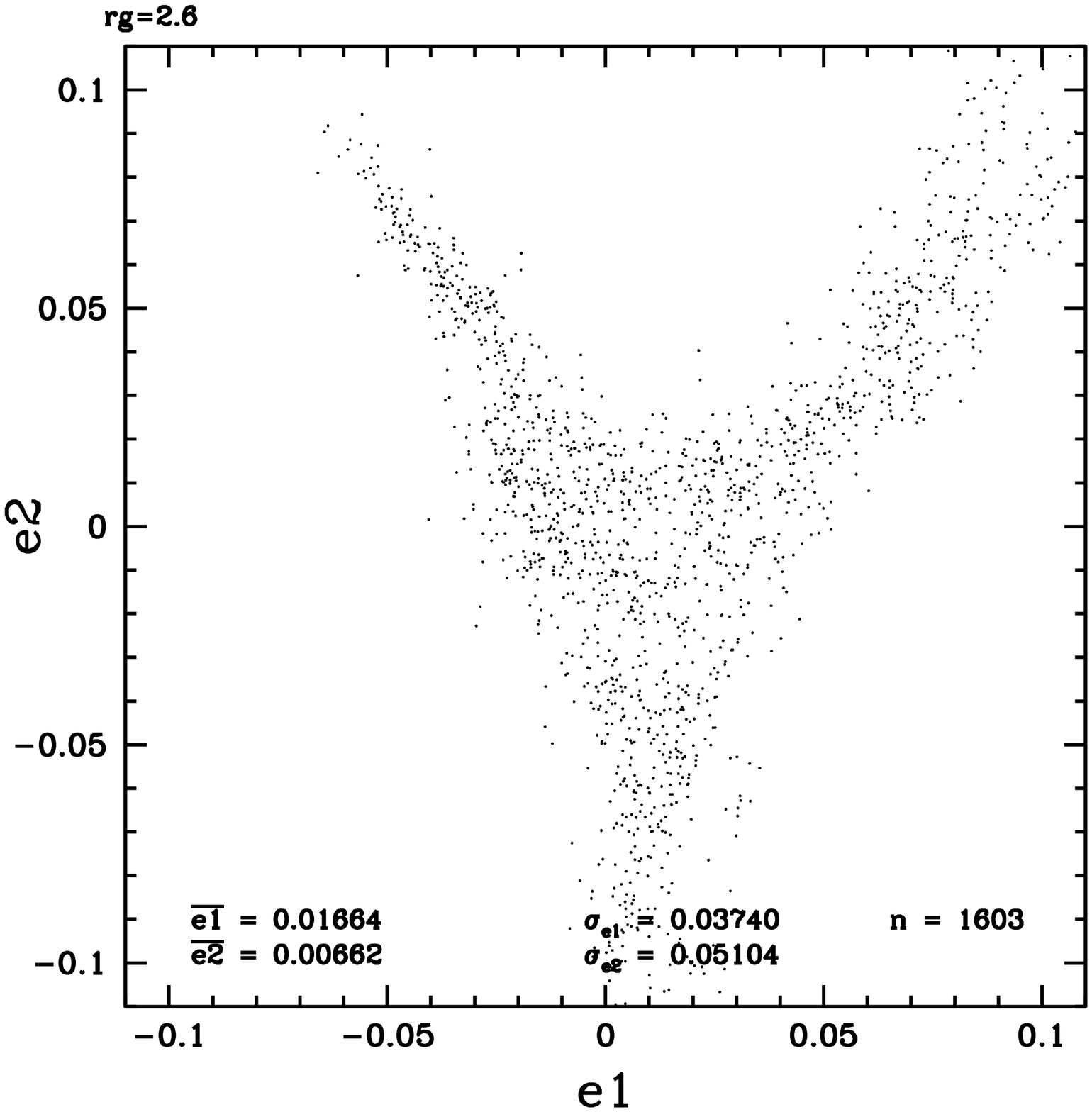}
\includegraphics[width=8.8cm]{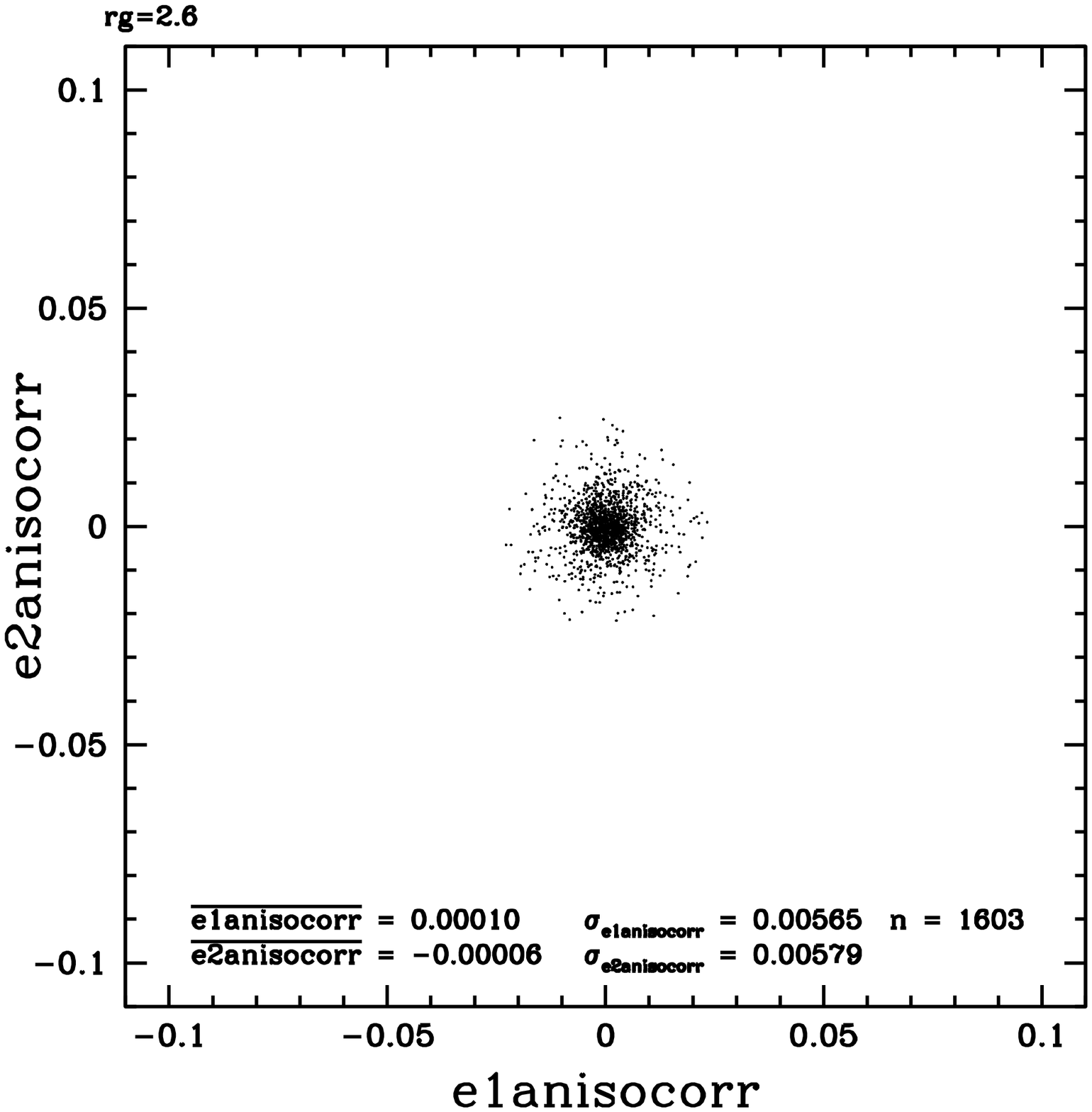}
\caption{PSF anisotropy for an early (pre-lens flip) CFHTLS field (W1p2p3). The upper left panel shows the observed PSF anisotropy pattern, the upper right panel shows the residual stellar anisotropy after modelling the two-dimensional PSF anisotropy pattern with a fifth order polynomial over the whole field of view. The amplitude of the anisotropy is given by showing the length of a 10 per cent anisotropy on the upper left of the two upper sub panels.
The lower left panel shows the stellar ellipticity distribution before the PSF anisotropy correction, the lower right panel shows the stellar ellipticity distribution after the correction. The mean stellar ellipticity components, their dispersions, and the number of stars used for the modelling are also added in these lower sub panels.}
\label{fig:W1p2p3.e1e2}
\end{figure*}
\\
\begin{figure*}
\centering
\includegraphics[width=8.8cm]{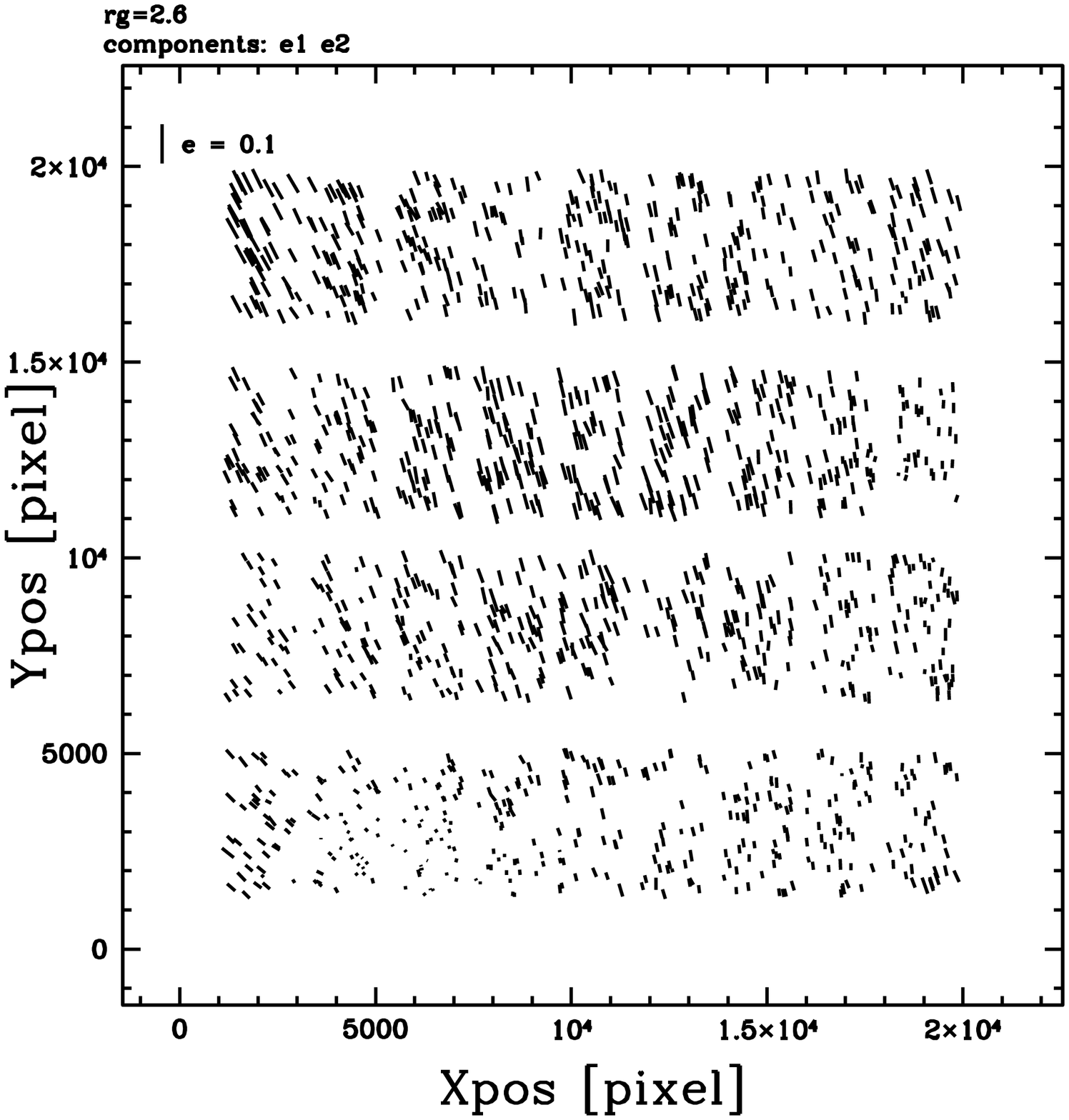}
\includegraphics[width=8.8cm]{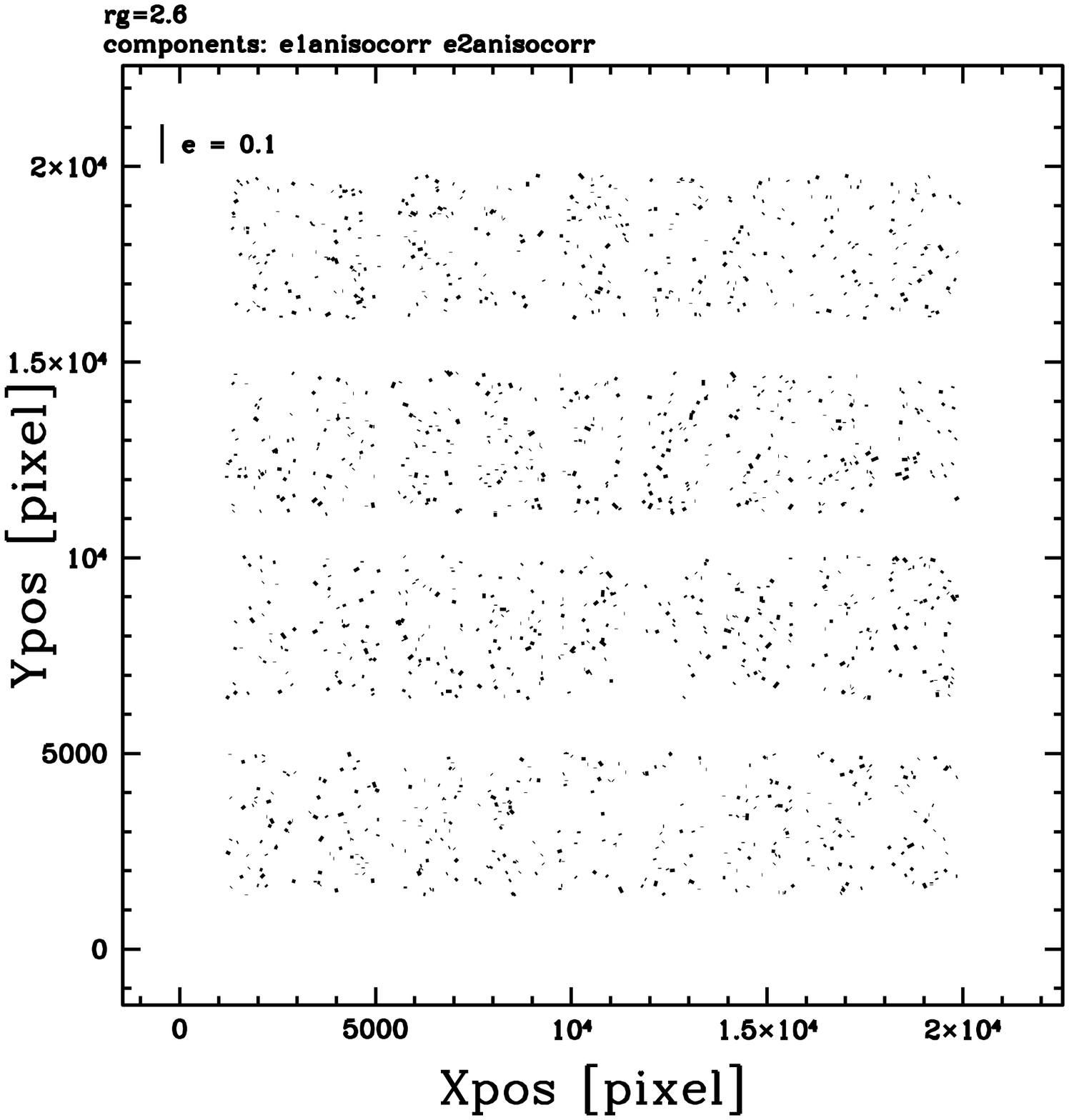}
\includegraphics[width=8.8cm]{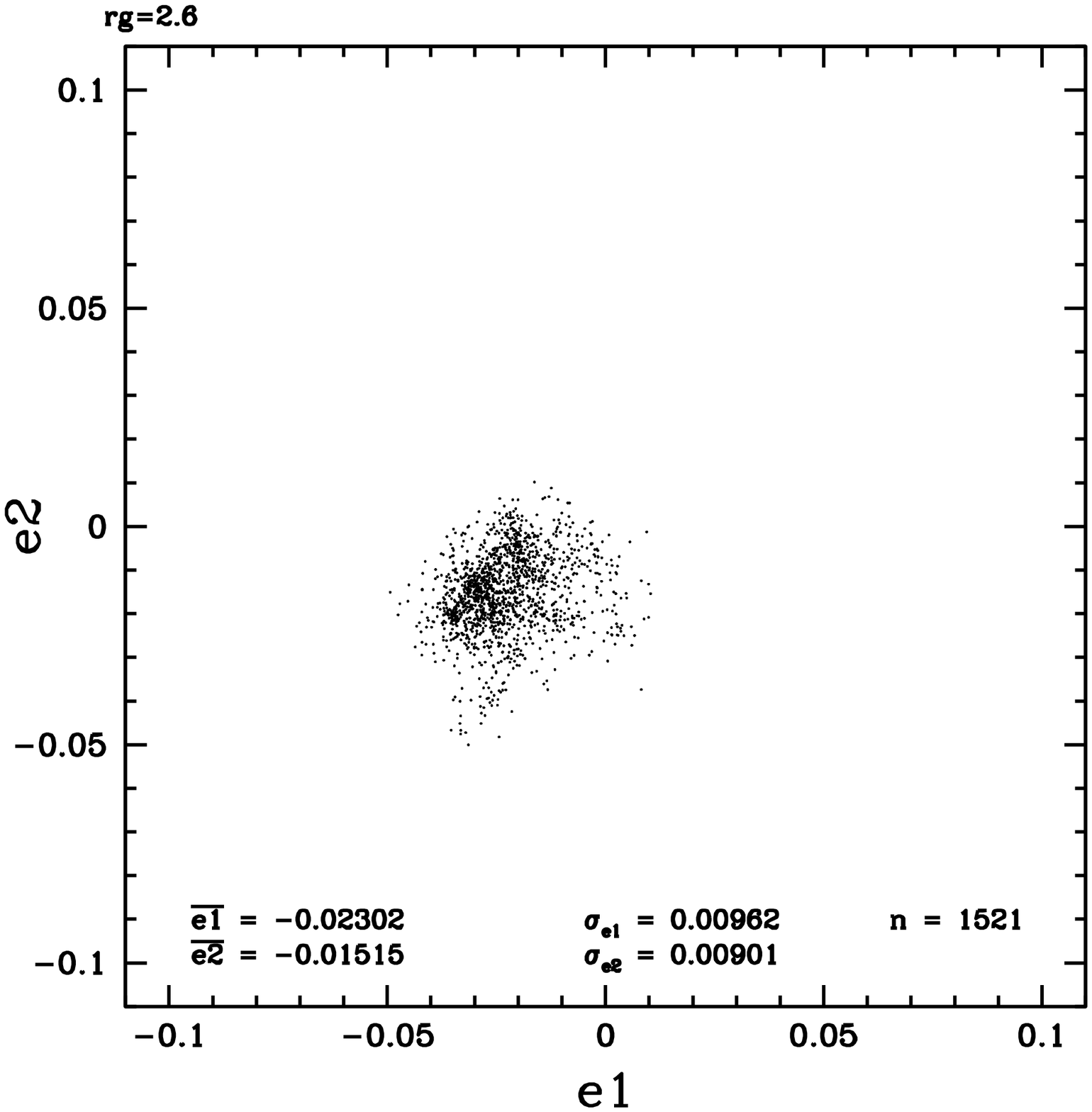}
\includegraphics[width=8.8cm]{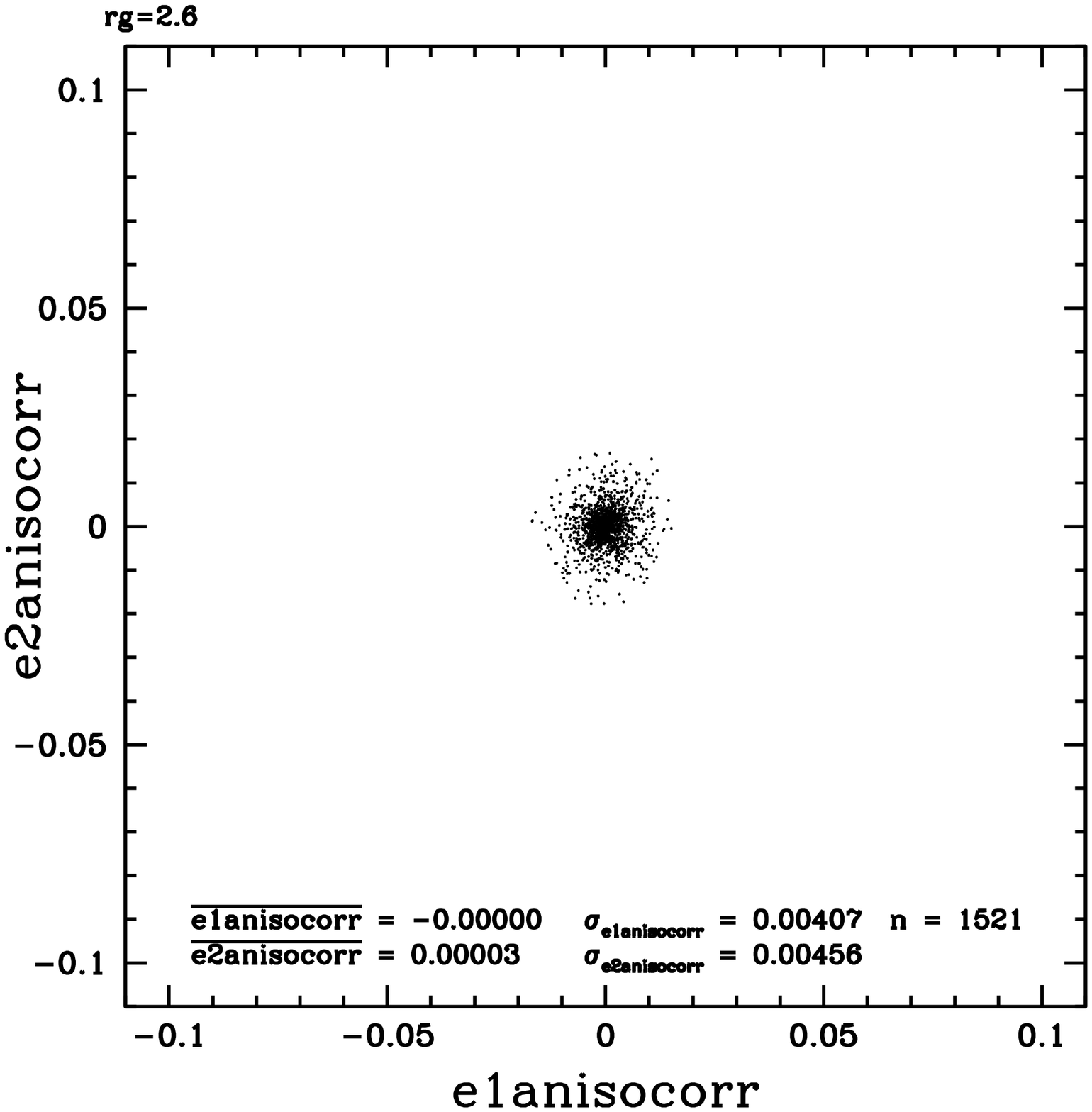}
\caption{Same as Fig.~\ref{fig:W1p2p3.e1e2} but this time using data for the field 
W1m1m2 which was observed after the lens flip. The PSF anisotropy pattern is more regular and also has a smaller amplitude.}
\label{fig:W1m1m2.e1e2}
\end{figure*}
To obtain the shape catalogue we run the \SExtractor software with a signal to noise detection threshold of 3$\sigma$ for four contiguous pixels. We then measure the objects' shapes and their smear and shear tensors.  Then bright but unsaturated stars with  magnitudes in the range of $18 < i' < 22$ are selected, where the exact limits depend on the S/N and PSF of each CFHTLS-field. Further we require a \SExtractor star classification of greater than 0.96. These stars are taken to measure the PSF anisotropy pattern. Examples are shown in the upper panels of  Figs.~\ref{fig:W1p2p3.e1e2} and \ref{fig:W1m1m2.e1e2}, where we have drawn a line at each location of a star, with length proportional to the anisotropy and the orientation along the major axis of the anisotropy ellipsoid (a so-called whisker plot). As can be seen in these Figures we do not use stars (and galaxies) which come from different chips in the individual exposures.
This results in 36 distinct regions from which shape estimates are used. The gaps between these regions depend on the dither pattern which was originally meant to fill the chip gaps. Since the camera is an array of 9 $\times$ 4 CCDs, and gaps in vertical direction are small, one can hardly see the vertical gaps in Figs.~\ref{fig:W1p2p3.e1e2} and \ref{fig:W1m1m2.e1e2}. The reason why we ignore coadded data that comes from different chips is because there are discontinuities in the PSF anisotropy fields over chip borders. The coadded data in these regions are a superposition of two or more anisotropy patterns. It is impossible to obtain a meaningful PSF anisotropy pattern from stars in these regions and  to correct the objects accordingly. Our procedure reduces the number of galaxies in our sample, but also the remaining systematic errors of the PSF anisotropy correction.
\\
Ignoring these regions we then take a fifth order polynomial to describe the PSF anisotropy pattern over the one square degree tiles. The fifth order is flexible enough for the complex PSF anisotropy patterns. We have also experimented with third order polynomials for subfields of the one square degree fields but obtained larger systematic errors in terms of PSF anisotropy model residuals and B-modes in the two-point shear correlation function (for a definition and measurements of B-modes see e.g. \citealt{vanwaerbeke00,vanwaerbeke01} or \citealt{fu08}). A sophisticated analysis that provides an objective measure whether to use lower or higher order polynomials in the PSF anisotropy correction is described in \citet{rowe10}.
\\
We show the residuals of the PSF anisotropy model and the data in the upper right panels of  Figs.~\ref{fig:W1p2p3.e1e2} and \ref{fig:W1m1m2.e1e2}. In the lower left and right panels the ellipticity components $\epsilon_1$ and $\epsilon_2$ of the stars before and after correction are shown. Their average values and their dispersion as well as the number of stars used for these models (1\ 603 and 1\ 521) is given in these Figures as well.  
The data shown in Figs.~\ref{fig:W1p2p3.e1e2} and \ref{fig:W1m1m2.e1e2} have been taken before and after a lens flip (the orientation of lens L3 has been inverted) at the \MegaCam camera optics\footnote{http://www.cfht.hawaii.edu/Science/CFHTLS-DATA/cfhtlsgeneralnews.html\#0007}. The lens flip significantly changed the image characteristics of CFHTLS observations. 
\\
This is why the anisotropy pattern is so remarkably different. In general, the large anisotropy in the early data is more difficult to correct for than the less complex anisotropy pattern in the later data, and several frames of the pre-lens flip data have a much worse anisotropy residual than the example shown in Figs.~\ref{fig:W1p2p3.e1e2}. For this reason we decided to discard 35 CFHTLS fields which performed distinctly worse after visual inspection and only consider 89 fields for our analysis.
\\
After the PSF anisotropy modelling is done the objects are PSF anisotropy corrected. For this we calculate the stellar anisotropy correction quantities for a weight function with a width  of $r_{\rm g}$ that equals the half light radius of the object, \mbox{$r_{\rm g}=r_{\rm h} $}, where the half light radius $r_{\rm h}$ comes from \SExtractor. 
\\
\begin{figure}
\centering
\includegraphics[width=8.8cm]{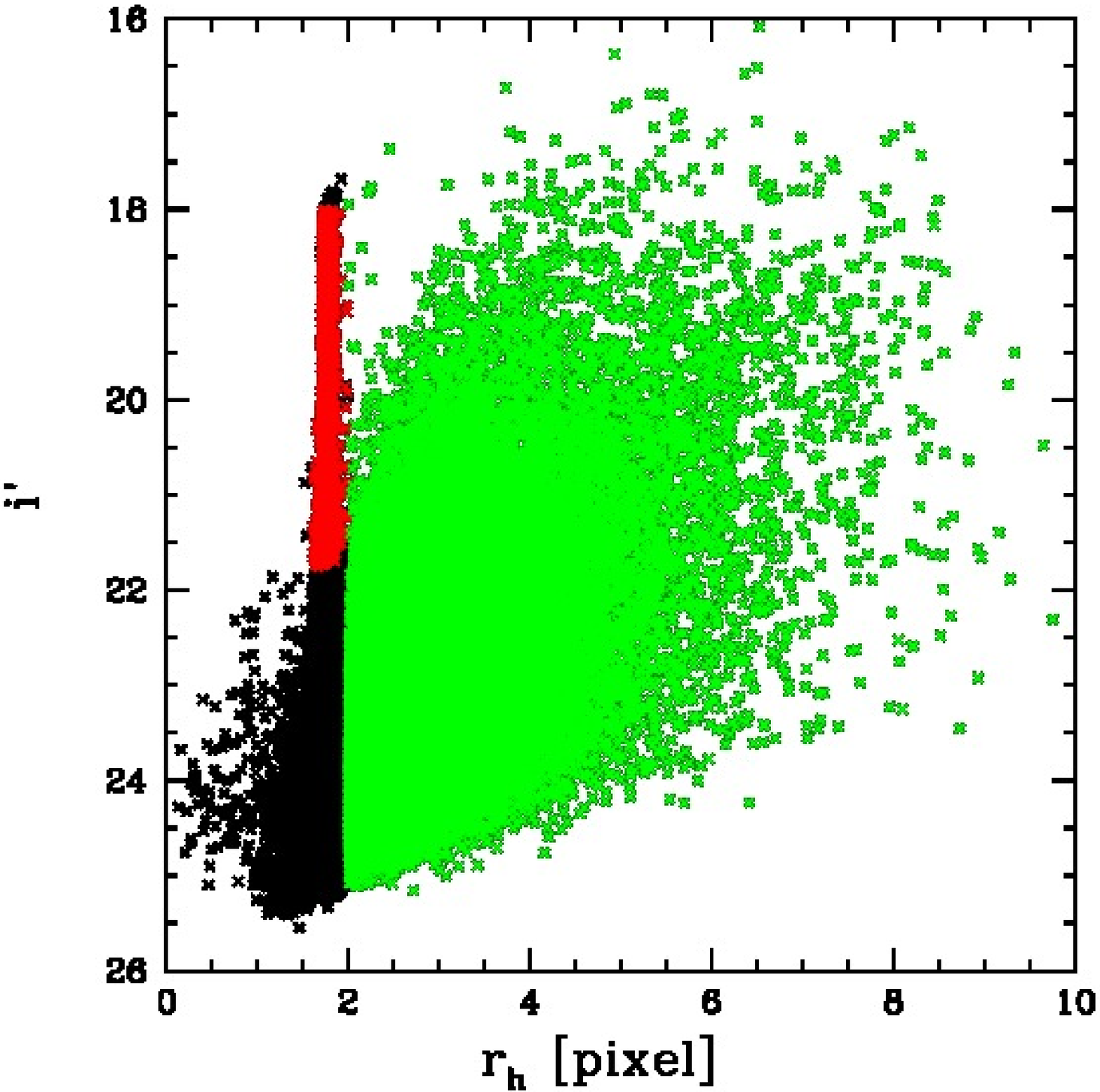}
\caption{Distribution of the $i'$-magnitude vs. \SExtractor half light radius for objects detected with \SExtractor for shape measurements in the W1m1m0-field. We mark stars in red and objects (galaxies) that enter our shape catalogue in green.} 
\label{fig:W1m1m0.rhmag}
\end{figure} 
\begin{figure}
\centering
\includegraphics[width=8.8cm]{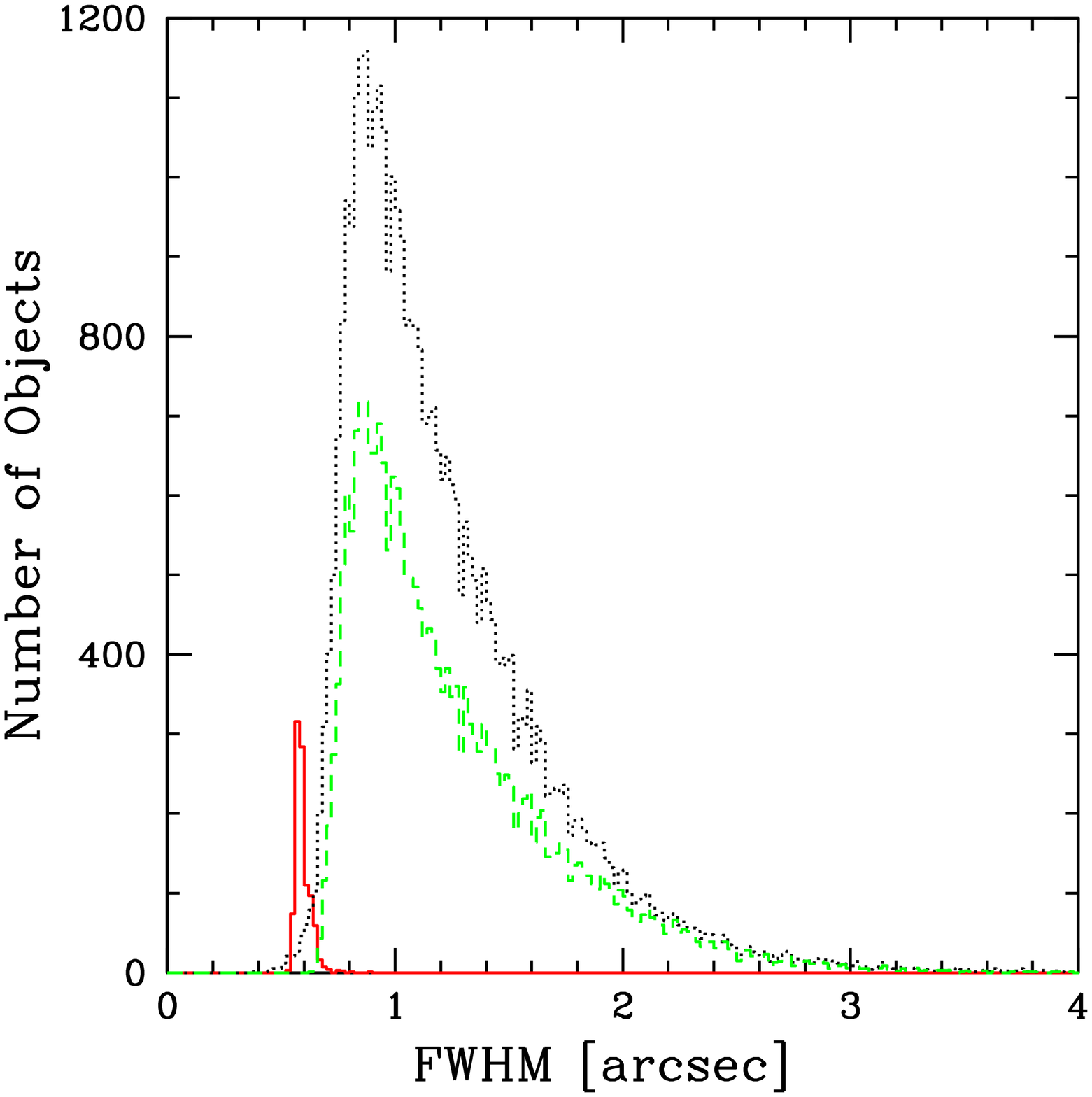}
\caption{Distribution of the FWHM of stars (red) and galaxies (dashed green) detected for shape measurements in the W1m1m0-field. The black dotted line shows the complete galaxy sample, while the dashed line shows the cleaned and selected galaxy sample for shape measurements (S/N$\ge$ 5 and $z _{\rm phot} \le 2$). Most of the galaxies in this field have a  FWHM that is larger than $1.4$ times the FWHM of stars (0.6 arcsec). This has been shown to allow a bias free shear estimate in GREAT08 (\citealt{bridle10}). After the photometric redshift and shape catalogue are merged there is another selection made which ensures that the FWHM of `lensed' objects is larger than of high S/N stars in each of the CFHTLS frames.}
\label{fig:W1m1m0.fwhm}
\end{figure} 
We show the selection of stars (red) and galaxies (green) in the magnitude-half light radius plane for the field W1m2m2 for illustration in Fig.~\ref{fig:W1m1m0.rhmag}. The resulting histograms for the FWHM of stars and galaxies can be seen in Fig.~\ref{fig:W1m1m0.fwhm}, which shows  that most of galaxies obey \mbox{${\rm FWHM}_{\rm gal}>1.4 \times {\rm FWHM}_{\rm star}\approx$ 0.84 arcsec}, which is the limit where no bias for the shear measurement signal was found in the KSB-implementation of Heymans according to GREAT08 (\citealt{bridle10}). In addition when the shape and photometric redshift catalogue are merged to obtain the background galaxy sample all objects with small FWHM are eliminated (see dashed histogram in Fig.~\ref{fig:W1m1m0.fwhm}).
\\
Finally, we  exclude objects  which have 
$\nu \equiv \left(S/N\right)_{\rm KSB}\le 5$ because for those shape measurements become very noisy and shear estimates are potentially biased low. We'll investigate later when measuring the GGL signal whether this cut is justified.
There is no common agreement above which S/N galaxies should be used for shear measurements. It certainly depends on the science case as well. 
Umetsu (priv. communication) suggests in the context of measuring the weak shear profile in the outskirts of clusters to cut all objects with S/N smaller than 7.
\\
We call the catalogue that remains after the KSB pipeline's and the S/N and size cut the `shape catalogue'. The distributions of the two ellipticity components $\epsilon_1$ and $\epsilon_2$ are shown in Figs.~\ref{fig:WIDE.sample.ellipticity_1} and \ref{fig:WIDE.sample.ellipticity_2} as green solid lines. The distributions are very similar and have widths of $\sigma_{\eps_1}=\sigma_{\eps_2}=0.29$. We also show the distributions for those objects in the shape catalogue that also have a photometric redshift estimate as green dashed lines. These distributions have the identical widths of again $0.29$; these values are used for the error analysis later on. The galaxy ellipticity distributions however have broader wings than Gaussians. In Fig.~\ref{fig:WIDE.sample.eps} we show the distribution of the absolute value of ellipticities as solid (shape catalogue) and dashed (after merging shape with photometric redshift catalogue) curves; its medians are  $\abs{\eps}_{\rm med}=0.31$. The distributions of the half light radii and the S/N of the galaxies can be seen in Figs.~\ref{fig:WIDE.sample.rh} and \ref{fig:WIDE.sample.sn}. The median source half light radius is  0.6 arcsec and the median S/N-ratio for the objects in the shape catalogue is $S/N_{\rm med}=11.90$. Also in this case the numbers remain the same when the shape catalogue is merged with the photometric redshift catalogue. This is because most of the objects in the shape catalogue do have a photometric redshift, whereas there are many  objects in the photometric redshift catalogue that have a too low signal to noise to enter the shape catalogue. 
\begin{figure}
\centering
\includegraphics[width=8.8cm]{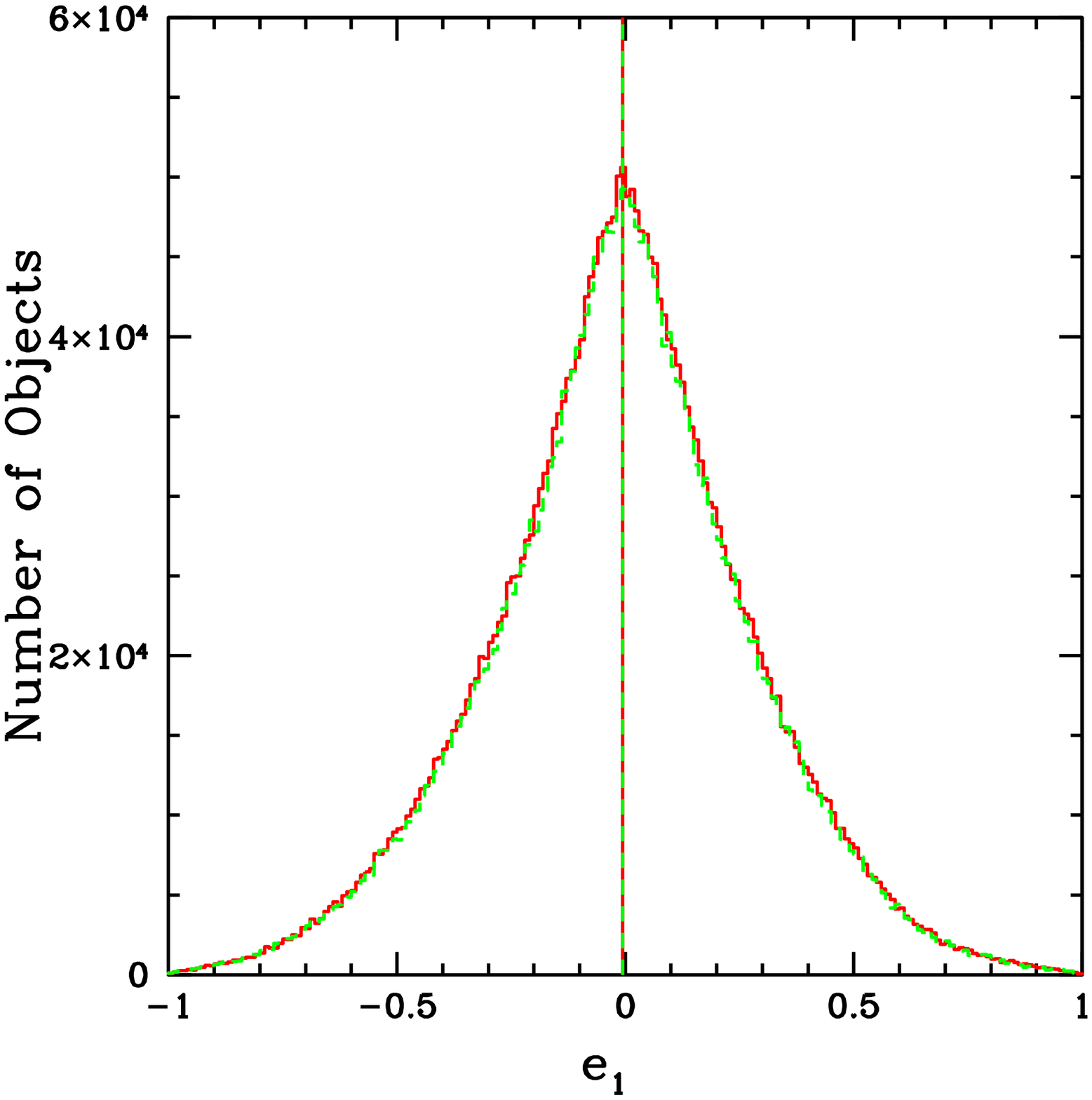}
\caption{Distribution of the $\epsilon_1$  component of galaxies in the shape catalogue (solid green) and after merging the shape catalogue with the photometric redshift catalogue (dashed red). The rms-widths in both cases are  equal to $\sigma_{\eps_1}=0.29$.}
\label{fig:WIDE.sample.ellipticity_1}
\end{figure}
\begin{figure}
\centering
\includegraphics[width=8.8cm]{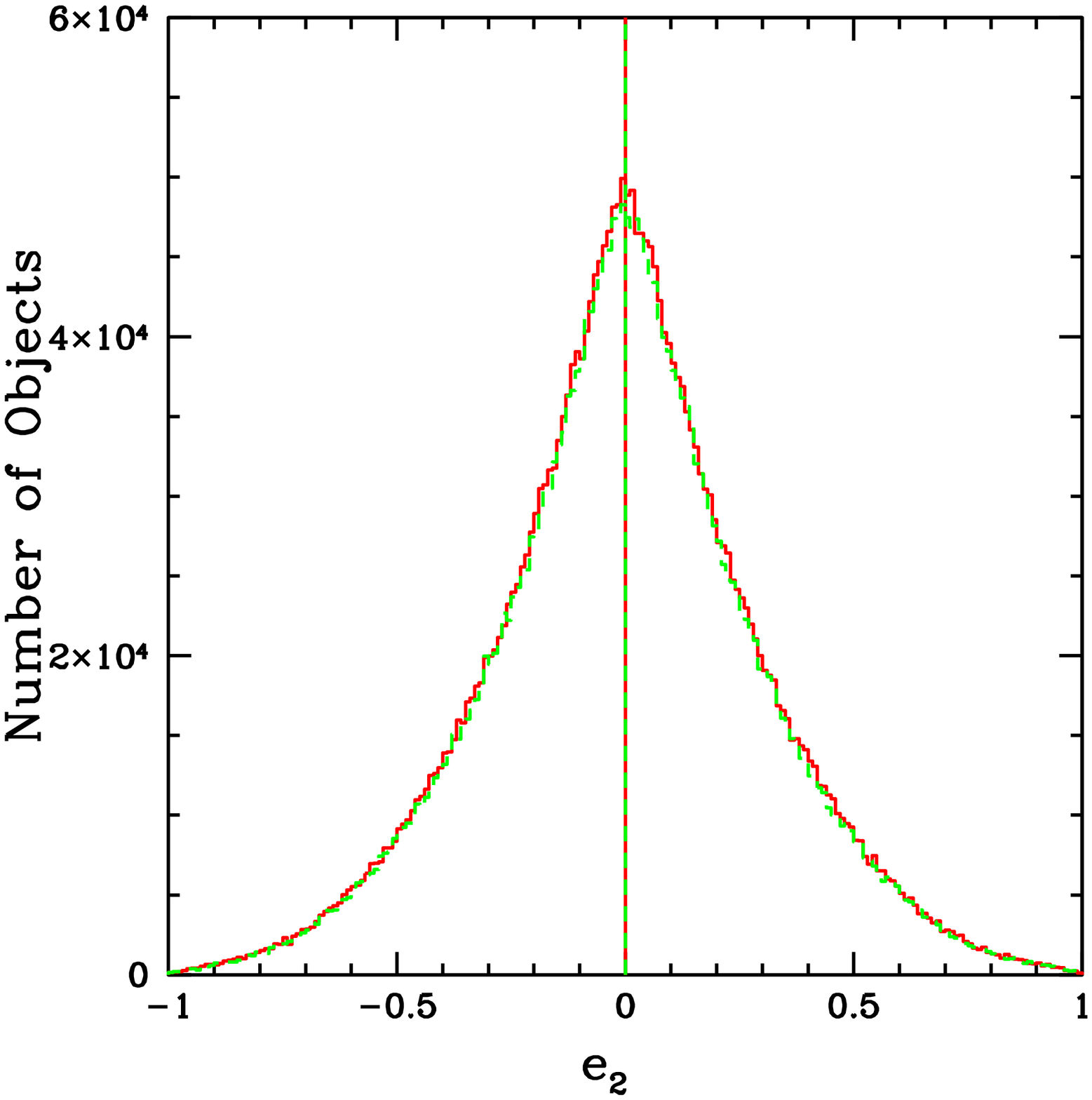}
\caption{Distribution of the $\epsilon_2$  component of galaxies in the shape catalogue (solid green) and after merging the shape catalogue with the photometric redshift catalogue (dashed red). The rms-widths in both cases are  equal to $\sigma_{\eps_2}=0.29$.}
\label{fig:WIDE.sample.ellipticity_2}
\end{figure}
\begin{figure}
\centering
\includegraphics[width=8.8cm]{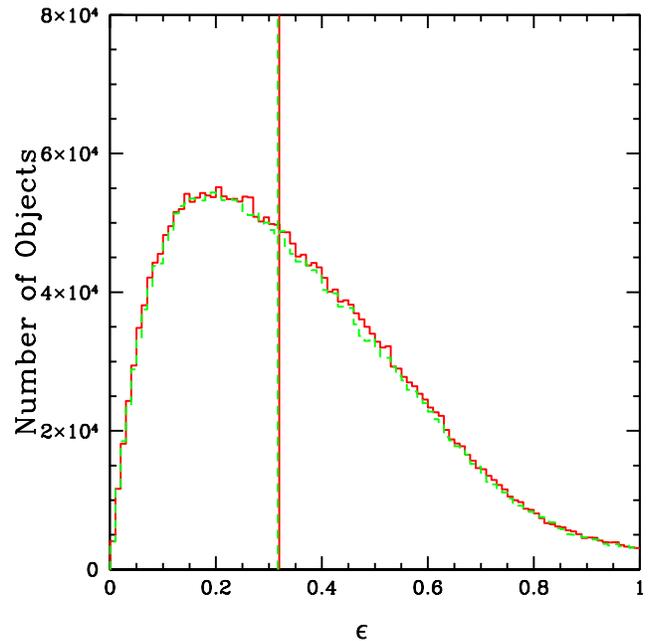}
\caption{Distribution the absolute seeing-corrected ellipticities $\abs{{\epsilon}}$ of objects in the shape catalogue (solid green) and after merging the shape catalogue with the photometric redshift catalogue (dashed red). The median value equals 0.31 in both cases.}
\label{fig:WIDE.sample.eps}
\end{figure}
\begin{figure}
\centering
\includegraphics[width=8.8cm]{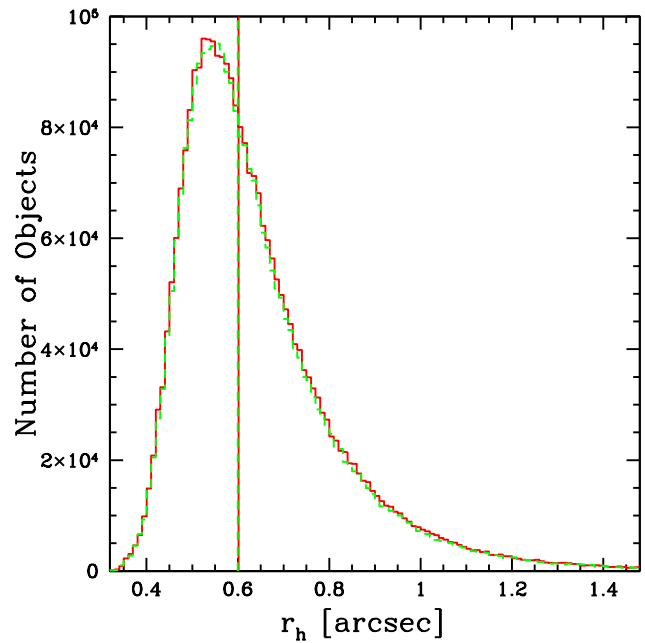}
\caption{Distribution for galaxies in the shape catalogue (solid green) and the shape catalogue merged with the photometric redshift catalogue (dashed red); the vertical line indicates the median of $r_{\rm h,med}=$0.6 arcsec.}
\label{fig:WIDE.sample.rh}
\end{figure}
\begin{figure}
\centering
\includegraphics[width=8.8cm]{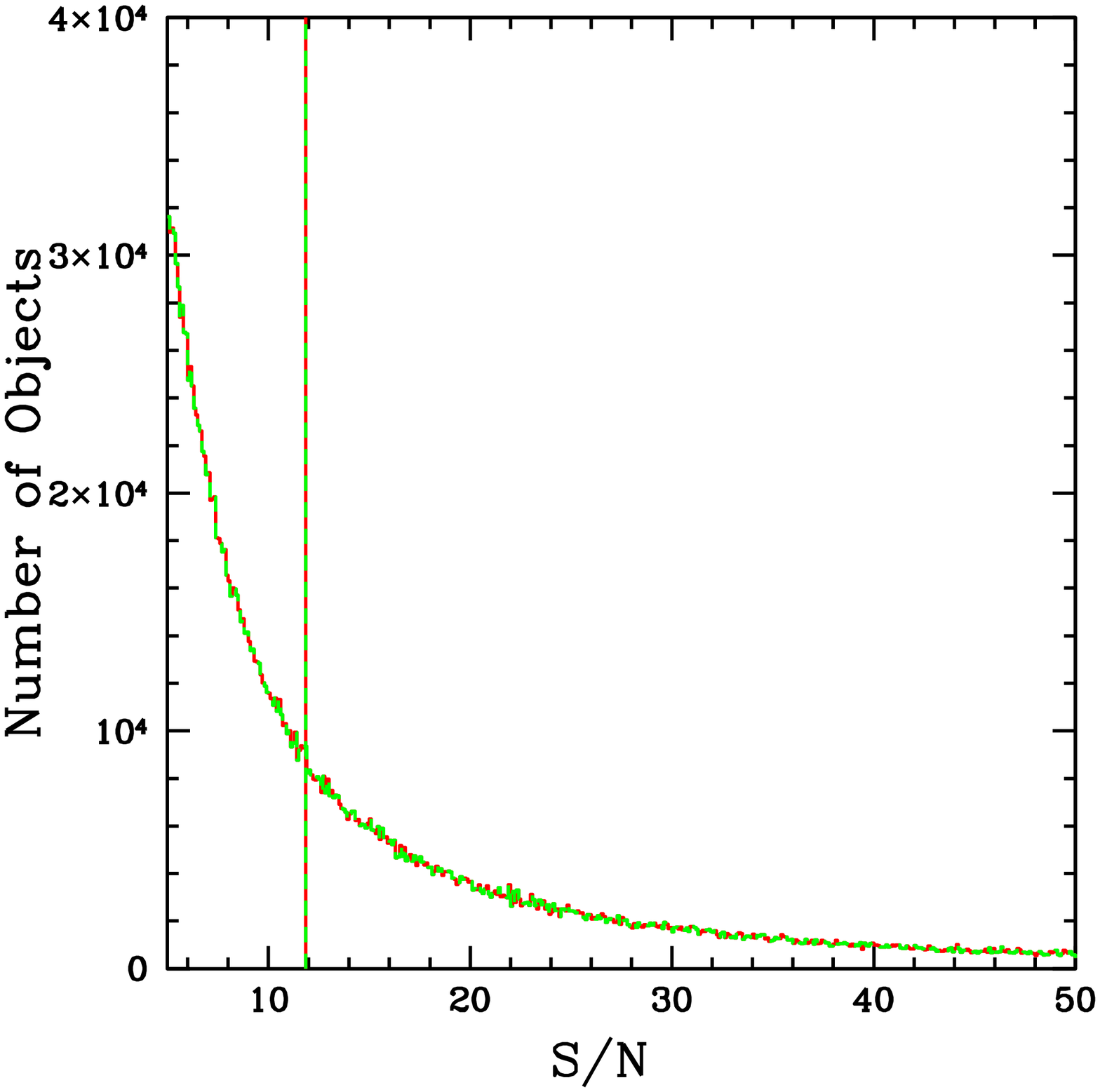}
\caption{S/N-ratio-distribution (we take the $\nu$-value of the KSB-pipeline as the S/N estimate) of objects in the shape catalogue and in the shape catalogue merged with the photometric redshift sample. The vertical line indicates the median of $S/N_{\rm med}=11.9$. Note that the shape catalogue itself only includes objects with a $S/N>5.$ }
\label{fig:WIDE.sample.sn}
\end{figure}
%
%
%
%
\subsection{Properties of galaxies in the photometric redshift and shape catalogues}
\label{sec:properties_catalogues}
\begin{figure}
\centering
\centering
\includegraphics[width=8.8cm]{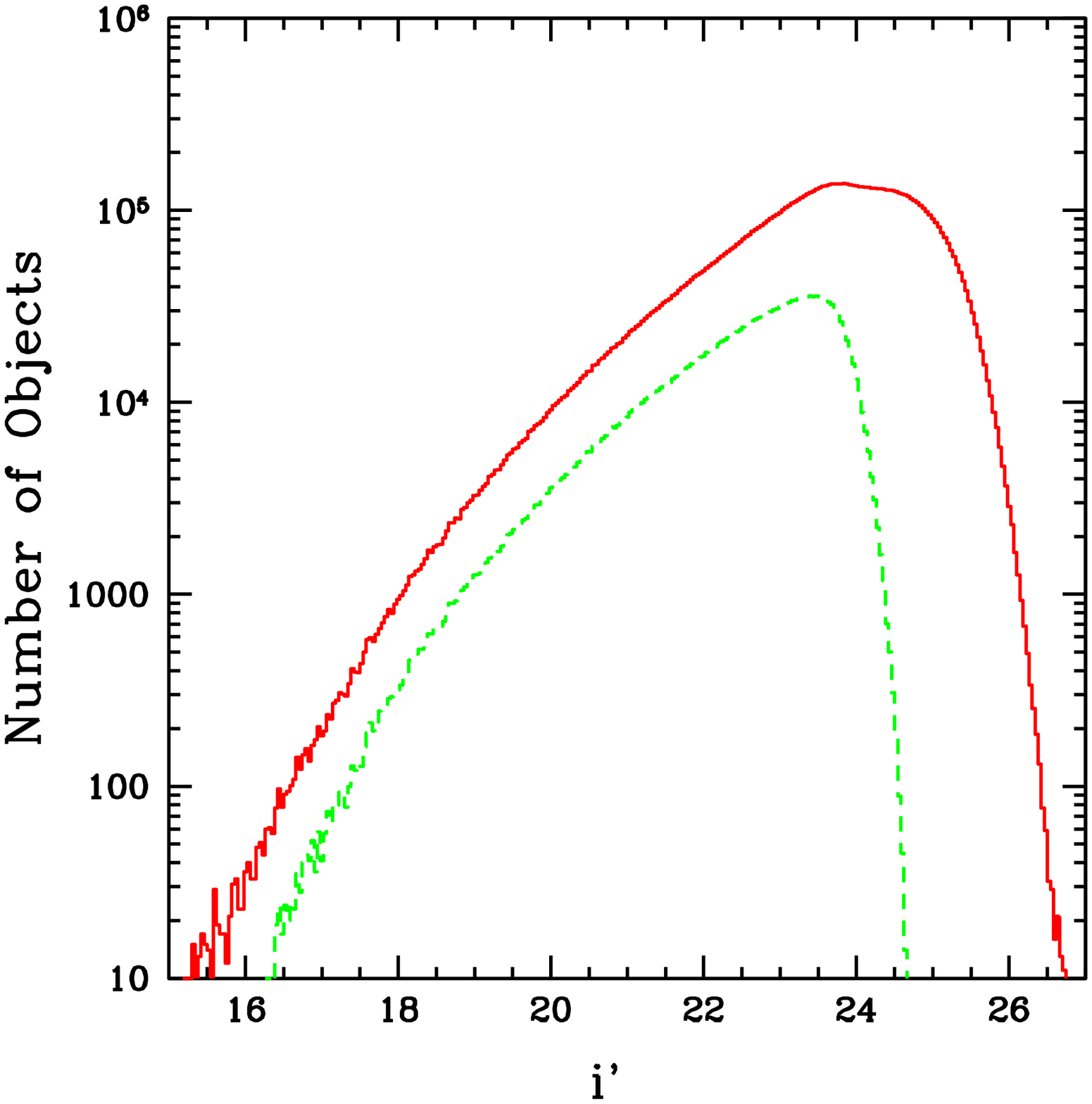}
\caption{Histogram of the $i'$-band magnitude for all {8\ 315\ 162} objects in our cleaned photometric redshift catalogue (red solid line) and for those in the photometric redshift catalogue which are not eliminated in the shape estimation procedure (green dashed line), i.e. for all 2\ 416\ 426 objects which enter our weak lensing catalogue. Objects that are large and bright and those which have low signal to noise do not enter the shape catalogue.
}
\label{fig:WIDE.imag_histo}
\end{figure}
\begin{figure}
\centering
\includegraphics[width=8.8cm]{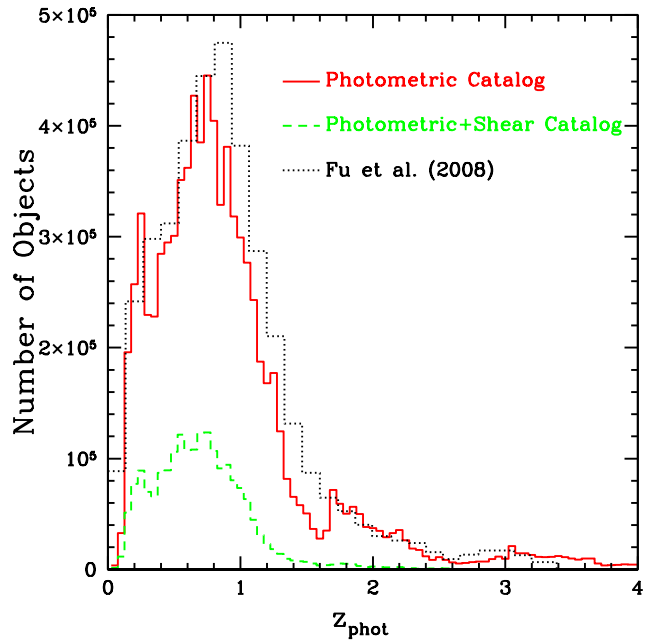}
\caption{Histogram of the cleaned photometric redshift sample. The red solid line shows the full photometric redshift sample, the green dashed line shows the photometric redshift distribution of our shear catalogue. The gap and bump at $z \sim 1.6$ are a feature of very faint objects ($i \ge 24.5$) and do neither affect our background sample (we have no background with $i' \ge 24$) nor our foreground (we only consider lenses with $z \le 1$). For comparison we show the photometric redshift distribution of \citet{fu08} (black dotted line).}
\label{fig:WIDE.zhisto}
\end{figure}
\begin{figure}
\centering
\includegraphics[width=8.8cm]{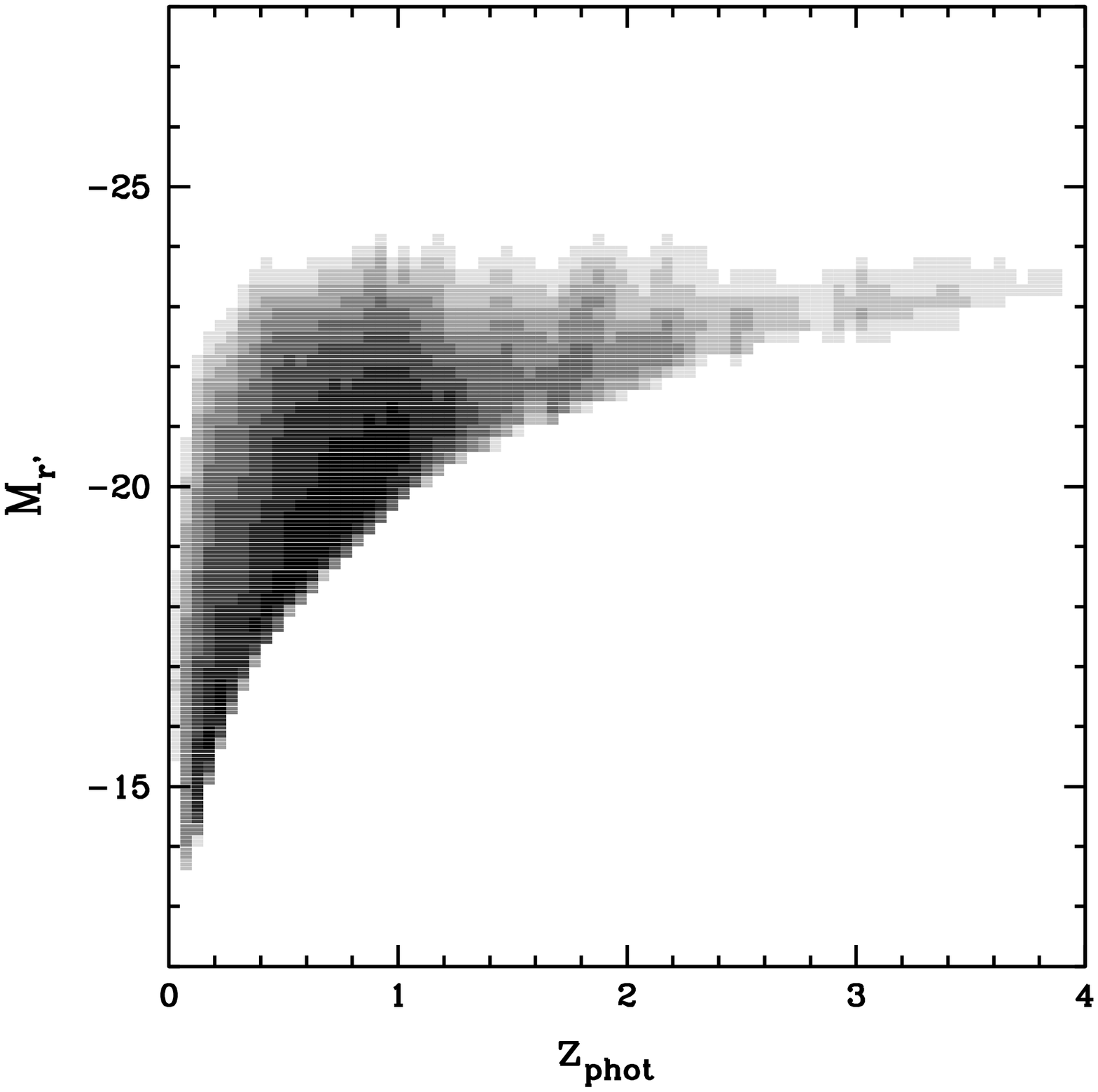}
\caption{Density plot for the absolute luminosity ($M_{r'}$) redshift ($z$) relation for the cleaned photometric redshift catalogue. The cleaning regards masked area, stellar contamination and non precise photometric redshifts and is explained in the text. The number density of very bright galaxies at low redshifts is non-zero, but rather low, leading to an apparent gap in this region in the plot.}
\label{fig:WIDE.Mag_abs_z}
\end{figure}
The $i'$-band detection based, PSF-matched total multi-colour aperture photometry catalogue for the 89  $\rm{deg^{2}}$ analysed comprises 12\ 060\ 104 objects.
The number of galaxies in our photometric redshift catalogue is reduced relative to the full photometric catalogue: a fraction of objects is eliminated because not all area is appropriate for precise photometry (areas around bright stars, satellite tracks and image artefacts have to be flagged), another fraction of objects is identified as stars by their morphology (\SExtractor classification) or by their colours (SED match to stellar library).
\\
Our photometric redshift catalogue contains 8\ 315\ 162  galaxies. Their number counts are shown in Fig.~\ref{fig:WIDE.imag_histo}. They become incomplete at $i'\approx 23.5$. The galaxies' photometric redshift distribution is shown in Fig.~\ref{fig:WIDE.zhisto}.  The majority of galaxies is below redshift one. The galaxy density strongly declines towards z=2, and has a long redshift tail up to $z=4$. A few of these high redshift galaxies are also spectroscopically verified as can be seen in Fig.~\ref{fig:WIDE.photoz_deep}. Our photometric redshift distribution looks similar to that of \citet{brimioulle08} and \citet{erben09}, who  used CFHTLS-Wide data as well, however with a smaller field (about $37$  $\rm{deg^{2}}$) and a slightly different selection criteria for the galaxies in these previous works. 
\\
Those galaxies which have photometric redshifts and shape measurements define our maximum background sample. They are 2\ 960\ 048  altogether, where 2\ 416\ 426 reside in areas that are not flagged as bad area. These are the galaxies that we take for the weak lensing analysis in the following. We include the number counts and photometric redshift distributions for galaxies with photometric redshifts and shape estimates in Figs.~\ref{fig:WIDE.imag_histo} and \ref{fig:WIDE.zhisto} as dashed curves. The differences to the (cleaned) photometric redshift sample are explained as follows: Very bright and large and thus nearby objects do not end up in shape catalogues which reduces the galaxy density at low redshifts in shape catalogues. On the other hand, very small (close to point like) objects are eliminated as well as faint objects because galaxies have to be larger than stars in terms of half light radius and one needs a higher S/N for shape measurements than for photometry. This removes $70$ per cent of all objects in the photometric redshift catalogue from the shape catalogue. Faint objects are spread over a broad redshift range. Fig.~14 in \citet{ilbert06}, for instance, shows that the redshift distribution of $23<i'<23.5$ and $23.5<i'<24$ galaxies has a similar shape for $z<2$. This implies that the shape of the redshift probability distribution is hardly altered by our $S/N>5$ cut in the galaxy shape catalogue relative to the photometric redshift catalogue although of course the number of galaxies is drastically reduced by this cut. For comparison we also have added the photometric redshift distribution for galaxies in the weak lensing sample (galaxies with shape and photometric redshift information) of \citet{fu08}. The redshift distributions look rather similar in general, but due to stricter magnitude requirements for our source sample our number counts are slightly lower above $z \sim 0.8$.
\\
The SED template fitting provides an absolute magnitude estimate for each object at its photometric redshift. We show the $M_{r'}-z$ relation for the  photometric redshift sample in Fig.~\ref{fig:WIDE.Mag_abs_z}. Galaxies fainter than $M_{r'}=-18$ are limited to $z<0.6$, whereas galaxies with $M_{r'}\la -20$ can be found out to $z=1$.
\subsection{Defining lens and background galaxy samples}
\label{sec:lenssample}
\begin{figure}
\includegraphics[width=8.8cm]{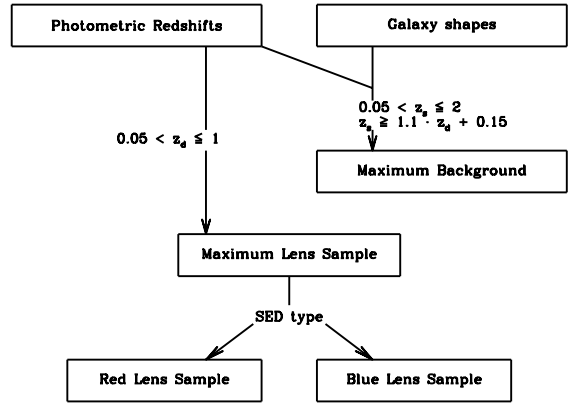}
\caption{Flowchart for our lens and background definition.}
\label{fig:lenssample}
\end{figure}

The procedure for defining  a `foreground' and `background' galaxy sample or a lens and lensed object sample is guided by the following facts: We want to study the mass distribution of galaxies as a function of redshift, rest frame luminosities, SED-types and later on also stellar masses. Hence we can only consider galaxies for which we have photometric redshifts. Among those we restrict the analysis to those galaxies from the photometric redshift sample with 
\begin{equation}
M_{\textit{r'} ,\rm{lens}} < -17 \ \ \text{and} \ \  z_{\rm lens}\le1 \ . 
\label{eq:lens_select}
\end{equation}
This defines our maximum lens sample. We  can study the properties of any lens subsample, provided we have enough galaxies with shape measurements behind the lens sample considered.
\\
We construct our maximum background sample from those galaxies which have reliable photometric redshifts and shape measurements. We ignore sources with $z_{\rm phot} > 2$ as the number counts strongly drop for higher redshifts (see Fig.~\ref{fig:WIDE.zhisto}). Another reason is that the accuracy of photometric redshifts deteriorates for $2 < z_{\rm phot} < 3$ without NIR information at the depth of CFHTLS-Wide and that the  fractional contribution for $z_{\rm phot} > 3$, where the photometric redshift accuracy improves due to u-band drop-outs, is negligible. We also ignore objects with large photometric redshift errors.
In summary we require
\begin{equation}
 \Delta_{z_{\rm phot \  source/lens }} < 0.25 \ (1 + z_{\rm phot \  source/lens }) \  \ \text{and} \  \  z_{\rm  source} \le 2.0  \  .
\label{eq:source_select}
\end{equation}
This provides us with  4\ 942\ 433 galaxies in the maximum lens and 1\ 684\ 290 galaxies in the maximum background sample.
\\
In the following we always assign a background sample dynamically to the foreground sample: for a fixed foreground sample we analyse the shapes of all galaxies  within an angular or physical radius in the background of every lens galaxy. As minimum and maximum angular scale we choose 5 arcsec and 15 arcmin. This corresponds to  $3.3\ h^{-1}$ kpc and  \mbox{$600\ h^{-1}$ kpc} at $z=0.05$, $20\ h^{-1}$ kpc and  \mbox{$3.8\ h^{-1}$ Mpc} at $z=0.5$, $28\ h^{-1}$ kpc and  $5.0\ h^{-1}$ Mpc  at $z=1$. The GGL signal will only be evaluated out to a distance of \mbox{$2\ h^{-1}$  Mpc} later on. The outer angular cut-off of $15'$ is chosen for computational reasons. It equals a transverse separation of  \mbox{$2.7\ h^{-1}$ Mpc} at a redshift of $0.3$.
\\
As the criterion for a galaxy to be a background galaxy we require that the photometric redshift estimate for the galaxy from the maximum background sample exceeds that of the lens by the $z_{\rm s}-z_{\rm d} \ge \sqrt{4\Delta_{\rm z_{\rm d}}^2 + 4 \Delta_{\rm z_{\rm s}}^2}\approx\sqrt{8} \Delta_{\rm z}\ $. This photometric redshift requirement becomes equal to $z_{\rm s}-z_{\rm d} \ge 0.1 $ for our photometric redshift errors of $\Delta_{\rm z} \approx 0.04$.
\\
On the other hand, even if the photometric redshifts were extremely precise one would not consider galaxies in the ultimate background of a lens as they only carry a small lensing signal. The gravitational shear observed in the background of a singular isothermal sphere scales as $D_{\rm ds}/D_{\rm s}$, where $D_{\rm ds}$ and $D_{\rm s}$ are the diameter distances from the lens to the source and from the observer to the source. This ratio approaches zero for $z_{\rm s}\approx z_{\rm d}$, rises steeply for $z_{\rm s}> z_{\rm d}$ and then flattens for larger redshifts. To exclude galaxies that carry no signal but only shape noise, for any given foreground sample we reject galaxies for which $D_{\rm ds}/D_{\rm s}>0.1$ for a given foreground sample does not hold for any of the galaxies in their foreground (cf. Fig.~\ref{fig:dd.error}).
Finally, when the measured shear is translated into a mass density, the fractional error  is proportional to the ratio of the error of the critical surface mass density and the critical surface mass density (see equation~\ref{eq:err2}). For Gaussian photometric redshifts errors with a width of $0.05\ (1+z)$ this ratio is smaller than $0.3$ if 
$z_{\rm s}> 1.1\ z_{\rm d}+0.15 \ \   {\rm and } \  \  z_{\rm d}>0.05  \ $ 
(see Fig.~\ref{fig:sigmac.error}).
This mass accuracy condition is slightly more conservative than  $z_{\rm s}>z_{\rm d}+0.1$. Furthermore, this criterium implies  $D_{\rm ds}/D_{\rm s} \approx 0.75 $ for $z_{\rm d}\approx 0.05$ and  $D_{\rm ds}/D_{\rm s}>0.15$ for $z_{\rm d}\approx 1.0$  which is the largest lens galaxy redshift that we consider in this work. 
In summary, our selection criterium for the lens and source redshifts are (see also Fig.~\ref{fig:lenssample}):
\begin{equation}
0.05 < z_{\rm d} \le  1.0, \ z_{\rm s} \le  2.0 \ \ \text{and} \ \ z_{\rm s} \ge  1.1 \ z_{\rm d} + 0.15 \ .
\label{eq:zlens-zsource-diff}
\end{equation}
These criteriae insure that foreground and background galaxies can be disentangled, that ${{D_{\rm ds}}/{D_{\rm s}}}>0.1$ holds  and that the fractional statistical error for the critical surface density due to photometric redshift errors is smaller than $0.3$ per foreground-background pair (see Fig.~\ref{fig:sigmac.error}).
\\
The definition of our maximum lens and background samples implies that the maximum lens sample is larger than the maximum background sample. It allows situations where the apparent magnitude of a foreground galaxy is larger than the magnitude of a background object tracing its gravitational shear. This is not the case in GGL analyses like that of \citet{parker07}, 
and de facto (although not by definition) not the case for the work of \citet{fischer00}, \citet{mckay01} or \citet{mandelbaum06a,mandelbaum06b}, due to the extremely bright foreground sample ($r'<17.77$ for the main spectroscopic sample, see \citealt{strauss02} and $r'<19.5$ for the spectroscopic LRG sample, see \citealt{eisenstein01}). Also, in our analysis the  redshift distribution of lenses and lensed objects overlap, although for each lens only galaxies in its background are considered. 
\\
Similar to \citet{mandelbaum06b} we split the maximum lens sample in absolute magnitude intervals of one magnitude width from  $M_{\rm r'}=-17$ to $M_{\rm r'}<-24$.
\subsection{Lens mass errors from photometric redshifts and shape measurement errors}
\label{sec:lensmassbias}
\begin{figure}
\includegraphics[width=8.8cm]{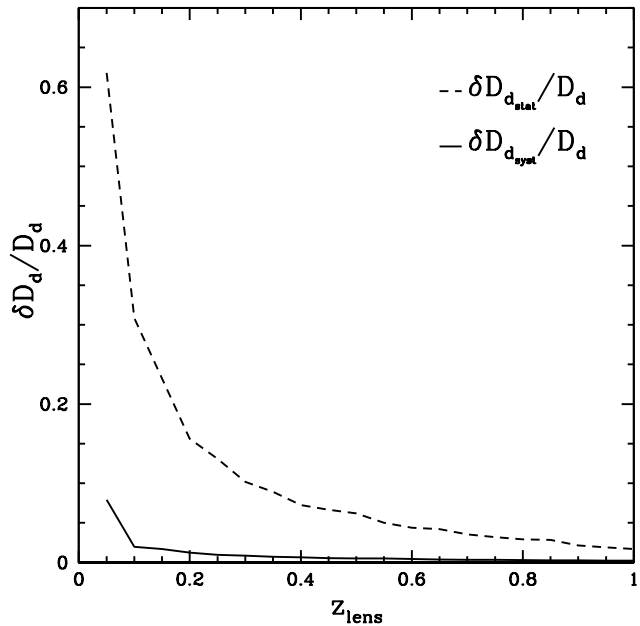}
\caption{Fractional systematic and statistical error $\delta D_{\rm{d}}/D_{\rm{d}}$ of the angular distance of the lens $D_{\rm{d}}$ in presence of photometric redshift errors. The dashed line shows the statistical and the solid line the systematic error for a Gaussian redshift error distribution with a scatter of $0.05 \ (1+z)$. The systematic error in $D_{\rm{d}}$ is below 10 per cent for $z \ge 0.05$ and well below 5 per cent for $z \ge  0.1$.}
\label{fig:dd.error}
\end{figure}

\begin{figure}
\includegraphics[width=8.8cm]{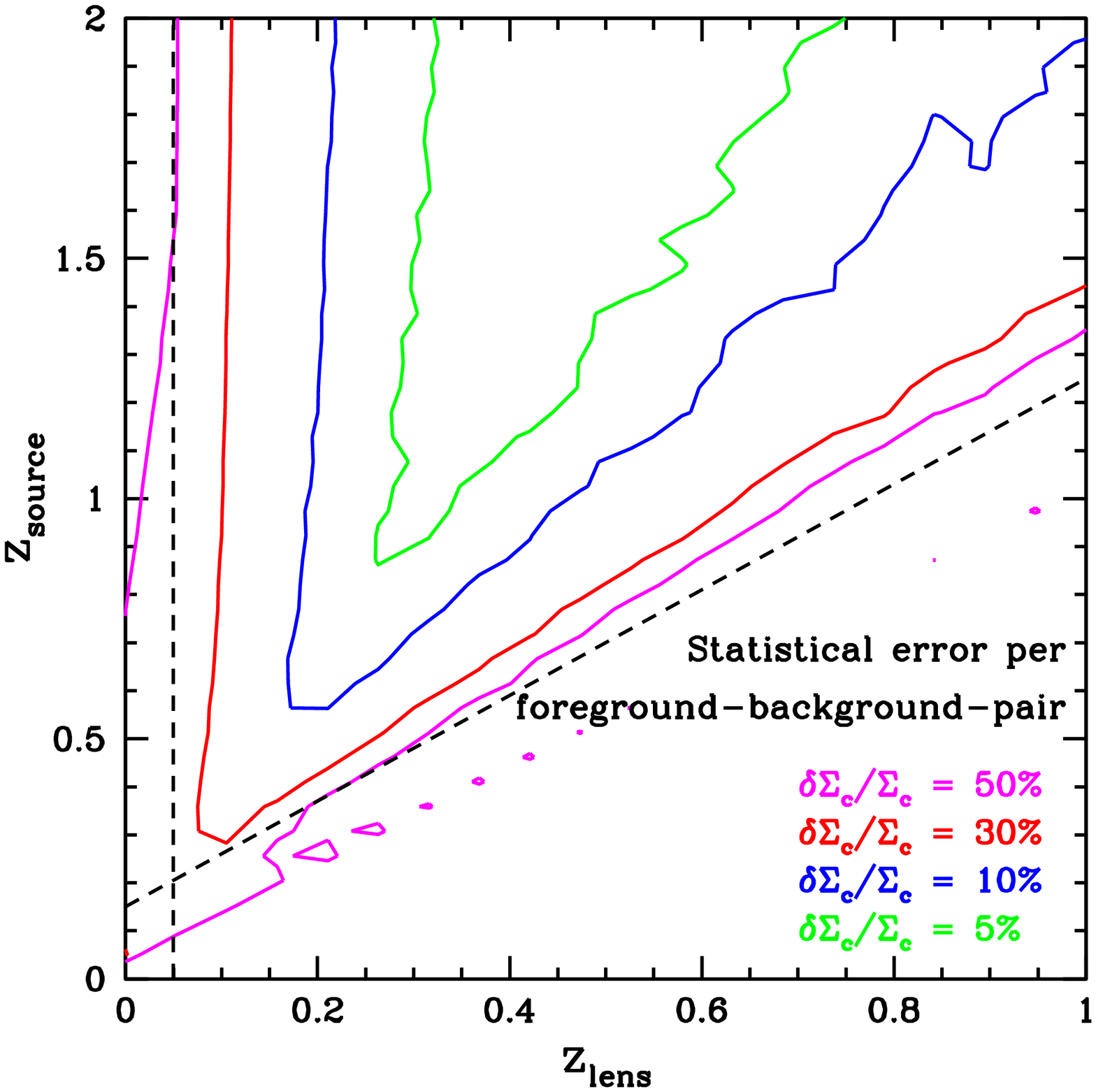}
\includegraphics[width=8.8cm]{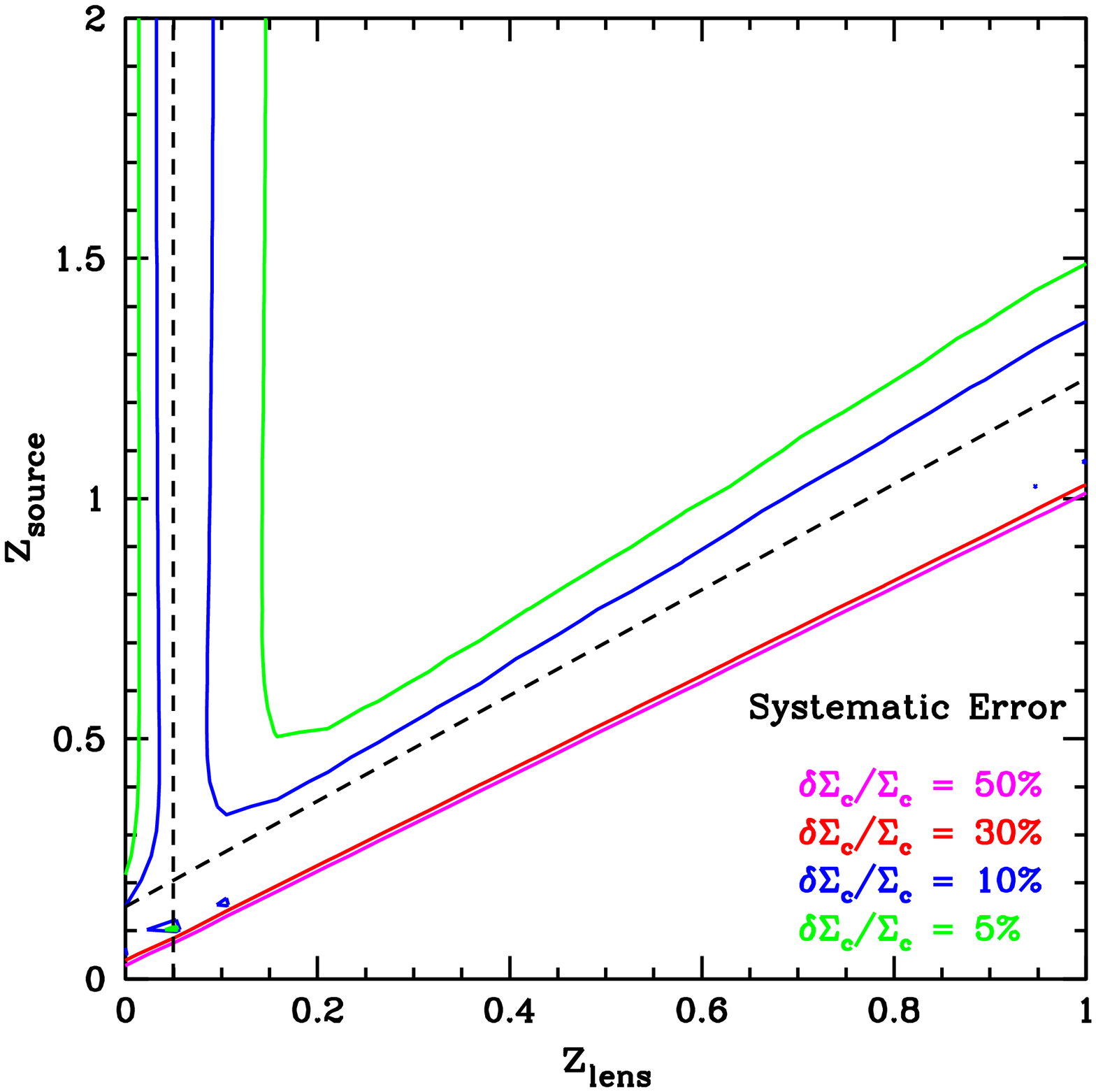}
\caption{Fractional error of the critical surface density, $\delta \Sigma_{\rm c}/\Sigma_{\rm c}$, in the presence of photometric redshift errors as a function of the lens and source redshift. The upper panel shows the statistical error per foreground-background pair, the lower panel the systematic error. The photometric redshift errors in this simulation are Gaussian with a scatter of $0.05 \ (1+z)$. The magenta and the red contours show $\delta \Sigma_{\rm c}/\Sigma_{\rm c}=0.5$ and $\delta \Sigma_{\rm c}/\Sigma_{\rm c}=0.3$ levels, respectively. Blue and green contours show the $\delta \Sigma_{\rm c}/\Sigma_{\rm c}=0.1$ and $\delta \Sigma_{\rm c}/\Sigma_{\rm c}=0.05$ levels. The fractional systematic error is below $0.3$ if  $z_{\rm source}= 1.1 \ z_{\rm lens} + 0.15$ (dashed black line) and $z_{\rm lens}>0.05$, which is the source-lens redshift requirement set in equation~(\ref{eq:zlens-zsource-diff}). The errors are in general rather overestimated than underestimated. This is especially true for very low redshift sources, where the scattering during randomisation of the lens redshift can lead to negative lens redshifts, which are not realised in the observational data.}
\label{fig:sigmac.error}
\end{figure}

As an estimator for equation~(\ref{eq:etan2}) we use the foreground-background pair average, i.e.  
\begin{equation}
\begin{array} {ll}
 \hat{\Delta\Sigma} (R) & = \ave{ \hat \Sigma_{\rm c} \ \hat \gamma_{\rm t } ( R) }_{\rm fg-bg-pairs}    \  .
\end{array}
\label{eq:etan4}
\end{equation}
In the upper equation the estimates carry a `hat'. $\hat \Sigma_{\rm c }$ is obtained from the photometric redshift estimates of the background-foreground pair and the shear estimate $\hat \gamma_{\rm t}(\hat R)$ is estimated from the background shapes and the lens photometric redshift because the observed angular scale $\theta$ needs to be translated into a length scale with $\hat R=\theta \hat D_{\rm d}$:
The relation between the true gravitational shear $\gamma_{\rm t}(R)$ at radius $R$ and its estimate $\hat \gamma_{\rm t}(R)$ at radius $R$ reads as
\begin{equation}
\begin{array} {cl}
\hat \gamma_{\rm t} (R)&= \gamma_{\rm t}(R)+ \left[\gamma_{\rm t}(\hat R) - \gamma_{\rm t} (R)\right] + \delta\gamma_{\rm shape} \\
& \equiv \gamma_{\rm t} (R) + \Delta\gamma_{\rm t}(\hat R, R) + \delta\gamma_{\rm shape}  \\ 
\end{array} \  .
\label{eq:err1}
\end{equation}
In equation~(\ref{eq:err1}) we have introduced a radius independent shear estimation error $ \delta\gamma_{\rm shape}$ that comes from shape measurement error, intrinsic shape noise and potential shape estimate bias.
\\\
The other quantity which we have introduced in equation~(\ref{eq:err1}) is the `profile error'  
\begin{equation}
\Delta\gamma_{\rm t}(\hat R, R)= \gamma_{\rm t}(\hat R) - \gamma_{\rm t} (R) = \gamma_{\rm t}(\theta \hat D_{\rm d}) - \gamma_{\rm t}(\theta D_{\rm d}) \ ,
\label{eq:Delta_gam}
\end{equation}
which comes from mixing physical scales when translating angles into length scales. The latter is zero when a spectroscopic lens sample is used. In general, the  profile error depends on the profile steepness. For a power law with $\gamma_{\rm t}(R) \propto R^{-\alpha }$ it equals
\begin{equation}
\Delta\gamma_{\rm t}(\hat R,R) = \gamma_{\rm t} (R) \left(\frac{R^\alpha-\hat R^\alpha}{\hat R^\alpha}\right) \  .
\label{eq:deltagam_alpha}
\end{equation}
Thus the profile error is the more important the steeper the shear profile is. 
For an isothermal profile with $\alpha=1$ equation~(\ref{eq:deltagam_alpha}) becomes 
\begin{equation}
\Delta\gamma_{\rm t}^{\rm SIS}(\hat R,R) = \gamma_{\rm t}(R)  \left(\frac{D_{\rm d}-\hat D_{\rm d}}{\hat D_{\rm d}} \right) \  .
\label{eq:deltagam_SIS}
\end{equation}
For small redshifts the profile error equals
\begin{equation}
\left(\frac{\Delta\gamma_{\rm t}^{\rm SIS} (\hat R, R)}{\gamma_{\rm t}(R)}\right)_{\rm z_{\rm d}\approx 0} \approx \frac{z_{\rm d} -\hat z_{\rm d}}{\hat z_{\rm d}} = \frac{\Delta z_{\rm d}}{z_{\rm d}} \  .
\label{eq:deltagam_SIS_0}
\end{equation}
It is obvious that photometric redshift errors without a bias can already impose a bias on the shear profile estimate.
\\
To understand this we consider foreground-background pairs with a transverse distance estimate $\hat{R}$ in the lens plane and a redshift estimate of $\hat{z_{\rm d}}$. If the redshift estimate is too low, i.e. $\hat{z_{\rm d}} < z_{\rm d}$ and thus $\hat{R} < R$, sources are scattered from larger to smaller transverse separations. This leads to an underestimate of the gravitational shear signal. Analogously if the true redshift is lower, the gravitational shear signal is overestimated. Therefore a possible bias depends on the redshift (and thus projected separation) distribution of the analysed lens sample. For a flat redshift distribution about the same number of galaxies are scattered down from higher and scattered up from lower redshifts into the bin of interest. In this case the mean projected distance of galaxies within a ring of a diameter of $\hat{R}$ equals the true value $R$. If the lens number increases as a function of redshift more galaxies with $z_{\rm d} > \hat{z_{\rm d}}$ get scattered down to a lower redshift than galaxies with $z_{\rm d} < \hat{z_{\rm d}}$ get scattered up. This can lead to $\langle\hat{z_{\rm d}}\rangle < \langle z_{\rm d} \rangle$ and thus $\langle\hat{R}\,\rangle < \langle R\,\rangle$. We performed a lensing signal simulation, scattering our lens redshifts by adding a Gaussian redshift error distribution of 0.03 (1+z), to estimate the maximal bias in the measurement of the velocity dispersion due to lens redshift errors. Even in the most extreme scenario (lowest redshift lenses, i.e. steep rise in redshift counts and asymmetric redshift scattering, as there are no lower redshift galaxies that can be scattered up) the bias in $\sigma$ is below 4 per cent. For higher redshift lenses the bias rapidly decreases. As the number of low redshift lenses is low compared to the total lens number we make no attempt to correct for this bias.
\\
For a given background-foreground distance distribution the bias is the larger, the larger the photometric redshift scatter is. For flatter profiles the bias is smaller (see equation~\ref{eq:deltagam_alpha}). 
\\
The ratio of the estimator and the true contribution for each pair in equation~(\ref{eq:etan4}) is in linear order:
\begin{equation}
\frac{\hat \Sigma_{\rm c} \ \hat \gamma_{\rm t } ( R)}{\Sigma_{\rm c} \gamma_{\rm t}(R) }   = 
1+ \frac{\delta \Sigma_{\rm c}}{\Sigma_{\rm c}} + 
\frac{\Delta\gamma_{\rm t}(\hat R, R)}{\gamma_{\rm t} (R)} + 
\frac{\delta\gamma_{\rm shape}}{\gamma_{\rm t}(R)} \  . 
\label{eq:err2}
\end{equation}
We now define
$\hat{{\Delta\Sigma}} = \Delta\Sigma +\delta \Delta\Sigma$,
and 
$\hat{\Sigma}_{\rm c} = \Sigma_{\rm c}  + \delta \Sigma_{\rm c } $, insert equation~(\ref{eq:err2}) into equation~(\ref{eq:etan4}) and obtain to linear order for the error of the estimator:
\begin{equation}
\begin{array} {ll}
& \delta \Delta\Sigma(R)  = \ave{  \gamma_{\rm t}(R) \Sigma_{\rm c} \left[ \frac{\delta \Sigma_{\rm c}}{\Sigma_{\rm c}} + \frac{\Delta\gamma_{\rm t} (\hat R, R)}{\gamma_{\rm t}(R) } + \frac{\delta\gamma_{\rm shape}}{ \gamma_{\rm t} (R)}  \right] }_{\rm fg-bg-pair} \\
& = 
\ave{  \gamma_{\rm t}(R) \  \delta \Sigma_{\rm c}
      +\Delta\gamma_{\rm t} (\hat R, R)  \Sigma_{\rm c}
      + \delta\gamma_{\rm shape}  \Sigma_{\rm c}  }_{\rm fg-bg-pair} \ .
\\
\label{eq:err3}
\end{array}
\end{equation}
Equation~(\ref{eq:err3}) can be used to obtain the error estimates in presence of scatter and biases for photometric redshifts and shape estimates. 
Equations~(\ref{eq:deltagam_alpha}), (\ref{eq:deltagam_SIS}) and (\ref{eq:deltagam_SIS_0}) explain together with equation~(\ref{eq:err3}) why lens redshift errors are more severe in GGL than source redshift errors, in particular if the lens is very close. In this case, for $\Delta z_{\rm d}\approx 0.03 $ and $z_{\rm d}\approx 0.1 $ this fractional error can easily approach 30 per cent. For larger redshifts the dependency of $D_{\rm d}$ on redshift weakens on, so does the associated profile error. The term $\delta\gamma_{\rm shape}$ includes the intrinsic shape noise, the shape measurement error and potential systematics. The value of $\delta\gamma_{\rm shape}/ \gamma_{\rm t}$ is in order of $0.3/0.002\approx 150$ per foreground background pair. So, the relative shape error exceeds the relative profile error by more than a factor of 500. 
\\
We now calculate the errors due to shape noise and photometric redshift errors more accurately.
\\
Regarding the shape measurement noise one can use the estimate of the tangential shear relative to a random foreground sample to estimate the statistical error of $\gamma_{\rm t}(R)$ and $\Delta\Sigma(R)$ estimates. This is shown in more detail in Appendix~\ref{sec:gamma_t random}. For a given lens sample as defined in Section~\ref{sec:lenssample} the corresponding background sample is specified by the selection criterion defined in equation~(\ref{eq:zlens-zsource-diff}).
We do the same estimate of $\gamma_{\rm }(R)$ and $\Delta\Sigma(R)$ as we do for real data but use the shape of the background object with a randomised phase or the shape of another object in the shape catalogue instead of the true tangential ellipticity.
\\
In addition we simulate photometric redshift errors by adding a Gaussian distributed error of $0.05\ (1+z)$. We calculate the lens distance $D_{\rm d}$ and the critical surface mass density $\Delta\Sigma_{\rm c}$ and estimate the propagated errors for both quantities due to the photometric redshift uncertainties. The simulation is designed to rather overestimate the estimated errors than to underestimate them. This is especially true for very low lens redshifts, where the scattering of photometric redshifts can lead to negative lens redshifts, which are not realised in observational data.
\\
The propagated errors comprise a systematic and a statistical part and read as
\begin{equation}
\delta \Sigma_{\rm c} = \sqrt{{\delta \Sigma_{\rm c}}_{\rm syst}^2 + \frac{{\delta \Sigma_{\rm c}}_{\rm stat}^2}{n}} \ .
\end{equation}
In order to disentangle their contributions we consider two samples of different sizes (\mbox{100\ 000} and \mbox{1\ 000} objects) and solve the resulting system of equations (see also \citealt{gruen10}). In both cases the systematic error is significantly smaller than the corresponding statistical error. For a redshift of $z=0.05$ the fractional systematic error in $\delta D_{\rm{d}}/D_{\rm{d}}$ is lower than 10 per cent, dropping below 5 per cent for higher redshifts (see Fig.~\ref{fig:dd.error}). The fractional error of the critical surface density $\delta \Sigma_{\rm c}/\Sigma_{\rm c}$ is larger than for $D_{\rm{d}}$ due to multiple dependencies on the redshift ($D_{\rm{d}}$, $D_{\rm{s}}$ and therefore $D_{\rm{ds}}$). However by applying the selection criterion from equation~(\ref{eq:zlens-zsource-diff}) we ensure that the systematic errors in unfortunate cases are at most slightly larger than 10 per cent (see Fig.~\ref{fig:sigmac.error}).
\\
We now estimating the systematic errors for extreme cases. For a lens at $z \sim 0.7$ we expect that about 25 per cent of the background galaxies are close enough to the lens to have an average bias of about $\sim 20$ per cent and that the $\Sigma_{\rm c}$ estimate is nearly unbiased for the remaining 75 per cent. This gives an average systematic error of $0.75 \times 1 + 0.25 \times 1.2 = 1.05$, i.e. an overestimation of $\delta \Sigma_{\rm c}/\Sigma_{\rm c} \sim 5$ per cent. For the other extreme at low redshifts the fraction of lens-source pairs with a bias of $\sim 20$ per cent is only 5 per cent. This leads to a fractional systematic error of $0.95 \times 1 + 0.05 \times 1.2 = 1.01$, i.e. $\delta \Sigma_{\rm c}/\Sigma_{\rm c} = 1$ per cent. For the whole sample we expect a systematic bias of not more than $0.88 + 0.12 \times 1.2 = 1.02$, i.e. a maximal bias of $\delta \Sigma_{\rm c}/\Sigma_{\rm c} = 2$ per cent due to photometric redshift inaccuracies. We therefore conclude that the expected systematic errors due to photometric redshift uncertainties are small enough to be neglected.
\subsection{Systematic errors from IA}
\label{sec:IA}
Galaxies which are in the same structure and thus physically connected are not randomly distributed in orientation but rather intrinsically aligned (\citealt{hirata04}), for instance satellite galaxies tend to be radially aligned relative to their central galaxies. This is why intrinsic alignment (IA) is a major issue in the interpretation of cosmic shear data (see e.g. \citealt{mandelbaum06c} or \citealt{bridle07}). The observed two-point correlation function for the ellipticity of galaxy pairs is 
\mbox{
$\langle \epsilon_i \epsilon_j^{*} \rangle = \langle \gamma_i \gamma_j^{*} \rangle + \langle \epsilon_{i}^{S} \epsilon_{j}^{S*} \rangle + \langle \gamma_i \epsilon_{j}^{S*} \rangle + \langle \epsilon_{i}^{S} \gamma_j^{*} \rangle$,
}
where $\epsilon^S$ and $\epsilon$ are the unlensed and lensed ellipticities and $\gamma_i$ is the cosmic shear at redshift $z_i$ (see e.g. \citealt{joachimi08}). The first term on the right hand side is the desired cosmic shear signal, the second term (called II) describes the intrinsic alignment of two galaxies. Unless the two galaxies are physically associated (i.e. they are required to be at same redshift) this term is zero. The third term describes (for $z_i \le z_j$) the correlation of a foreground gravitational shear with the intrinsic ellipticity of a background galaxy and is zero. The fourth term (called GI) describes the correlation between the intrinsic ellipticity of a foreground galaxy and the gravitational shear acting on a background galaxy.
\\
In GGL, however, one measures the tangential alignment, i.e. the cross-correlation of a background galaxy shape and the foreground lens position. Therefore intrinsic alignment theoretically should not be an issue at all. This situation is different in case of a foreground-background mismatch due to photometric redshift errors, where the photometric redshift of the assumed background object is overestimated and the galaxy actually is embedded in the foreground structure. If the falsely assumed background galaxy is randomly oriented relative to the foreground galaxy considered, then the shear signal is just diluted and our error considerations from Section~\ref{sec:lensmassbias} apply. If however the background galaxy has a preferred direction to the foreground an additional source of systematic error arises. If these false `background' galaxies are fainter than the foreground galaxies, they will likely be their satellites (if associated to the foreground structure) and thus will on average be radially aligned (see \citealt{agustsson06}). This then leads to a false detection of the GI-signal.
\\
The separation of GGL and intrinsic alignment is investigated in detail by \citet{blazek12}. To isolate IA from the lensing signal, they exploited the fact, that the contamination of the background galaxy sample with foreground galaxies should decrease if a more distant background slice is considered. They measured the excess surface mass density $\Delta\Sigma$ associated with SDSS-LRGs using two source subsamples in two redshift slices behind the lens.
\\
They conclude that the size of IA for their lens sample is small (and consistent with zero, see their fig.~3). Their fig.~2 shows that for all scales larger than $100\ h^{-1}$ kpc the signal extracted for two redshift subsets agrees which implies that the imprint of IA on $\Delta \Sigma$ can be neglected.
\\
In our case we can infer the potential error due to IA from the upper left panel of Fig.~\ref{fig:WIDE.ds.syst} in Appendix~\ref{sec:ds.syst}. The magenta and green points show the $\Delta \Sigma$ values obtained for foreground lenses with $0.05 < z \le 0.5$ using the shear signal from galaxies in the redshift slices of $0.6 \le z \le 0.73$ and $1.01 \le z \le 2$. Since the contamination of the 
$z = 1-2$ sample should be zero, the difference between the green and magenta points quantifies the maximal error due to IA in the low $z$ background sample with $0.6 \le z \le 0.73$. All values agree within $2\sigma$. We therefore conclude that systematic errors due to IA are small enough to be neglected.
\subsection{Defining lens subsamples}
\label{sec:subsamples}
\begin{figure}
\includegraphics[width=8.8cm]{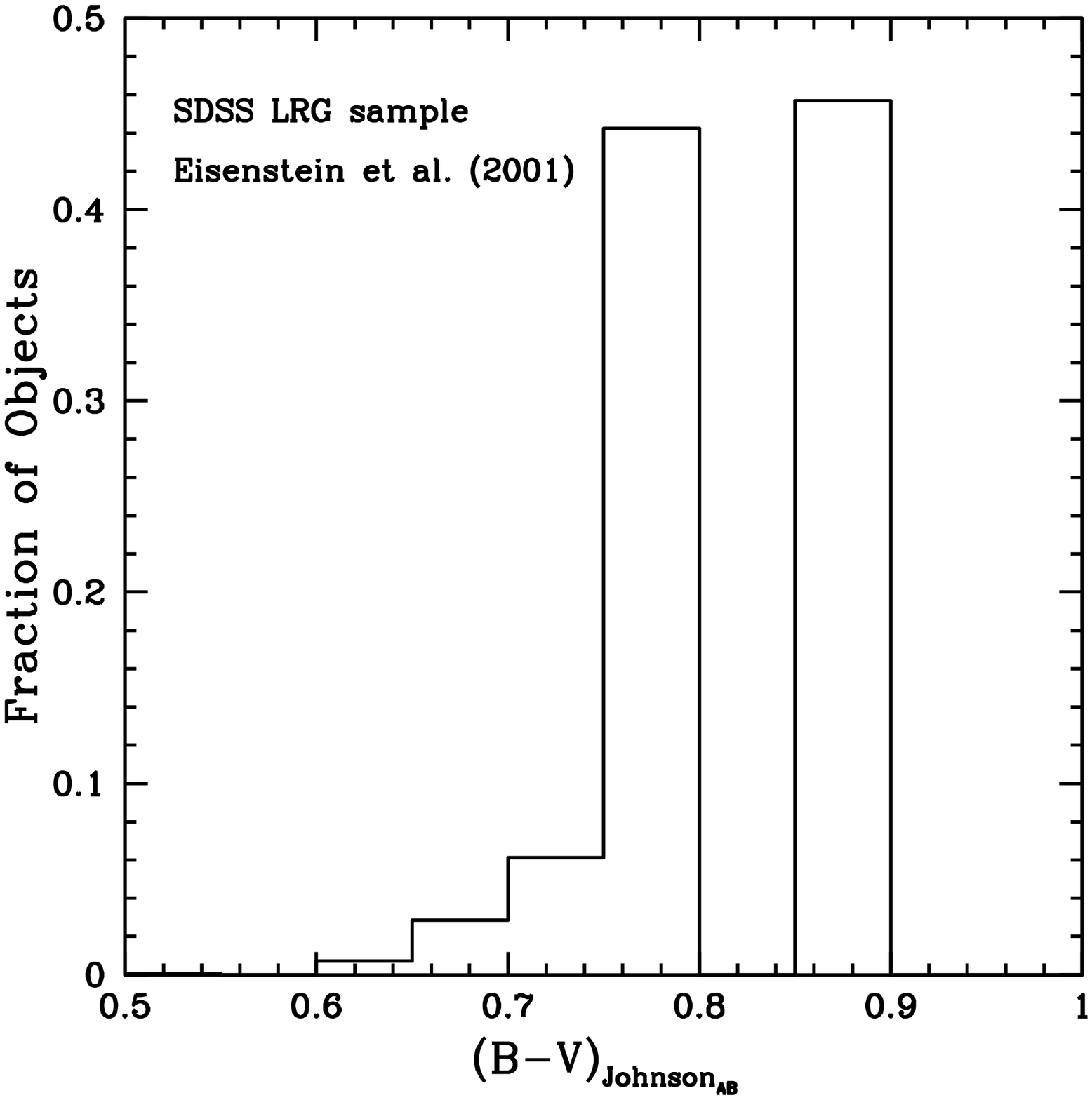}
\caption{$(B-V)$-rest frame colour histogram for the SDSS LRG sample (\citealt{eisenstein01}) in the AB-system. The $(B-V)$-colours are taken from our SED-templates that best match the SDSS photometry at the spectroscopic redshift of the LRG. For almost all objects the assumption $(B-V) > 0.7$ holds, justifying the chosen galaxy classification for red and blue galaxies (see  \citealt{dahlen05}).}
\label{fig:B-V.histo}
\end{figure}

\begin{figure*}
\includegraphics[width=18.5cm]{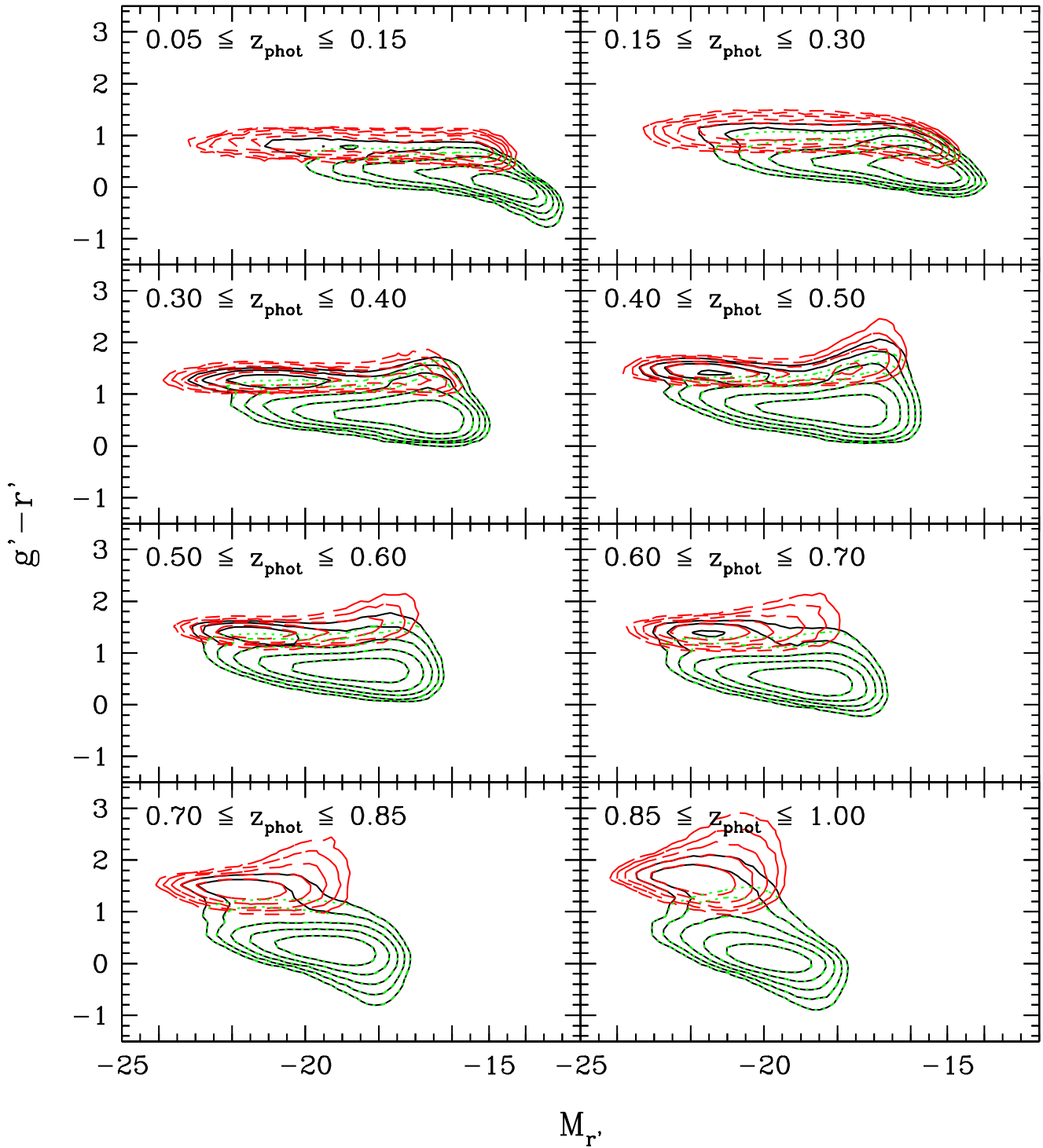}
\caption{Density contours for the colour-absolute magnitude-relation of our red and blue galaxies. The black solid lines show the distribution of all galaxies, the red dashed lines the distribution of red galaxies and the green dotted lines the distribution of blue galaxies. As we can see red and blue galaxies populate distinct regions in the colour-magnitude space, especially for higher redshifts. For low redshifts, there is a small overlap of red and blue galaxies in the colour-magnitude-plane.}
\label{fig:R-gr}
\end{figure*}

\begin{figure*}
\includegraphics[width=18.5cm]{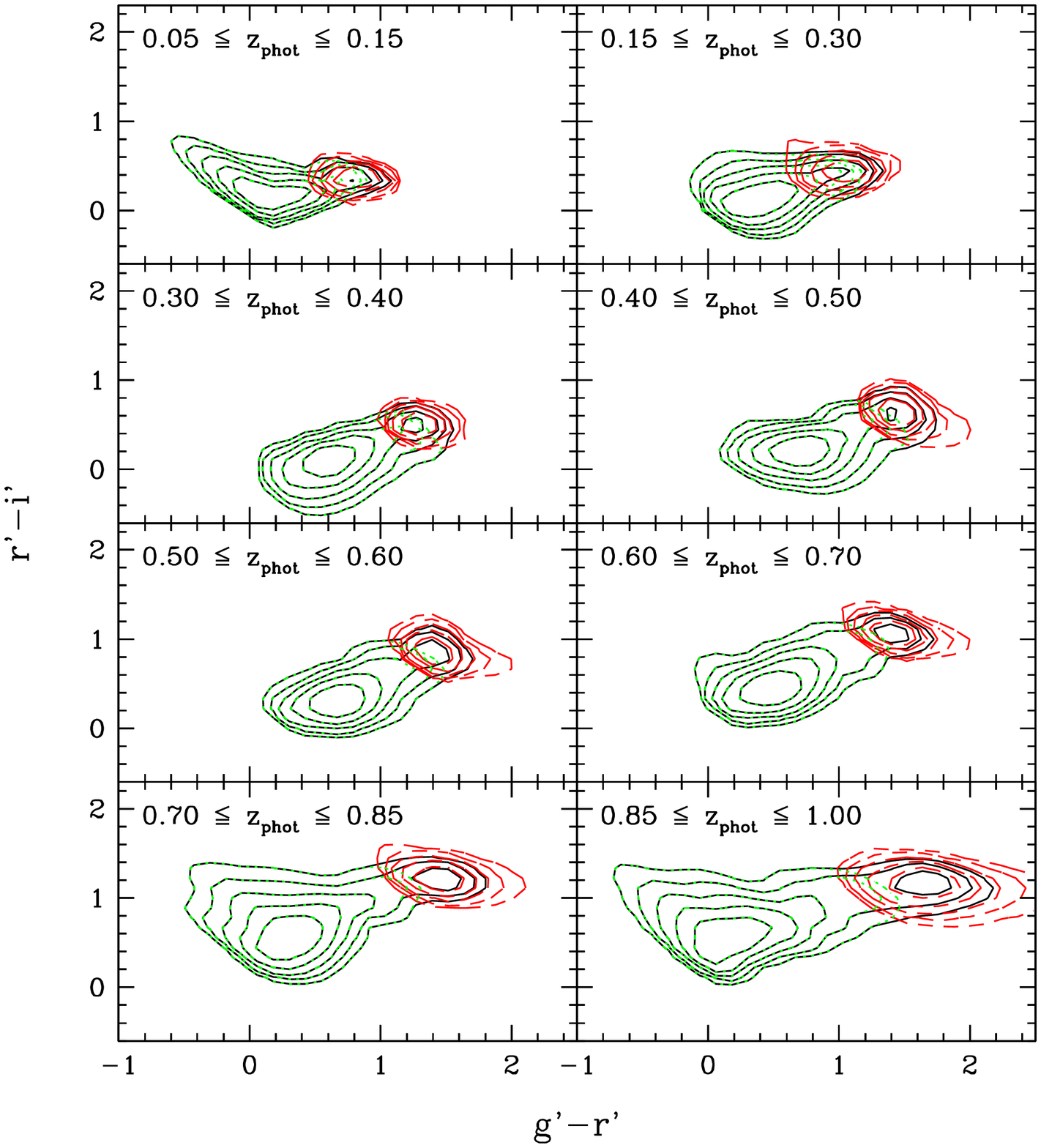}
\caption{Density contours for the apparent colour-colour-relation. The black solid lines show the distribution of all galaxies, the red dashed lines the distribution of red galaxies and the green dotted lines the distribution of blue galaxies. In colour-space our defined red and blue galaxy samples populate distinct regions when considering redshift $z \ge 0.4$, in a agreement with \citet{tojeiro12}. For redshifts lower than $z \approx 0.4$ the $(g'-r')$ vs. $(r'-i')$-colours of red and blue galaxies overlap.}
\label{fig:gr-ri}
\end{figure*}

As we do not only want to analyse the properties of the mean lenses as a function of luminosities, but also want to investigate the differences based on SED and environment we define several lens subsamples. We follow the approach of \citet{dahlen05} who used the rest frame $(B-V)$-colours to classify red and blue galaxy types. We therefore estimate the $(B-V)$-colours for our objects from the best-fitting templates in the photometric redshift estimation in the AB-system and define our red galaxy sample as all galaxies with $(B-V) > 0.7$ and consequently our blue galaxy sample as all galaxies with $(B-V) \le 0.7$. We test this selection by measuring the (rest frame) $(B-V)$-colours of the best-fitting SED (at the spectroscopic redshift) for the SDSS Luminous Red Galaxy (LRG) sample (see \citealt{eisenstein01}). The $(B-V)$-colour histogram of the best-fitting SDSS-LRG templates (Fig.~\ref{fig:B-V.histo}) shows that this assumption excellently holds for the LRG sample.
\\
We further verify our galaxy classifications by plotting the apparent galaxy colours and rest frame luminosities. We use the absolute $(g'-r')$-colour  vs. the rest frame $r'$-band luminosity to discriminate red and blue galaxies (cf. \citealt{loveday12}). As we can see in Fig.~\ref{fig:R-gr} red and blue galaxy populate distinct regions in the magnitude-colour space, especially considering higher redshifts, indicating a low contamination rate of our red and blue lens samples. In contrast to \citet{loveday12} we do not have individual rest frame colours for our lens galaxies. We thus consider the apparent colours  vs. the absolute magnitudes. However, this is at least for very low redshifts a valid approximation. Further we plot the $(g'-r')$-  vs. $(r'-i')$-colours (cf. \citealt{tojeiro12} for their red galaxy sample at moderate redshifts between 0.5 and 0.7). We see in Fig.~\ref{fig:gr-ri} that for redshifts $z \ge 0.4$ our red and blue galaxy samples are clearly separated and also for lower redshifts our red galaxy sample shows few overlap in this colour-colour-plane.
\\
\\
In addition we measure a relative local galaxy density for all galaxies in our lensing sample. For this we consider all galaxies within an interval of $\Delta z = \pm0.2$ in front and behind the investigated lens galaxies. We measure the galaxy density (galaxies per area) within an angular distance of 30 arcsec around a lens and on the entire field in the previously defined redshift interval and divide the quantities to derive the relative local galaxy density. We then define following environmental lens samples:
\ben
\item a very dense environment lens sample (10 per cent of galaxies populating the densest environments),
\item a dense environment lens sample (50 per cent of galaxies populating denser environments),
\item a low density environment lens sample (50 per cent of galaxies populating lower density environments)
\item and a very low density environment lens sample (10 per cent of galaxies populating the lowest density environments).
\een
%
\section{Weak lensing analysis}
\label{sec:lensing}
\begin{figure}
\includegraphics[width=8.8cm]{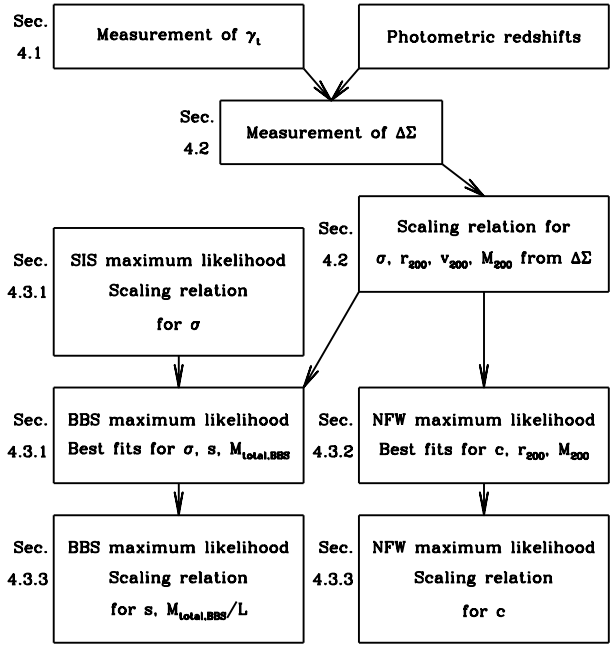}
\caption{Flow chart for our lensing analysis.}
\label{fig:lensing}
\end{figure}

The structure of this section is as follows. First we measure the tangential shear signal $\gamma_{\rm t}(R)$ as a function of luminosity, galaxy SED and environment density to obtain a qualitative overview about the lensing signal strength. Using photometric redshifts we calculate the excess surface mass density $\Delta\Sigma(R)$. We finally use the results for different luminosity bins to derive the scaling relations for various halo profile parameters (e.g. velocity dispersion $\sigma$ or virial radius $r_{200}$) as a function of the galaxies' absolute luminosity. In a first step we only consider $\Delta\Sigma(R)$ on small scales and treat each galaxy halo as being isolated and derive the luminosity scaling relations. In the second step we account for associated haloes by explicitly modelling nearby haloes traced by galaxies and analyse the $\Delta\Sigma(R)$-signal on larger scales using a maximum likelihood analysis that parameterises the scaling relations as well. We confirm the scaling relations obtained in the first step.  Then we fix the luminosity scaling relations and obtain the halo parameters for the `reference galaxy' as a function of SED-type and galaxy environment density. Finally we use the obtained values for the halo parameters and thus the masses to estimate the remaining scaling relations for the halo parameters, i.e. the scaling of the BBS truncation radius $s$ and the NFW concentration parameter $c$ as a function of luminosity. The structure in our lensing analysis is also illustrated in Fig.~\ref{fig:lensing}.
\subsection{Tangential shear}
\label{sec:alignment}
\begin{figure*}
\centering
\includegraphics[width=8.8cm]{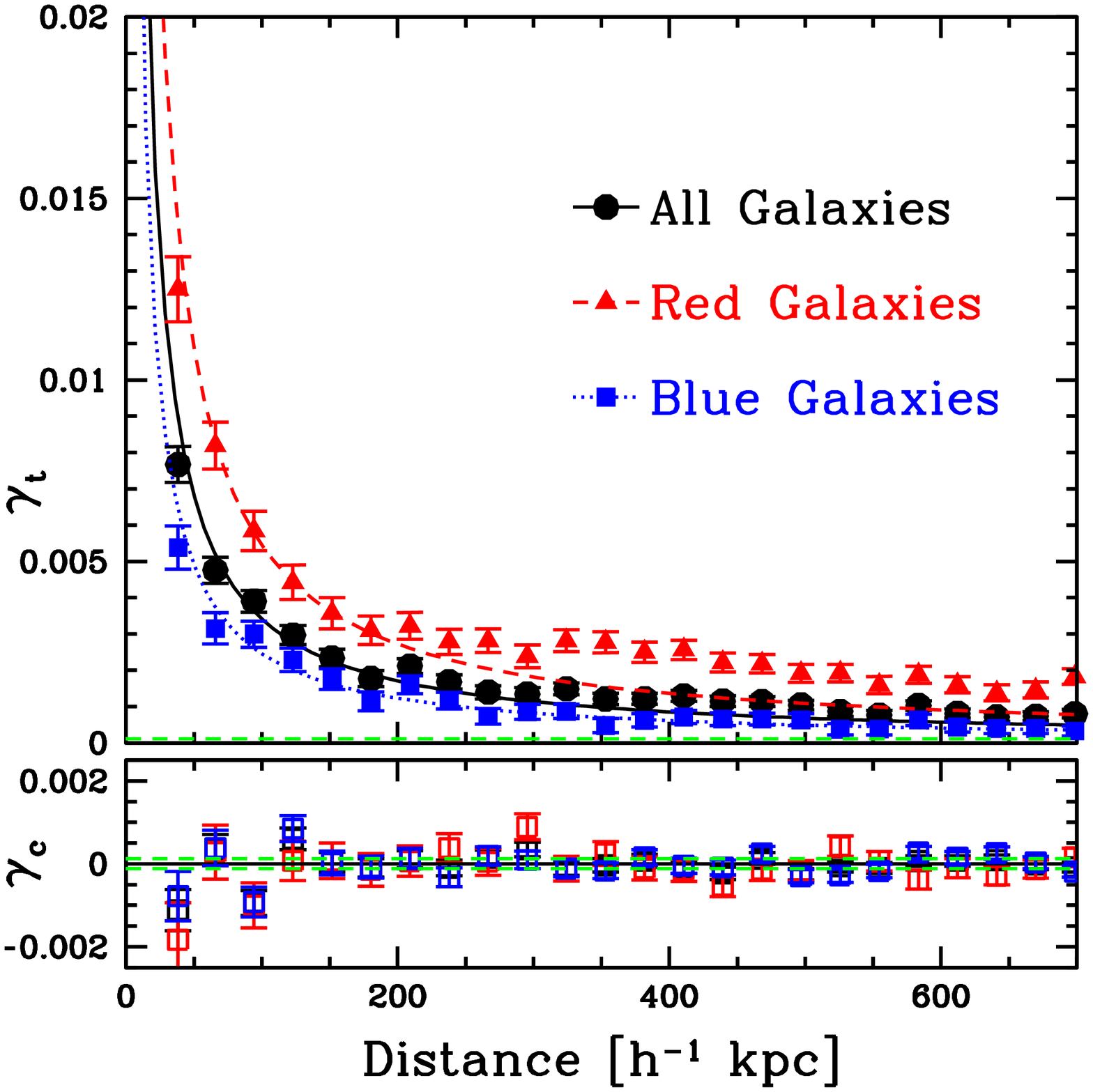}
\includegraphics[width=8.8cm]{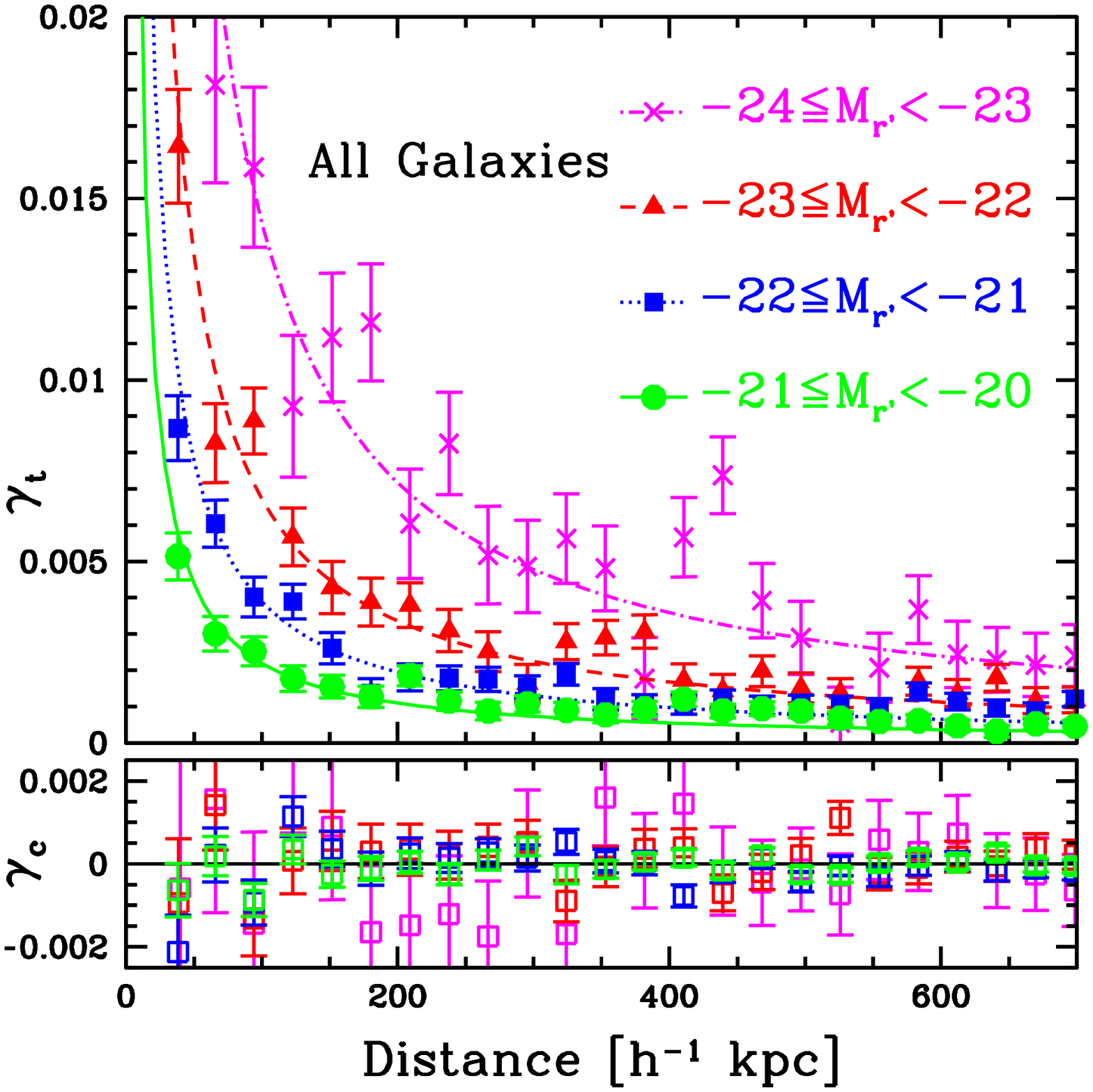}
\includegraphics[width=8.8cm]{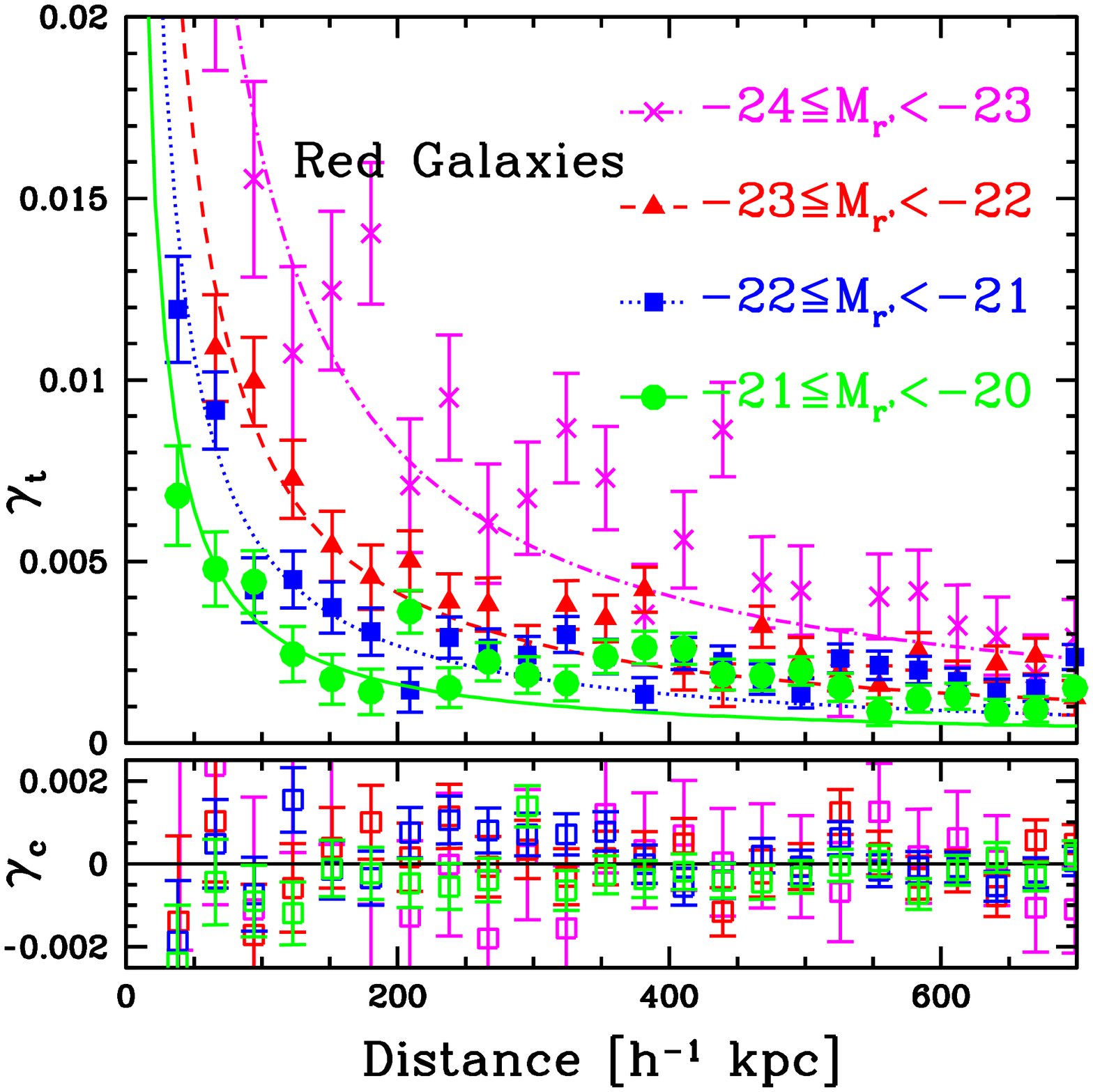}
\includegraphics[width=8.8cm]{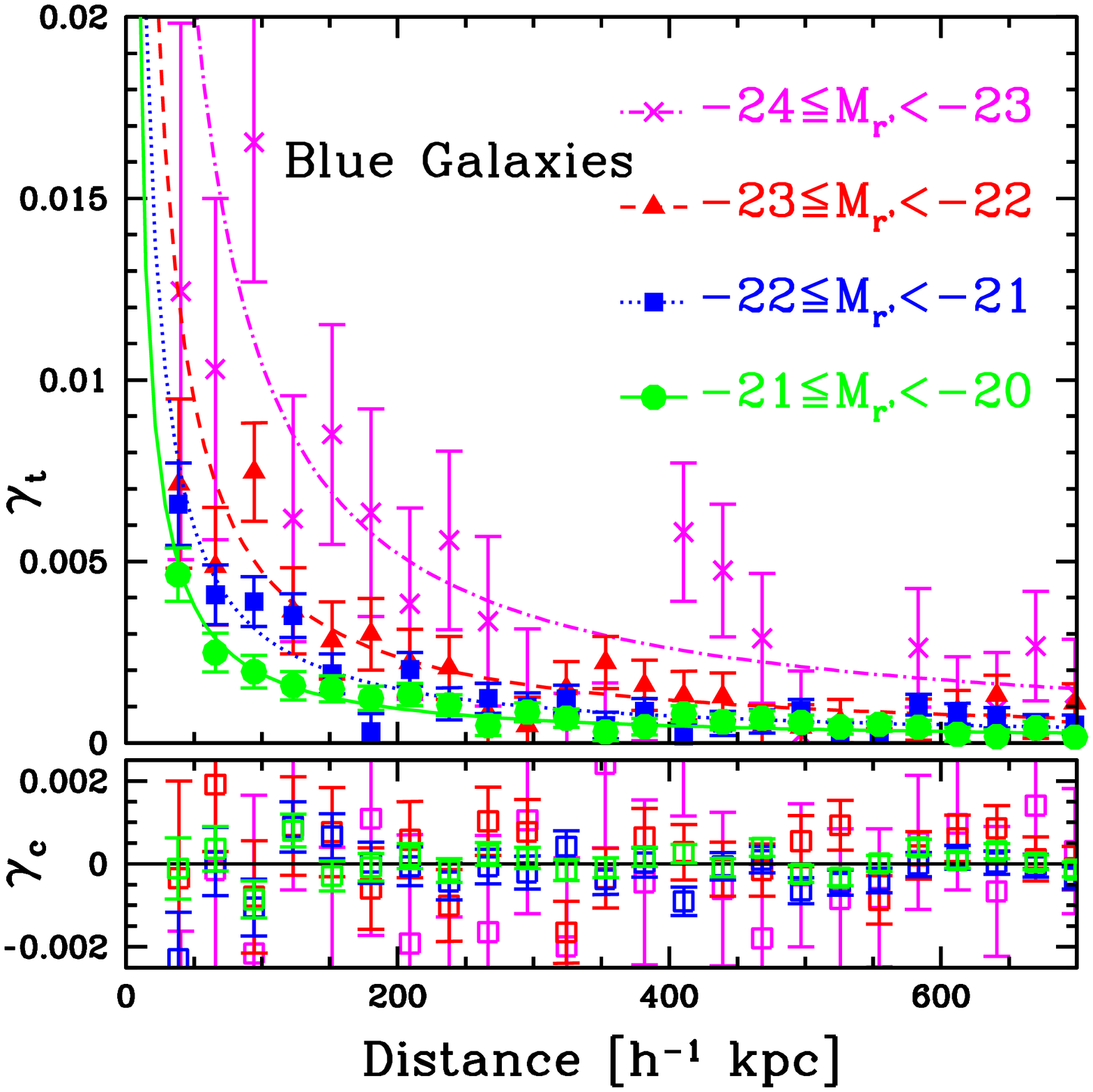}
\caption{Tangential shear signal (filled circles) for different lens samples. The cross shear (systematic error) is shown with empty squares. It is consistent with zero at all radii and in all samples. The green dashed lines in the upper left panel show the $1-\sigma$-level of the remaining systematic error. The weighting scheme is given in the text. Obviously, the fainter the subsample, the less distant it is on average.  We use the weighted mean redshift of the galaxies to fit the amplitude of an isothermal sphere on scales \mbox{$R \le 200 \ h^{-1}$ kpc}. The corresponding effective velocity dispersion of the sample is then derived as well. The shear of the `effective' isothermal sphere is added. The measured gravitational shear signal is isothermal for $R \le 200 \ h^{-1}$ kpc.
The upper left panel shows the tangential shear signal in the luminosity bin $-24\le M_{r'}<-20$ for red galaxy lenses as filled red triangles (dashed fit-line), blue galaxy lenses as filled blue squares (dotted fit-line) and for all galaxy lenses as filled black circles (solid fit-line). The deviations from a single halo signal  due to the neighbouring galaxy haloes on scales $R > 200\ h^{-1}$ kpc are most prominent for the red galaxy lens sample as this galaxy type is mainly populating higher density regions such as galaxy clusters. The gravitational shear caused by the red galaxy lens sample is significantly higher than for blue galaxies. 
The upper right panel shows the combined lens sample signal separated into four luminosity bins of $-24\le M_{r'}<-23$ in magenta (crosses, dashed-dotted fit-line), $-23\le M_{r'}<-22$ in red (filled triangles, dashed fit-line), $-22\le M_{r'}<-21$ in blue (filled squares, dotted fit-line) and $-21\le M_{r'}<-20$ in green (filled circles, solid fit-line). As expected, the signal strength and so the effective velocity dispersion of the sample decreases with decreasing brightness of the lenses. The fractional decrease of the signal for the fainter samples is of the expected amplitude if Tully-Fisher or Faber-Jackson luminosity-velocity scaling relations are assumed (more details in the text). 
The lower panels show the red galaxy lens sample (left) and the blue galaxy sample (right) split into the same luminosity bins with the same colour scheme as for the combined sample before. Obviously the observed lensing signal and therefore galaxy mass is higher in every single luminosity bin for red galaxies than for blue galaxies.}
\label{fig:WIDE.etc.lum}
\end{figure*}
Measuring the gravitational shear as a function of angular scale with respect to the foreground galaxies is the most direct measurement of GGL. 
The use of photometric redshifts is in principle not required. Instead one can define magnitude intervals for foreground and background galaxies and then interpret the shear measurements using an assumption on the foreground and background redshift distribution (see \citealt{hoekstra03,hoekstra04} or \citealt{parker07}).
 However, if galaxies are spread over a broad redshift range, then physical scales are mapped into various angular scales and the observed signal (as a function of angle) is a mix of various physical scales. 
We therefore make use of the photometric redshift estimates for the lenses and measure the mean gravitational shear as a function of the projected distance to the foreground lenses at their redshifts. The  quantitative interpretation is not the major goal of this analysis, but to see qualitatively to which distance the signal is dominated by the halo the galaxy is centred in, and when the contribution of nearby galaxy haloes becomes visible. 
As lenses we either consider all galaxies (in our photometric redshift sample)  or we consider subsamples  in absolute $r'$-band magnitudes. For each radius $R$ the mean gravitational shear is obtained by a weighted mean of shear estimates for all foreground-background pairs at this radius. The weighting factors for the shear estimates are as introduced by \citet{hoekstra00}
\begin{equation}
w=\frac{1}{\sigma_{g}^2}=\frac{\left(P^g\right)^2}{\left(P^g\right)^2 \sigma_{\epsilon}^2+\langle \Delta \epsilon^2 \rangle} ,
\label{eq:weight-hoekstra}
\end{equation}
where $\sigma_{\epsilon}$ stands for the scatter of the intrinsic ellipticities and $\langle \Delta \epsilon ^2 \rangle^{1/2}$ for the uncertainty in the ellipticity measurement. Applying these weight factors we are able to reduce the contribution of background galaxies with large uncertainties in the ellipticity measurement and to enhance the signal-to-noise ratio.
\\
\\
Weak lensing measurements also provide a simple and comfortable possibility to identify systematic errors, the cross-shear, also called B-modes. Being a conservative force and therefore not producing curls, the tangential shear should vanish if all background objects are rotated by 45 degrees. Vanishing B-modes do not absolutely guarantee the absence of systematics, but their presence is an indicator for remaining systematic effects.
\\
\\
In Fig.~\ref{fig:WIDE.etc.lum} the gravitational shear estimate is shown as a function of the projected distance $R$ to the lens when we average over all galaxies and over all red and blue galaxies, respectively. Singular isothermal shear fits are added as dashed curves, where the fit  has been obtained by using separations smaller than $R=200 \ h^{-1}$ kpc only. For the distance ratio $D_{\rm ds}/D_{\rm s}$ required for translating the shear into a velocity dispersion we use the foreground-background pair weighted average of the individual $D_{\rm ds}/D_{\rm s}$ ratios for each foreground-background pair. The fit shows that the signal approximately follows an isothermal profile out to  \mbox{$R=200 \ h^{-1}$ kpc}. For scales larger than $R=200 \ h^{-1}$ kpc the deviations are stronger for the red galaxies than for the overall sample, which can be explained by red galaxies being more strongly correlated with each other, and by the fact that the fraction of red galaxies residing in denser dark matter environments (groups and clusters of galaxies) is higher than for non-red galaxies. The observed B-modes are consistent with zero.
In the other three panels of  Fig.~\ref{fig:WIDE.etc.lum} the signals are evaluated for four absolute magnitude intervals of $M_{r'}$ between -24 and -20 with interval widths of one magnitude. As expected we see a clear sequence from the brightest lens sample to the faintest. Assuming a Faber-Jackson \citep{faberjackson} or Tully-Fisher \citep{tullyfisher} relation, respectively, we observe a picture consistent with the decrease of velocity dispersion for fainter lens samples. The expectation that red galaxies are more massive than blue galaxies for given $r'$-luminosity is confirmed by the investigation of the individual luminosity bins. The gravitational shear for red galaxies is significantly higher than for their blue counterparts. The gravitational shear of the combined galaxy sample is between that of the red and blue sample for every luminosity bin as expected. The B-modes are consistent with zero for all luminosity bins.
\begin{figure}
\includegraphics[width=8.8cm]{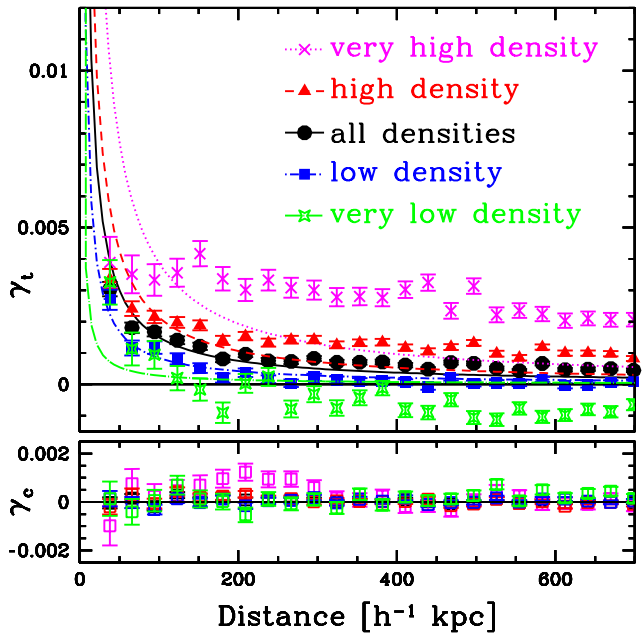}
\caption{Tangential shear signal for lens samples with luminosities $-24 \le M_{r'} < -17$ in different environments. The SIS-fits are obtained within a separation of $R=200\ h^{-1}$ kpc. The definitions of the environments are given in Section~\ref{sec:subsamples}. The black circles (solid fit-line) show the complete lensing sample (see also Fig.~\ref{fig:WIDE.etc.lum} upper left panel), blue (filled squares, dashed-dotted fit-line) and green (diamonds, dashed-dotted fit-line with longer dashes) samples show lenses in low density environments. It becomes apparent that in dense environments (red filled triangles, dashed fit-line) and especially for very dense environment (magenta crosses, dotted fit-line) the observed gravitational shear signal cannot be explained with only one lens.}
\label{fig:WIDE.total.etc.env}
\end{figure}
\\
We now also analyse the tangential shear as a function of the environment density defined in Section~\ref{sec:subsamples}. Fig.~\ref{fig:WIDE.total.etc.env} shows the gravitational shear around galaxies with luminosities of  $-24  < M_{r'} < -17$ for the four different environment classes defined by us. The denser the environment, the higher the signal. In principle the higher signal could be due to a higher mean luminosity in the considered luminosity interval. But this is not only reason as can be seen if the behavior as a function of distance to the galaxies is analysed: The gravitational shear becomes approximately equal at the smallest scale of $R = 30 - 40\ h^{-1}$ kpc. This implies that, on average, the galaxies in the four environment samples have roughly the same central density profiles but differ in their outer halo profiles and experience a different impact of nearby haloes. As expected, the gravitational shear is lowest for galaxies in the lowest environment density: it rapidly drops down on very sort scales and becomes negative for larger scales, which means that the average convergence at radius $R$  is higher than the averaged convergence within radius $R$, or that the larger scale environment is denser than the density around the galaxy within $R \approx\ 100-200\ h^{-1}$ kpc itself. For the low density environment (blue points in Fig.~\ref{fig:WIDE.total.etc.env}) the shear signal follows an isothermal profile (blue dashed curve) out to $R=200\ h^{-1}$ kpc, and approaches zero on larger scales. There is hardly any impact of nearby haloes visible in the signal (as expected if the environment is poor). For the high density sample the signal decreases similar to an isothermal sphere out to $R=200\ h^{-1}$ kpc and then stays constant. For the highest environment density sample this happens already at scales of approximately \mbox{$R=100\ h^{-1}$ kpc} and the shear stays high at a level of $0.003$.
At scales of $400\ h^{-1}$ kpc this signal of $0.003$ is the same as that of a group with a dark matter halo with  velocity dispersion of \mbox{400 $\rm{km\ s^{-1}}$} if the typical distances of sources and lenses are used. The shape of the signal is, due its `flatness', not in agreement with being caused by a single group halo, where the majority of the considered galaxies are central. This flat behavior is also confirmed in our 3D-LOS-projected lensing signal simulations (see Fig.~\ref{fig:WIDE.total.veryhigh.ds.sim} in Appendix~\ref{sec:simulations}), where we see that this flatness originates in the multiple gravitational deflections on brighter nearby galaxies in the close environment.
%
%
%
\subsection{Measurements of the excess surface mass density $\Delta\Sigma (R)$}
\label{sec:DeltaSigma}
\begin{figure*}
\includegraphics[width=18.5cm]{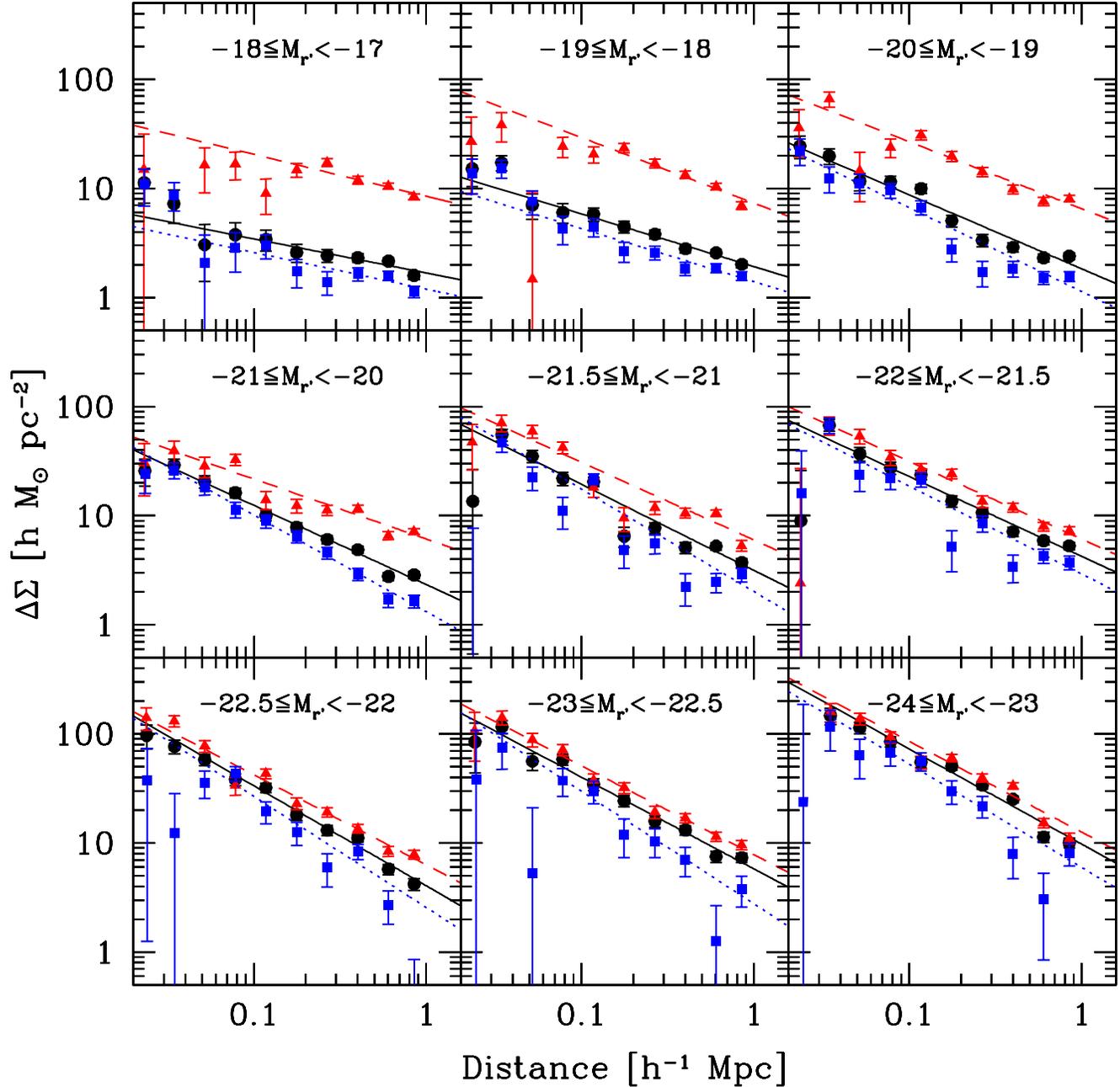}
\caption{Excess surface mass density $\Delta\Sigma$ for individual luminosity bins. The fainter luminosity bins are dominated by blue lens galaxies (blue squares, dotted fit-line), outnumbering red galaxies (red triangles, dashed fit-line) in those redshift bins. The red galaxy sample shows a significantly higher signal, both due to the higher mass of red galaxies for given luminosity and the denser environment red galaxies populate in general. For brighter luminosity bins the fraction of red galaxies increases. }
\label{fig:WIDE.total.ds.bins}
\end{figure*}

The measured gravitational shears and the photometric redshifts of the lenses and sources are now combined to  obtain the excess surface mass density $\Delta\Sigma (R)$ (see equation~\ref{eq:etan2}) according to the estimator in equation~(\ref{eq:etan4}), further using the weighting factor from equation~(\ref{eq:weight-hoekstra}).
We fit the excess surface density with a power law,
\begin{equation}
\Delta\Sigma = A[R/1 \ \rm{Mpc}]^{-\alpha} \quad,
\label{eq:deltasigma-fit}
\end{equation}
where an exponent of $\alpha=1$ corresponds to an isothermal profile. The fits include scales out to a projected distance of $1 \  h^{-1}$ Mpc. 
A power law fit is a too simple model to properly describe the excess surface mass density, however, the profile steepness of a power law is a measure of the relative importance of the halo hosting the galaxy itself (slope close to -1), and the haloes in which the galaxy host halo resides (flatter slope). We split our galaxy sample samples into several absolute luminosity intervals in $M_{r'}$ and analyse them separately (see Fig.~\ref{fig:WIDE.total.ds.bins}). We find that the slope increases from $\alpha \sim 0.3$ for luminosities $-17 > M_{r'} > -18$ to values of $\alpha \sim 0.9$ for galaxies with $-23 > M_{r'} > -24$. The amplitude of the excess surface mass density increases with luminosity as seen for the tangential shear signal before (Fig.~\ref{fig:WIDE.etc.lum}). For the fainter luminosity bins the signal of the total galaxy sample is dominated by blue lens galaxies. The excess surface mass density for faint red galaxies is for all scales significantly higher than for blue galaxies, suggesting that the red galaxies in this luminosity range are significantly more massive than their blue counterparts and that faint red galaxies are more likely in denser environments, where in addition larger scale haloes from galaxies, groups and clusters contribute. With increasing luminosity the signal amplitude increases and the difference between red and blue lens galaxies decreases. For the brightest luminosity interval the excess surface mass density of the combined sample is dominated by the bright red galaxies as in this luminosity range red galaxies outnumber their blue counterparts.
\\
\\
We compare our results regarding $\Delta\Sigma$ to \citet{mandelbaum06b}, who analysed the excess surface mass density for luminosities $-17 \ge  M_{\textit{r'},\rm{SDSS,\rm{AB}}} > -22.5$ for red and blue galaxy samples using SDSS data (spectroscopic redshifts for the lens sample, photometric redshifts for bright source galaxies with $r'_{\rm SDSS,\rm{AB}} < 21$ and statistical redshift distributions for fainter sources). 
In contrast to \citet{mandelbaum06b} our rest frame magnitudes are not given in the AB but in the vega system and are not calculated for a Hubble parameter of $H_0 = 100\ \rm{km\ s^{-1}\ Mpc^{-1}}$ but for $H_0 = 72\ \rm{km\ s^{-1}\ Mpc^{-1}}$. To adjust for these differences we need to apply a magnitude offset of roughly $\Delta \rm mag=-0.55$ to our rest frame magnitudes.
Compared to \citet{mandelbaum06b} we use a much smaller area but significantly deeper data. We choose the corresponding luminosity bins and investigate the individual results for $\Delta\Sigma$.  The signals for the blue galaxy lens sample are in good agreement, yet fairly noisy.
\\
This is not the case for red galaxies. While for the faintest luminosity bins we obtain higher amplitudes up to luminosities of $M_{r'} \sim -21$ we measure lower amplitudes for brighter galaxies.
\\
However, in their later work \citealt{mandelbaum08b} (see also \citealt{dutton10}) do find a higher signal for faint galaxies, which agrees with our results. A very good illustration of the \citet{mandelbaum06b}, \citet{mandelbaum08b} and \citet{schulz10} results can be found in \citet{dutton10}, fig.~1. It shows that the mass-to-light (or mass-to-stellar mass) ratio of red galaxies increases relatively moderately with stellar mass (or light) in contrast to the earlier results of \citet{mandelbaum06b} (which had still relatively large uncertainties). The compilation of \citet{dutton10}, Fig.1,  shows that the mass-to-light or mass-to-stellar mass ratio increases rather slowly for stellar masses with $\log M_{\rm star}$=10.5 to 11.3 according to the \citet{schulz10} results. This fully supports our findings.
\\
We further compare our measurements to \citet{vanuitert11} who measured the GGL signal using the overlapping region between the SDSS-R7 and the RCS2 survey. While the lens sample is chosen spectroscopically, the shear estimates are derived on the deeper RCS2 $r'$-bands from CFHT. Comparing the corresponding luminosity bins we observe a good agreement for the excess surface mass density in our measurements and the results of \citet{vanuitert11} up to a luminosity of $M_{r',\rm{AB}} \sim$ -23 to -24 (cf. their fig.~8). 
\\
\\
The dependence of the measured excess surface mass density on the environment density is shown in Appendix~\ref{sec:sim.ds} (see Figs.~\ref{fig:WIDE.red.ds.sim}-\ref{fig:WIDE.total.verylow.ds.sim}). At this point we then also show how well the observations (for either red or blue galaxies residing in different environments) are described by the best fitting BBS- and NFW halo parameters derived for red and blue galaxies later on in this Section.
\subsubsection{L-$\sigma$-Scaling Based on Fits to $\Delta\Sigma$}
\label{sec:ds scaling sigma}
\begin{figure}
\includegraphics[width=8.8cm]{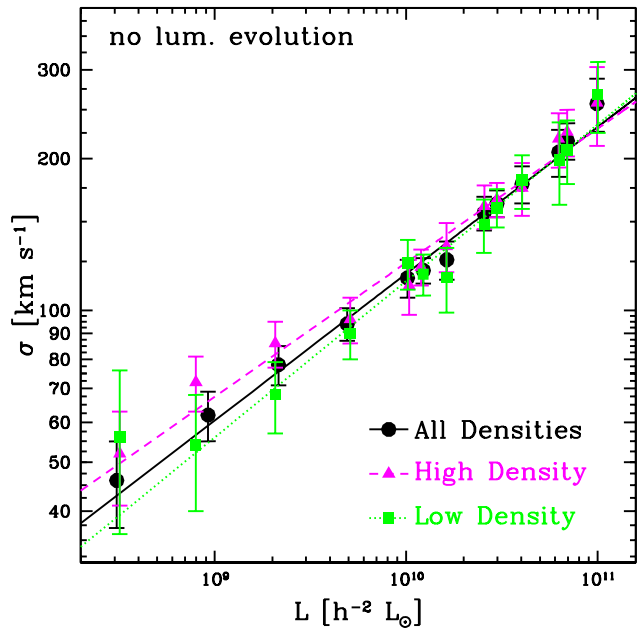}
\caption{Velocity dispersion $\sigma$  as a function of absolute luminosity for galaxies in different environments. We determine the velocity dispersion by fitting an SIS out to a scale of $100\ h^{-1}$ kpc to the excess surface mass density $\Delta\Sigma$ in separate luminosity bins (see Fig.~\ref{fig:WIDE.total.ds.bins}). We show the results for the combined (red and blue galaxies) sample for all environments (black circles, solid fit-line), for dense environment (magenta triangles, dashed fit-line)  and environment with low density (green squares, dotted fit-line).}
\label{fig:lsigma.env}
\end{figure}

We now fit an isothermal sphere to the $\Delta\Sigma$-profile for the previously defined magnitude intervals (see Fig.~\ref{fig:WIDE.total.ds.bins}). The corresponding best fit velocity dispersions are shown in Fig.~\ref{fig:lsigma.env} as a function of the weighted mean (see equation~\ref{eq:weight-hoekstra}) luminosities of galaxies from the respective magnitude interval. The black data points show the complete sample (all galaxies in all density environments). The results for galaxies in high and low density environments are shown in magenta and green.
To avoid contamination by neighbouring galaxy haloes, especially in the fainter luminosity bins, we only consider $\Delta\Sigma$ data points out to $\sim 100\ h^{-1}$ kpc. 
For bright galaxies the velocity dispersions agree irrespective of the galaxy environment density.
In high density environments faint galaxies show a mildly higher signal than their counterparts in low density environments (see Fig.~\ref{fig:lsigma.env}), increasing in difference with decreasing luminosities. 
The interpretation for this is that for faint galaxies the environment could impact the $\Delta\Sigma$ signal already on scales smaller than 100 $h^{-1}$ kpc, but more likely it is the change of fraction of blue and red galaxies: in dense environment the red galaxies are relatively more abundant and thus increase the $\Delta\Sigma$-amplitude of the combined sample.
Therefore the scaling relation of the velocity dispersion becomes slightly steeper with decreasing environment density, being $\sigma \propto  L_{r'}^{0.31\pm0.03}$ in low density environments, $\sigma \propto  L_{r'}^{0.29\pm0.02}$ in all density environments and finally $\sigma \propto  L_{r'}^{0.27\pm0.02}$ in high density environments. 
With these power law slopes we obtain for a galaxy with luminosity $\rm{L^{*}=1.6 \times 10^{10}\ h^{-2}\ L_{r',\odot}}$, which corresponds to an absolute magnitude of $M_{r'}^{*} \sim -21.7$ in the vega system, a velocity dispersion of $\sigma^{*}=135\pm2\ \rm{km\ s^{-1}}$ if all lenses in all environments are considered, $\sigma^{*}=141\pm2\ \rm{km\ s^{-1}}$ for all lenses in dense environments and $\sigma^{*}=132\pm2\ \rm{km\ s^{-1}}$ for all lenses in low density environments.
\begin{figure*}
\includegraphics[width=8.8cm]{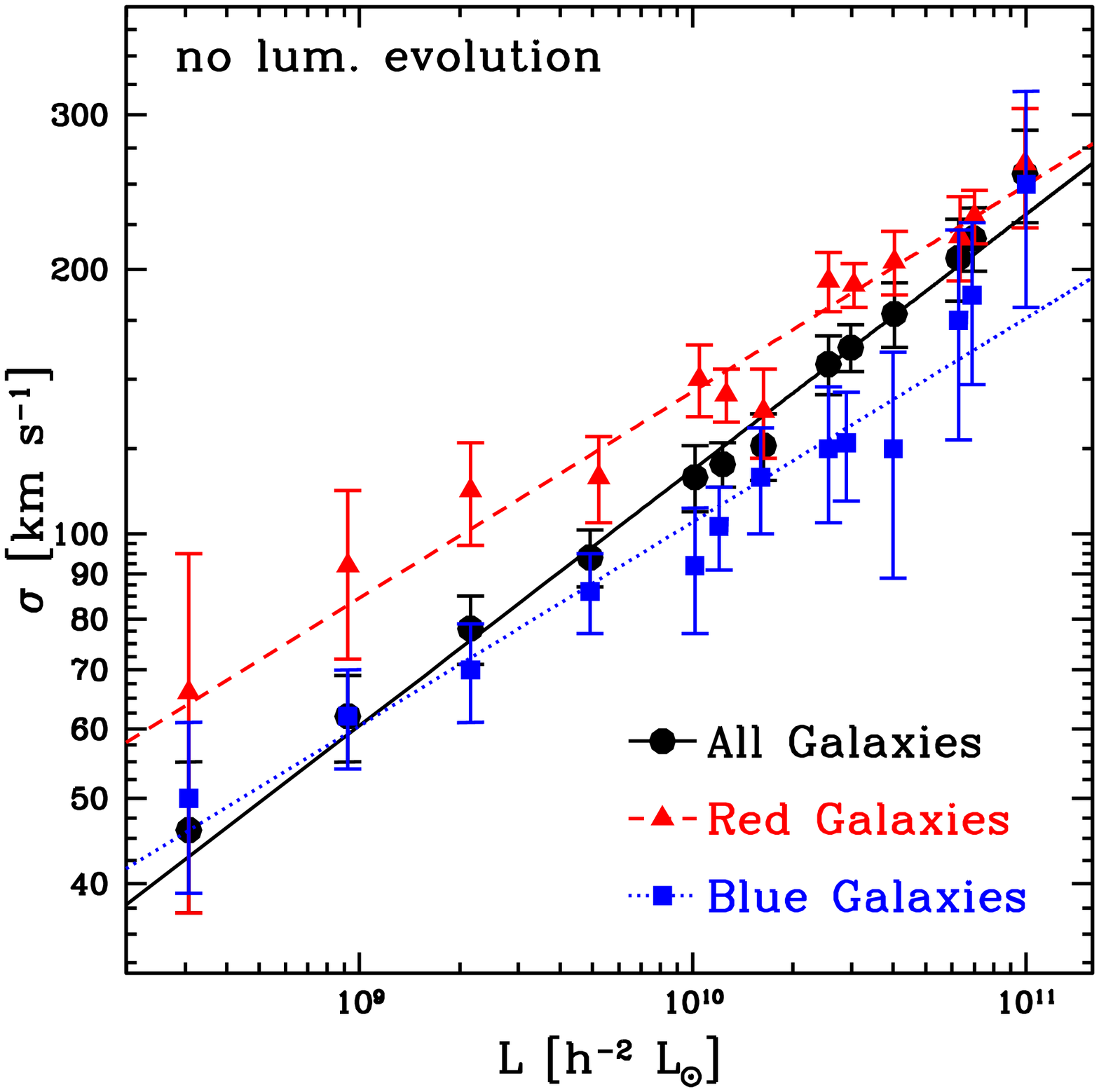}
\includegraphics[width=8.8cm]{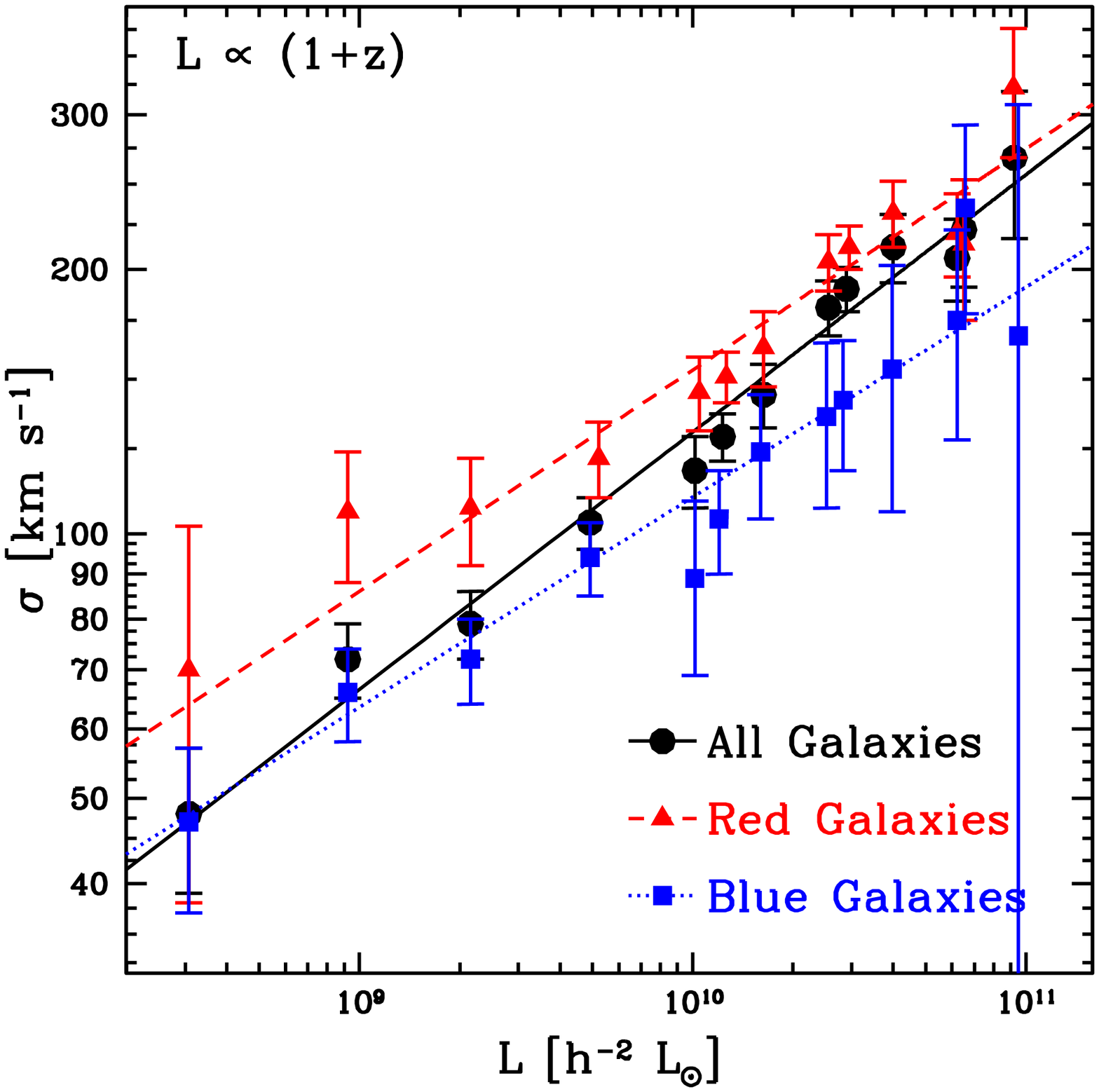}
\caption{Velocity dispersion $\sigma$ as a function of luminosity. We determine the velocity dispersion by fitting an SIS out to a scale of $100\ h^{-1}$ kpc to the excess surface mass density $\Delta\Sigma$ in separate luminosity bins (see Fig.~\ref{fig:WIDE.total.ds.bins}). The left panel shows the results for the combined lens sample (black circles, solid fit-line) and for red (red triangles, dashed fit-line) and blue (blue squares, dotted fit-line) lenses separately. The slopes are $(0.24\pm0.03)$ and $(0.23\pm0.03)$ for the red and blue sample. Due to the excess of elliptical galaxies at high luminosities and the excess of spiral galaxies at low luminosities this combines to a slope of $0.29\pm0.02$ (black dashed line) for the complete lens sample.
The right panel shows the corresponding values for $\sigma$ assuming a luminosity evolution according to $L \propto (1+z)$. The amplitude is increased in comparison to the values without evolution, however, the slope and thus the scaling relation between $\sigma$ and luminosity remain almost unaffected ($0.30\pm0.02$ for all galaxies, $0.26\pm0.03$ for red and $0.25\pm0.03$ for blue galaxies). The parameterisations for the derived scaling relations are shown in Table~\ref{tab:scaling relations}.}
\label{fig:lsigma}
\end{figure*}
\\
As next step we analyse the lens sample for red and blue lens galaxies separately without further discriminating the environment density. Both galaxy types, red and blue, show the same scaling behavior, $\sigma_{\rm red} \propto  L_{r'}^{0.24\pm0.03}$ and $\sigma_{\rm blue} \propto  L_{r'}^{0.23\pm0.03}$, respectively, consistent with the Faber-Jackson relation \citep{faberjackson}, as can be seen in the left panel of Fig.~\ref{fig:lsigma} (red and blue dashed lines). However, red galaxies show a significantly higher amplitude than their blue counterparts over the whole luminosity range. The transition from red SED-dominated galaxies at high luminosities to blue SED-dominated galaxies at fainter luminosities leads to the steeper scaling relation of $\sigma \propto  L_{r'}^{0.29\pm0.02}$ for the complete lens sample (see black solid line in the left panel of Fig.~\ref{fig:lsigma}).
\\
Based on our fit-values for $\sigma$ in the different luminosity bins and the obtained scaling relations, we obtain a velocity dispersion of $\sigma^{*}=135\pm2\ \rm{km\ s^{-1}}$ for the combined lens sample, $\sigma^{*}_{\rm red}=162\pm2\ \rm{km\ s^{-1}}$ for red lenses and a velocity dispersion of $\sigma^{*}_{\rm blue}=115\pm3\ \rm{km\ s^{-1}}$ for blue galaxies with luminosity $L=L^{*}$.
\\
Until now we have grouped galaxies together that have the same rest frame luminosity, ignoring the fact that their luminosity evolves with lookback time and thus redshift. If we want to group galaxies together that have the same present day luminosity we have to account for their luminosity evolution from the redshift of observation to $z=0$. In the following we assume that galaxy luminosities evolve as $L \propto (1+z)$. This assumption is correct for red galaxies (see e.g. \citealt{saglia10} or also \citealt{bernardi10}). Blue galaxies evolve more rapidly. For an estimate we could take their SED-types and plausible star formation histories to obtain their luminosity dimming. This will not be extremely precise, and therefore in the following we take the evolution of red galaxies as a lower limit.
\\
Accounting for luminosity evolution increases the velocity dispersion for a galaxy with present day reference luminosity to $\sigma^{*}=150\pm2\ \rm{km\ s^{-1}}$ for all galaxies, $\sigma^{*}_{\rm red}=173\pm2\ \rm{km\ s^{-1}}$ for red and $\sigma^{*}_{\rm blue}=123\pm3\ \rm{km\ s^{-1}}$ for blue galaxies. The scaling relations, however, are nearly unaffected. We obtain slopes of $\sigma \propto L^{0.29\pm0.02}$ for all galaxies, $\sigma \propto L^{0.25\pm0.03}$ and $\sigma \propto L^{0.24\pm0.03}$ for galaxies with red and blue SEDs (see also Table~\ref{tab:scaling relations}).
The closer inspection of red and blue data points in both panels of Fig.~\ref{fig:lsigma} shows that for a luminosity of $L \sim 6-7 \times 10^{10}\ h^{-2}\ L_{\odot}$ two red data points are decreased relative to the red SED linear fit and that one blue data point is increased relative to the blue SED linear fit. This could point to a problem in contamination of the red and blue samples with blue and red galaxies at this luminosity. We will see this feature (that apparently two data points for the red galaxies are biased low) in most of the following Figures.
\\
\\
The velocity dispersion $\sigma_{\rm halo}^{\rm WL}$ obtained from the weak lensing analysis (out to $100\ h^{-1}$ kpc) describes the circular velocity $v_{\rm circ,halo} = \sigma_{\rm halo} \sqrt{2}$ of the dark matter halo assuming an SIS-profile (see equation~\ref{eq:rho-SIS}).
\citet{gerhard01} (see their fig.~2) have studied the circular velocity curves of local ellipticals with stellar dynamics out to a few ($\le 3$) effective radii. They constrained the anisotropy profiles $\beta(r)$ (see \citealt{binney87}) of the stellar orbits and obtained that the mean values for $\beta$ are typically between 0.2 and 0.4. The detailed dynamical models yield a relation between the central stellar velocity dispersion and the maximal rotation velocity profile of
\begin{equation}
  \sigma_{\rm star}= 0.66\ v_{\rm max}^{\rm dyn}\ .
\end{equation}
The radii where these maximal velocities are reached are of order of 0.5 times the effective radii. The rotation velocities for larger radii ($> R_e$) are flat and have values of $\approx 0.9\ v_{\rm max}^{\rm dyn}$.
\\
If one sets these `asymptotic values' equal to the halo circular velocity we obtain
\begin{eqnarray}
  v_{\rm circ,halo} = \sqrt{2} \ \sigma_{\rm halo}^{\rm WL} &=& 0.9\ v_{\rm max}^{\rm dyn} = 0.9 \times 1/0.66\ \sigma_{\rm star} \nonumber \\
  &\text{or}& \nonumber \\
  \sigma_{\rm halo}^{\rm WL} &=& 0.96\ \sigma_{\rm star} \ .
\end{eqnarray}
If one sets the maximal circular velocity equal to the halo circular velocity one obtains
\begin{equation}
  \sigma_{\rm halo}^{\rm WL} = 1.07\ \sigma_{\rm star} \ .
\end{equation}
In Fig.~\ref{fig:WIDE.red.sigma} we compare how the measured velocity dispersion $\sigma_{\rm star}$ of LRGs compare with predictions from our WL-analysis for red galaxies, based on \citet{eisenstein01} and \citet{gallazzi06}, i.e. we add the best-fitting lines for the $\sigma_{\rm halo}^{\rm WL}$-luminosity relation, rescaled with 1/0.96 (in magenta) and 1/1.07 (in red). We have added the relation between the $\sigma_{\rm star}$ and evolution corrected luminosities of SDSS-LRGs (\citealt{eisenstein01}) obtained from \citet{gallazzi06} as green dashed line. This relation is however obtained from fitting a linear relation of velocity dispersions  vs. absolute magnitude to the overall LRG sample. In the lensing analysis we first average the signal within some (small) luminosity bin and then study the signal. To treat the LRG-galaxies in a similar way we have obtained the $\sigma_{\rm star}$-values from the SDSS data base and estimated the luminosity evolved redshift zero absolute magnitudes in the r-band  (from SED-fits and a luminosity evolution proportional to $1+z$) and obtained the mean stellar velocity dispersion within equidistant luminosity intervals. For this we only include galaxies with redshifts between 0.05 and 0.3 and with secure velocity dispersion estimates $0.03 < d\sigma_{\rm star}/\sigma_{\rm star} < 0.1$. The results are plotted with filled black circles, whereas the density contours for all considered galaxies are shown in black. We see that $0.96\ \sigma_{\rm star} \le \sigma_{\rm halo}^{\rm WL} \le 1.07\ \sigma_{\rm star}$ holds at least for for luminosities above $10^{10}\ h^{-2}\ L_{\odot}$. Therefore the halo velocity is between the maximal circular velocity found around 0.5 $R_e$ and 90 per cent of this value which equals the velocity of galaxies at a few effective radii. This indicates that at least for galaxies above this luminosity threshold the halo indeed is isothermal on scales out to $100\ h^{-1}$ kpc.
\begin{figure}
\centering
\includegraphics[width=8.8cm]{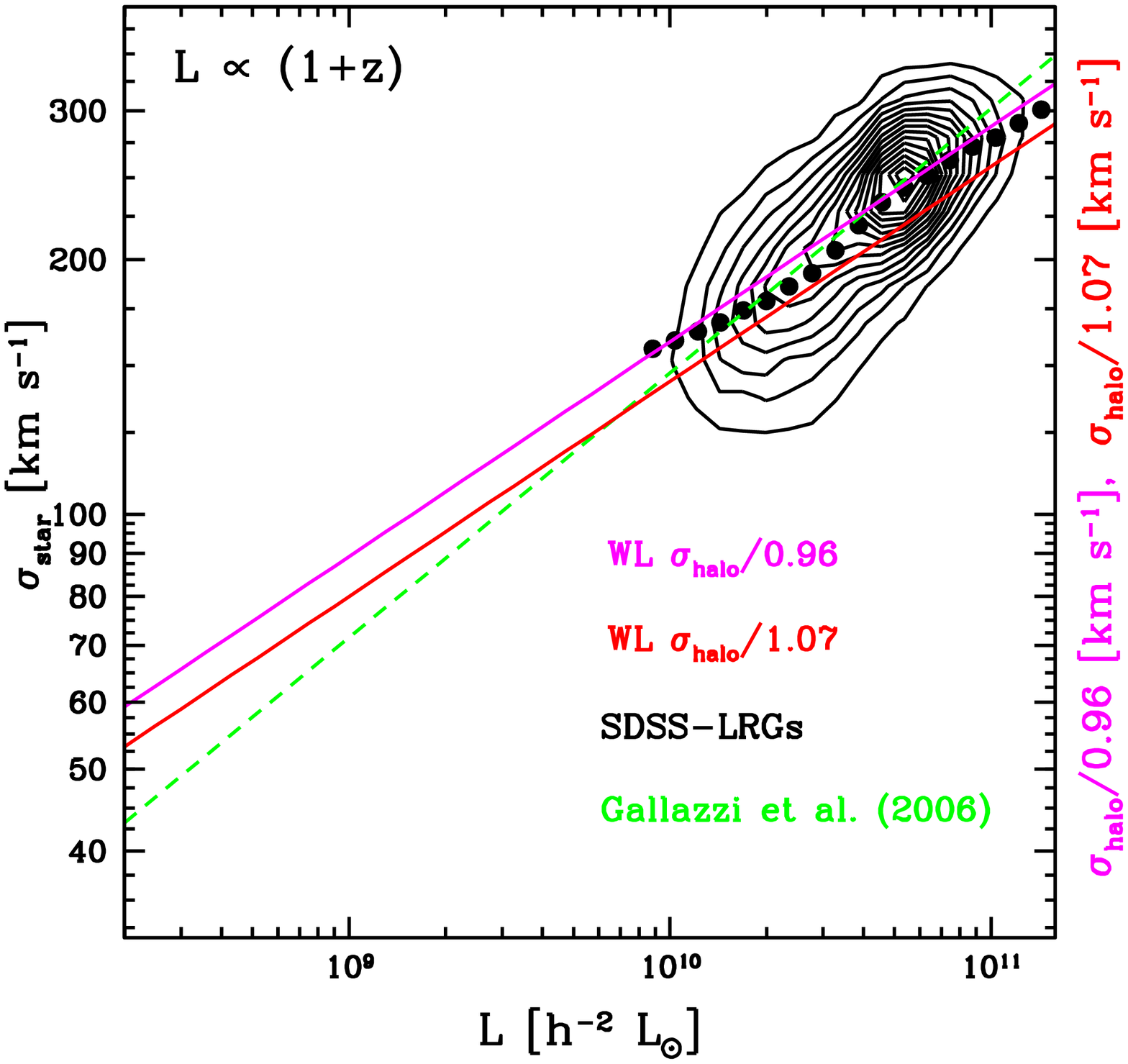}
\caption{Velocity dispersion for our red galaxy sample, rescaled by 1/0.96 (in magenta) and 1/1.07 (in red) as a function of absolute luminosity. We compare our values to the ones of \citet{gallazzi06} (green dashed fit-line) and a SDSS-LRG sample based on \citet{eisenstein01} (black circles and contours), only considering LRGs with redshifts of $0.05 < z < 0.3$ and velocity dispersion uncertainties of $0.03 < d\sigma_{\rm star}/\sigma_{\rm star} < 0.1$.}
\label{fig:WIDE.red.sigma}
\end{figure}
\subsubsection{L-$r_{200}$- and L-$M_{200}$-Scaling Based on Fits to $\Delta\Sigma$}
\label{sec:ds scaling r200}
\begin{figure*}
\includegraphics[width=8.8cm]{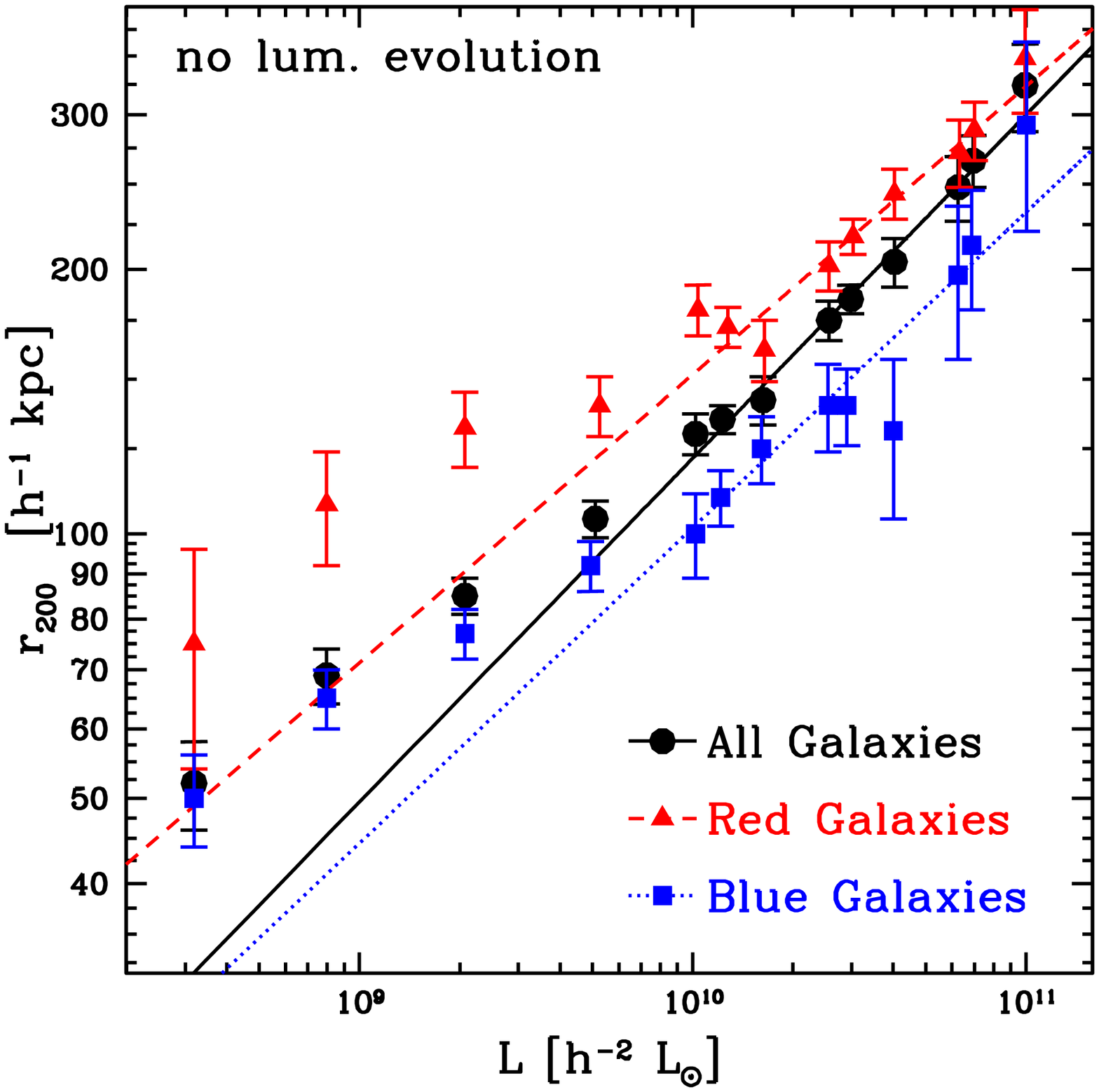}
\includegraphics[width=8.8cm]{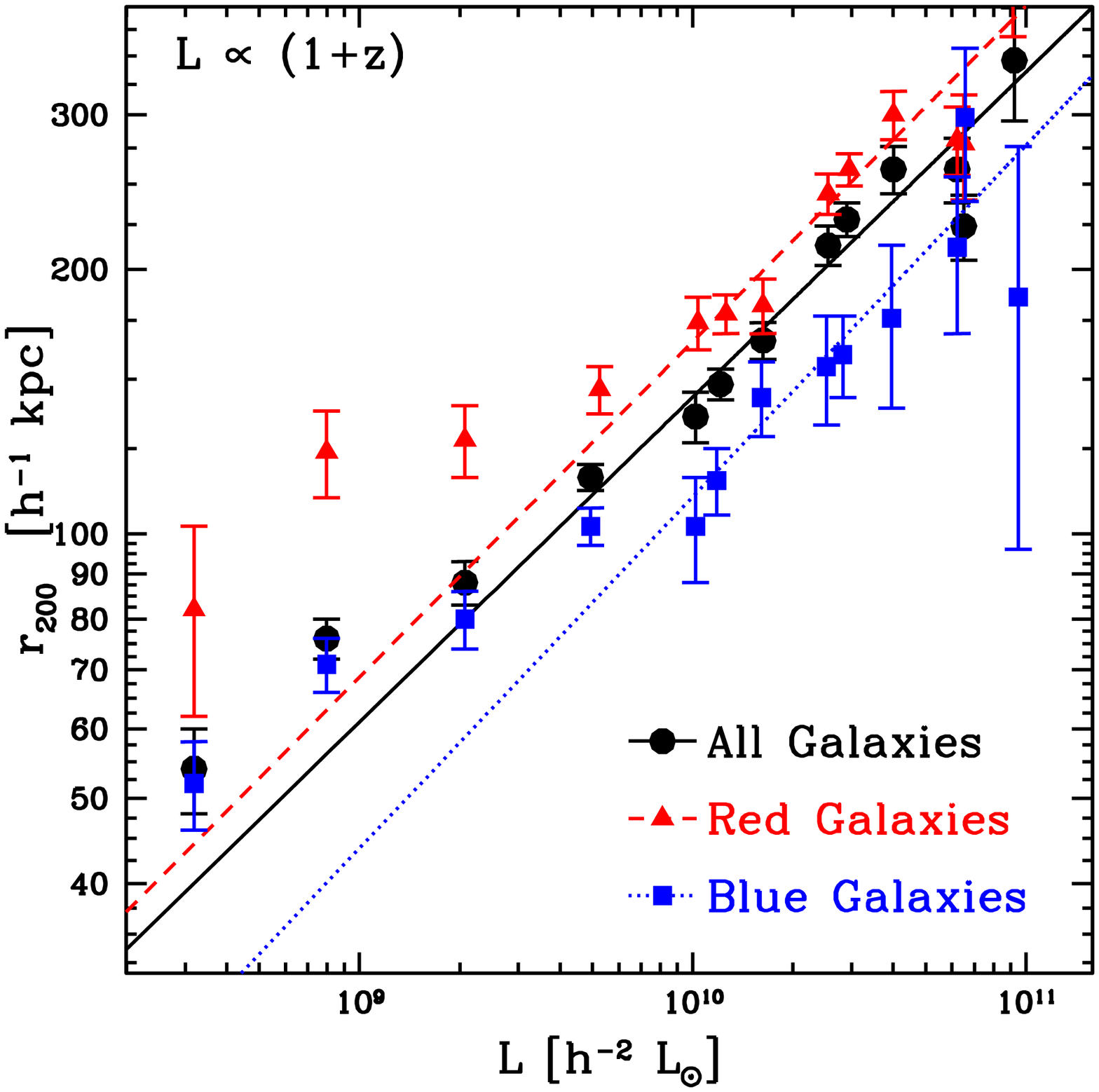}
\caption{Virial radius $r_{200}$ as a function of absolute luminosity. The left panel shows the result without, the right panel with luminosity evolution $L \propto (1+z)$. The red triangles and dashed fit-lines denote red galaxies, the blue squares and dotted fit-lines blue galaxies and the black cirlces and solid fit-lines all galaxies. We only use galaxies with $L > 10^{10}\ h^{-2}\ L_{\odot}$ for the determination of the scaling relation. For the combined lens sample the $r_{200}$ scales with $L^{0.39\pm0.04}$ ignoring, and with $L^{0.37\pm0.04}$ including luminosity evolution. The parameterisations for the derived scaling relations are shown in Table~\ref{tab:scaling relations}.}
\label{fig:L-r200}
\end{figure*}

\begin{figure*}
\centering
\includegraphics[width=8.8cm]{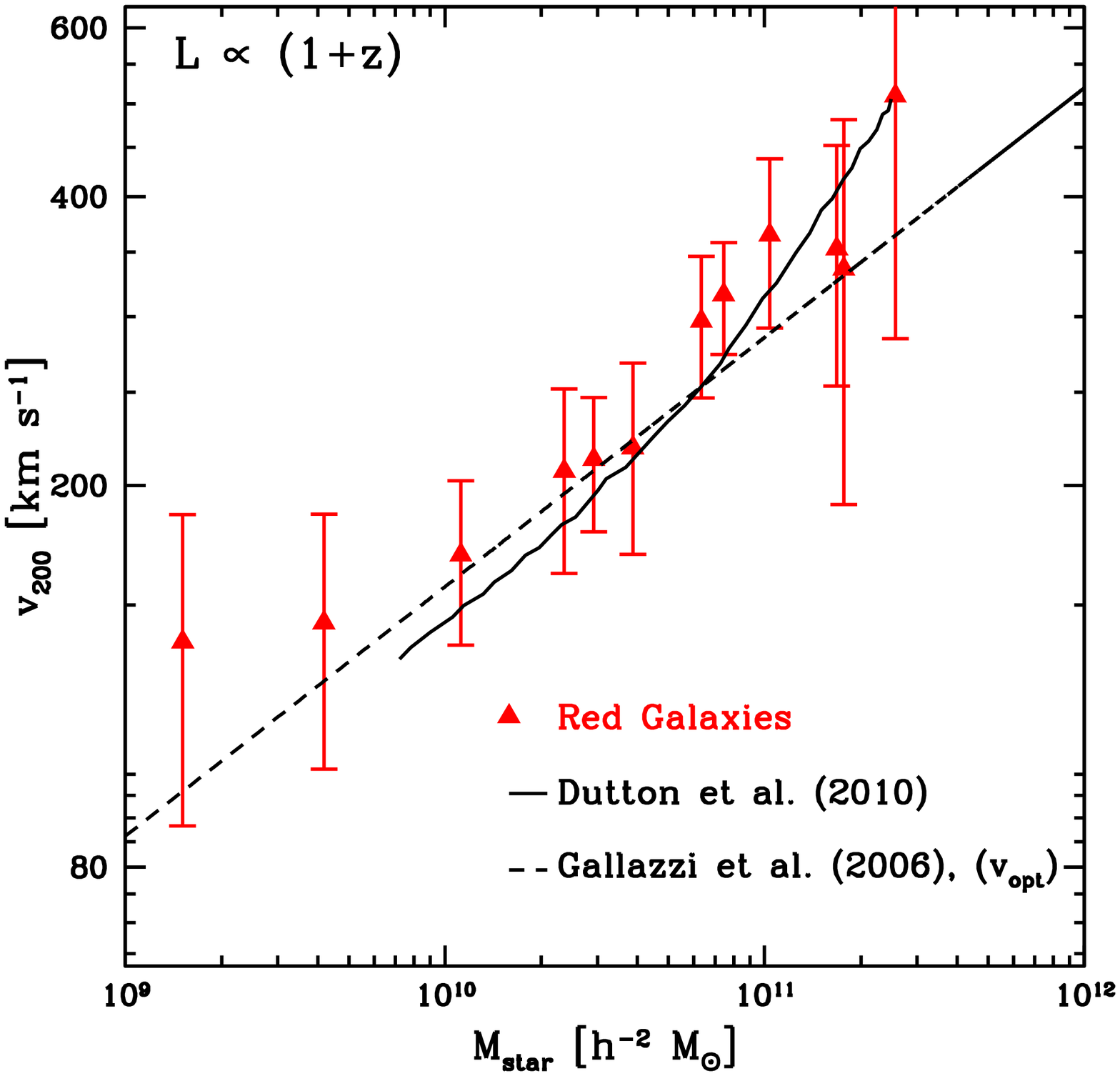}
\includegraphics[width=8.8cm]{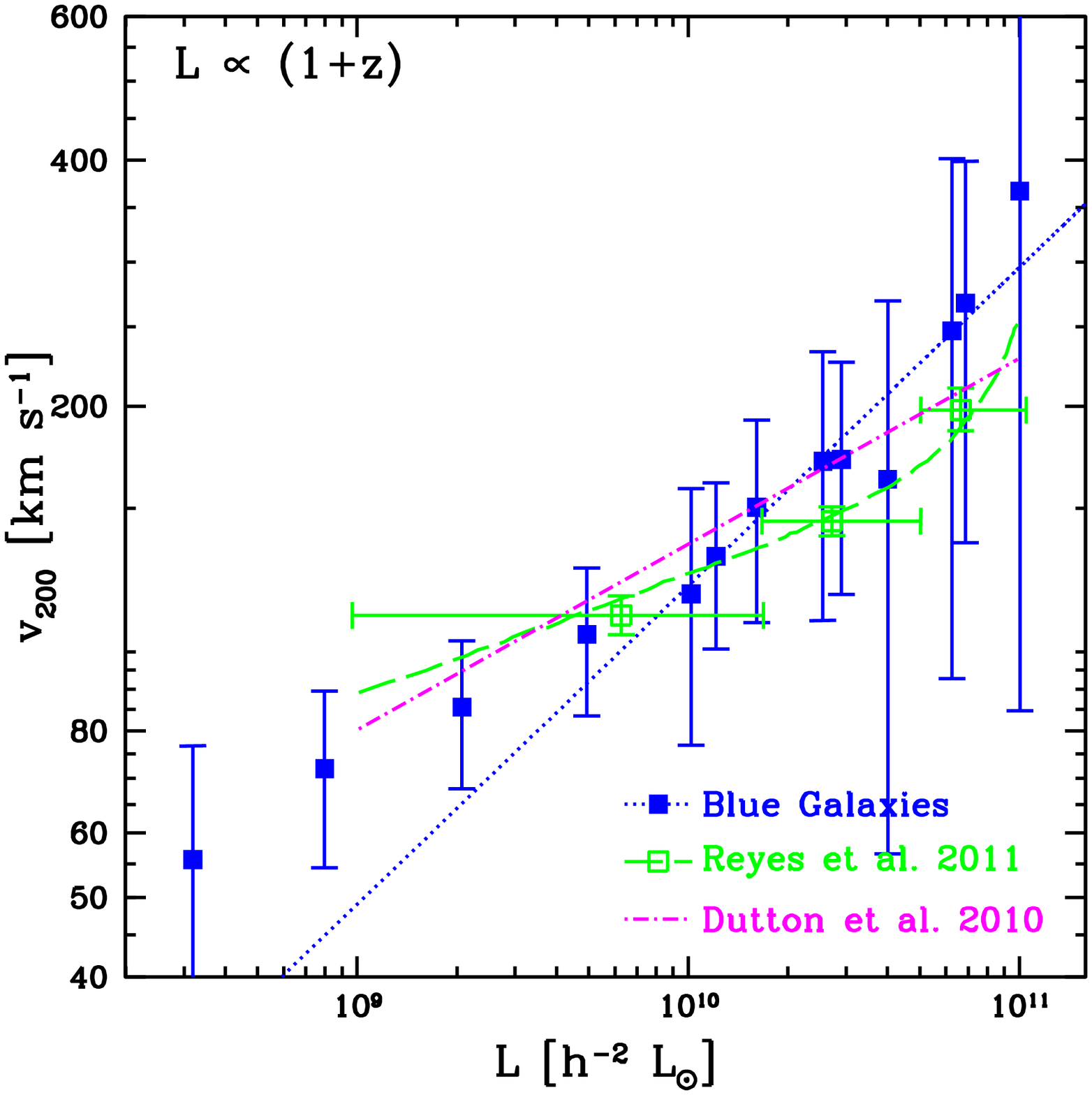}
\caption{Circular velocity $v_{200}$ as a function of absolute luminosity for red (left panel) and blue galaxies (right panel). Analogously to the fit of the $r_{200}$ we only use data points with $L > 10^{10}\ h^{-2}\ L_{\odot}$ for the determination of the scaling relation fit. In the right panel we have added the results of \citet{reyes11} and the model of \citet{dutton10}, after translating their stellar masses to luminosities for blue galaxies (see text). In the left panel we have added the results of \citet{dutton10} and \citet{gallazzi06} using the relation of \citet{dutton10} (see text).
}
\label{fig:L-v200}
\end{figure*}

\begin{figure*}
\includegraphics[width=8.8cm]{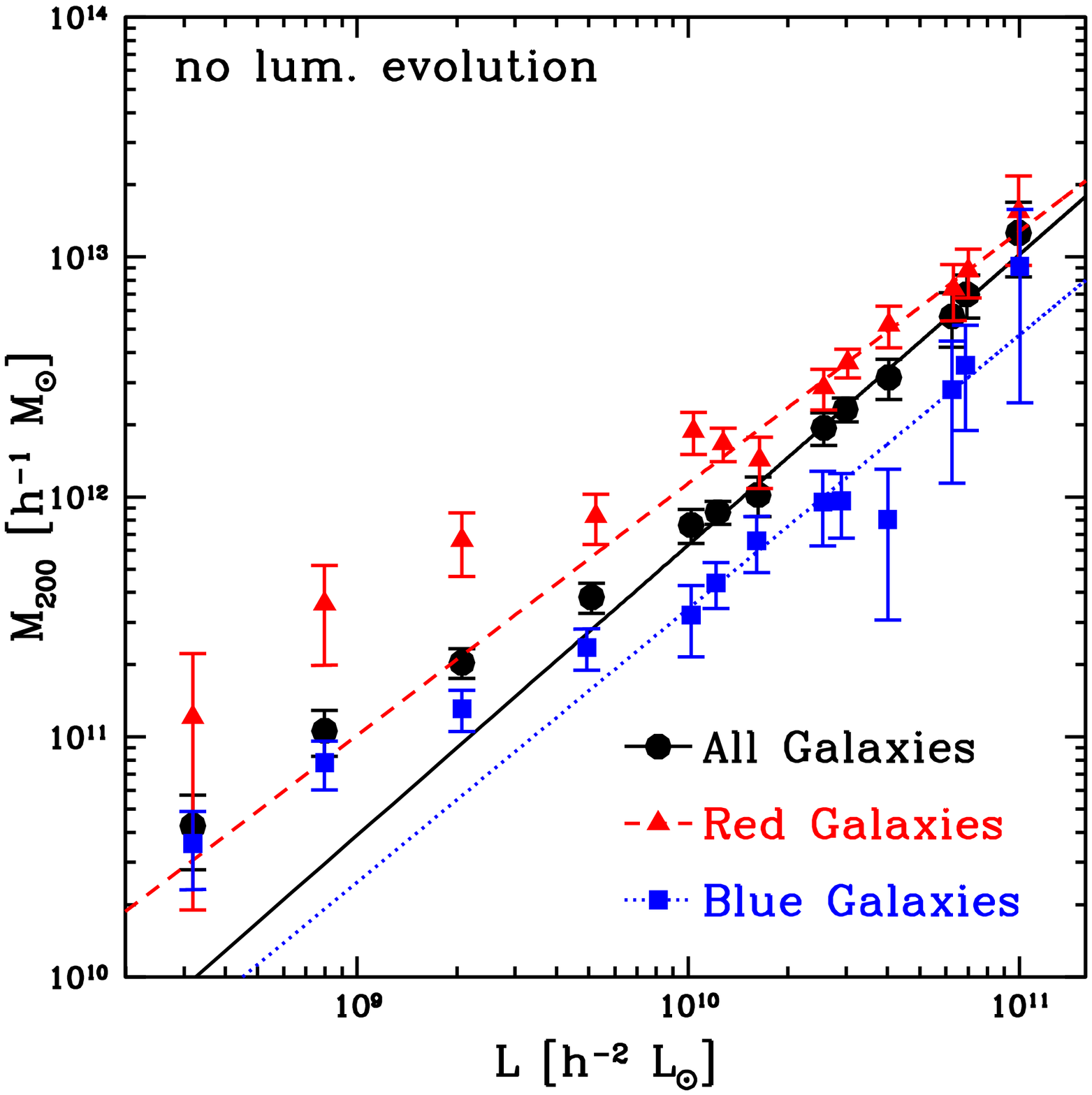}
\includegraphics[width=8.8cm]{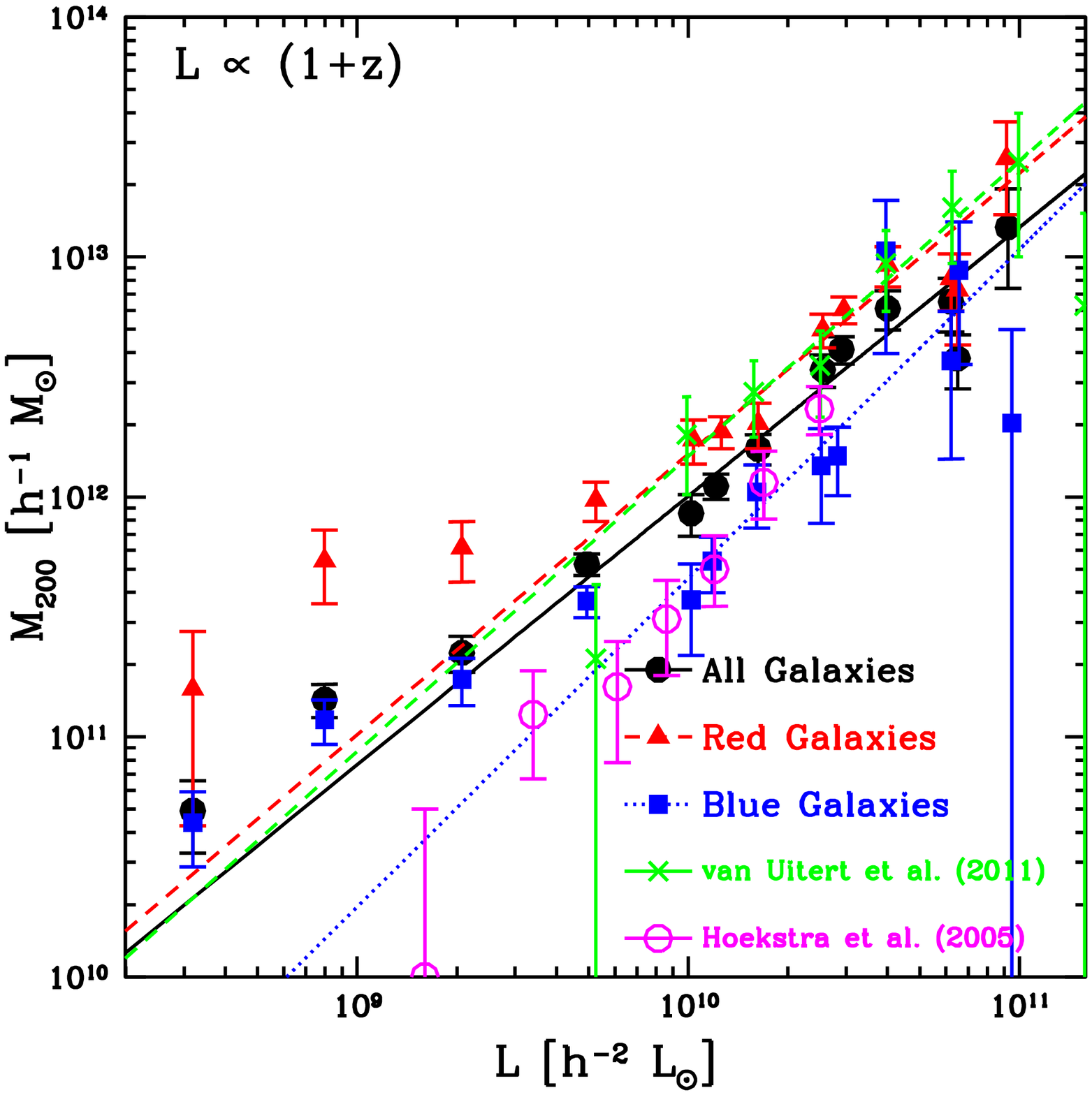}
\caption{Virial mass $M_{200}$ as a function of absolute luminosity. The left panel shows the result without, the right panel with luminosity evolution $L \propto (1+z)$. The red triangles and dashed fit-lines denote red galaxies, the blue squares and dotted fit-lines blue galaxies and the black circles and solid fit-lines all galaxies. We see as expected the same scaling behavior as for the $r_{200}$ (see Fig.~\ref{fig:L-r200}). Analogously to the fits of the $r_{200}$ we only use data points with $L > 10^{10}\ h^{-2}\ L_{\odot}$ for the determination of the scaling relation. For the complete lens sample the $M_{200}$ scales with $L^{1.21\pm0.10}$ ignoring, and with $L^{1.12\pm0.11}$ including luminosity evolution. We included the results from \citet{vanuitert11}  for a red SED-dominated lens sample in the right panel (green crosses) and from \citet{hoekstra05} for an isolated and thus blue SED-dominated galaxy sample (open magenta circles). The parameterisations for the derived scaling relations are shown in Table~\ref{tab:scaling relations}.}
\label{fig:L-m200}
\end{figure*}

\begin{figure}
\centering
\includegraphics[width=8.8cm]{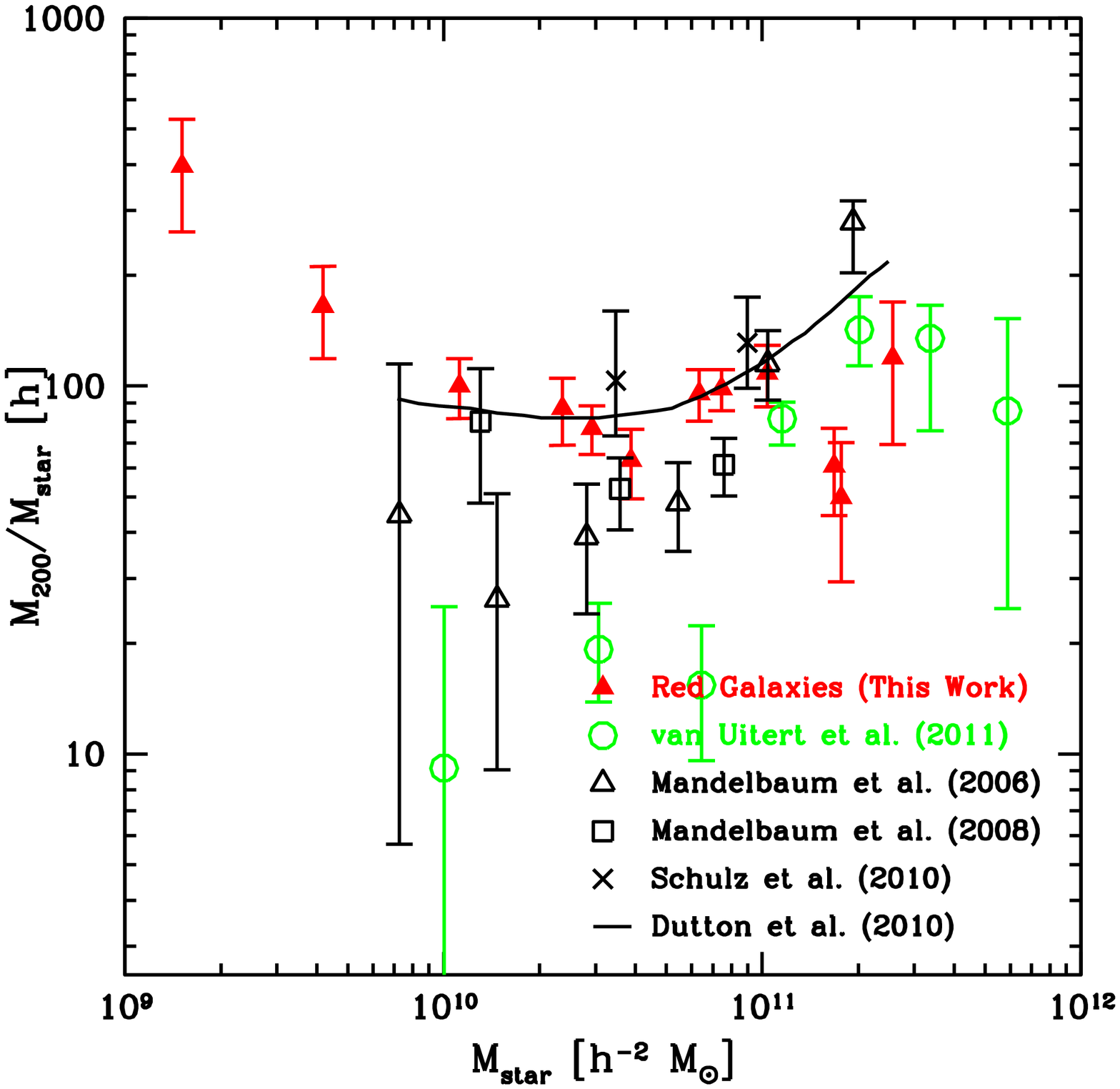}
\caption{Stellar Mass  vs. $M_{200}/M_{\rm star}$ ratio for red galaxies converted to $z=0$. The red triangles denote our red galaxies. We have added the results of \citet{mandelbaum06b} (open triangles), \citet{mandelbaum08b} (open squares), \citet{schulz10} (black crosses) and \citet{dutton10} (black solid line), see also fig.~1 in \citet{dutton10}. We further include the results of \citet{vanuitert11} from their fig.~14 as green open circles.}
\label{fig:mstar-m200}
\end{figure}

We now assume that the concentration-mass-relation $c \propto M^{-0.084}$ (see equation~\ref{eq:duffy}) of \citet{duffy08} holds which reduces the NFW-profile to a one-parametric profile. We fit NFW-profiles to the $\Delta\Sigma$-profiles (shown in Fig.~\ref{fig:WIDE.total.ds.bins}) and thus obtain the virial radius for each luminosity interval. The $r_{200}$  vs. luminosity relations (with and without luminosity evolution) are shown in Fig.~\ref{fig:L-r200}. They imply that if the assumed concentration-mass-relation is correct then the $r_{200}$  vs. luminosity relation cannot be described by a single-power law anymore, but instead with double-power laws and a break at around $L=10^{10}\ h^{-2}\ L_{\odot}$. In this case, the mean mass-to-light ratio of  galaxies within a luminosity interval  would indeed be minimal at this break luminosity luminosity. This is in agreement with results from abundance matching (AM) techniques and some satellite kinematic results (see fig.~1 \citealt{dutton10}), in particular with the results of \citet{more11} (see their fig.~5) who also obtained a change of slope for the red galaxies'  $M_{200}$  vs. luminosity relation at a luminosity of about $10^{10}\ h^{-2}\ L_{\odot}$. We would like to point out however, that the result in Fig.~\ref{fig:L-r200} only holds if the concentration is only weakly changing with virial mass. One could instead approximately reconcile a single-power law $r_{200}$-luminosity relation if one required the concentration to rise steeply for luminosities smaller than $10^{10}\ h^{-2}\ L_{\odot}$.  We will investigate these two alternatives more in Section~\ref{sec:scaling relations}. Because of the apparently broken $r_{200}$-luminosity scaling relation we measure the power law slope only for galaxies brighter than $10^{10}\ h^{-2}\ L_{\odot}$. We obtain power laws of $r_{200}^{\rm red} \propto L^{0.33\pm0.04}$ and $r_{200}^{\rm blue} \propto L^{0.36\pm0.07}$ for red and blue galaxies without luminosity evolution and of $r_{200}^{\rm red} \propto L^{0.38\pm0.04}$ and $r_{200}^{\rm blue} \propto L^{0.40\pm0.08}$ for luminosities evolving with $(1+z)$. If galaxies are not separated into blue and red SED types we obtain (for the combined sample) $r_{200} \propto L^{0.39\pm0.03}$ and $r_{200} \propto L^{0.37\pm0.04}$ ignoring luminosity evolution and assuming a $(1+z)$ scaling. As before the steeper scaling is due to the fact that the amplitudes for the $r_{200}$  vs. luminosity scalings are different for red and blue galaxies and the fractional mix of red and blue galaxies changes as a function of luminosity.
\\
In Fig.~\ref{fig:L-v200} we translated the result for $r_{200}$ to the virial velocity $v_{200}$ using equation~(\ref{eq:v200}). In the right panel of Fig.~\ref{fig:L-v200} we show $v_{200}$  vs. luminosity for our blue sample (blue data points) and the power law fit for $L>10^{10}\ h^{-2}\ L_{\odot}$ (blue dotted line). \citet{reyes11} have measured $v_{200}$  for SDSS disc galaxies as function of stellar mass. We try to compare their result to ours by translating their stellar mass estimate (back) to luminosity. It seems that for local disc galaxies (the \citealt{reyes11} disc galaxies have redshifts between 0.02 and 0.1) an on average mass-to-light ratio of $M_{\rm star}/L_{r} = 1\ M_{\odot}/L_{\odot}$  is a good description. This can be seen on one hand in fig.~1 of \citet{vanuitert11} by comparing their blue histograms on the vertical to the horizontal axis showing the luminosity distribution and stellar mass distribution of blue SDSS-galaxies. This estimate is in agreement with \citet{bell03}, if one takes into account that our local (see Fig.~\ref{fig:R-gr}) galaxies have a $(g-r)$-colour of approximately $0.3-0.4$ at the bright end (which are the galaxies in common with \citealt{reyes11}). The same result is obtained from \citet{kauffmann03}, fig.~14, upper right panel, if one takes into account that our local blue galaxies have mostly absolute magnitudes fainter than $M_{r'}=-21$. For the three luminosity intervals provided by \citet{reyes11} their data points (translated to luminosity) agree well with ours (see Fig.~\ref{fig:L-v200}, right panel).  We have a larger dynamical range and can extend our analysis down to to a few times $10^9\ L_{\odot}$. In an analogous way we have translated the \citet{dutton10} model for the $v_{200}$-stellar mass relation to the $v_{200}$-luminosity relation. It agrees very well with our result, but might have a slightly shallower slope.
\\
For the red galaxies we have translated our luminosities into stellar mass estimates, because in this case we are confident that we can do so for the absolute magnitude and redshift range considered. We use $\log_{10}(M_{\rm star}) = 1.093\ \log_{10}L_{r} - 0.573$ (which was used by \citealt{dutton10}  and derived from \citealt{gallazzi06}), and insert luminosity evolution corrected luminosities. Our results for $v_{200}$ are shown in red in Fig.~\ref{fig:L-v200}, together with the model of \citet{dutton10}. They are the same to a remarkable level. Only the results for the second and third brightest luminosity interval lie below, for a reason we speculated about
already.  On top we have added the result for $v_{\rm opt}$ as obtained from \citet{gallazzi06} velocity dispersion  vs. luminosity relation, using the prefactors of \citet{dutton10} for the relation between velocity dispersion and rotation velocity. (Using our translation factor the \citealt{gallazzi06} curve would be slightly below).  We concluded that for luminosities between $10^{10}$ and $6 \times10^{10}\ h^{-2}\ L_{\odot}$ the mass density profile of ellipticals is not
only isothermal out to $100\ h^{-1}$  kpc (as shown before), but out to the virial radius. For higher luminosities the virial velocity exceeds the optical velocity.
\\
\\
Finally we translate our $r_{200}$  vs. luminosity results into virial masses and show results with and without luminosity evolution correction in the left and right panels of Fig.~\ref{fig:L-m200}. For the power law fits (added as red dashed and blue dotted lines) we have again used only galaxies with $L>10^{10}\ h^{-2}\ L_{\odot}$.  For the combined sample we obtain $M_{200} \propto L^{1.21\pm0.10}$ and $M_{200} \propto L^{1.12\pm0.11}$ for the case without and with luminosity evolution correction. This scaling agrees with the results of \citet{guzik02} within their larger uncertainties ($M \propto L_{r'}^{1.34\pm0.17}$).
We have added the results of \citet{hoekstra05} as magenta points, which agree well with our blue sample. This agreement appears reasonable since the \citet{hoekstra05} sample contains isolated galaxies which mostly consist of blue galaxies. In addition we have considered the excess surface mass density profile of \citet{vanuitert11} (see their fig.~8), and translated them into virial mass estimates in the same way as we did for our work. These estimates are shown as green points. They agree well with our red sample results, which again is reasonable since the \citet{vanuitert11} sample is dominated by red galaxies. All results obtained for $r_{200}$ and $M_{200}$ are summarised in Table 2.
\\
At last we translate our $M_{200}$  vs. luminosity relation from the right panel of Fig.~\ref{fig:L-m200} into the $M_{200}$  vs. stellar mass relation (MSR) using the relation $\log_{10}(M_{\rm star}) = 1.093\ \log_{10}L_{r} - 0.573$ as above. The result is shown in Fig.~\ref{fig:mstar-m200}. The virial-to-stellar mass ratio (shown as red points) is almost constant (at 100) for a decade in stellar mass ($10^{10}$ to $10^{11}\ h^{-2}\ M_{\odot}$), and increases for lower stellar masses. In the mass range of $10^{10}$ to $10^{11}\ h^{-2}\ M_{\rm star}$ our result precisely agrees with the \citet{dutton10} model shown as black solid curve. At the high stellar mass end the MSR seems to increase (if at all) only slowly with stellar mass. This saturation is in agreement with the results of \citet{vanuitert11} (green points, taken from their fig.~14, and converting their stellar masses to $H_0=100\ \rm{km\ s^{-1}\ Mpc^{-1}}$, as in this Figure the stellar masses are given for $H_0=70\ \rm{km\ s^{-1}\ Mpc^{-1}}$ and the virial masses are given for $H_0=100\ \rm{km\ s^{-1}\ Mpc^{-1}}$ according to van Uitert, private communication) which also saturates at a value of about 100 to 150. The \citet{vanuitert11} points for low stellar masses are however even below the early \citet{mandelbaum06b} results and seem very low. Since the \citet{vanuitert11}  $M_{200}$  vs. luminosity relation derived by us from their $\Delta\Sigma$ results agree well with ours, the difference can only be due to a different relation for the stellar masses (Note, that \citealt{vanuitert11} aim to add up the total stellar mass, i.e. not only that of the central galaxy but also that of its satellites).
\begin{table*}
\centering
\caption{}
Without luminosity evolution, $L^{*}_{r'}=1.6\times10^{10}\ h^{-2}\  L_{r',\odot}$
\\
\begin{tabular}{c|c|c|c|c|c|c}
\hline \hline
type &  $\sigma^{*}\ \rm{[km\ s^{-1}}]$ & $\eta_{\sigma}$ & $r^{*}_{200}\ [h^{-1}\ \rm{kpc}]$ & $\eta_{r_{200}}$ & $M^{*}_{200}\ [10^{11}\ h^{-1}\ \rm{M_{\odot}}]$ & $\eta_{M_{200}}$ \\
\hline \hline
All & $135\pm2$ & $0.29\pm0.02$ & $146\pm2$ & $0.39\pm0.03$ & $11.1\pm0.4$ & $1.21\pm0.10$ \\ 
Red & $162\pm2$ & $0.24\pm0.03$ & $177\pm3$ & $0.33\pm0.04$ & $18.6\pm0.8$ & $1.05\pm0.12$ \\
Blue & $115\pm3$ & $0.23\pm0.03$ & $120\pm2$ & $0.36\pm0.07$ & $5.8\pm0.5$ & $1.14\pm0.20$ \\
\hline \hline
\end{tabular}
\\
With luminosity evolution, $L^{*}_{r'}=1.6\times10^{10}\ h^{-2}\  L_{r',\odot} \times (1+z)$
\\
\begin{tabular}{c|c|c|c|c|c|c}
\hline \hline
type &  $\sigma^{*}\ \rm{[km\ s^{-1}}]$ & $\eta_{\sigma}$ & $r^{*}_{200}\ [h^{-1}\ \rm{kpc}]$ & $\eta_{r_{200}}$ & $M^{*}_{200}\ [10^{11}\ h^{-1}\ \rm{M_{\odot}}]$ & $\eta_{M_{200}}$ \\
\hline \hline
All & $150\pm2$ & $0.29\pm0.02$ & $170\pm2$ & $0.37\pm0.04$ & $17.0\pm0.6$ & $1.12\pm0.11$ \\
Red & $173\pm2$ & $0.25\pm0.03$ & $198\pm3$ & $0.38\pm0.04$ & $26.1\pm1.1$ & $1.17\pm0.13$ \\
Blue & $123\pm3$ & $0.24\pm0.03$ & $133\pm3$ & $0.40\pm0.08$ & $8.7\pm0.6$ & $1.37\pm0.25$ \\
\hline \hline
\end{tabular}
\label{tab:scaling relations}
\\
Exponents for the scaling relations of the velocity dispersion $\sigma$, assuming an SIS and for the $r_{200}$ and $M_{200}$, assuming an NFW profile without and with luminosity evolution. The SIS fits have been extracted from all all luminosity bins, the fits for the NFW profiles only include luminosities brighter than $L=10^{10}\ h^{-2}\ L_{\odot}$.
\end{table*}
%
\subsection{Maximum likelihood analysis}
In the previous subsection we have analysed the lensing signal on small scales, ignoring the impact of further galaxy haloes. We now analyse the signal out to scales \mbox{$R\le 2\ h^{-1}$ Mpc} in the case of BBS. Since the integrated mass for the NFW profile diverges for infinite radii and the integrated mass value within a radius of $1\ h^{-1}$ Mpc already exceeds the total BBS mass, assuming reference halo parameters of $\sigma^{*} \sim 130\ \rm{km\ s^{-1}}$, $s^{*} \sim 200\ h^{-1}$ kpc, $c^{*} \sim 6$ and $r_{200}^{*} \sim 130\ h^{-1}$ kpc, we limit the NFW maximum likelihood analysis to a maximum distance of $R\le 400\ h^{-1}$ kpc. A projected separation of $R=400\ h^{-1}$ kpc corresponds to $\approx 1 \ r_{200}$ for bright and $\approx 5\ r_{200}$ for faint galaxies. Since at these scales the original NFW profile hardly differs from a truncated NFW profile (see \citealt{baltz09}) the not-finite total NFW masses do not affect our results. At scales of $400 \ h^{-1}$ kpc the signal of other haloes becomes comparable to the haloes of interest. This means, we have to model all the haloes simultaneously. We do this in a maximum likelihood analysis (see Section~\ref{sec:likelihood theory}), where we measure the (two) parameters of a reference galaxy (with a reference luminosity) and describe haloes of galaxies with different luminosities by scaling relations. As reference luminosity we choose $L_{r'} = 1.6 \times 10^{10}\ h^{-2} \ \rm{L_{r',\odot}}$ for the rest of this work.
\label{sec:likelihood}
\subsubsection{Truncated isothermal sphere (BBS)}
\label{sec:BBS}
\begin{figure*}
\includegraphics[width=16.5cm]{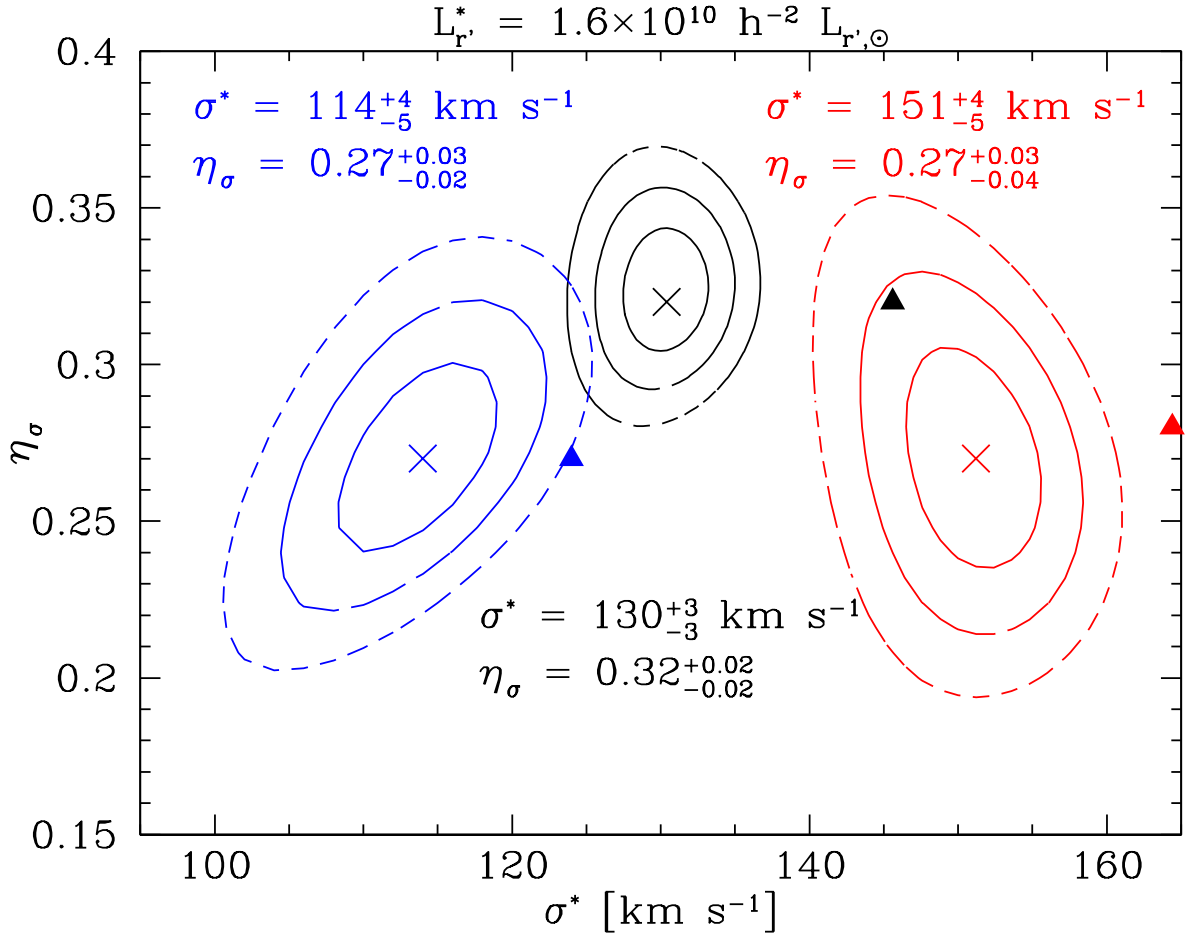}
\caption{Constraints on velocity dispersion $\sigma^{*}$ and the velocity dispersion  vs. luminosity scaling index $\eta_{\sigma}$ for a fiducial galaxy with luminosity $L_{r'}^{*} = 1.6 \times 10^{10} h^{-2} \ L_{r',\odot}$. Blue galaxies are shown in blue in the left, red galaxies in red in the right and the combined galaxy sample in black in the middle. The contours indicate the 68.3 per cent (solid), the 95.4 per cent (dashed) and the 99.7 per cent (dotted) confidence levels for both parameters combined, the crosses indicate the best-fit values, the filled triangles indicate the best-fit values assuming luminosity evolution with $(1+z)$.}
\label{fig:WIDE.redblue.likely.eta}
\end{figure*}
\begin{figure*}
\includegraphics[width=18.5cm]{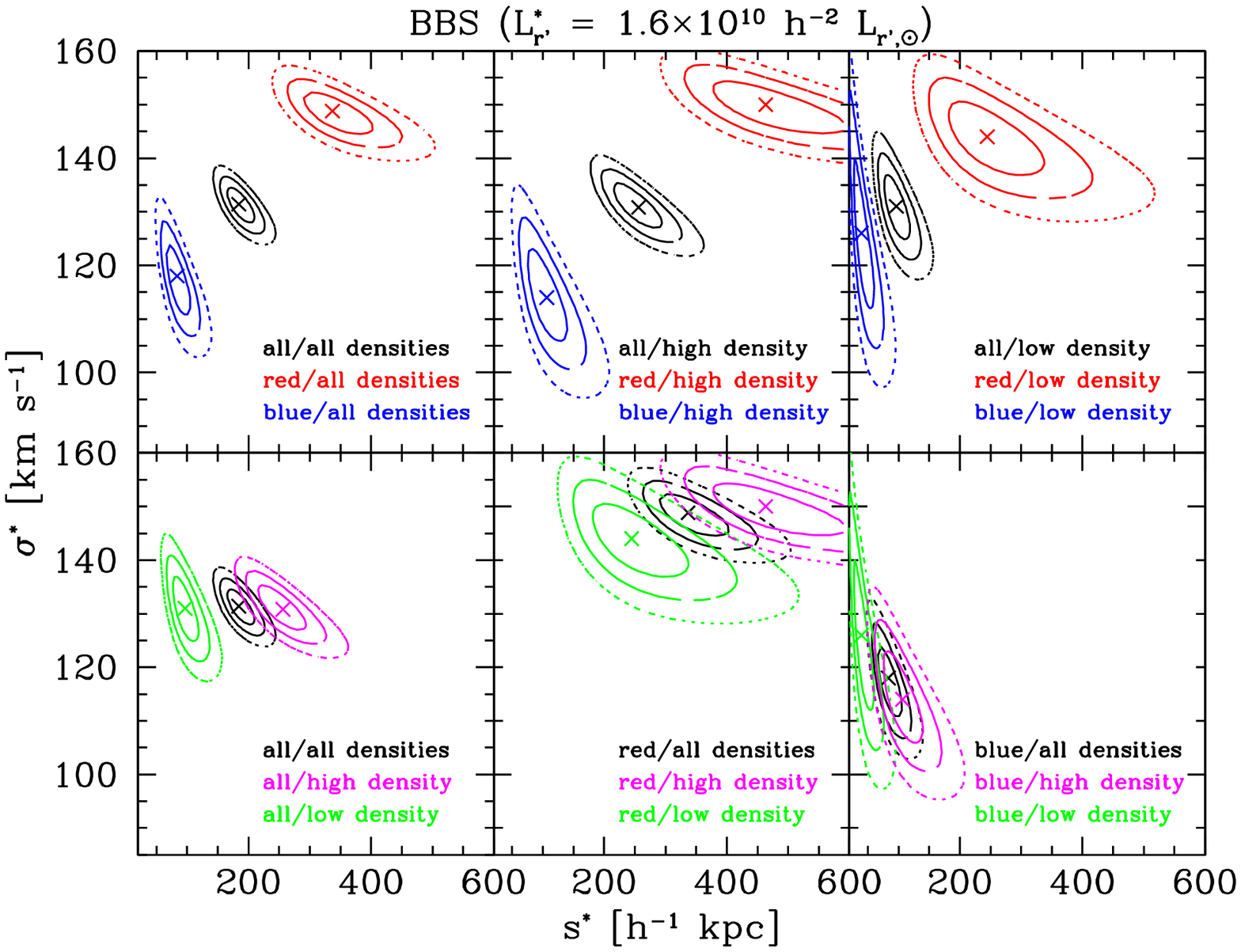}
\caption{Maximum likelihood results for the BBS profile. The truncation radius $s$ is shown on the x-axis, the velocity dispersion $\sigma^{*}$ is shown on the y-axis, the contours show the 68.3 per cent, the 95.4 per cent and the 99.7 per cent confidence levels.
In the upper row the results for different galaxy types are compared for one environment in each panel. Black contours show the combined galaxy lens samples, red contours the red and blue contours show the blue galaxy lens samples. The left panel shows the galaxies in all environments, the middle one shows the galaxies in high density environment and the right panel shows the galaxies in low density environment. 
In the lower row we compare the results for identical galaxy types in different environments, black is all environments, magenta is high density environment and green is low density environment. The left panel shows all galaxy types, the middle one shows red galaxies and the right panel shows blue galaxies. We see that red galaxies in general have larger velocity dispersions and halo sizes than blue galaxies, leading to higher masses. This holds for all environments. Looking at the dependence of the individual galaxy profiles with respect to their environmental conditions no galaxy type shows significant evolution in velocity dispersion, but the truncation radius clearly increases with environment density.}
\label{fig:WIDE.likely.bbs}
\end{figure*}
In Section~\ref{sec:ds scaling sigma} we had analysed the combined sample fitting an SIS to the excess surface mass density $\Delta\Sigma$ and obtained a scaling relation of $\sigma \propto L^{0.29\pm0.02}$ for the combined sample in Section~\ref{sec:ds scaling sigma}. As a cross-check we now analyse the luminosity  vs. velocity dispersion scaling with a maximum likelihood analysis. We use an SIS and analyse foreground-background pairs only to a distance of \mbox{$200 \ h^{-1}$ kpc}, since in this region the difference between the singular and the truncated isothermal sphere is negligible. In this way all (visible) nearby haloes are accounted for, in particular the environment of satellite galaxies sitting in more massive haloes traced by another galaxy. This analysis yields an exponent of $\eta_{\sigma} = 0.31\pm0.02$ (see black contours in Fig.~\ref{fig:WIDE.redblue.likely.eta}), in agreement with the results from Section~\ref{sec:ds scaling sigma}. The inclusion of luminosity evolution with $L \propto (1+z)$ increases the values for $\sigma^{*}$ (see also Section~\ref{sec:ds scaling sigma}), but basically does not change the slope of the scaling relation, as the best fit values (triangles in Fig.~\ref{fig:WIDE.redblue.likely.eta}) show. Therefore we now fix this scaling to  $\sigma \propto  L^{0.3}$ for the combined sample. In Section~\ref{sec:ds scaling r200} we have found $M_{200} \propto L^{1.2}$ for the combined sample. Although NFW and BBS masses are not the same, we expect similar luminosity scalings to hold. We therefore, in agreement with \citet{hoekstra04}, assume $M_{\rm BBS} \propto L^{1.2}$ to hold. This then (using equation~\ref{eq:eta_m}) implies $\eta_s = 0.6$.
\\
With these assumptions and if we ignore luminosity evolution we obtain for the reference $\rm{L^{*}}$-galaxy a velocity dispersion of $\sigma^{*} = 131^{+2}_{-2}$ km\ $\rm{s^{-1}}$ and a truncation radius of $s^{*} = 184^{+17}_{-14} \ h^{-1}$ kpc, implying a BBS mass of $M^{*}_{\rm total,BBS} = 2.4^{+0.3}_{-0.2} \times 10^{12} \ h^{-1} \ \rm{M_{\odot}}$ (measuring the haloes out to scales of $2\ h^{-1}$ Mpc). The results for the case of luminosity evolution are given in Table~\ref{tab:maximum_likelihood}.
\\
The left panel of Fig.~\ref{fig:lsigma} showed that the luminosity  vs. $\sigma$ scaling of the combined lens samples differs from that of pure red or blue galaxy samples, as it additionally traces the transition from massive red SED-dominated galaxies to low-mass blue SED-dominated galaxies. We therefore separately investigate the scaling relations for red and blue galaxies. In the analysis of red and blue galaxies, only for one galaxy type at a time the best-fitting parameters are examined, while for the other galaxy type the lensing signal is calculated with adjusted values for the fiducial quantities, as e.g. the velocity dispersion $\sigma$ and the scaling parameter $\eta_{\sigma}$. I.e. we assume fixed values $\sigma^{*}$ and $\eta_{\sigma}$ for the blue galaxies and run a maximum likelihood for the corresponding values for red galaxies, and vice verse. These results then are each fed into a further iteration for the maximum likelihood analysis, repeating the procedure until the calculations converge and do not deliver results deviating from the previous iteration step. We obtain values of $\eta_{\sigma}^{\rm red}=0.27^{+0.03}_{-0.04}$ for red and $\eta_{\sigma}^{\rm blue}=0.27^{+0.03}_{-0.02}$ for blue lenses (see Fig.~\ref{fig:WIDE.redblue.likely.eta}).
\\
\\
We now fix the slope of the red and blue samples to  $\eta_{\sigma}^{\rm red} = \eta_{\sigma}^{\rm blue} = 0.25$ and measure the velocity dispersion and truncation radius of the red and blue reference galaxies. For this we keep the assumption regarding size  vs. velocity dispersion scaling of $ s \propto \sigma^2$. Using $\eta_s = 0.5$ in equation~(\ref{eq:eta_m}) then gives a M/L-ratio independent from luminosity for both the red and blue galaxy samples.
\\
The analysis of the red lens sample yields a velocity dispersion of $\sigma^{*}_{\rm red} = 149^{+3}_{-3}$ km\ $\rm{s^{-1}}$ and a truncation radius $s^{*}_{\rm red} = 337^{+43}_{-37} \ h^{-1}$ kpc, implying a total mass of $M^{*,\rm red}_{\rm total,BBS} = 5.5^{+0.9}_{-0.8} \times 10^{12} \ h^{-1} \ \rm{M_{\odot}}$. For the blue lens sample we obtain a velocity dispersion of \mbox{$\sigma^{*}_{\rm blue} = 118^{+4}_{-5}$ km\ $\rm{s^{-1}}$} and a truncation radius $s^{*}_{\rm blue} = 84^{+13}_{-14} \ h^{-1}$ kpc and a total mass of $M^{*,\rm blue}_{\rm total,BBS} = 8.6^{+1.9}_{-2.2} \times 10^{11} \ h^{-1} \ \rm{ M_{\odot}}$ These results are also visualised in the upper left panel of Fig.~\ref{fig:WIDE.likely.bbs}.
\\
We measure the same quantities for galaxies in high and low density environments. It can be seen in the lower panels of Fig.~\ref{fig:WIDE.likely.bbs}  that the environment hardly influences the velocity dispersion of the galaxies.
This means that the central matter density of galaxies mostly depends on their luminosities and hardly on their environment. The truncation radius however, increases significantly with the density of the environment  (see Fig.~\ref{fig:WIDE.likely.bbs} as well) for galaxies of both SED-types. The best-fitting values for the different samples are shown in Table~\ref{tab:maximum_likelihood}.
\\
In general dense environments make it more difficult to measure the truncation radius $s$, because at distances where the `end' of the halo becomes measurable many other haloes then contribute to the combined signal. Further we would like to point out that the interpretation of the BBS-mass (and thus the size $s$) is that it gives the total mass for a halo hosting a galaxy with a central density profile with amplitude $\sigma$. Since in dense environments there is also dark matter on scales of groups and clusters the mean mass (and thus the truncation radius) associated to galaxies grows.
\subsubsection{Universal density profile (NFW)}
\label{sec:NFW}
\begin{figure*}
\includegraphics[width=18.5cm]{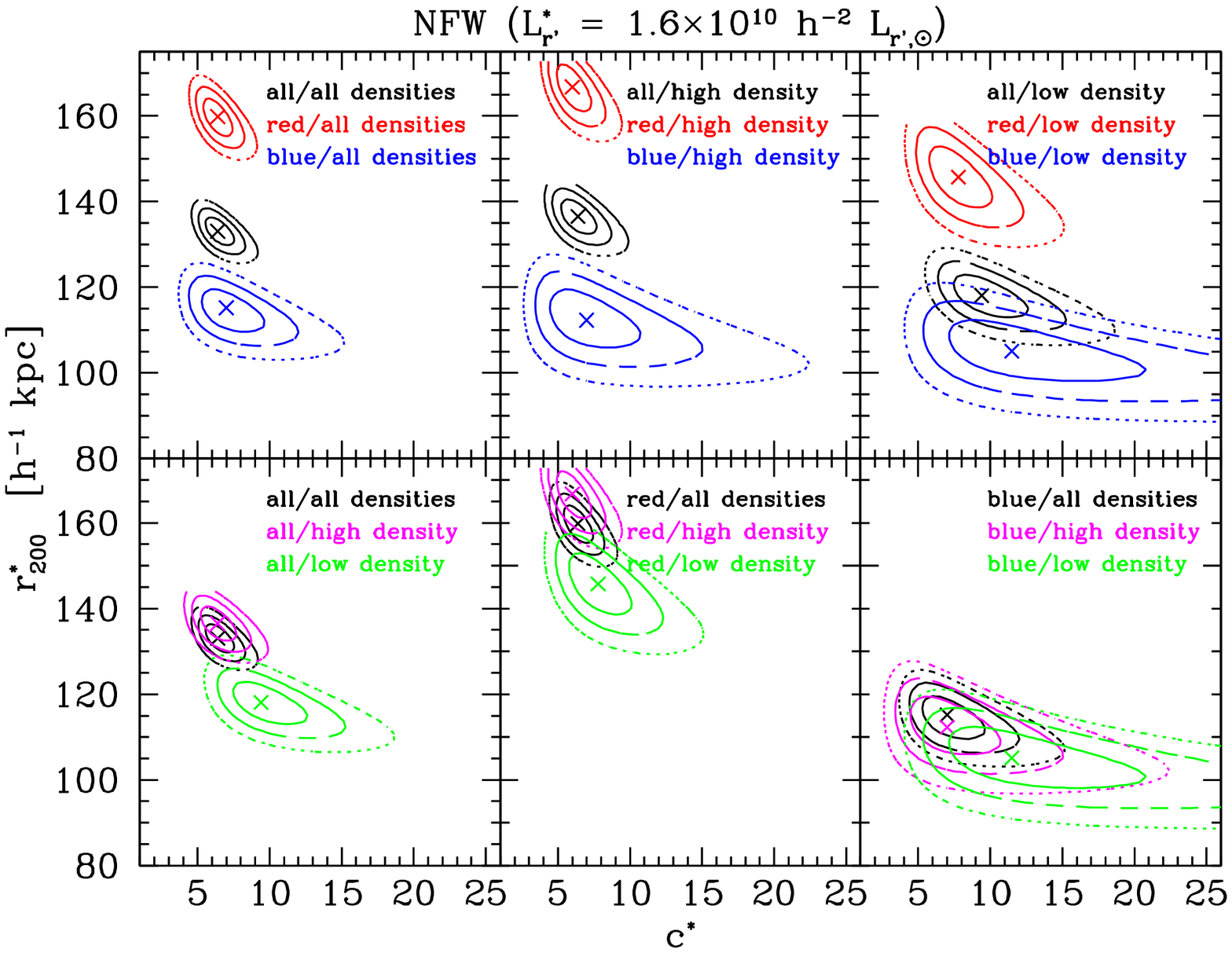}
\caption{Maximum likelihood results for  the NFW profile. The concentration parameter $c^{*}$ is shown on the x-axis, the $\rm{r^{*}_{200}}$ is shown on the y-axis. The contours show the 68.3 per cent, the 95.4 per cent and the 99.7 per cent confidence levels. 
In the upper row different galaxy types are compared for the one environment in each panel: black contours show the combined galaxy lens sample, red contours the red and blue contours show the blue galaxy lens sample. The left panel shows the galaxies in all environments, the middle one shows the galaxies in high density environment and the right panel shows the galaxies in low density environment.
In the lower row we compare the results for identical galaxy types in different environments, black is all environments, magenta is high density environment and green is low density environment. The left panel shows all galaxy types, the middle one shows red galaxies and the right panel shows blue galaxies. The concentration shows no significant difference comparing the different galaxy types at the reference luminosity, but the $r_{200}$ is significantly higher for red galaxies than for blue galaxies, also increasing with environmental density for all SED types.}
\label{fig:WIDE.likely.nfw}
\end{figure*}
\begin{figure*}
\includegraphics[width=18.5cm]{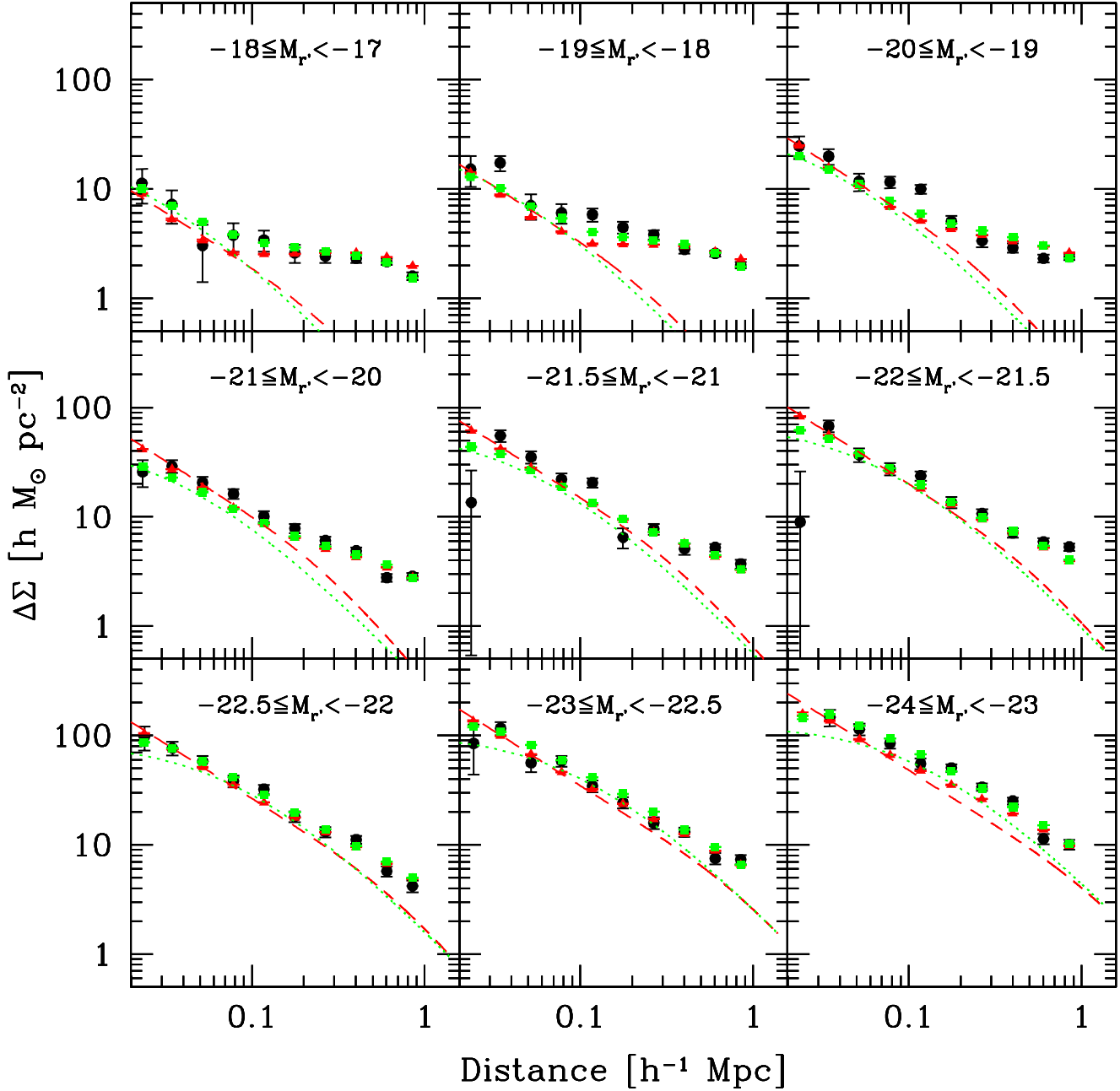}
\caption{Excess surface mass density $\Delta\Sigma$ for different luminosity bins. Black circles are the observational data points, red triangles result from a BBS simulation, green squares stand for the NFW simulation, based on maximum likelihood analyses (see Section~\ref{sec:likelihood theory}). The exact procedure for the simulation is described in Appendix~\ref{sec:simulations}. The dashed red and dotted green lines show the analytic single dark matter haloes scaled to corresponding luminosity for the BBS and NFW profile, respectively.}
\label{fig:WIDE.total.ds.sim}
\end{figure*}

In Section~\ref{sec:ds scaling r200} we carried out an NFW fit to the $\Delta\Sigma$-profiles on scales $R\le 100\ h^{-1}$ kpc and obtained a scaling compatible with $r_{200} \propto L^{0.4}$ for the combined galaxy sample and $r_{200} \propto L^{1/3}$ for the red and blue galaxy samples. We now fix these slopes, i.e. we assume the mass-luminosity relation to scale like this for the overall luminosity range (single power law) and we also assume the concentration-luminosity relation of \citet{duffy08} (see equation~\ref{eq:eta_c}) to hold. The maximum likelihood analysis for the combined sample yields $c^{*} = 6.4^{+0.9}_{-0.7}$ and a virial radius of $r^{*}_{200} = 133^{+3}_{-2} \ h^{-1}$ kpc.  This translates into a virial mass of $M_{\rm 200} = 7.6^{+0.5}_{-0.3} \times 10^{11} \ h^{-1} \ \rm{M_{\odot}}$. 
\\
As next step we measure the parameters for the SED types separately. The procedure is the same as for the BBS profile: we fix one galaxy type with adjusted fiducial parameter values (assuming \mbox{M/L=constant}) and obtain the best-fitting parameters for the other galaxy type, iterating the analyses until convergence.
For red SED lenses we measure \mbox{$c^{*,\rm red} = 6.4^{+0.7}_{-0.8}$} and a virial radius of \mbox{$r^{*,\rm red}_{200} \ = 160^{+3}_{-4} \ h^{-1}$ kpc}, leading to a virial mass of \mbox{$M^{*,\rm red}_{\rm 200} = 1.2^{+0.1}_{-0.1} \times 10^{12} \ h^{-1} \ \rm{M_{\odot}}$}. For blue galaxies, we measure \mbox{$c^{*,\rm blue} = 7.0^{+1.9}_{-1.6}$} and a virial radius \mbox{$r^{*,\rm blue}_{200} = 115^{+4}_{-5} \ h^{-1}$ kpc}, resulting in a virial mass of $M^{*,\rm blue}_{\rm 200} = 5.0^{+0.5}_{-0.6} \times 10^{11} \ h^{-1} \ \rm{M_{\odot}}$. The virial radii of the red galaxies are significantly larger than for the blue galaxies, leading to the significant mass difference, as already measured for the BBS profile. This also can be seen in the upper row of Fig.~\ref{fig:WIDE.likely.nfw}. For the (same) fiducial luminosity the halo concentration thus is similar for red and blue galaxies. But since the virial radii and thus the virial masses for the red galaxies are significantly larger than for blue galaxies (of the same luminosity), the concentration is higher for massive red galaxies, if galaxies with the same masses are compared. If we scale the mass of a blue galaxy to a value of $M^{*, \rm red}_{200}=12.4 \times 10^{11}\ h^{-1}\ M_{\odot}$, we obtain a concentration parameter of $c^{\rm blue} = 3.1^{+0.9}_{-0.7}$ for this galaxy. For less massive galaxies ($M_{200} < M^{*,\rm blue}_{200}$) this relations inverts. The results of the maximum likelihood analyses are summarised in Table~\ref{tab:maximum_likelihood}.
\\
Investigating the dependence of NFW haloes on the environment (see also the lower panels of Fig.~\ref{fig:WIDE.likely.nfw}) shows that galaxies in high density environments hardly differ in concentration and only have a slightly higher virial radius than galaxies in average density environments, but that galaxies in low density environments have a significantly lower virial radius and also show a tendency of having a higher concentration (especially blue lens galaxies) than in denser environments.  The best-fitting values for the NFW-likelihood analyses are shown in Table~\ref{tab:maximum_likelihood}.
\begin{table*}
\centering
\caption{}
Without luminosity evolution, $L^{*}_{r'}=1.6\times10^{10}\ h^{-2}\  L_{r',\odot}$
\begin{tabular}{c|c|c|c|c|c|c|c|c|c} 
\hline \hline
type & density &  $\sigma^{*}$ & $s^{*}$ & $c^{*}$ & $r^{*}_{200}$ & $M^{*}_{\rm total,BBS}$ & $M^{*}_{\rm 200}$ & $M^{*}_{\rm BBS}(r^{*}_{200})$ & $M^{*}_{\rm total,BBS}/L^{*}$ \\ 
 & & $[\rm{km\ s^{-1}}]$ & $[h^{-1} \ \rm{kpc}]$ & & $[h^{-1} \ \rm{kpc}]$ & $[10^{11} \ h^{-1} \ \rm{M_{\odot}}]$ & $[10^{11} \ h^{-1} \ \rm{M_{\odot}}]$ & $[10^{11} \ h^{-1} \ \rm{M_{\odot}}]$ & $[h \ \rm{M_{\odot}/L_{r',\odot}}]$ \\
\hline \hline
all  & all & $131^{+2}_{-2}$ & $184^{+17}_{-14}$ & $6.4^{+0.9}_{-0.7}$ & $133^{+3}_{-2}$ & $23.2^{+2.8}_{-2.5}$ & $7.6^{+0.5}_{-0.3}$ & $9.2^{+1.1}_{-0.9}$ & $178^{+22}_{-19}$\\
all  & high & $131^{+3}_{-3}$ & $256^{+24}_{-26}$ & $6.4^{+1.0}_{-1.0}$ & $137^{+3}_{-3}$ & $32.2^{+4.5}_{-4.8}$ & $8.3^{+0.5}_{-0.5}$ & $10.1^{+1.4}_{-1.4}$ & $248^{+35}_{-37}$ \\
all  & low & $131^{+4}_{-5}$ & $96^{+15}_{-15}$ & $9.4^{+2.4}_{-1.7}$ & $118^{+4}_{-4}$ & $12.1^{+2.6}_{-2.8}$ & $5.3^{+0.5}_{-0.5}$ & $6.8^{+1.4}_{-1.5}$ & $93^{+20}_{-22}$ \\
\hline
red  & all & $149^{+3}_{-3}$ & $337^{+43}_{-37}$ & $6.4^{+0.7}_{-0.8}$ & $160^{+3}_{-4}$ & $54.9^{+9.2}_{-8.2}$ & $12.4^{+0.7}_{-0.9}$ & $15.5^{+2.6}_{-1.7}$ & $422^{+71}_{-63}$ \\
red  & high & $150^{+3}_{-4}$ & $464^{+75}_{-68}$ & $6.0^{+1.0}_{-0.9}$ & $167^{+4}_{-5}$ & $76.6^{+15.4}_{-15.3}$ & $14.1^{+1.0}_{-1.3}$ & $16.8^{+3.4}_{-2.4}$ & $589^{+119}_{-118}$\\
red  & low & $144^{+5}_{-6}$ & $245^{+64}_{-52}$ & $7.8^{+1.6}_{-1.7}$ & $146^{+5}_{-6}$ & $37.3^{+12.3}_{-10.5}$ & $9.4^{+1.0}_{-1.2}$ & $12.8^{+4.0}_{-2.9}$ & $287^{+95}_{-85}$\\
\hline
blue  & all & $118^{+4}_{-5}$ & $84^{+13}_{-14}$ & $7.0^{+1.9}_{-1.6}$ & $115^{+4}_{-5}$ & $8.6^{+1.9}_{-2.2}$ & $5.0^{+0.5}_{-0.6}$ & $5.1^{+1.1}_{-1.3}$ & $66^{+15}_{-17}$ \\
blue  & high & $114^{+5}_{-6}$ & $107^{+22}_{-23}$ & $7.0^{+3.1}_{-2.3}$ & $112^{+6}_{-6}$ & $10.2^{+3.0}_{-3.3}$ & $4.6^{+0.7}_{-0.7}$ & $5.3^{+1.5}_{-1.7}$ & $78^{+23}_{-25}$ \\
blue  & low & $126^{+8}_{-9}$ & $40^{+11}_{-8}$ & $11.5^{+6.5}_{-3.9}$ & $105^{+6}_{-6}$ & $4.7^{+1.9}_{-1.6}$ & $3.8^{+0.6}_{-0.6}$ & $3.6^{+1.4}_{-1.3}$ & $36^{+14}_{-12}$ \\
\hline\hline
\end{tabular}
\newline
\newline
With luminosity evolution, $L^{*}_{r'}=1.6\times10^{10}\ h^{-2}\  L_{r',\odot} \times (1+z)$
\\
\begin{tabular}{c|c|c|c|c|c|c|c|c|c} 
\hline \hline
type &  density &  $\sigma^{*}$ & $s^{*}$ & $c^{*}$ & $r^{*}_{200}$ & $M^{*}_{\rm total,BBS}$ & $M^{*}_{\rm 200}$ & $M^{*}_{\rm BBS}(r^{*}_{200})$ & $M^{*}_{\rm total,BBS}/L^{*}$ \\ 
 & & $\rm{[km\ s^{-1}]}$ & $[h^{-1} \ \rm{kpc}]$ & & $[h^{-1} \ \rm{kpc}]$ & $[10^{11} \ h^{-1} \ \rm{M_{\odot}}]$ & $[10^{11} \ h^{-1} \ \rm{M_{\odot}}]$ & $[10^{11} \ h^{-1} \ \rm{M_{\odot}}]$ & $[h \ \rm{M_{\odot}/L_{r',\odot}}]$ \\
\hline \hline
all  & all & $144^{+3}_{-2}$ & $253^{+23}_{-20}$ & $5.4^{+0.8}_{-0.6}$ & $158^{+3}_{-2}$ & $38.5^{+5.1}_{-4.1}$ & $12.7^{+0.7}_{-0.5}$ & $13.7^{+1.8}_{-1.1}$ & $296^{+39}_{-32}$\\
red  & all & $161^{+3}_{-3}$ & $414^{+49}_{-48}$ & $6.2^{+0.8}_{-0.7}$ & $183^{+4}_{-4}$ & $78.7^{+12.3}_{-12.1}$ & $18.6^{+1.2}_{-1.2}$ & $20.9^{+3.2}_{-2.1}$ & $605^{+95}_{-93}$ \\
blue & all & $126^{+5}_{-5}$ & $108^{+19}_{-17}$ & $6.2^{+1.7}_{-1.4}$ & $135^{+4}_{-6}$ & $12.6^{+3.2}_{-3.0}$ & $8.0^{+0.7}_{-1.1}$ & $7.2^{+1.8}_{-1.5}$ & $97^{+25}_{-23}$ \\
\hline\hline
\end{tabular}
\newline
\label{tab:maximum_likelihood}
Best-fitting parameter values from the maximum likelihood analyses for the different galaxy SEDs and environmental densities, velocity dispersion $\sigma^{*}$ and truncation radius $s^{*}$ for a BBS profile, concentration parameter $c^{*}$ and virial radius $r^{*}_{200}$ for an NFW profile, the corresponding total mass (BBS) and virial mass (NFW) and the total M/L-ratio based on the total BBS mass. The upper table shows the values assuming no luminosity evolution with redshift, the lower table shows the values assuming luminosity evolution with $L \propto (1+z)$.

\end{table*}
\\
We finally check the consistency of our BBS and NFW analyses by comparing the corresponding (approximate) virial masses. We calculate the halo masses based on the profile parameters within the $r_{200}$ (obtained from the NFW fitting). For the maximum lens sample the BBS profile yields $M^{*}_{\rm BBS}(r^{*}_{200}) = 9.2^{+1.1}_{-0.9} \times 10^{12} h^{-1} \ \rm{M_{\odot}}$, which is only slightly higher than the corresponding $M^{*}_{200}$ from the NFW analysis. For the total lens sample in high and low density environment we measure a mass of \mbox{$M^{*}_{\rm BBS}(r^{*}_{200}) = 10.1^{+1.4}_{-1.4} \times 10^{11} h^{-1} \ \rm{M_{\odot}}$} and \mbox{$M^{*}_{\rm BBS}(r^{*}_{200}) = 6.8^{+1.4}_{-1.5} \times 10^{11} h^{-1} \ \rm{M_{\odot}}$}, respectively, which is still slightly higher than the corresponding NFW values, but agrees well within the uncertainties.
\\
The same comparison for the red and blue galaxy lens samples gives analogous results, as can be seen in Table~\ref{tab:maximum_likelihood}. The same is true for results obtained assuming $L \propto (1+z)$, see Table~\ref{tab:maximum_likelihood}.
In general the BBS profiles give slightly higher values for the masses than the NFW profiles. This is consistent with the results \citet{wright00} who obtained that fitting isothermal profiles to NFW profiles leads to higher mass estimates for the same given gravitational shear within the virial radius.
\\
In Fig.~\ref{fig:WIDE.total.ds.sim} we show how well the $\Delta\Sigma(R)$ predictions based on the best fit model parameters for the $L^*$-galaxies and the assumed scaling relations fit to the observed $\Delta\Sigma$-profile (black). Here the red symbols show the $\Delta\Sigma$-profiles based on the BBS simulation whereas the green symbols show the case for the NFW simulation. The dashed red and green curves show the analytic BBS and NFW single haloes scaled to correspond to the considered luminosity bins.
\subsubsection{Scaling relations for the M/L ratio and the concentration parameter $c$}
\label{sec:scaling relations}
\begin{figure*}
\includegraphics[width=1.02\textwidth]{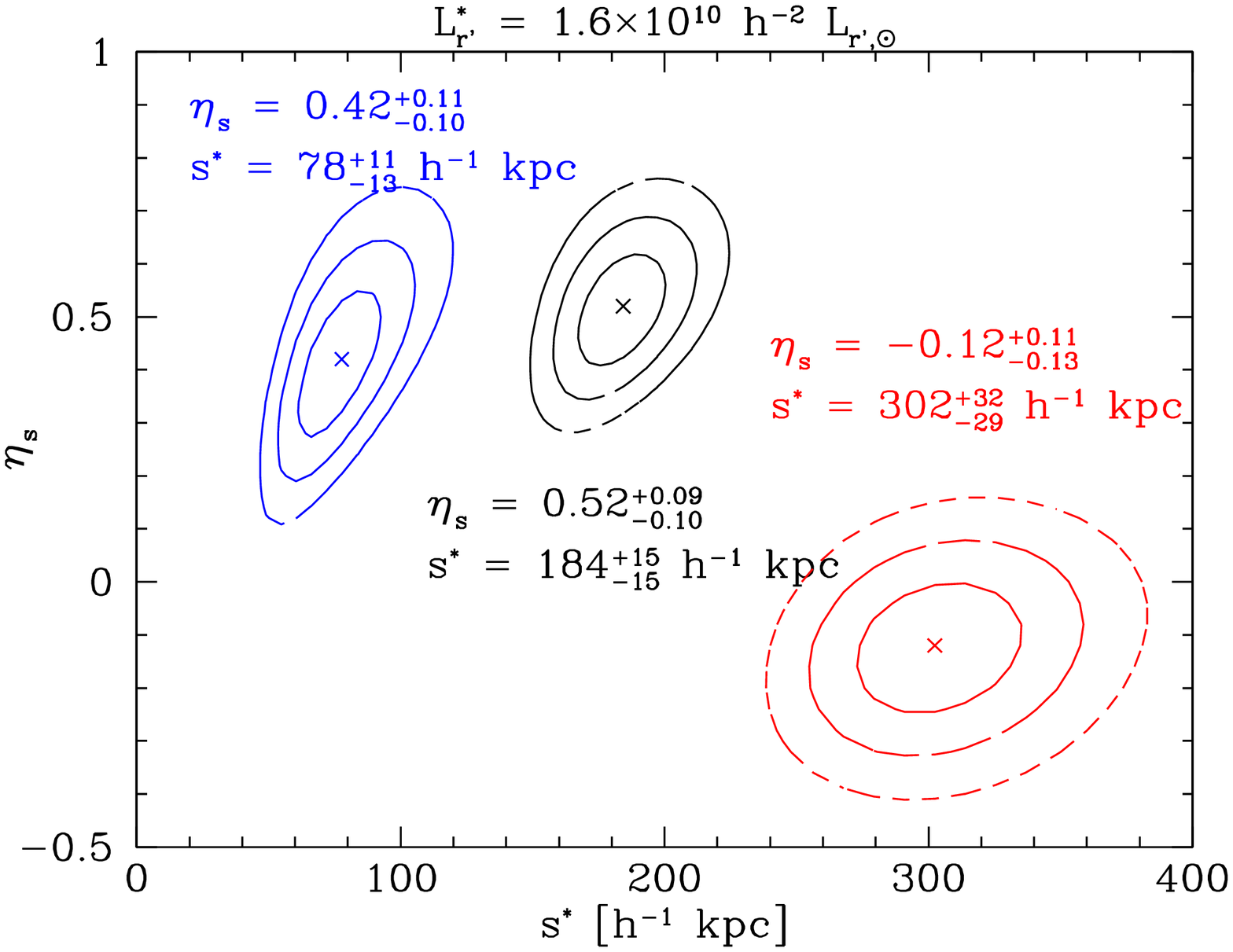}
\caption{Scaling relation for the BBS truncation radius $s$ with luminosity. The blue contours on the left show the constraints for blue galaxies, the red contours on the right show the constraints for red galaxies and the black contours in the middle show the combined galaxy sample. The contours indicate the 68.3 per cent (solid), the 95.4 per cent (dashed) and the 99.7 per cent (dotted) confidence levels, the cross indicates the best-fit value.}
\label{fig:WIDE.redblue.ml}
\end{figure*}

Until now in our analyses we assumed a scaling relation of $M_{\rm BBS}/L \propto  L^{0.2}$
(see equation~\ref{eq:scaling M/L})
. Given the stability of the $L-\sigma$-relation we now keep the slope $\eta_{\sigma}$ fixed and additionally fix the velocity dispersion to a value of \mbox{$\sigma^{*} = 131$ km\ $\rm{s^{-1}}$} for an $L^{*}$-galaxy and run a BBS maximum likelihood analysis to measure the scaling index $\eta_{s}$ for the truncation radius $s$. We find $\eta_{s} = 0.52^{+0.09}_{-0.10}$ for $s$ (see black contours in Fig.~\ref{fig:WIDE.redblue.ml}). As shown in equation~(\ref{eq:massBBStotal}) the total BBS-mass scales as $M \propto \sigma^2  s $, from one obtains the scaling relation for the mass-to-light ratio:
\begin{equation}
\eta_{M/L} = 2 \eta_{\sigma} + \eta_{s} - 1 \ ,
\end{equation}
(see also equation~\ref{eq:eta_m}).
Combined with the scaling behavior of the velocity dispersion $\eta_{\sigma} = 0.30\pm0.02$ we obtain a scaling of the mass-to-light ratio of 
\begin{equation}
\left(M/L\right) \propto  L^{0.12^{+0.10}_{-0.11}}
\end{equation}
for the combined sample.
\begin{figure*}
\includegraphics[width=1.02\textwidth]{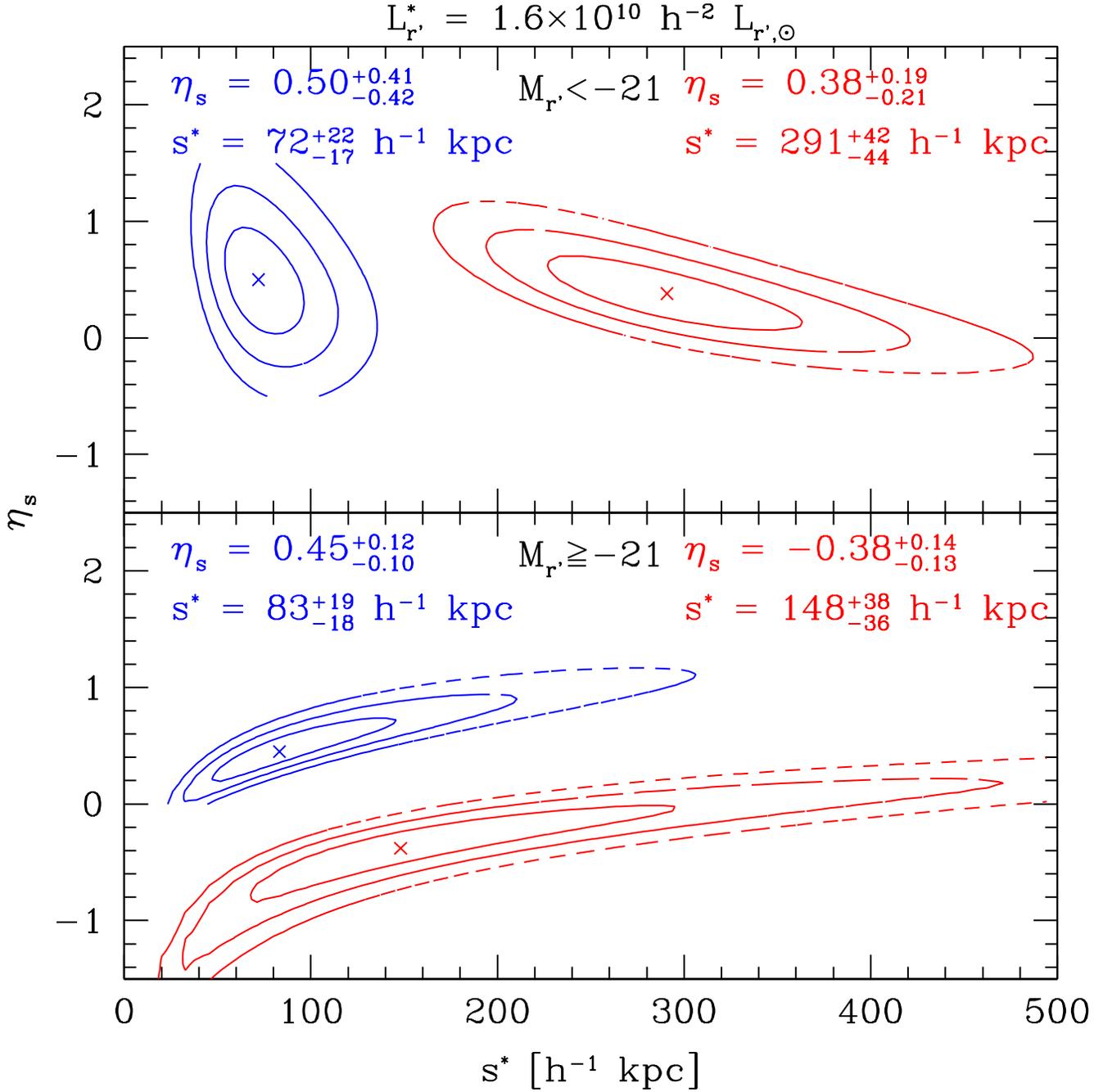}
\caption{Scaling relations for the BBS truncation radius $s$ distinguishing between bright and faint foreground lenses. Blue galaxies are shown as blue contours in the left side of both panels, red galaxies in the right in red contours. The upper panel shows the scaling behavior for bright galaxies with $M_{r'}<-21$, the lower panel the faint lens sample with $M_{r'}\ge-21$. We see that the results for bright ($\eta_s=0.50^{+0.41}_{-0.42}$) and faint ($\eta_s=0.45^{+0.12}_{-0.10}$) blue galaxies are consistent with each other, following the predicted scaling behavior of $s \propto L^{0.5}$. We also see that for bright red galaxies the scaling behavior ($\eta_s=0.38^{+0.19}_{-0.21}$) now also agrees with our expectations within the uncertainties, while the deviating scaling behavior of the complete red lens sample is caused by the faint lens fraction, as we see in the lower panel ($\eta_s=-0.38^{+0.14}_{-0.13}$).}
\label{fig:WIDE.redblue.ml.bf}
\end{figure*}
\begin{figure*}
\includegraphics[width=8.8cm]{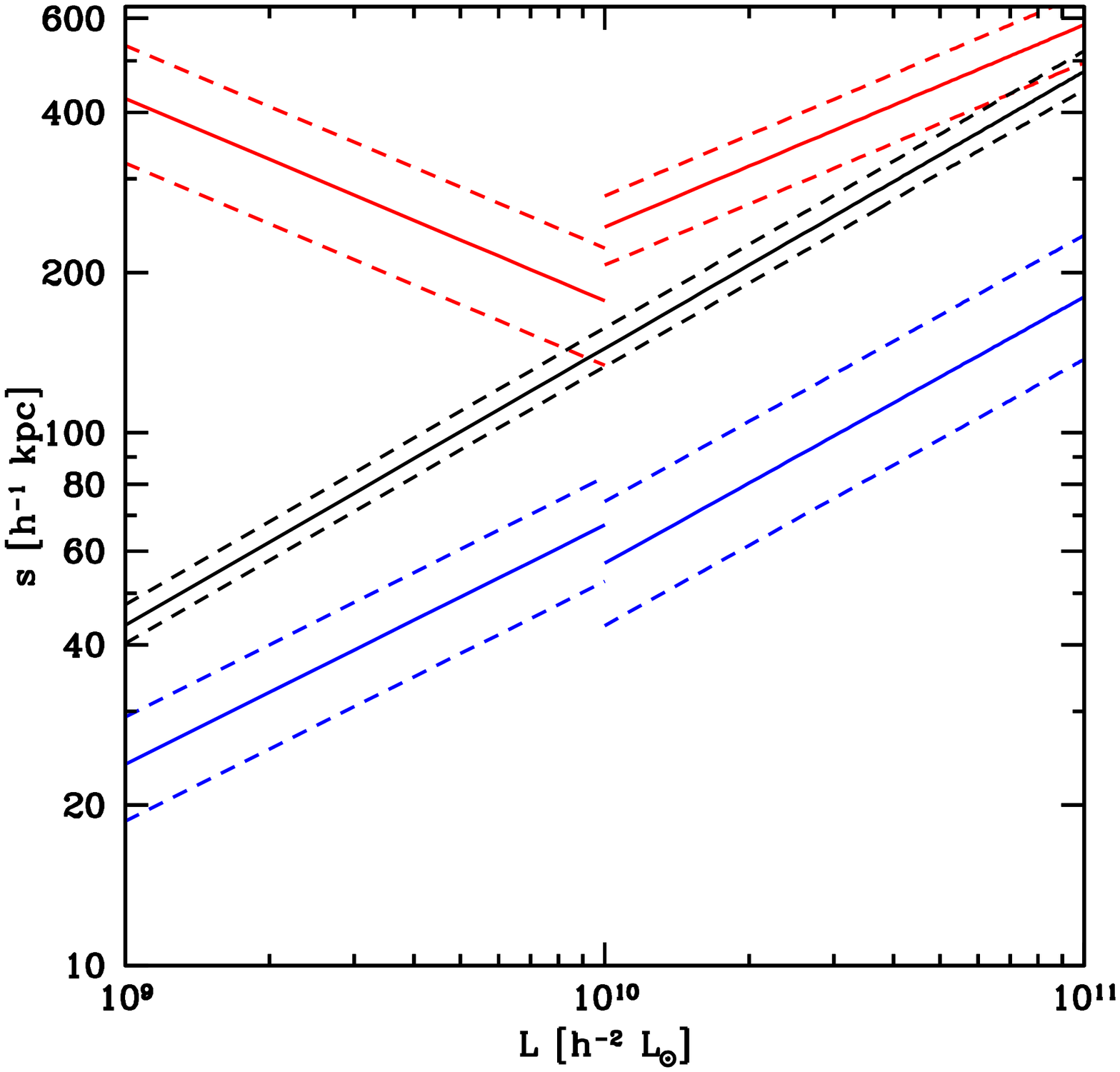}
\includegraphics[width=8.8cm]{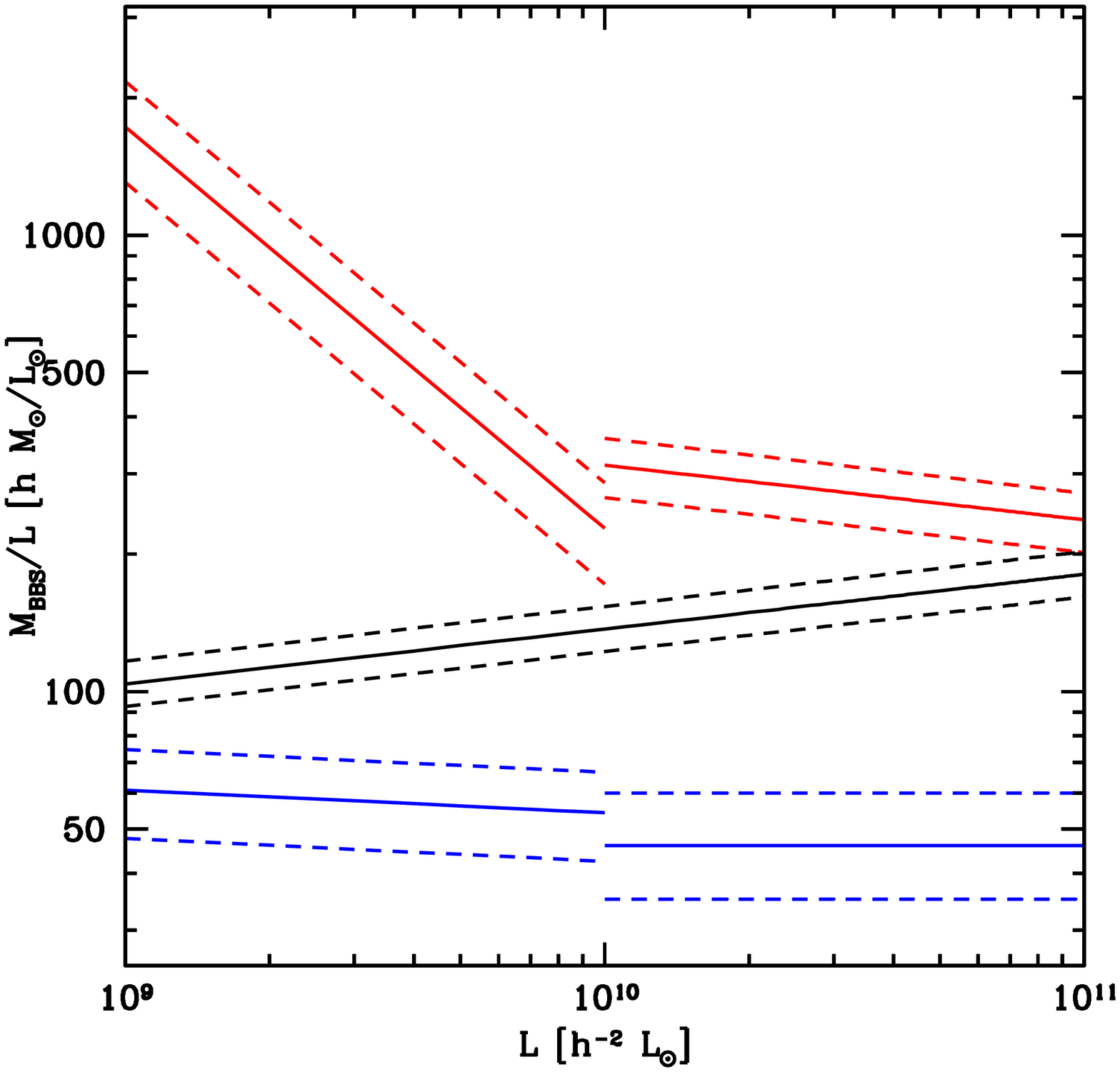}
\caption{Truncation radius s (left panel) and M/L-ratio (right panel), calculated for red (in red) and blue galaxies (in blue) according the scaling relation from Fig.~\ref{fig:WIDE.redblue.ml.bf}. The solid lines denote the best fit, the dashed lines denote the 68.3 per cent-confidence levels (based on the uncertainties of the halo parameters and not the scaling relation). The negative s-slope for faint and positive s-slope for blue galaxies lead to a strong decrease in the M/L-ratio for faint red galaxies and a weak further decrease for bright red galaxies, while the universally positive $s$-slope for blue galaxies lead to an approximately luminosity-independent M/L-ratio. Within its uncertainties the slope for the brighter galaxies is consistent with values slightly larger than $\eta_{\rm s}=0.5$, which would correspond to a slight increase of the M/L-ratio with luminosity. The discontinuity at $L=10^{10}\ h^{-1}\ L_{\odot}$ is a numerical artefact, since we allow for two different scaling relations in the two luminosity intervals. We do not require that the two relations agree at the point of connection (the 1 $\sigma$ error intervals are shown as dashed lines). We included for comparison the single-power scaling for the combined sample in both panels in black ($s \propto L^{0.52}$, $M/L \propto L^{1.12}$, see Fig.~\ref{fig:WIDE.redblue.ml}).}
\label{fig:ml-broken}
\end{figure*}
\\
The calculation of the scaling relation of $s$ for individual galaxy types is challenging and requires a large galaxy sample to be statistically significant. This holds especially for galaxies in dense environments such as red galaxies, because in this case the signal of the outer halo in the $\Delta\Sigma(R)$-profile is superposed by the imprint of neighbouring haloes. Since the truncation radius itself is already difficult to measure, it is hardly possible to constrain the scaling relation unless one considers a very large sample of galaxies which are approximately undisturbed by neighbouring haloes. We nevertheless try to analyse the scaling relation for $s$ for red and blue galaxies separately and show the results in Fig.~\ref{fig:WIDE.redblue.ml}. We see that the scaling relation for the blue galaxy samples follows our expectations based on the result that $\eta_{s}^{\rm blue} = 0.42^{+0.11}_{-0.10}$, which corresponds to a scaling relation of $M/L \propto L^{\eta_{M/L}^{\rm blue}}$ with $\eta_{M/L}^{\rm blue} = -0.08^{+0.12}_{-0.11}$ (which is within the errors consistent to a constant mass-to-light ratio). However, considering the red galaxy sample, the result of $\eta_{s}^{\rm red} = -0.12^{+0.12}_{-0.13}$ is surprising and difficult to understand at the first glance. To determine the reason we now further investigate our red and blue lens samples as a function of their rest frame luminosity, splitting the samples into bright ($M_{r'}<-21$) and faint lens samples ($M_{r'} \ge -21$). As we see in Fig.~\ref{fig:WIDE.redblue.ml.bf} the results for the blue galaxies remain unchanged, both luminosity samples showing the same scaling behavior, i.e. $\eta_{\rm s,bright}^{\rm blue} = 0.50^{+0.41}_{-0.42}$ for bright compared to $\eta_{\rm s,faint}^{\rm blue} = 0.45^{+0.12}_{-0.10}$ for faint blue galaxies, roughly corresponding to luminosity-independent mass-to-light-ratio. 
For the red sample we find that the deviating scaling behavior seen in Fig.~\ref{fig:WIDE.redblue.ml} has its origin in the faint part of our red lens sample. While the bright red galaxies in principle scale as expected ($\eta_{\rm s,bright}^{\rm red} = 0.38^{+0.19}_{-0.21}$, corresponding to $\eta_{M/L,\rm bright}^{\rm red} = -0.12^{+0.20}_{-0.22}$, consistent with a constant mass-to-light ratio), the scaling behavior for the fainter fraction is inverted ($\eta_{\rm s,faint}^{\rm red} = -0.38^{+0.14}_{-0.13}$). 
Since the velocity dispersion is decreasing with decreasing luminosity this implies that the halo masses are only slowly decreasing with luminosity. This is in agreement with the halo mass vs. stellar mass relation from \citet{guo10} (based on abundance matching). Their fig.~2 (upper panel) shows that the halo mass decreases only slowly with stellar mass once the stellar mass is below $\sim 10^{10}\ h^{-2} \ M_{\odot}$.
To exclude that environment structure causes this rise of the truncation radius with decreasing luminosity we have repeated this analysis for the low density environment galaxy sample. We have summarised our results in Fig.~\ref{fig:ml-broken}: the M/L-ratio for red galaxies steeply decreases for increasing luminosities for galaxies fainter than $L \sim 10^{10}\ h^{-2}\ L^{\odot}$ and then turns into a further, but shallow decrease for more massive galaxies. In contrast, for blue galaxies the M/L-ratio is approximately constant for the whole luminosity range considered. The confidence intervals shown are based on the uncertainties of the halo parameters (for the best fit luminosity scaling parameters) and do not include the uncertainties for the luminosity scaling parameters themselves. Within its uncertainties the slope of the scaling relation is consistent with values slightly larger than $\eta_{\rm s}=0.5$, which would correspond to a slight increase of M/L-ratio with luminosity.
\\
\\
Further we consider the scaling relation for the concentration parameter $c$ of the NFW profile with luminosity. We perform a maximum likelihood analysis with fixed fiducial virial radius $r^{*}_{200}$, obtained from the previous NFW analysis, using the concentration parameter $c$ and its scaling exponent $\eta_c$ as fit parameters. For the combined lens sample we find a scaling relation of $c \propto  L^{-0.07^{+0.11}_{-0.11}}$ (see the black contours in Fig.~\ref{fig:WIDE.cm}). For a scaling of the $M_{200}/L$-ratio with $L^{0.2}$ this leads to $c \propto  M_{200}^{-0.06^{+0.09}_{-0.09}}$, for $M_{200}/L \propto  L^{0.12}$ to $c \propto  M_{200}^{0.06^{+0.10}_{-0.10}}$, both values being in agreement with the results of \citet{duffy08}. We also investigate possible differences in the scaling behavior for red and blue lenses, running separate examinations, and repeat the calculations with the results from the previous iteration until convergence, as described in the previous subsections.
Indeed the scaling behavior differs. While the concentration parameter of red galaxies hardly depends on the luminosity ($c \propto  L^{-0.04^{+0.10}_{-0.13}}$), the best fit for the slope of the c-L-relation of blue galaxies is negative ($c \propto L^{-0.34^{+0.24}_{-0.26}}$, see Fig.~\ref{fig:WIDE.cm}). However, the uncertainties for $\eta_{\rm c}^{\rm blue}$ are quite large and thus the confidence intervals for $\eta_{\rm c}^{\rm red}$ and $\eta_{\rm c}^{\rm blue}$ overlap.
\\
The concentration-luminosity relation can be easily translated into a concentration-mass relation, for a given luminosity-mass relation. We assume $M_{200} \propto L^{1.2}$, but in any case the exact value for the scaling exponent is not important, as the relation only weakly depends on it and hardly changes if e.g. $M_{200} \propto L$ is assumed. Using the best-fit slope for the c-L-scaling for given luminosity the mass of red galaxies is higher than for blue ones. Therefore the fiducial luminosity for the lens subsamples splits into two different fiducial masses. For massive galaxies, the concentration of red galaxies exceeds the concentration of their blue counterparts with same mass, but due to the steeper concentration-mass relation for blue galaxies, considering low-mass galaxies, this relation turns into its opposite.
\\
\\
\begin{figure*}
\includegraphics[width=18.5cm]{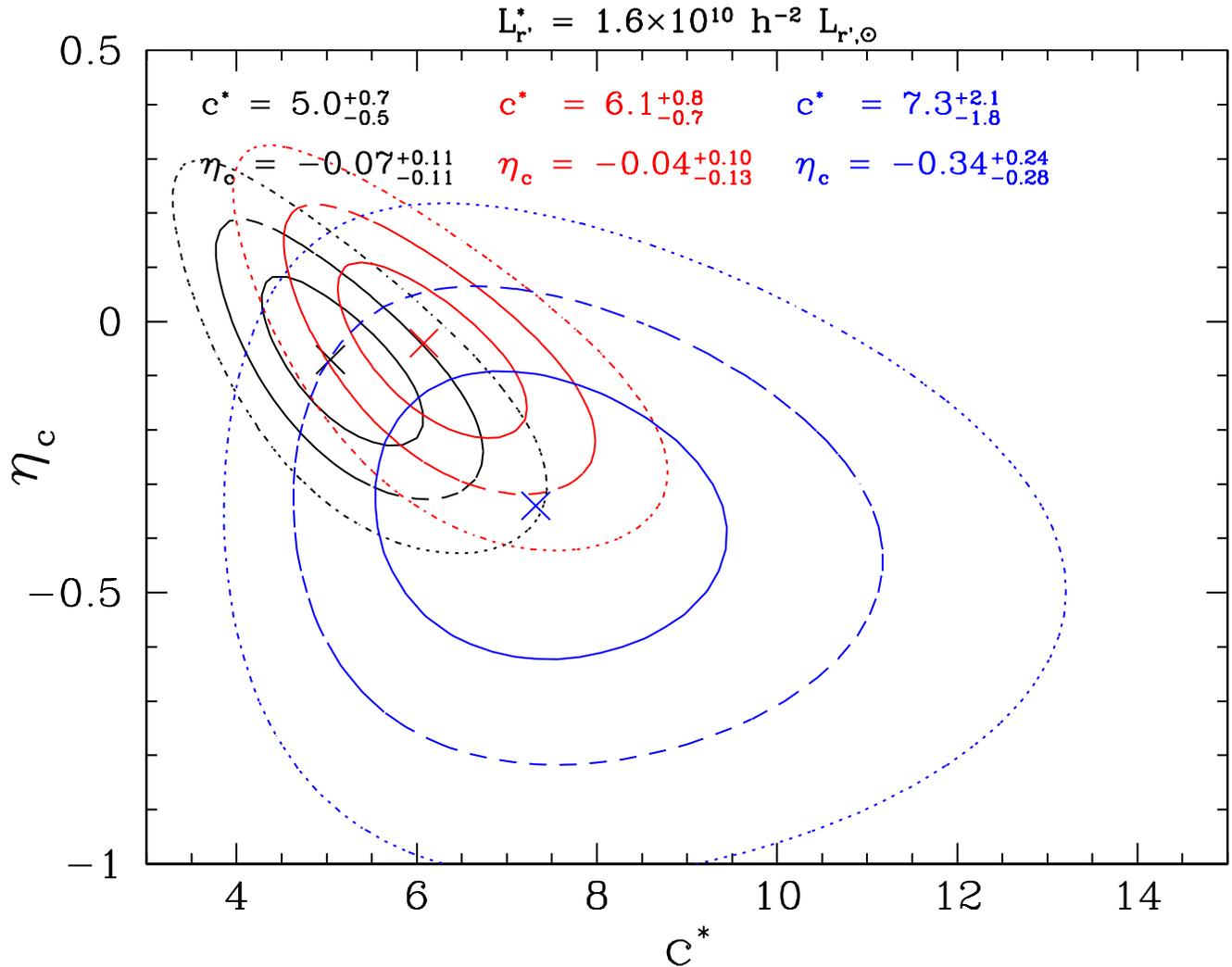}
\caption{Scaling relation for the concentration parameter $c$ with luminosity. The red contours show the scaling relation for red galaxies, the blue contours show the scaling relation for blue galaxies and the black contours for the combined galaxy sample. While for the combined and the red galaxy sample the result is consistent with a constant c-L-scaling, the best-fit value for the slope of the c-L-relation of blue galaxies is negative. However, the uncertainties are quite large and thus the confidence levels for $\eta^{\rm red}_{\rm c}$ and $\eta^{\rm blue}_{\rm c}$ overlap. The contours indicate the 68.3 per cent (solid), the 95.4 per cent (dashed) and the 97.3 per cent (dotted) confidence levels, the cross indicates the best-fit value.}
\label{fig:WIDE.cm}
\end{figure*}
As we have seen in Section~\ref{sec:DeltaSigma}, in Figs.~\ref{fig:L-r200} and \ref{fig:L-m200} the scaling behavior of $r_{200}$ deviated from a power law when including galaxies with luminosities $L<10^{10}\ h^{-1}\ L_{\odot}$. In principle there are two possibilities to explain this observation. Firstly one could assume a modification of the concentration-luminosity relation when moving to fainter galaxies, leading to increased values for $r_{200}$, when not accounted for. Secondly the scaling behavior of the mass-to-light ratio could change, leading to higher masses and thus $r_{200}$-values than expected with a single-power law. We will now further investigate both possibilities in more detail.
\\
\\
To investigate a change in the concentration-luminosity relation, explicitly assuming that the $M_{200}-L$ relation is described by a single-power law ($M_{200} \propto L$ for pure red and blue galaxies), we repeat our analysis and separately analyse the scaling behavior of the concentration parameter for galaxies brighter and fainter than $10^{10}\ h^{-2}\ L_{\odot}$. We fix the fiducial virial radius $r_{200}^{*}$ including its scaling relation with luminosity ($r_{200} \propto L^{1/3}$ for red and blue galaxies separately) and the fiducial concentration parameter $c^{*}$ and run maximum likelihood analyses, fitting independent slopes for the c-L relation for the brighter and for the fainter galaxies. Indeed we observe a different scaling behavior in both luminosity ranges. While for galaxies with $L > 10^{10}\ h^{-2}\ L_{\odot}$ the scaling corresponds to what we obtained for the all luminosities sample, for the fainter galaxies a value of $\eta_c < -1$ is indicated. In order to exclude that we observe an effect of remaining unproperly treated multiple deflections, we repeat the analysis for galaxies in low density environments, obtaining the same result. This means, if we assume that the $M_{200}/L$-ratio is described by a single-power law, the c-L relation comes significantly steeper for galaxies fainter than $L=10^{10}\ h^{-2}\ L^{\odot}$, leading to much higher values for the concentration.
\\
\\
We further consider the second possibility, assuming a change of slope for the $M_{200}/L$-ratio (see earlier in this section) and thus for $r_{200}$, as observed in Section~\ref{sec:ds scaling r200}.  We assume the concentration of \citet{duffy08} to hold and fit the $r_{200}$ and its scaling index $\eta_{r_{200}}$ for bright and faint galaxies independently) and obtain a significantly shallower \mbox{$r_{200}-L$ relation} for  galaxies fainter than $L_{r'} = 10^{10}\ h^{-2}\ L_{\odot}$, resulting in $r^{\rm red}_{200} \propto L^{0}$ for red and $r^{\rm blue}_{200} \propto L^{0.16}$ for blue galaxies). This implies an increase in mass-to-light ratio for faint galaxies with decreasing luminosities which is in consistence with the results of \citet{vandenbosch05} (see their fig.~3).
\subsubsection{Comparison to previous results}
\label{sec:disc}
\citet{parker07} analysed 22 $\rm{deg}^2$ of CFHTLS $i'$-band data, but without photometric redshift knowledge, choosing lens galaxies to have apparent magnitudes of $19 < i' < 22$ and source galaxies $22.5 < i' < 24$. Assuming a truncation radius of $s = 185 \pm 30 \ h^{-1}$ kpc,
based on the results of \citet{hoekstra04},
they found for a $\rm{L_{R_c}^{*}} = 1.3  \times 10^{10} \ h^{-2} \ \rm{L_{\odot}}$ galaxy a velocity dispersion of $\sigma^{*}=137 \pm 11$ km\ $\rm{s^{-1}}$ and a rest frame mass-to-light ratio of $173\pm34 \ h \ M_{\odot}/L_{\odot}$. Taking the higher value for our fiducial luminosity into account, our values of $\sigma^{*} = 131^{+2}_{-2}$ km\ $\rm{s^{-1}}$ and our mass-to-light ratio of $M_{\rm total,BBS}/L = 178^{+22}_{-19} \ h \ M_{\odot}/L_{\odot}$ are slightly lower, however consistent. For a luminosity evolved mainly red lens sample with luminosity $L_{R}=10^{10} \ h^{-1} \ L_{\odot}$ \citet{vanuitert11} found a virial mass of $M_{\rm vir} = 7.2 \ \pm 1.5 \times 10^{11}\ h^{-1} \ M_{\odot}$. If we convert our result for $M_{200}$ from the NFW likelihood for a pure red sample to their reference luminosity we obtain a value of \mbox{$M_{\rm vir} = 7.3 \ \pm 0.5\times 10^{11} \ h^{-1} \ M_{\odot}$} , being in good agreement with the result of \citet{vanuitert11}.
\subsubsection{Consistency of maximum likelihood and $\Delta\Sigma$ fit results}
\begin{figure*}
\includegraphics[width=8.8cm]{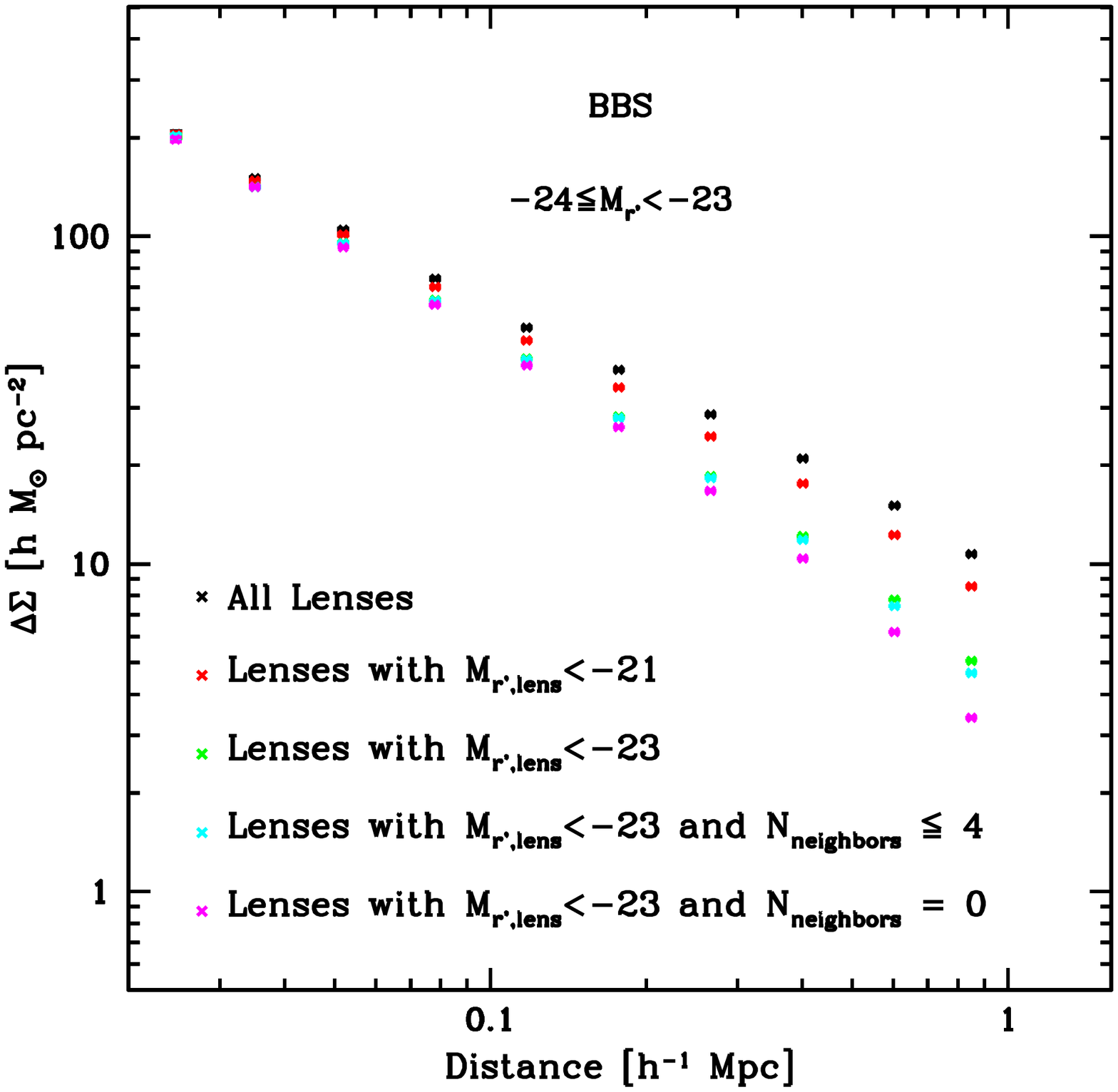}
\includegraphics[width=8.8cm]{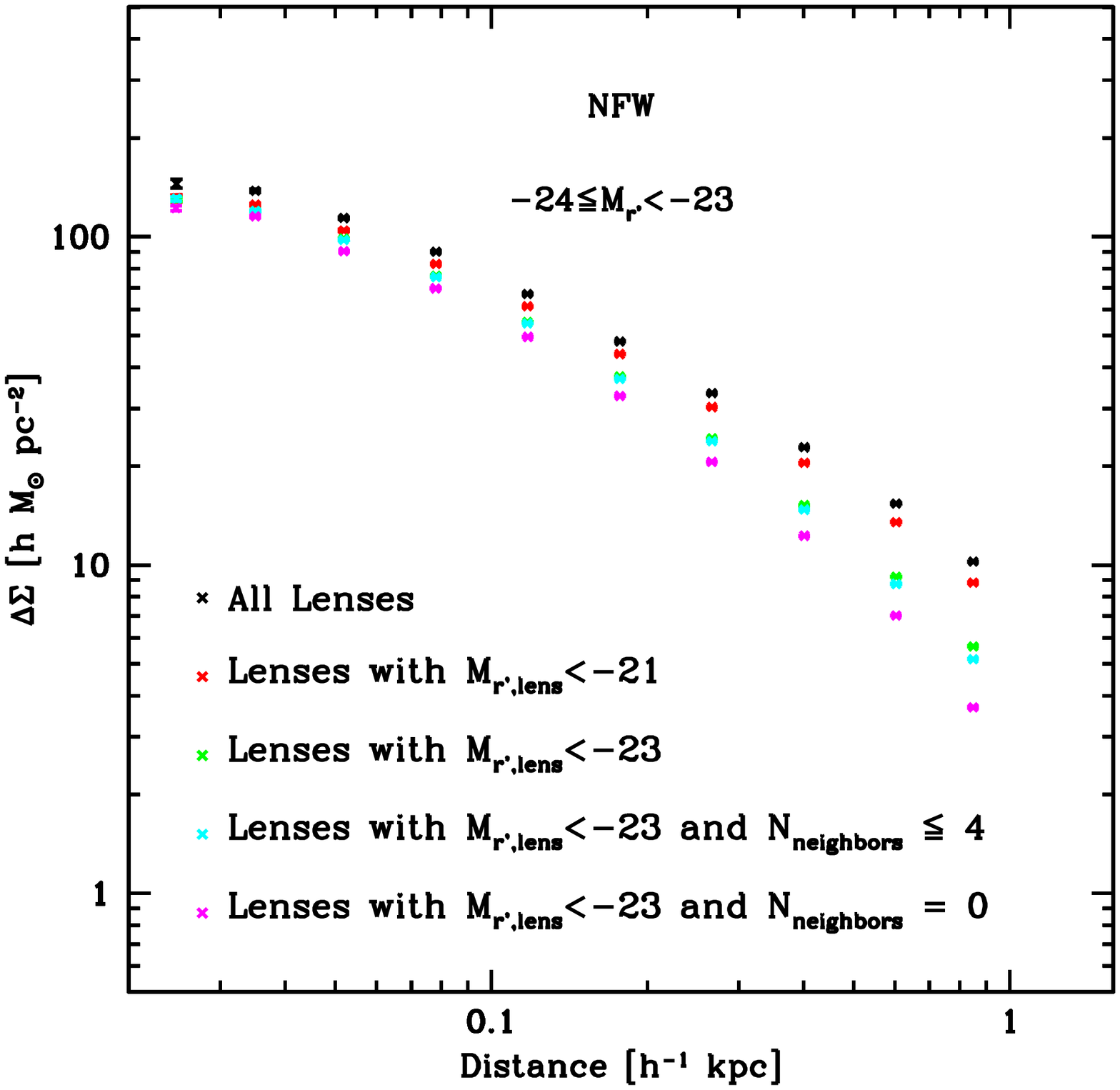}
\caption{Excess surface surface mass density $\Delta\Sigma$ extracted from lens galaxies with $-23 < M_{r'} \le -24$ calculated for simulated lens signals. The black dots show simulations including lenses with all magnitude, red dots only include lenses with $M_{r'}\le-21$, green dots only include lenses with $M_{r'}<-23$, cyan dots additionally restrict the lens sample to maximum neighbour number of 4 galaxies within a projected distance of $720\ h^{-1}$ kpc and finally the magenta plots only show those lenses without any neighbour within this projected distance. We see that in the case of a BBS profile (left panel) multiple deflection affect the signal amplitude only on larger scales, while assuming an NFW profile (right panel) the signal amplitude already at very low scale is biased high about 20 per cent.}
\label{fig:WIDE.ds.2423}
\end{figure*}

We compare our results from the maximum likelihood analyses and the measurements of the excess surface density $\Delta\Sigma$ (see Tables~\ref{tab:scaling relations} and \ref{tab:maximum_likelihood}) and find in general a good agreement. For an $L^{*}$-galaxy we obtain a velocity dispersion of $\sigma^{*}=135\pm2\ {\rm{km\ s^{-1}}}$ from $\Delta\Sigma$ and $\sigma^{*}=131^{+2}_{-2}\ {\rm{km\ s^{-1}}}$ from the maximum likelihood analysis for all galaxies. For blue galaxies we obtained values of  \mbox{$\sigma^{*}_{\rm blue}=115\pm3\ {\rm{km\ s^{-1}}}$} from $\Delta\Sigma$ and $\sigma^{*}_{\rm blue}=118^{+4}_{-5}\ {\rm{km\ s^{-1}}}$ from the likelihood analysis. The consistency for the red lens sample with \mbox{$\sigma^{*}_{\rm red}=162\pm2\ {\rm{km\ s^{-1}}}$} from $\Delta\Sigma$ and \mbox{$\sigma^{*}_{\rm red}=149^{+3}_{-3}\ {\rm{km\ s^{-1}}}$} from the maximum likelihood is only marginal. If we include luminosity evolution, the results from the maximum likelihood analyses and from the $\Delta\Sigma$-profiles in general also agree well. For all galaxies we obtained a velocity dispersion of \mbox{$\sigma^{*}=150\pm2\ {\rm{km\ s^{-1}}}$} from the $\Delta\Sigma$-analysis and $\sigma^{*}=144^{+3}_{-2}\ {\rm{km\ s^{-1}}}$ from the likelihood analysis, for blue galaxies we obtained \mbox{$\sigma^{*}_{\rm blue}=123\pm3\ {\rm{km\ s^{-1}}}$} from the $\Delta\Sigma$-analysis and $\sigma^{*}_{\rm blue}=126^{+5}_{-5}\ {\rm{km\ s^{-1}}}$ from the maximum likelihood analysis. For red galaxies the values are again only marginally consistent with \mbox{$\sigma^{*}_{\rm red}=173\pm2\ {\rm{km\ s^{-1}}}$} from the $\Delta\Sigma$-analysis and $\sigma^{*}_{\rm red}=161^{+3}_{-3}\ {\rm{km\ s^{-1}}}$ from the likelihood analysis.
For the virial radius \mbox{$r_{200}$}, the situation is different, as in this case the results are in general only marginally consistent. The maximum likelihood analysis yields a value of \mbox{$r^{*}_{200}=133^{+3}_{-2}\ h^{-1}$ kpc} for the combined lens sample while the amplitude of the $\Delta\Sigma$ lens signal leads to a value of \mbox{$r^{*}_{200}=146\pm2\ h^{-1}$ kpc}. Also for red (\mbox{$r^{*,\rm red}_{200}=177\pm3\ h^{-1}$ kpc} from $\Delta\Sigma$ and \mbox{$r^{*,\rm red}_{200}=160^{+3}_{-4}\ h^{-1}$ kpc} from maximum likelihood) and blue galaxies (\mbox{$r^{*,\rm blue}_{200}=120\pm2\ h^{-1}$ kpc} from $\Delta\Sigma$ and \mbox{$r^{*,\rm blue}_{200}=115^{+4}_{-5}\ h^{-1}$ kpc} from maximum likelihood) separately, the virial radii derived from the excess surface mass density are up to 10 per cent higher than in the maximum likelihood analysis. We obtain the same result when extracting the $r_{200}$ from a simulation-based $\Delta\Sigma$-profile.
\\
In order to understand the origin of this apparent discrepancy, we perform 3D-LOS-projected lensing signal simulations (BBS and NFW), initially including all lenses.  Incrementally we restrict the lens sample to $M_{r'}<-21$, then we only allow lenses with $M_{r}<-23$. We then further restrict the lens galaxies to less than five neighbours within a projected distance of $720\ h^{-1}$ kpc and finally only consider isolated lenses without neighbours within the same projected distance. In this way we incrementally reduce the influence of multiple deflections (mainly by galaxies in the lens environment but also by other lens galaxies in the line-of-sight). We use the simulated shape catalogues to calculate the corresponding $\Delta\Sigma$-profiles. For the analysis we focus on the lens luminosity bin with $-24 \le M_{r'} \ -23$ (see Fig.~\ref{fig:WIDE.ds.2423}). The profiles show for both models that on large scales the amplitude significantly decreases with decreasing importance of multiple deflections. This confirms the assumption that in this range the shear signal is significantly influenced by nearby galaxy haloes. Focussing on the BBS-based $\Delta\Sigma$-profiles, the results for the various lens samples hardly differ in the inner regions and only deviate on larger scales. This implies that velocity dispersions derived from a single-halo-SIS-fit to the $\Delta\Sigma$-profile on short scales instead of a full maximum likelihood analysis should be hardly biased. Surprisingly, the situation for the NFW-based profiles looks different. In this case multiple deflections lead to an increased signal amplitude also on short scales. I.e. virial radii derived from the $\Delta\Sigma$ in general are biased high about 20 per cent even on short scales, when not explicitly considering a lens sample in a low density environment. This can be explained, if we take into account, that the profile slopes for BBS and NFW behave differently. While the BBS profiles are isothermal on shorter scales and thus their slopes are independent of luminosity, this is not the case for NFW profiles. The brighter a lens, the larger its $r_{200}$ (and for similar concentration its scale radius $r_s$). For brighter galaxies, on short scales we basically only analyse regions within $r_{s}$ or a few scale radii, while for fainter galaxies the contribution of regions outside the $r_{s}$ (where the profile slope becomes much steeper) significantly increases. Therefore the slopes of the $\Delta\Sigma$-amplitudes might increase with increasing contribution of fainter lenses. I.e. NFW halo parameters should not be purely or only carefully extracted from $\Delta\Sigma$ profiles.
%
\section{Summary and conclusions}
\label{sec:conclusions}
We performed a GGL analysis on 89 pointings of the CFHTLS-WIDE. We used the $u^*g' r' i' z'$ photometry to estimate photometric redshifts with an accuracy of smaller than $0.04\ (1+z)$ for objects with $i' \le  24.0$. The background galaxy shapes were measured and corrected with the KSB+-method (see \citealt{kaiser95} and \citealt{hoekstra98}).
\subsection{Weak lensing results from analyses on shorter scales}
We measured the tangential gravitational shear signal on scales from 25\,$h^{-1}$ kpc out  to 700\,$h^{-1}$ kpc, examining all lens galaxies and subsequently red and blue galaxies separately. On average the measured velocity dispersions for red galaxies are about 25 per cent higher than for the average lens sample, while the velocity dispersions for blue galaxies are about 15 to 20 per cent lower. We also analysed different galaxy density environments. The tangential shear amplitude, especially on larger projected distances, shows higher values in denser environments, due to multiple deflection by nearby galaxy and by host haloes (see also \citealt{brainerd10}).
\\
\\
We measured the excess surface mass density $\Delta\Sigma(R)$ and compared our results for different luminosity intervals with previous work, finding partial consistence with \citet{mandelbaum06b}. The results for the blue galaxy sample are in good agreement. The measurements for red galaxies show some differences in amplitude depending on the luminosity bin. Our results however agree well with \citet{mandelbaum08b}. We further compared our measurements for red galaxies to \citet{vanuitert11}, finding good agreement for most investigated luminosity bins.
\\
\\
We investigated the luminosity scaling relations for the velocity dispersion $\sigma$, analysing $\Delta\Sigma$ in different luminosity bins for our red, blue and combined lens samples on scales out to $100\ h^{-1}$ kpc. We find $\sigma_{\rm red} \propto L^{0.24\pm0.03}$ for red and $\sigma_{\rm blue} \propto L^{0.23\pm0.03}$ for blue galaxies. For the same luminosities the red galaxies have a higher value of $\sigma$ than the blue ones. The combined galaxy sample is dominated by blue galaxies at low luminosities and red galaxies at high luminosities, which leads to a steeper relation $\sigma \propto L^{0.29\pm0.02}$. 
\\
\\
Our results from fitting isolated haloes to $\Delta\Sigma(R)$ are confirmed in a maximum likelihood analysis, thus properly treating multiple deflections by independent lens galaxies. For an SIS profile out to maximum separations of \mbox{$200\ h^{-1}$ kpc}, we obtain values for the $\sigma-L$ scaling of $\eta_{\sigma} = 0.31^{+0.02}_{-0.03}$ for all galaxies or $\eta_{\sigma}^{\rm red} = 0.27^{+0.03}_{-0.02}$ and $\eta_{\sigma}^{\rm blue} = 0.27^{+0.03}_{-0.04}$, when considering red and blue SED galaxies separately. Therefore we have fixed the scaling relations to $\sigma \propto L^{0.3}$ for the combined sample and $\sigma_{\rm red/blue} \propto L^{0.25}$ for red and blue galaxy samples for the rest of our analysis.
\\
\\
We analysed the scaling of the NFW-based halo parameters $r_{200}$, $M_{200}$ and $v_{200}$, assuming the concentration-mass-relation of \citet{duffy08}. We find that for galaxies with rest frame luminosities $L_{r'}>10^{10}\ h^{-2}\ L_{\odot}$ the scaling relations follow a power law. The power laws are $r_{200} \propto L^{0.33\pm0.04}$ and $r_{200} \propto L^{0.36\pm0.07}$ for red and blue galaxies. For the combined lens sample we find, analogously to the velocity dispersion $\sigma$ (and for the same reason), a steeper relation of $r_{200} \propto L^{0.39\pm0.04}$. The virial mass of the combined lens sample scales as $M_{200} \propto L^{1.21\pm0.10}$. However, for galaxies fainter than $L=10^{10}\ h^{-2}\ L_{\odot}$ the values of the NFW halo parameters remain significantly higher than suggested by a single-power law scaling relation. This either indicates a significantly slower decrease of galaxy radii and masses or a significant increase of the concentration parameter $c$ with decreasing luminosity.
\\
\\
We compared the results for the velocity dispersions of our red lens sample to kinematical measurements from \citet{gallazzi06} and LRGs based on \citet{eisenstein01}. We saw that $0.96\ \sigma_{\rm star} \le \sigma_{\rm halo}^{\rm WL} \le 1.07\ \sigma_{\rm star}$ holds. This means that the halo velocity is between the maximal circular velocity found around $\sim 0.5\ R_e$ and 90 per cent of this value which equals the circular velocity of galaxies at a few effective radii. Taken together this implies that haloes are isothermal from a few effective radii out to about $100\ h^{-1}$ kpc. Also, the values for the circular velocity $v_{200}$ of our blue galaxy sample agree well with the results of \citet{reyes11} and \citet{dutton10}.
\\
\\
We find a break in the luminosity scaling of the M/L-ratio (red galaxies have a minimum M/L-ratio for a stellar mass of $M_{\rm star} \sim 3-4 \times 10^{10}\ h^{-2}\ M_{\odot})$. This is in agreement with the result of \citet{dutton10}, who combined weak lensing measurements with satellite kinematics.
\subsection{Weak lensing results from analyses on larger scales}
We performed maximum likelihood analyses as described by \citet{schneider_rix97}. We investigated two different halo profiles, an NFW profile (out to $400\ h^{-1}$ kpc) and a truncated isothermal sphere (BBS, out to $2\ h^{-1}$ Mpc). We assumed the scaling relations of to \citet{guzik02} to scale the masses with luminosities and  \citet{duffy08} for the mass-concentration relation, while further applying the scaling relations we previously extracted from the $\Delta\Sigma$-profiles. For the maximum lens sample we find a virial radius of $r^{*}_{200} = 133^{+3}_{-2} \ h^{-1}$ kpc and a concentration parameter of $c^{*} = 6.4^{+0.9}_{-0.7}$ with a viral mass of $M^{*}_{200} = 7.6^{+0.5}_{-0.3} \times 10^{11} \ h^{-1} M_{\odot}$.
The virial radii are about 20 per cent higher for red galaxies and 15 per cent lower for blue galaxies. This results in virial masses about 60 per cent higher for the red galaxies and 35 per cent lower for blue galaxies.
For the BBS profile we find a velocity dispersion of \mbox{$\sigma^{*} = 131^{+2}_{-3}$ km\ $\rm{s^{-1}}$} and a truncation radius of $s^{*} = 184^{+17}_{-14} \ h^{-1}$ kpc. This corresponds to a total mass of $M^{*}_{\rm total,BBS} = 2.32^{+0.28}_{-0.25} \times 10^{12} \ h^{-1} \ M_{\odot}$. Both measurements are in agreement with literature (see \citealt{hoekstra04}).
\\
\\
Considering red and blue galaxies separately, we find that red galaxies show significantly higher values for the velocity dispersion and the truncation radius than the blue galaxies. Blue galaxies have (for an $L^{*}$-galaxy) basically the same concentration parameter as red galaxies but a significantly lower value for the virial radius. This leads to a significantly lower virial mass than for red galaxies (about a factor of 2.5). The values for the best fitting halo parameters for an $L^{*}$-galaxy are summarised in Table~\ref{tab:maximum_likelihood}. 
\\
\\
We analysed the properties of our lens galaxies as a function of their environment density. We split the galaxies into two halves relative to the median galaxy environment density.  This results in a modestly underdense field galaxy dominated galaxy sample and a modestly overdense galaxy sample, where more galaxies reside in groups.  We find that the velocity dispersion $\sigma$ of BBS haloes hardly depends on the environment, i.e. that the central galaxy halo profile mostly depends on the galaxy luminosity.  The BBS truncation and NFW virial radii for red galaxies are however significantly larger for galaxies in denser environments.  The increase of virial radii translates into 50 per cent higher virial masses for red galaxies in the moderately overdense relative to the moderately underdense sample.  We do not discriminate between central and satellite galaxies in this work, i.e. our analysis is insensitive to the level at which satellite haloes are stripped and central galaxies grow in mass. We instead measure the average mass for a given luminosity in different environments. The increase of average halo masses is expected if one takes into account that the total mass-to-light ratio increases from galaxies to groups (and galaxy clusters) (\citealt{vanuitert11}, \citealt{sheldon09}).
\\
\\
We performed a BBS maximum likelihood analysis, fixing $\sigma^{*}$ and its luminosity scaling $\eta_{\sigma}$ to fit the truncation radius $s^{*}$ and its luminosity scaling relation $s \propto L^{\eta_s}$. For the combined lens sample we find a value of $\eta_s=0.52^{+0.09}_{-0.10}$, leading to a M/L-scaling of $M/L \propto L^{0.12^{+0.10}_{-0.11}}$. When we analysed the blue and red sample separately we found that the blue and the bright red sample have a M/L-scaling consistent with a constant M/L-ratio. For the faint ($L<10^{10}\ h^{-2}\ L_{\odot}$) red galaxy sample we found $M/L \propto L^{-1}$, with a minimum at $\sim 10^{10}\ h^{-2}\ L_{\odot}$ (corresponding to $M_{\rm star} \approx 10^{10}\ h^{-2}\ M_{\odot})$. The existence and location of this minimum is consistent with results from abundance matching (see \citealt{guo10} and \citealt{dutton10}).
\\
\\
We analysed the concentration-luminosity relation as a function of the combined lens sample and for red and blue galaxies separately. We find a scaling of $c \propto  L^{-0.07^{+0.11}_{-0.11}}$ for the lens sample as a whole, consistent with \citet{duffy08}. The scaling of red galaxies is similar, but slightly flatter ($c \propto  L^{-0.04^{+0.10}_{-0.13}}$), showing hardly any evolution with luminosity or mass. For blue galaxies we find $c \propto  L^{-0.34^{+0.24}_{-0.26}}$. The best-fit value would imply a significant increase of the concentration with decreasing luminosity. This would mean, that at the high luminosity end of our lens sample the concentration of red galaxies exceeds the concentration of blue galaxies of same luminosity while for their faint counterparts the opposite is the case. However, the uncertainties for the measured slopes for  blue galaxies are quite large. Therefore the values for $\eta^{\rm red}_{\rm c}$ and $\eta^{\rm blue}_{\rm c}$ still are consistent with each other within the uncertainties.
\\
\\
For galaxies fainter than $L=10^{10}\ h^{-2}\ L_{\odot}$ we measured higher values for $r_{200}$ from $\Delta\Sigma$ than expected for a single-power scaling with luminosity. We investigated two possibilities to fix this. On one hand we analysed the implications of a modified concentration-luminosity-relation in comparison to \citet{duffy08}. The analysis of the modified c-L scaling relation yields comparable values for $\eta_c^{\rm bright}$ as for the case where we considered all luminosities. For the faint luminosity range a slope of $\eta_c^{\rm faint} < -1$ is suggested, leading to significantly higher concentration values. On the other hand we assumed a double-power-law scaling relation for the $r_{200}$ with luminosity. When allowing two different slopes for the $r_{200}$-L-relation for $L>10^{10}\ h^{-2}\ L_{\odot}$ and $L<10^{10}\ h^{-2}\ L_{\odot}$, the resulting boost of the fainter lenses' $\Delta\Sigma$ profiles makes an additional modification of the c-L scaling unnecessary and also gives better agreement to the observed $\Delta\Sigma$-profiles for faint galaxies. The change of slope in $M_{200}/L$ is also predicted by \citet{vandenbosch05}.
\subsection{Comparison with 3D-LOS-projected lensing signal simulations}
Finally we created two sets of synthetic lensing observations. The locations, the redshifts, the SED types and the luminosities of the lens and source galaxies were chosen to be identical to the true observations. The lensing properties of the lenses were described with either BBS or NFW profiles with parameters obtained from the maximum likelihood analyses regarding the $L^{*}$-halo parameters and the luminosity scaling relations.
\\
\\
We analysed the simulated observations in the same way as the true observations and measured $\gamma_{\rm t}$- and $\Delta\Sigma$-profiles as a function of SED type and absolute luminosity and compared the results of both simulations with the observational data. The $\Delta\Sigma$-profiles derived from the simulated observations mostly agree well with the observed profiles. More precisely if the NFW simulation is based on a single-power law scaling for $r_{200}$ and assuming the concentration-mass relation of \citet{duffy08}, it predicts a too low lensing signal for the fainter luminosity bins. This can be fixed by either assuming a steeper concentration-luminosity scaling or a shallower scaling relation for the $r_{200}$ of faint galaxies. The NFW simulation based on a double-power law scaling relation for the $r_{200}$ ($r_{200}^{\rm red/blue} \propto L^{1/3}$ for $L>10^{10}\ h^{-2}\ L_{\odot}$, $r_{200}^{\rm red} \propto L^{0}$ and $r_{200}^{\rm blue} \propto L^{0.21}$ for  $L<10^{10}\ h^{-2}\ L_{\odot}$) indeed leads to predictions agreeing significantly better with the observations.
\section*{Acknowledgements}
We are grateful to the CFHTLS survey team for conducting the observations and the TERAPIX team for developing software used in this study. 
We acknowledge use of the Canadian Astronomy Data Centre, which is operated by the Dominion Astrophysical Observatory for the National Research Council of Canada's Herzberg Institute of Astrophysics.
This research has made use of the NASA/IPAC Extragalactic Database (NED) which is operated by the Jet Propulsion Laboratory, California Institute of Technology, under contract with the National Aeronautics and Space Administration.
\\
We would like to thank T. Erben and T. Schrabback for kindly introducing and providing the \THELI - and the KSB+-pipeline. 
\\
We would like to thank the anonymous referee for helpful comments on our manuscript.
\\
This work was supported by the the DFG priority programme 1177, the DFG Cluster of Excellence `Origin and Structure of the Universe', the TRR33 `The Dark Universe' and the Marie Curie research training network `DUEL'.
\\
M. L. thanks the European Community for the `DUEL' doctoral fellowship MRTN-CT-2006-036133.\\
\bibliographystyle{mn2e}
\bibliography{mybib}
\appendix
\section{Systematics}
In this section we verify the integrity of our weak lensing analysis. We demonstrate that there is no weak lensing signal if we replace foreground galaxies by stellar objects or random points and that our measured lensing signal amplitudes do not show a significant dependence on source magnitude, S/N-ratio or size. We also show, that misassignment of foreground objects to the background does not introduce a bias of more than $\sim 1$ per cent to the measurements of velocity dispersions. The contribution of faint not detected galaxies in the neighbourhood of more massive galaxies does not significantly affect  the shear measurement of the massive galaxies and can be neglected.
\subsection{Gravitational shear estimates relative to stars and a random foreground distribution}
\label{sec:gamma_t random}
\begin{figure*}
\includegraphics[width=8.8cm]{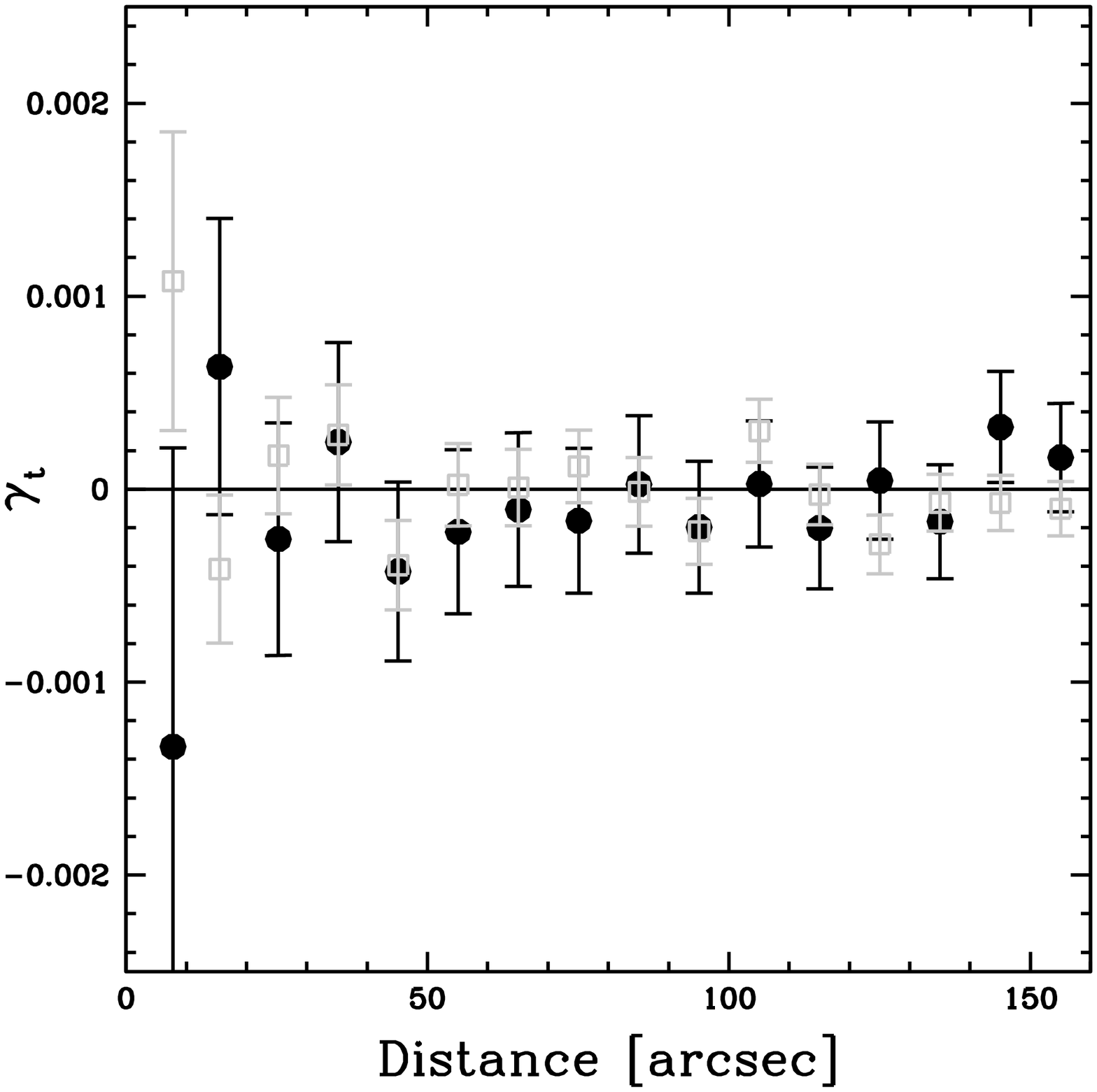}
\includegraphics[width=8.8cm]{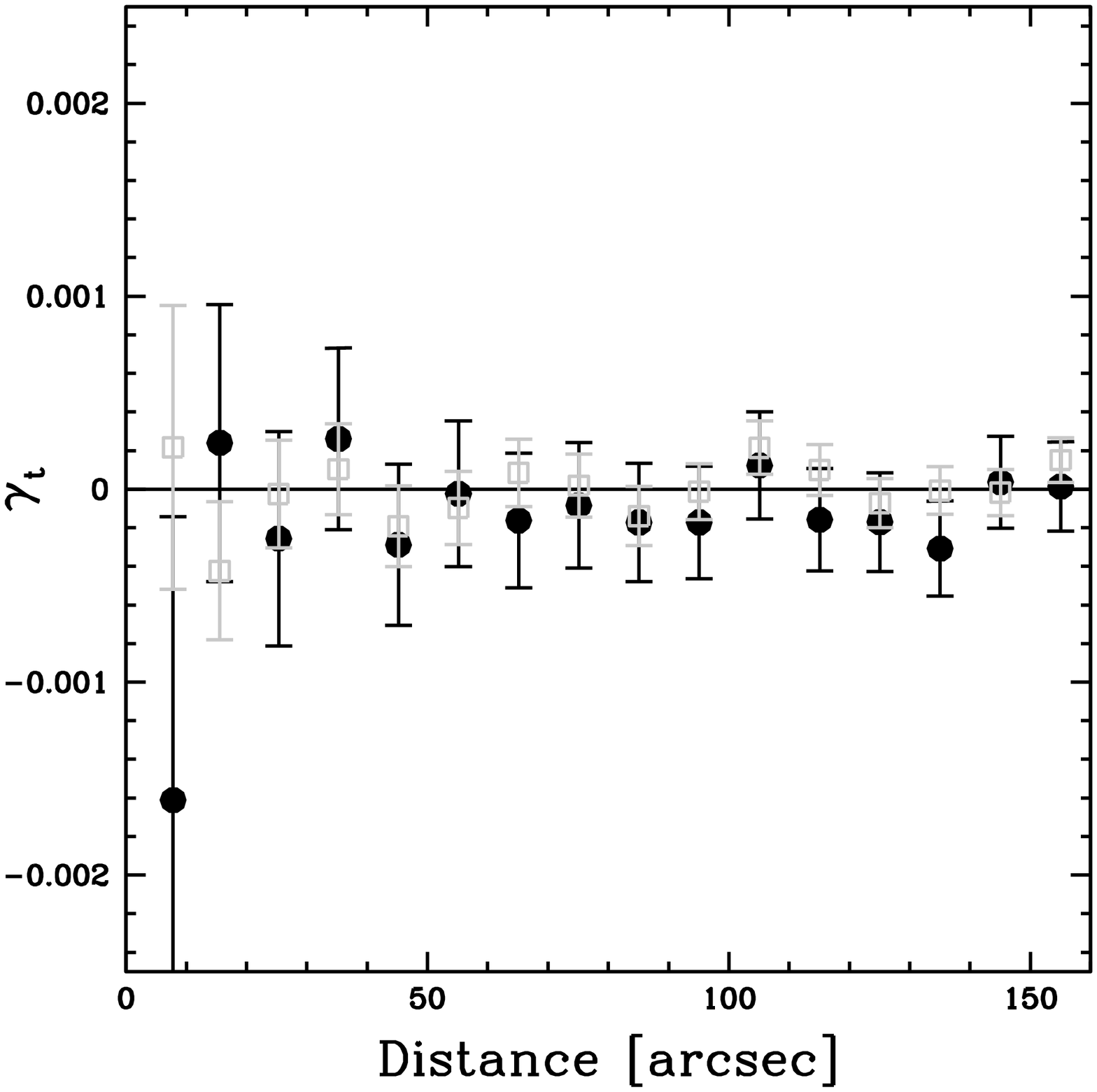}
\caption{Tangential gravitational shear (black symbols) and cross-shear (grey symbols) relative to a stellar foreground sample. Both the E-modes and B-modes are consistent with zero. The left panel shows the stellar selection performed with magnitude cut, \SExtractor classification and the KSB-pipeline, the right panel shows the stellar selection from the original photometric catalogues by selecting via magnitudes and \SExtractor classification.}
\label{fig:WIDE.syst.stars}
\end{figure*}

\begin{figure*}
\centering
\includegraphics[width=8.8cm]{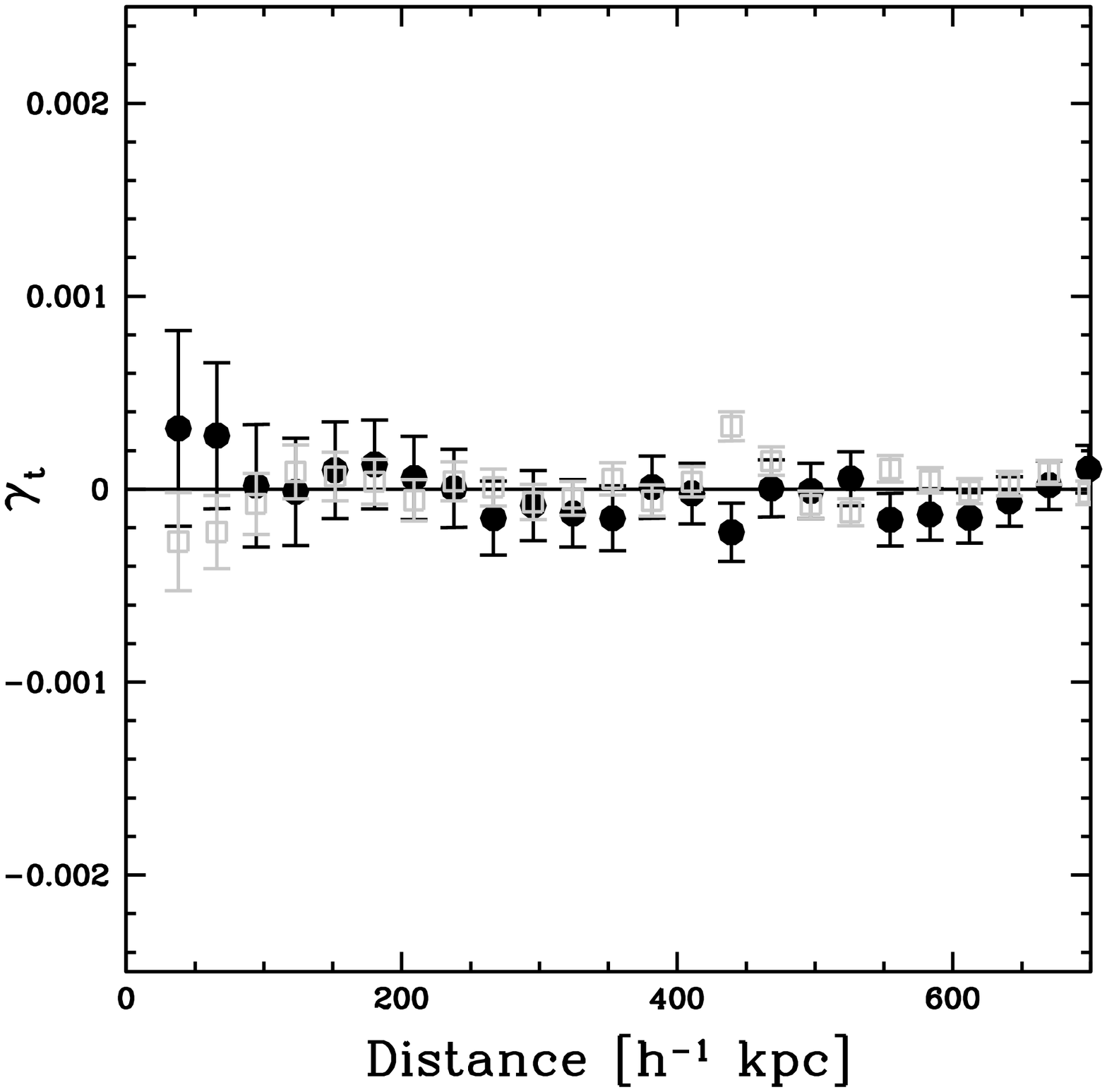}
\includegraphics[width=8.8cm]{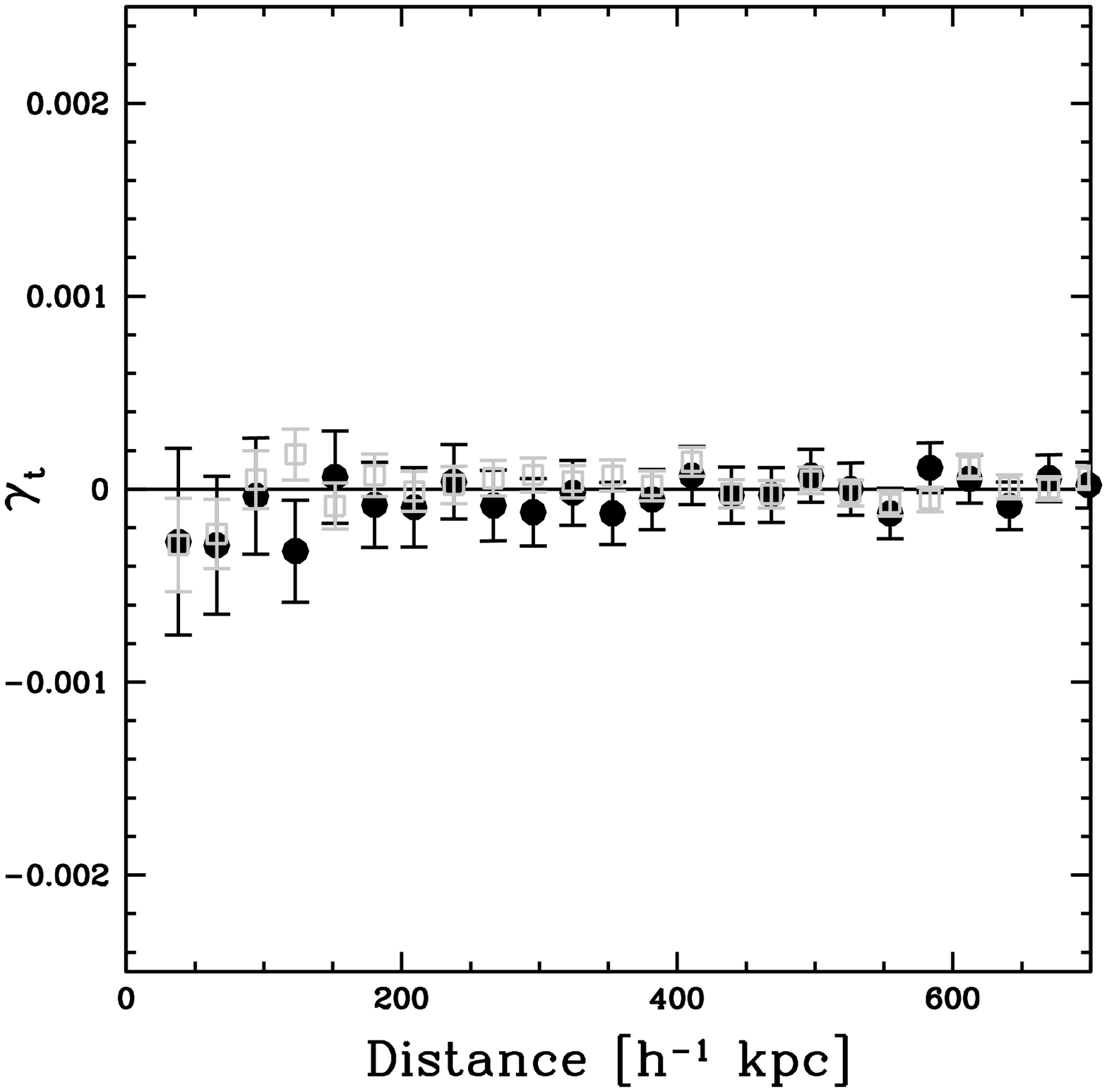}
\caption{Tangential alignment of galaxies from the maximum background catalogue relative to random foreground positions (left panel) and tangential shear of background with permuted shears relative to original foreground objects (right panel). The E-mode signal shown in black and the B-mode estimate in grey, are both consistent with zero.}
\label{fig:WIDE.syst.random}
\end{figure*}

In absence of systematic errors the background galaxies must not show any systematic alignment relative to foreground stars or a random distribution of points. We therefore take the stars used for our KSB PSF anisotropy correction. These are \mbox{274\ 589} stars with magnitudes of about $18 < i' < 22$ and a \SExtractor star classification of greater than 0.96. We treat this stellar sample as foreground sample and analyse the lensing signal for our complete and unaffected background sample. As Fig.~\ref{fig:WIDE.syst.stars} (left panel) shows, both the tangential shear $\gamma_{\rm t}$ and the cross-shear $\gamma_{\rm c}$ are consistent with zero. In the second step we take $all$ stars from our photometric catalogue with magnitudes $18\lesssim i' \lesssim 22$ and a \SExtractor star classification of greater than 0.96 leading to total `foreground' sample of 471\ 066 objects. Also in this case E-modes and B-modes are consistent with zero as can be seen in the right panel of Fig.~\ref{fig:WIDE.syst.stars}.
\\
We then dice random points over the full CFHTLS area and then only consider random points outside the masked area. In this way the randomly positioned `foreground' objects sample the same area as the original (photometric) foreground catalogue. We again keep the original background shape catalogue and measure E- and B-modes consistenct with zero for both quantities on all scales (see left panel of Fig.~\ref{fig:WIDE.syst.random}).
Finally we take the original photometric foreground sample and modify our background shape catalogue. We keep the original background positions but randomly reassign the measured shears by permutation, conserving the ellipticity distribution. In this case the tangential shear and the cross-shear are in agreement with zero as well, see right panel of Fig.~\ref{fig:WIDE.syst.random}.
%
%
\subsection{Estimate of the signal dilution from contamination of the background shape samples with foreground galaxies}
\label{sec:boost}
\begin{figure*}
\includegraphics[width=18.25cm]{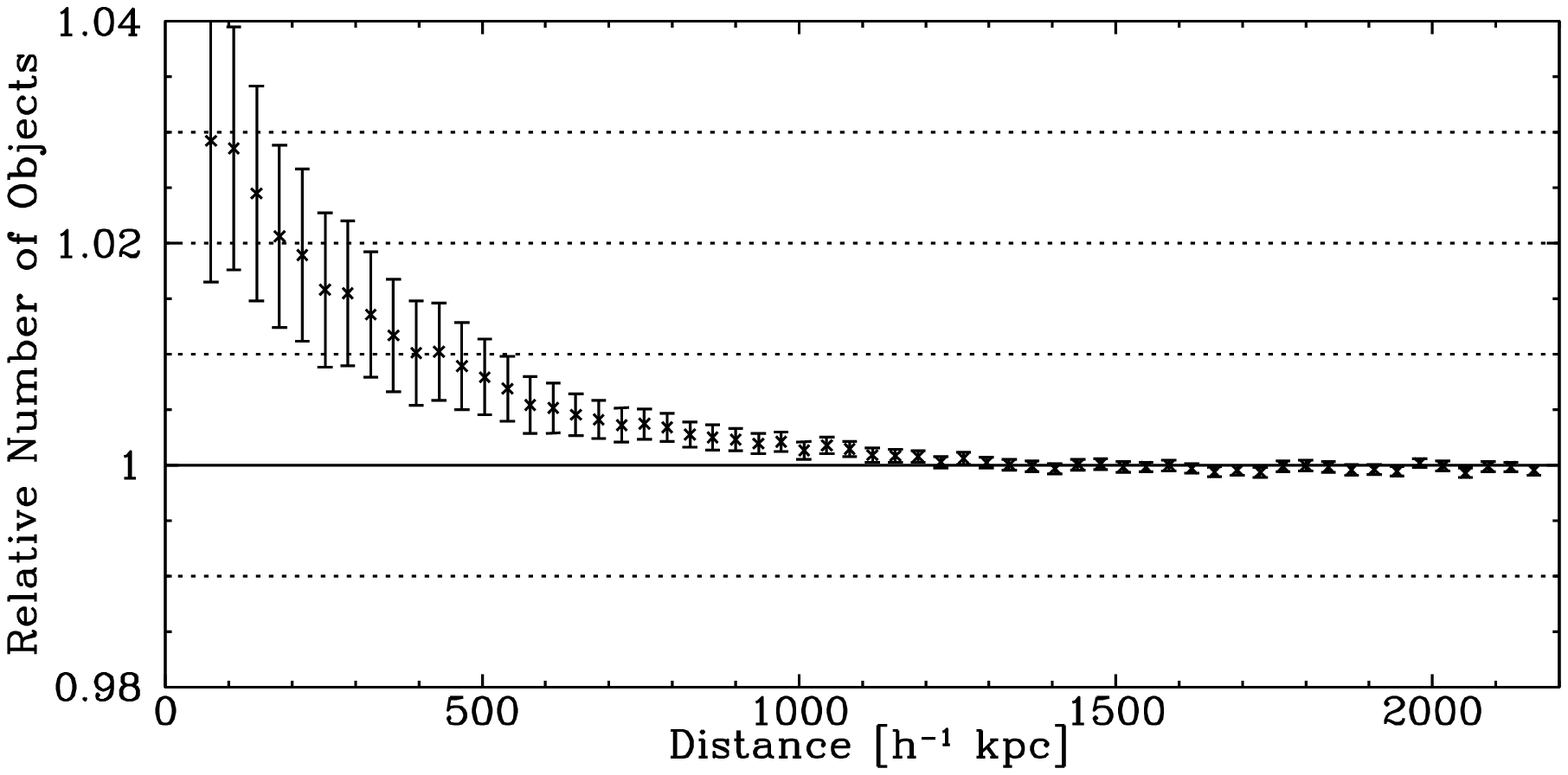}
\caption{Relative galaxy background density around galaxies in the lens sample (relative to random foreground points). The excess background density is between 2 and 3 per cent on very short scales, drops below 2 per cent for scales larger than $\sim 200\ h^{-1}$ kpc and is smaller than 1 per cent for scales larger than $\sim450\ h^{-1}$ kpc.}
\label{fig:WIDE.background.density}
\end{figure*}

The lensing signal can be diluted if galaxies which are physically related to the lenses are accidentally  assigned as background objects (e.g. due to photometric redshift mismatch). To quantify this effect we measure the density of background objects around foreground objects within a radius of \mbox{$\sim2\ h^{-1}$ Mpc} as a function of the projected separation (see also \citealt{mandelbaum05a}). On short scales the number density of background objects is slightly increased (see Fig.~\ref{fig:WIDE.background.density}) by up to 3 per cent. Therefore we either have indeed a dilution of the lensing signal by 1 to 3 per cent on scales $\le 500\ h^{-1}$ kpc or the increase in background galaxies is caused by the magnification effect. Using that $\gamma_t \propto \sigma^2$ we conclude that in the worst case the velocity dispersions can be systematically underestimated by $\sim 1.5$ per cent on very short scales and less than $\sim 1$ per cent on larger scales due to contamination of background by foreground objects.
\subsection{Signal contribution of undetected low mass nearby galaxies}
\begin{figure*}
\includegraphics[width=8.8cm]{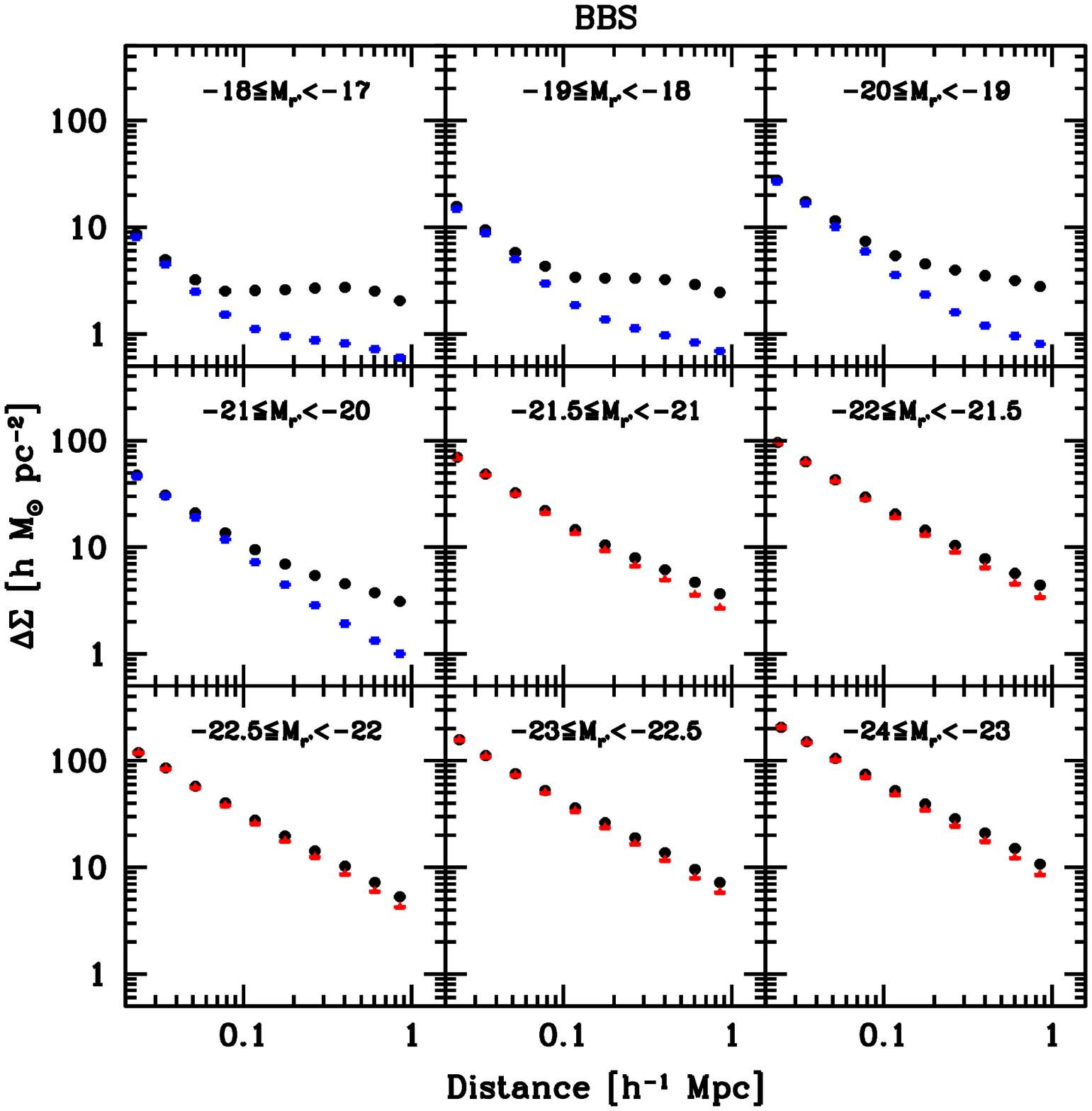}
\includegraphics[width=8.8cm]{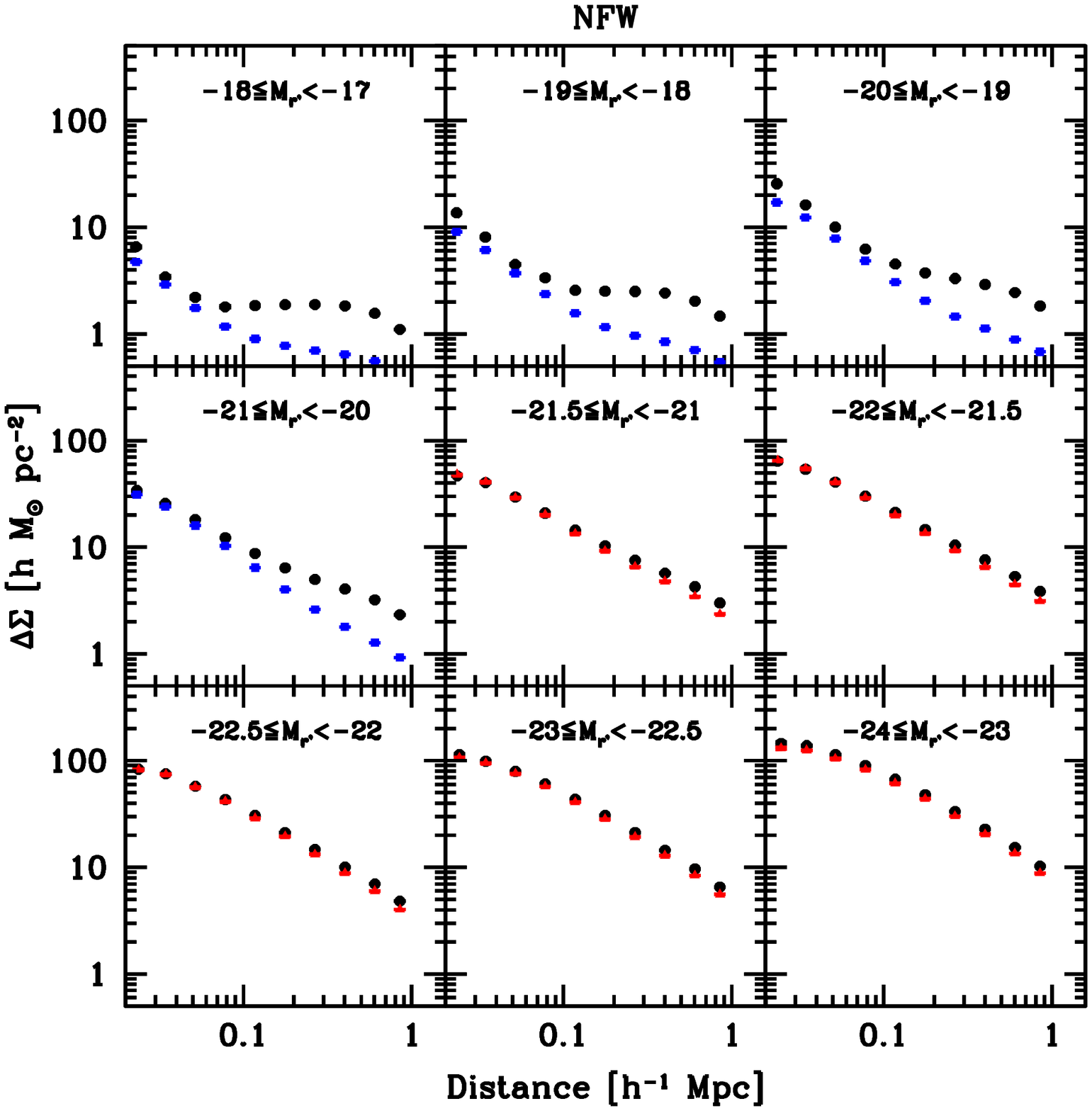}
\caption{Contribution of bright and faint galaxies to the excess surface mass density $\Delta\Sigma$. The left panel shows the BBS simulations, the right panel shows the NFW simulations. The red triangles show the simulated signal contribution only for lenses with $M_{r'} < -21$, the black circles show the results for simulations including lenses down to $M_{r'} = -17$. As can be seen their contribution especially on short scales is negligible. The blue squares show the results for simulations only including lenses with $M_{r'} \ge -21$. The signal amplitude for faint galaxies on large scales is dominated by mass associated with brighter galaxies in the environment.}
\label{fig:WIDE.ds.21}
\end{figure*}

We now measure the contribution of low-mass objects to the lensing signal of the central galaxy. For larger redshifts these galaxies cannot be detected anymore and their halo mass is assigned to the central halo. To estimate the amount we run simulations (see Appendix~\ref{sec:simulations} for details on the simulations). In the first setup we only consider bright galaxies ($M_{r'}<-21$) at lower redshifts ($z _{\rm phot} \le 0.5$) with faint companions ($M_{r'} > -21$, maximal separation $720\ h^{-1}$ Mpc in projected distance and 0.15 in redshift) but ignore the contribution of the companions in the lensing signal simulation. In the second step we also consider the lensing contribution of the companions. We then calculate the excess surface mass density $\Delta\Sigma$ for both cases. We see that the contribution of the low-luminosity companions is negligible (see Fig.~\ref{fig:WIDE.ds.21}, red points for the bright galaxy signal vs. black points for the combined signal). For comparison we also investigate the complementary lens sample, only comprising lenses with $M_{r'}>-21$ (see Fig.~\ref{fig:WIDE.ds.21}, blue points for the faint galaxy signal vs. black points for the combined signal). We conclude that for faint galaxies $\Delta\Sigma$ on large scales is dominated by mass associated with brighter galaxies in the neighbourhood.
\subsection{Scaling of the lensing signal with redshift}
\label{sec:ds.syst}
\begin{figure*}
\centering
\includegraphics[width=8.8cm]{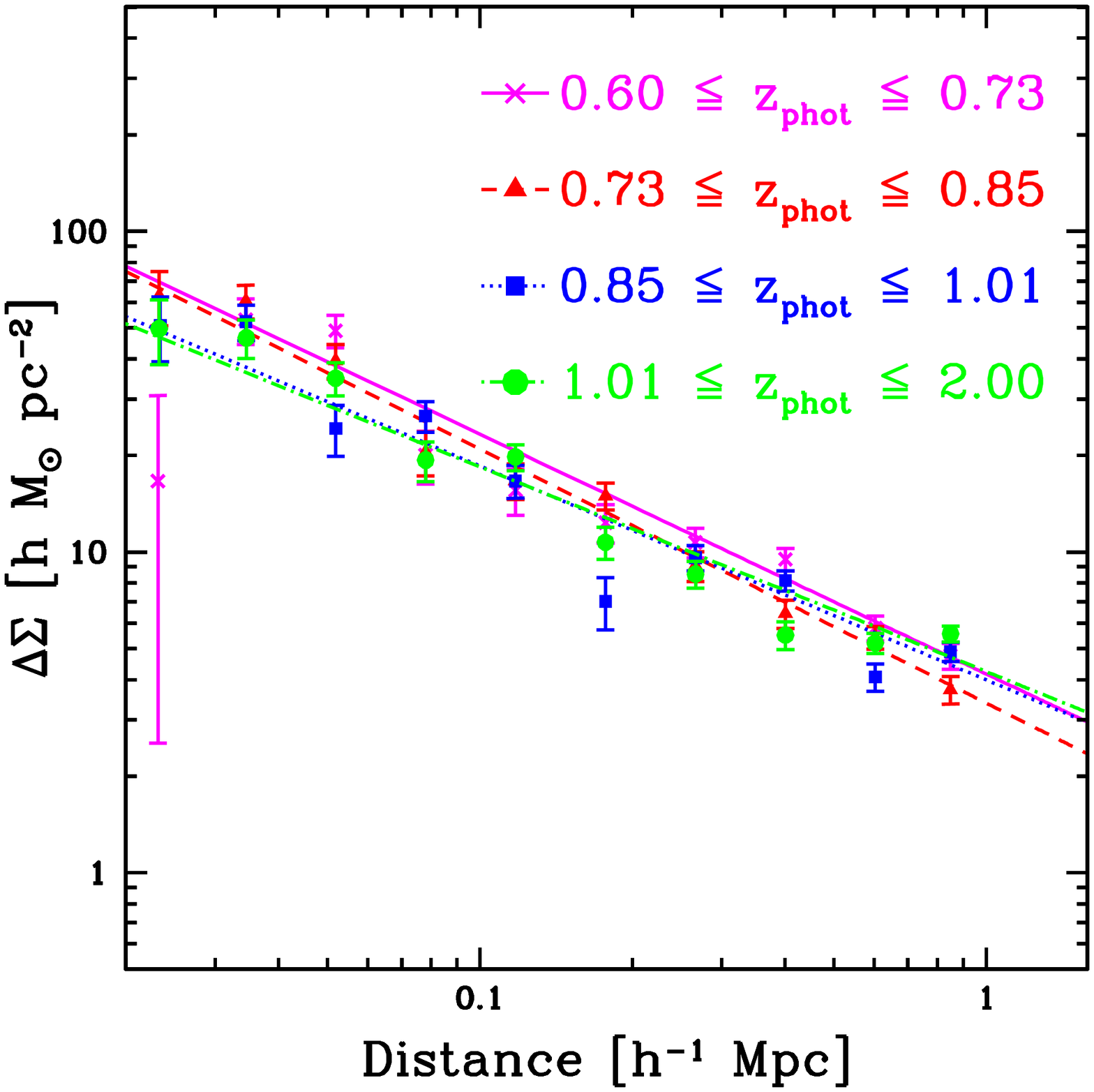}
\includegraphics[width=8.8cm]{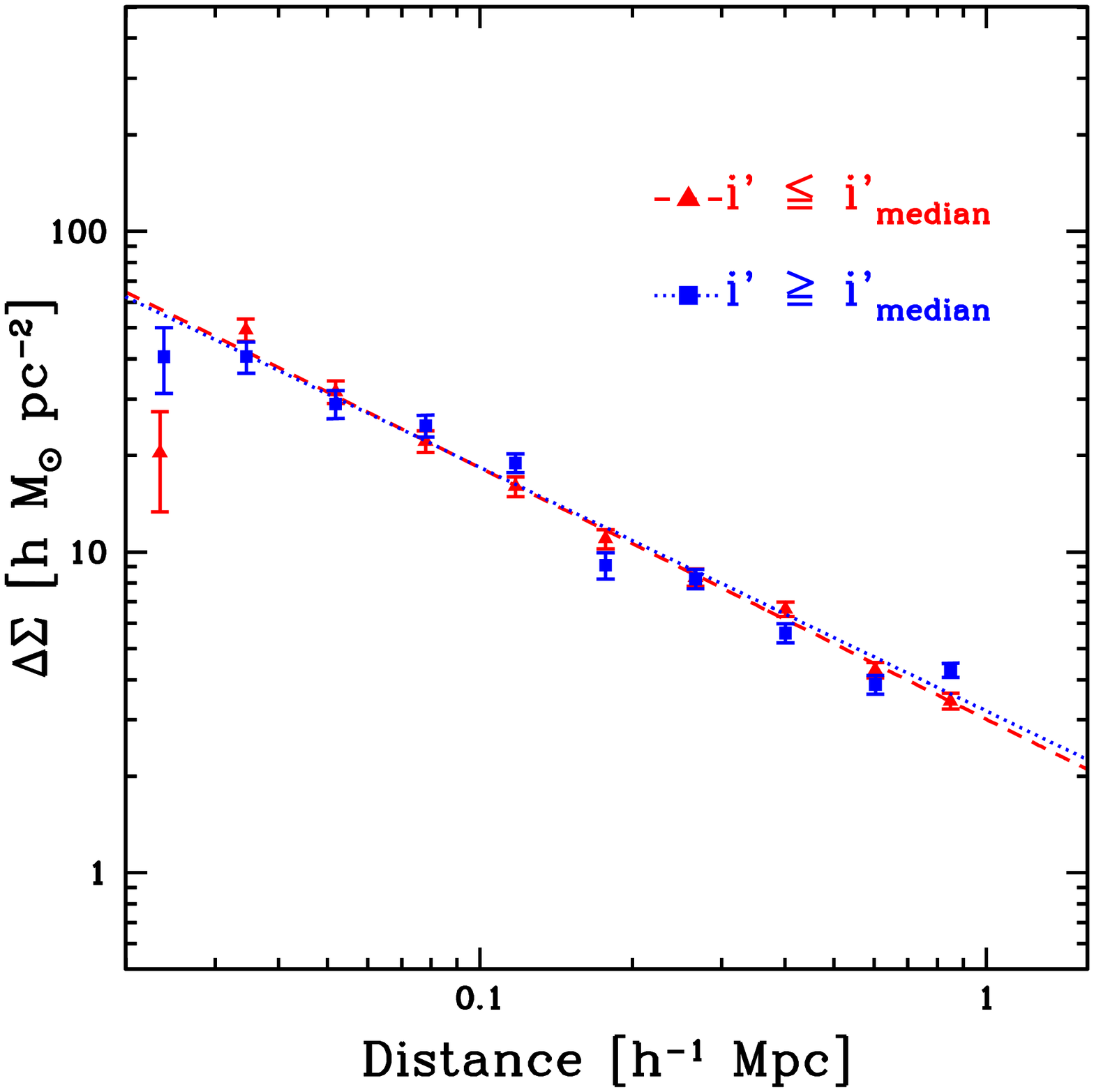}
\includegraphics[width=8.8cm]{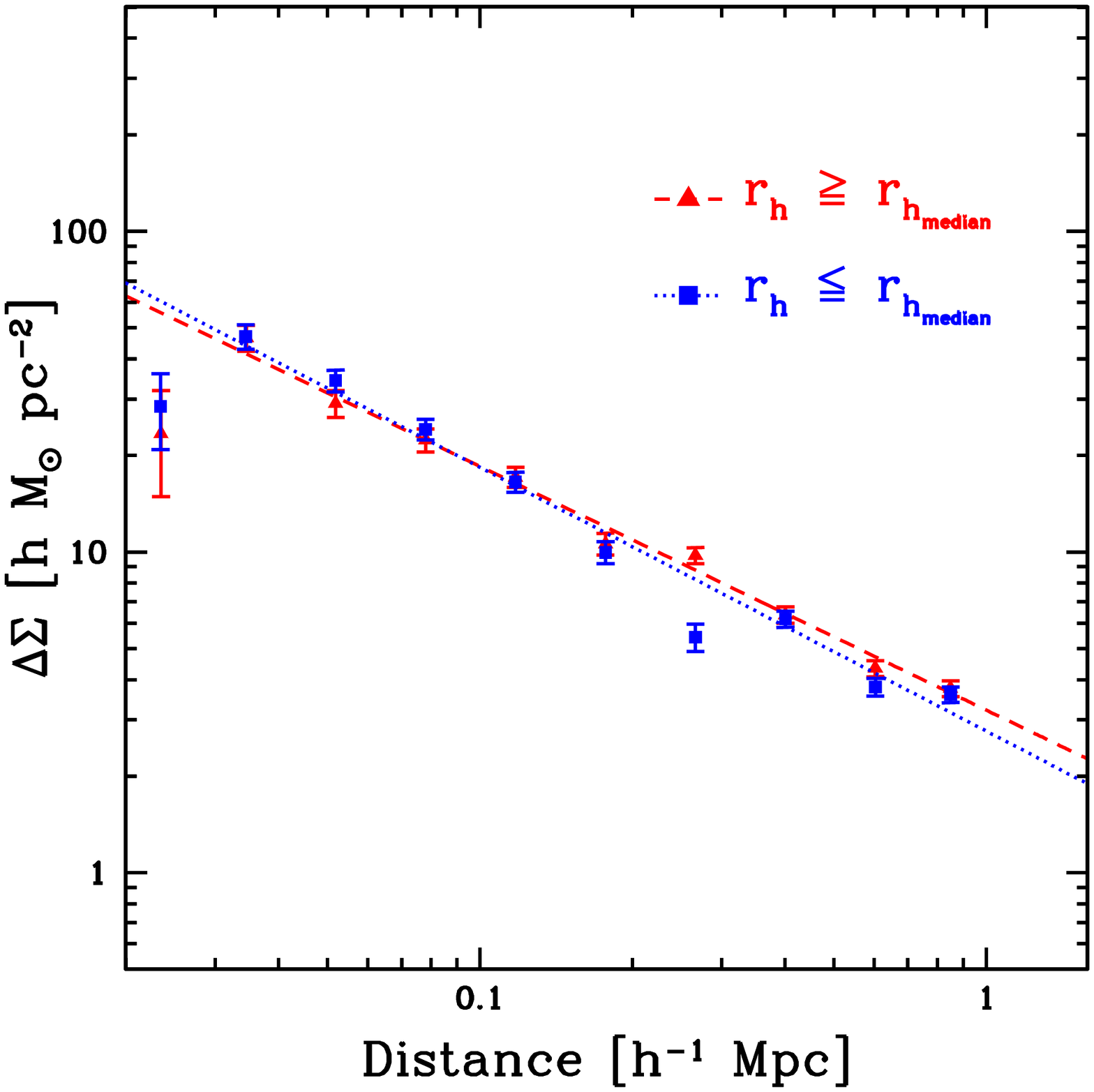}
\includegraphics[width=8.8cm]{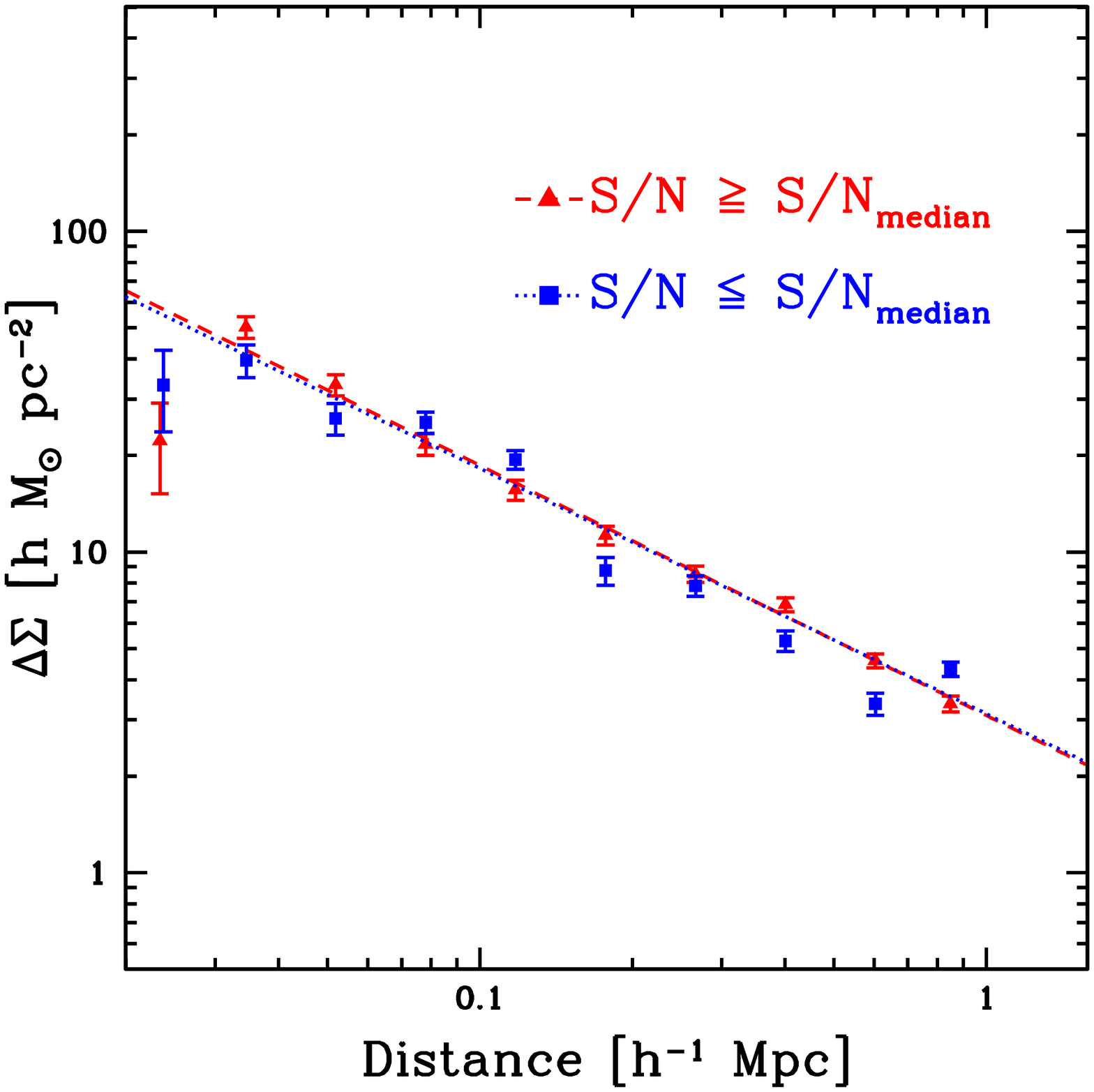}
\caption{Systematic check. We split our samples into different subsamples and compare the results. In the upper left panel we measure the excess surface mass density $\Delta\Sigma$ in for lenses with $0.05 < z_{\rm d} \le  0.50$ in four redshift bins. In the other three panels we consider the complete foreground sample. The upper right panel shows $\Delta\Sigma$ for bright (red symbols) and faint (blue symbols) sources, the lower left panel shows $\Delta\Sigma$ for large (red symbols) and small (blue symbols) sources and the lower right panel shows $\Delta\Sigma$ for sources with high (red symbols) and low (blue symbols) signal-to-noise. The measurements are in good agreement.}
\label{fig:WIDE.ds.syst}
\end{figure*}
We now analyse the scaling behavior of the gravitational lens signal with redshift. We consider all source galaxies with $0.6 \le  z _{\rm phot} \le  2.0$ and divide them into four different redshift bins of same size. We select lenses with of $0.05 < z _{\rm phot} \le  0.5$ to ensure the equality of the lens sample for independent background bins and avoid differences in the examined redshift range due to our dynamical lens-source assignment. We then measure both the tangential shear signal and the excess surface mass density and observe good agreement for all four subsamples (see upper left panel in Fig.~\ref{fig:WIDE.ds.syst}). The velocity dispersions $\sigma$ for all four redshift bins, fitted within a projected separation of $200\ h^{-1}$ kpc (see also Section~\ref{sec:alignment}), are in agreement. The results are summarised in Table~\ref{tab:systematics}.
\subsection{Shear calibration biases for small and low-S/N objects}
We now use the complete foreground lens sample again. We split our background sources each into two subsamples of same number in terms of following properties:
\ben
\item apparent magnitude, defining  a bright and faint source sample, 
\item object size, defining a large and a small object sample,
\item signal-to-noise ratio, defining a sample with high and low photometric signal-to-noise.
\een
We measure the gravitational shear and $\Delta\Sigma$ for all subsamples and obtain an estimate of the velocity dispersion $\sigma$ from an SIS fit to the shear.
For the bright and for the faint background samples we measure a velocity dispersion of \mbox{$\sigma = 115\pm2$ km\ $\rm{s^{-1}}$} and \mbox{$\sigma = 119\pm1$ km\ $\rm{s^{-1}}$}, respectively. For the large and for the small source samples we measure $\sigma = 117\pm2$ km\ $\rm{s^{-1}}$ and \mbox{$\sigma = 116\pm2$ km\ $\rm{s^{-1}}$}, respectively, and for the high and the low signal-to-noise samples we measure velocity dispersions of $\sigma = 115\pm2$ km\ $\rm{s^{-1}}$ and $\sigma = 119\pm1$ km\ $\rm{s^{-1}}$. The $\Delta\Sigma$ estimates are shown in Fig.~\ref{fig:WIDE.ds.syst} to agree well with each other, see also Table~\ref{tab:systematics}.
\begin{table*}
\begin{center}
\caption{}
\begin{tabular}{c|c|c|c|c|c|c|c|c|c} 
\hline \hline
name & $z_{\rm{fg}}$ &  Background \ selection & $\ave{M_{r'}}$ & $\ave{z_{\rm fg}}_{\rm pair}$ & $\ave{z_{\rm bg}}_{\rm pair}$ & $N_{\rm fg}$ & $N_{\rm bg}$ & $N_{\rm pair}$ & $\sigma \ [\rm{km\ s^{-1}}]$\\ 
\hline \hline
$L_{\rm bg1     }$   & $0.05 <  z_{\rm{fg}} \le  0.5$   & $0.60 \le  z_{\rm bg} \le  0.73$ & $-21.0$  & $0.29$ & $0.67$ & $277\ 219  $ & $292\ 900  $ & $17\ 230\ 313 $ & $ 121 \pm 5 $ \\
$L_{\rm bg2     }$   & $0.05 <  z_{\rm{fg}} \le  0.5$   & $0.73 \le  z_{\rm bg} \le  0.85$ & $-21.0$  & $0.32$ & $0.78$ & $277\ 577  $ & $283\ 754  $ & $19\ 174\ 161 $ & $ 124 \pm 4 $ \\
$L_{\rm bg3     }$   & $0.05 <  z_{\rm{fg}} \le  0.5$   & $0.85 \le  z_{\rm bg} \le  1.01$ & $-21.0$  & $0.32$ & $0.92$ & $277\ 513  $ & $273\ 113  $ & $18\ 325\ 822 $ & $ 118 \pm 3 $ \\
$L_{\rm bg4     }$   & $0.05 <  z_{\rm{fg}} \le  0.5$   & $1.01 \le  z_{\rm bg} \le  2.00$ & $-21.0$  & $0.32$ & $1.20$ & $277\ 598  $ & $260\ 978  $ & $17\ 531\ 415 $ & $ 124 \pm 3 $ \\
$L_{\rm bgbright}$   & $0.05 <  z_{\rm{fg}} \le  1.0$   & $r' \le 23.06                 $ & $-21.0$  & $0.38$ & $0.81$ & $1\ 711\ 502 $ & $1\ 035\ 270 $ & $67\ 033\ 496 $ & $ 115 \pm 2 $ \\
$L_{\rm bgfaint }$   & $0.05 <  z_{\rm{fg}} \le  1.0$   & $r' >   23.06                 $ & $-21.0$  & $0.48$ & $1.04$ & $1\ 811\ 810 $ & $626\ 348  $ & $67\ 025\ 398 $ & $ 119 \pm 1 $ \\
$L_{\rm rhbig   }$   & $0.05 <  z_{\rm{fg}} \le  1.0$   & $r_{h} \le 3.17               $ & $-21.0$  & $0.40$ & $0.87$ & $1\ 772\ 769 $ & $890\ 184  $ & $67\ 031\ 724 $ & $ 117 \pm 2 $ \\
$L_{\rm rhsmall }$   & $0.05 <  z_{\rm{fg}} \le  1.0$   & $r_{h} >   3.17               $ & $-21.0$  & $0.42$ & $0.92$ & $1\ 791\ 460 $ & $771\ 434  $ & $67\ 027\ 170 $ & $ 116 \pm 2 $ \\
$L_{\rm snhigh  }$   & $0.05 <  z_{\rm{fg}} \le  1.0$   & $S/N \ge  9.34                $ & $-21.0$  & $0.37$ & $0.80$ & $1\ 695\ 097 $ & $1\ 037\ 322 $ & $67\ 029\ 067 $ & $ 115 \pm 2 $ \\
$L_{\rm snlow   }$   & $0.05 <  z_{\rm{fg}} \le  1.0$   & $S/N <  9.34                  $ & $-21.0$  & $0.48$ & $1.04$ & $1\ 819\ 346 $ & $624\ 296  $ & $67\ 029\ 827 $ & $ 119 \pm 1 $ \\
\hline\hline
\end{tabular}
\label{tab:systematics}
\end{center}
Results for the systematics check by splitting the background sample (redshift, apparent luminosity, object size and S/N-ratio) and analysing the subsamples with respect to the same foreground sample. The table shows effective lens luminosities and foreground/background-redshifts, number of lenses, sources and lens-source pairs and the velocity dispersions $\sigma$ from the SIS-fit to the inner region ($R \le 200\ h^{-1}$ kpc.)
\end{table*}
\section{3D-LOS-projected lensing signal simulations}
\label{sec:simulations}
The 3D-LOS-projected lensing signal simulations are based on our maximum likelihood results for the best-fit values of the halo parameters and their scaling relation with galaxy luminosity.
\\
\\
For these shape-noise-free simulations we keep positions, luminosities and redshifts from our previously defined lens and source samples and calculate the tangential shear signal $\gamma_{\rm t}$ and the excess surface mass density $\Delta\Sigma$ induced by the lens sample to the source sample, in analogy to the maximum likelihood analysis, but with fixed values for the halo parameters. We use two different fiducial parameter sets for red and blue lens galaxies in our simulations. This takes the higher red galaxy mass for a given luminosity into account. Firstly assuming a truncated isothermal sphere (BBS) (see chapters \ref{sec:BBS_theory} and \ref{sec:BBS}), we adopt our measured values for the velocity dispersion $\sigma^{*}_{\rm red} = $149 km\ $\rm{s^{-1}}$ and the truncation radius $s^{*}_{\rm red} = 337 \ h^{-1}$ kpc for red galaxies and \mbox{$\sigma^{*}_{\rm blue} = 118$ km}\ $\rm{s^{-1}}$ and $s^{*}_{\rm blue} = 84 \ h^{-1}$ kpc for blue galaxies. Secondly assuming an NFW profile (see Sections \ref{sec:NFW_theory} and \ref{sec:NFW}), we adopt a concentration parameter $c^{*}_{\rm red} = 6.4$ and a virial radius $r^{*}_{200,\rm red} = 160 \ h^{-1}$ kpc for red galaxies and $c^{*}_{\rm blue} = 7.0$ and $r^{*}_{200,\rm blue} = 115 \ h^{-1}$ kpc for blue type galaxies. Since the simulations with a single-power law fit for the $r_{200}-L$ scaling relation lead to an underestimation of the excess surface mass density $\Delta \Sigma$ for the faintest luminosity bins based on the NFW profile, we assume a double-power law scaling relation for the $r_{200}$ with a change in slope at $L=10^{10}\ h^{-2}\ L_{\odot}$ (\mbox{$\eta_{r_{200},\rm{bright}}^{\rm red}=\eta_{r_{200},\rm{bright}}^{\rm blue}=1/3$}, $\eta_{r_{200},\rm{faint}}^{\rm red}=0$, $\eta_{r_{200},\rm{faint}}^{\rm blue}=0.21$). In addition we assume a change in slope for red galaxies for the $s-L$ scaling behavior when considering a BBS profile ($\eta_{s,\rm{faint}}^{\rm red}=0$). We then calculate the gravitational shears for the galaxies expected from GGL, set them equal to the `measured' ellipticities and rerun our analysis based on these artificial catalogues.
\subsection{Predicted tangential alignment}
\label{sec:sim.etc}
\begin{figure*}
\centering
\includegraphics[width=8.8cm]{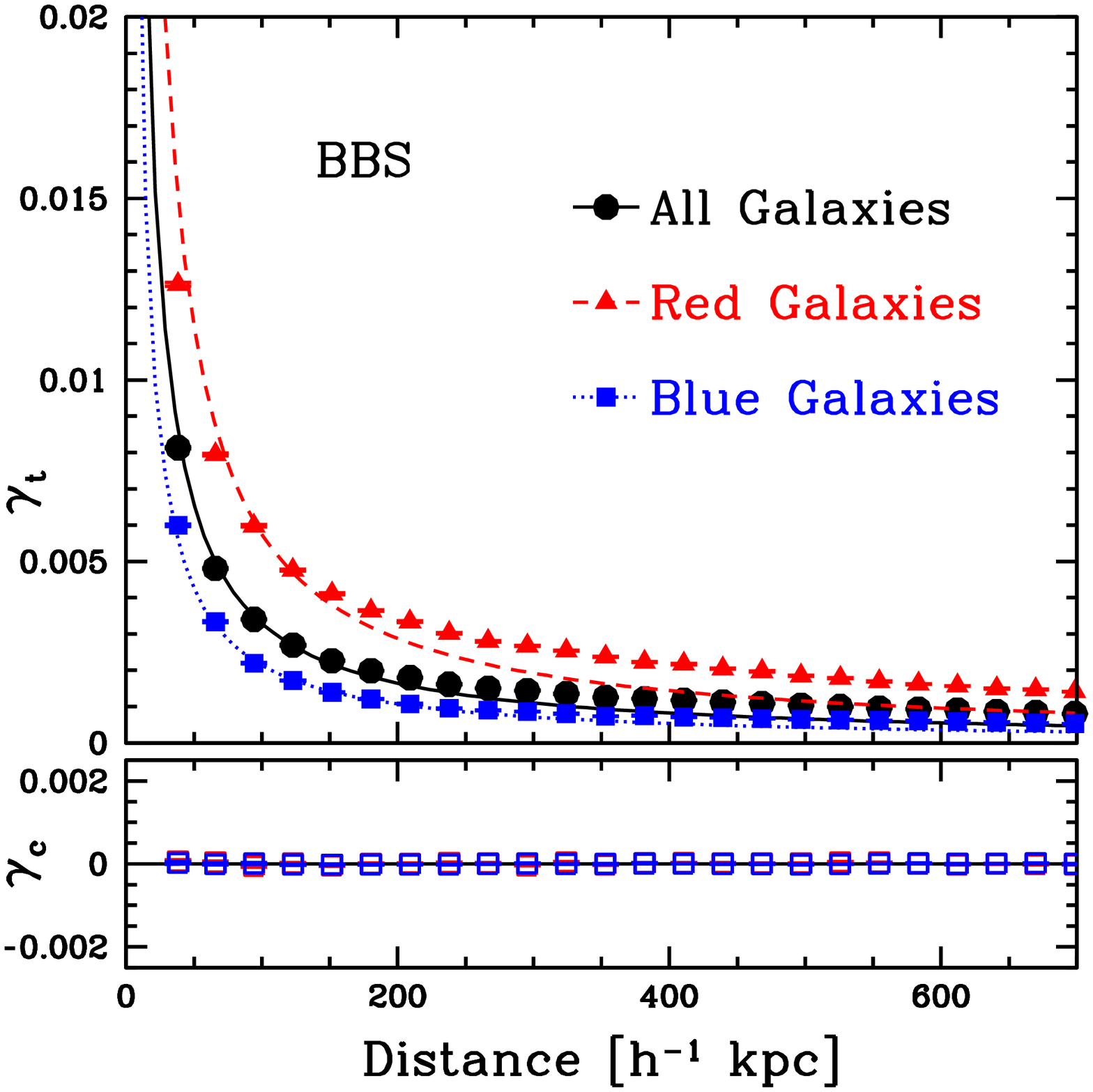}
\includegraphics[width=8.8cm]{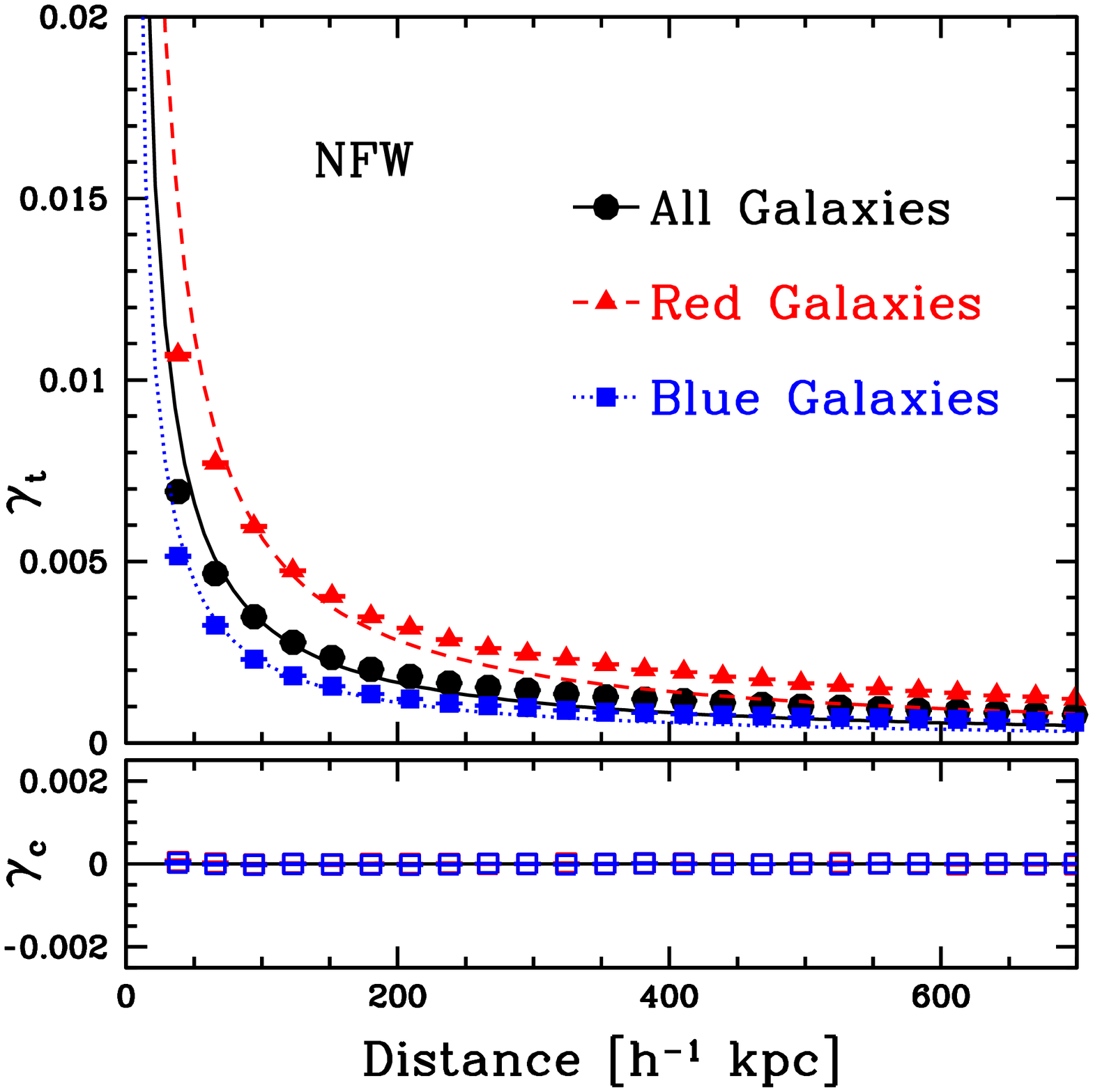}
\caption{Tangential shear signal for simulated sample, BBS (left) and NFW (right) for galaxies with $-24 \le  M_{r'} < -20$. The colour code is the same as before, black stands for all galaxies, red for red galaxies and blue for blue galaxies.}
\label{fig:WIDE.total.etc.simulations}
\end{figure*}

\begin{figure*}
\centering
\includegraphics[width=8.8cm]{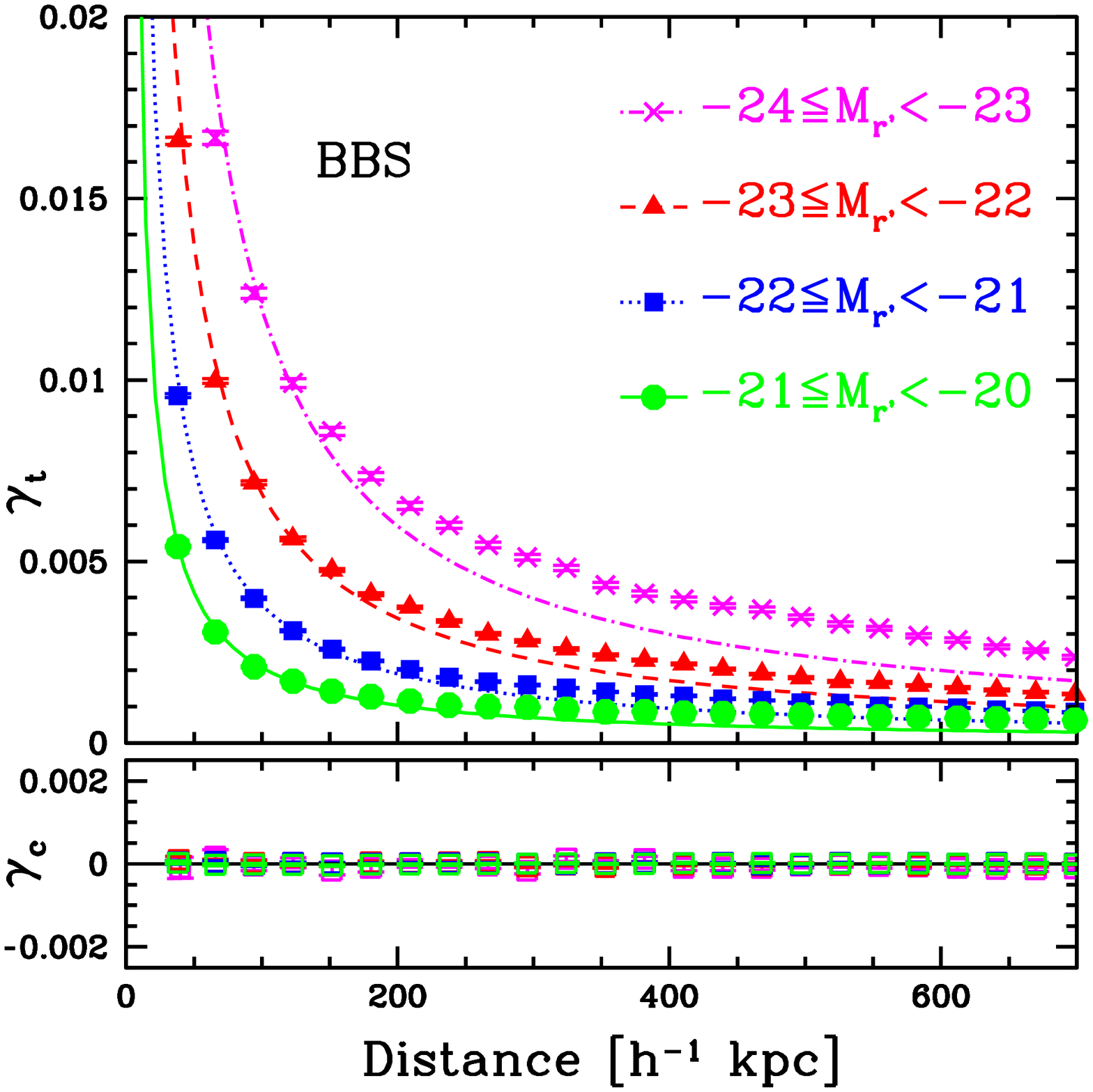}
\includegraphics[width=8.8cm]{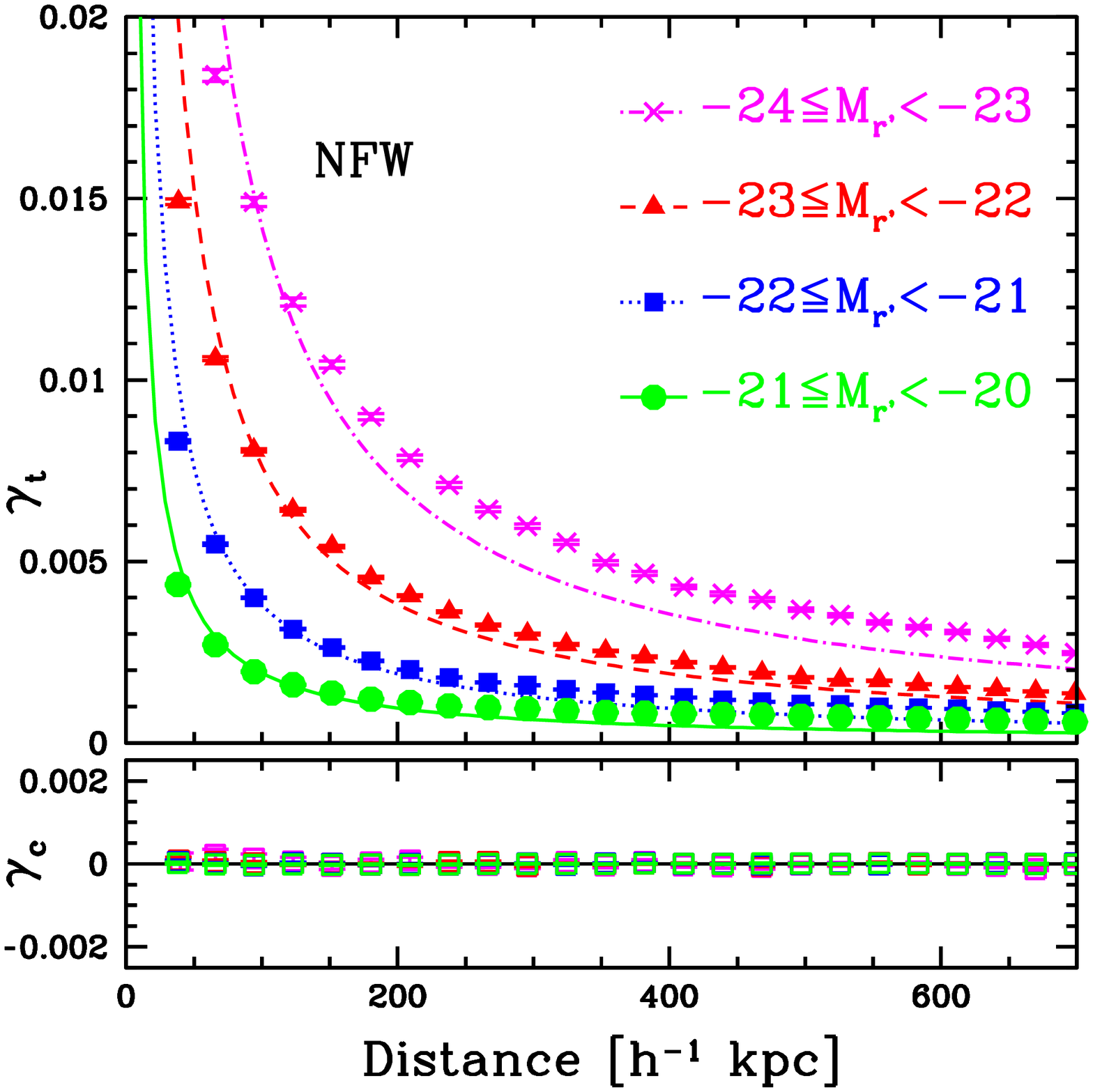}
\caption{Tangential shear signal for simulated samples, BBS (left) and NFW (right), looking at four individual one-magnitude luminosity bins with $-24 \le  M_{r'} < -20$.}
\label{fig:WIDE.total.etc.lbins.simulations}
\end{figure*}

We repeat the analysis as described in Section~\ref{sec:alignment} and measure the tangential shear $\gamma_{t}$-profile for our simulated shear catalogues. The qualitative results are in agreement to the results of \citet{brainerd10}, as on short scales we see an excellent agreement of the profile to an SIS or BBS profile (the difference on scales significantly shorter than the truncation radius is negligible). On larger scales we observe a higher signal than expected from an SIS, due to multiply deflected sources by nearby galaxy haloes (see Figs.~\ref{fig:WIDE.total.etc.simulations} and \ref{fig:WIDE.total.etc.lbins.simulations}). We observe that the reliability of the estimate for the velocity dispersion $\sigma$ depends on the luminosity range of the investigated lens sample. While for the brighter luminosity bins the results of the simulated tangential shear analyses agree quite well with the observations, for the fainter luminosity bins the simulated signal amplitude is slightly lower than the observed one. The reasons for this behavior can either be caused by not considered host haloes such as group or cluster haloes which are not sufficiently taken into account by only considering galaxy haloes. Another reason can be that the assumed scaling relations which are assumed to be power laws in luminosity are not universally valid but only a justified approximation depending on the investigated luminosity range.
\subsection{Predicted versus measured excess surface mass density}
\label{sec:sim.ds}
After the tangential shear profile analysis of the gravitational shear simulations we extend our investigation and calculate the excess surface mass density $\Delta\Sigma$-profile (see Section~\ref{sec:DeltaSigma}) based on our simulations.
The results confirm the results from the tangential shear analysis: for bright galaxies the signals from observational and simulated data agree very well down to a lens rest frame magnitude of $M_{r'} \sim  -20$. For fainter (particularly BBS) lenses the amplitudes of the simulated sample starts to slightly underestimate the observational signal increasing with decreasing luminosity, especially on short scales. This effect appears to be weaker on scales larger than $\sim 200 \ \rm{h^{-1}}$ kpc, see also Fig.~\ref{fig:WIDE.total.ds.sim}. In general the simulated BBS profiles fit the observational data better than the NFW profiles, when assuming single-power law fits for the scaling relations (not shown).  However, when assuming double-power law fits for the scaling of the virial radius $r_{200}$ and the truncation radius $s$ the NFW profile provides the better fit.
\begin{figure*}
\includegraphics[width=18.5cm]{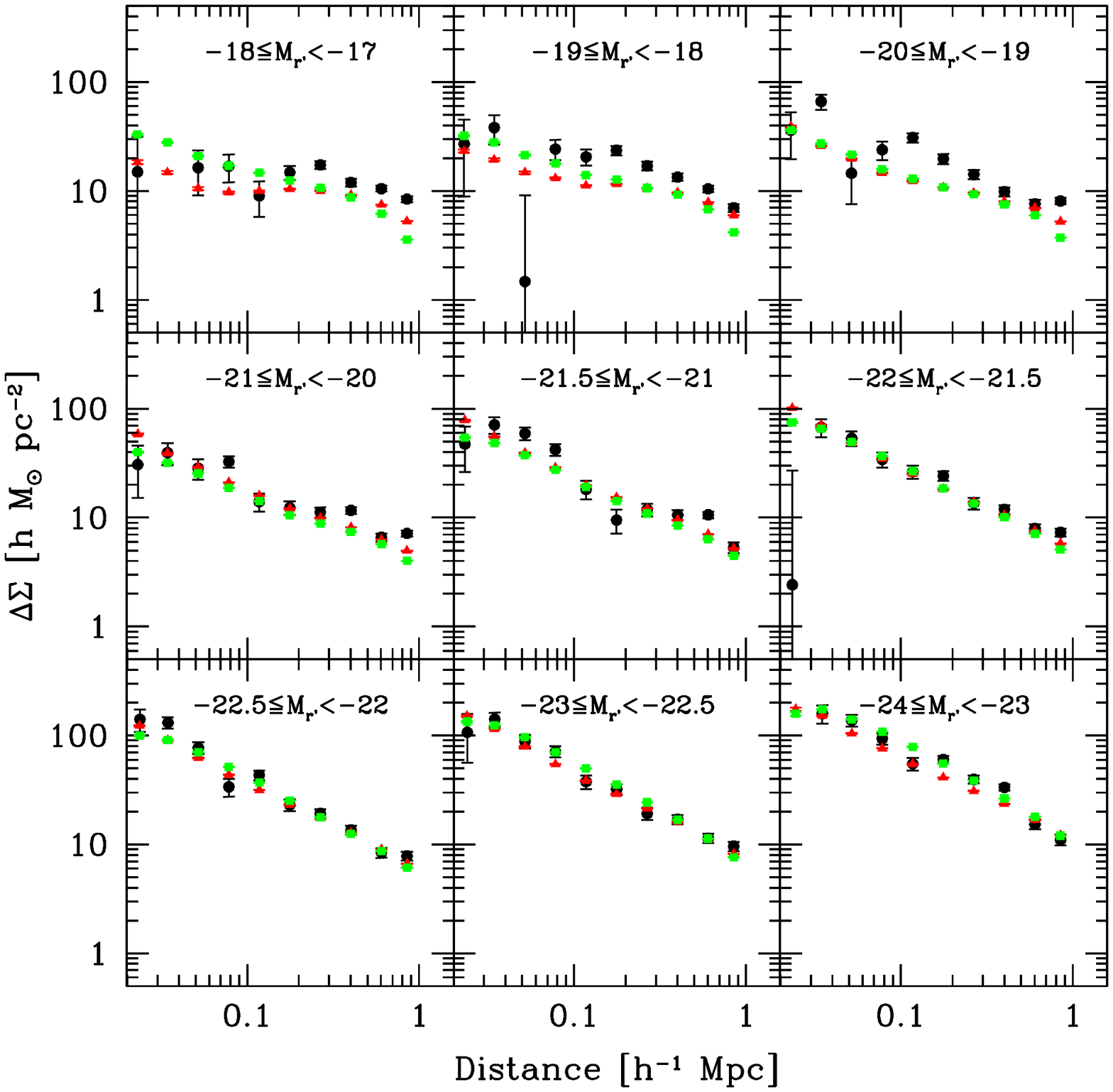}
\caption{Excess surface mass density $\Delta\Sigma$ for different luminosity bins only considering red galaxies. Black circles are the observational data points, red triangles come from the BBS simulation and green squares from the NFW simulation.}
\label{fig:WIDE.red.ds.sim}
\end{figure*}
\begin{figure*}
\includegraphics[width=18.5cm]{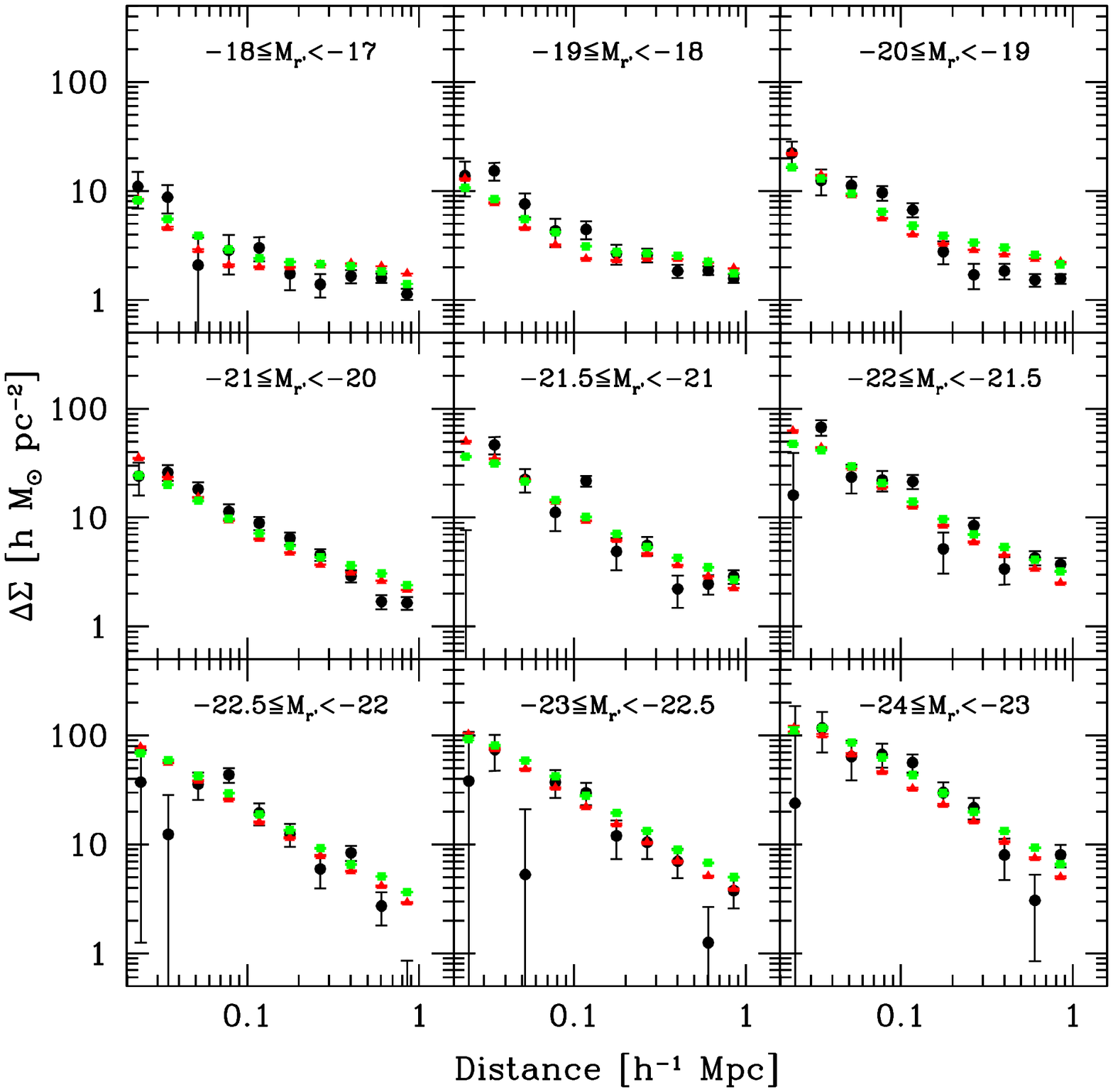}
\caption{Excess surface mass density $\Delta\Sigma$ for different luminosity bins only considering blue galaxies. Black circles are the observational data points, red symbols come from the BBS simulation and green squares from the NFW simulation.}
\label{fig:WIDE.blue.ds.sim}
\end{figure*}
\\
If red and blue galaxies are considered separately, the simulations do not appear to fit equally well, in particular in the fainter luminosity bins. While the simulations of the blue and bright red ($M_{r'}<-20$) galaxies describe the observational results fairly well, this is not the case for the fainter red galaxies, where the red lens simulations seem to partially underestimate the observed $\Delta\Sigma$-profiles (see Figs.~\ref{fig:WIDE.red.ds.sim} and \ref{fig:WIDE.blue.ds.sim}). 
\begin{figure*}
\includegraphics[width=18.5cm]{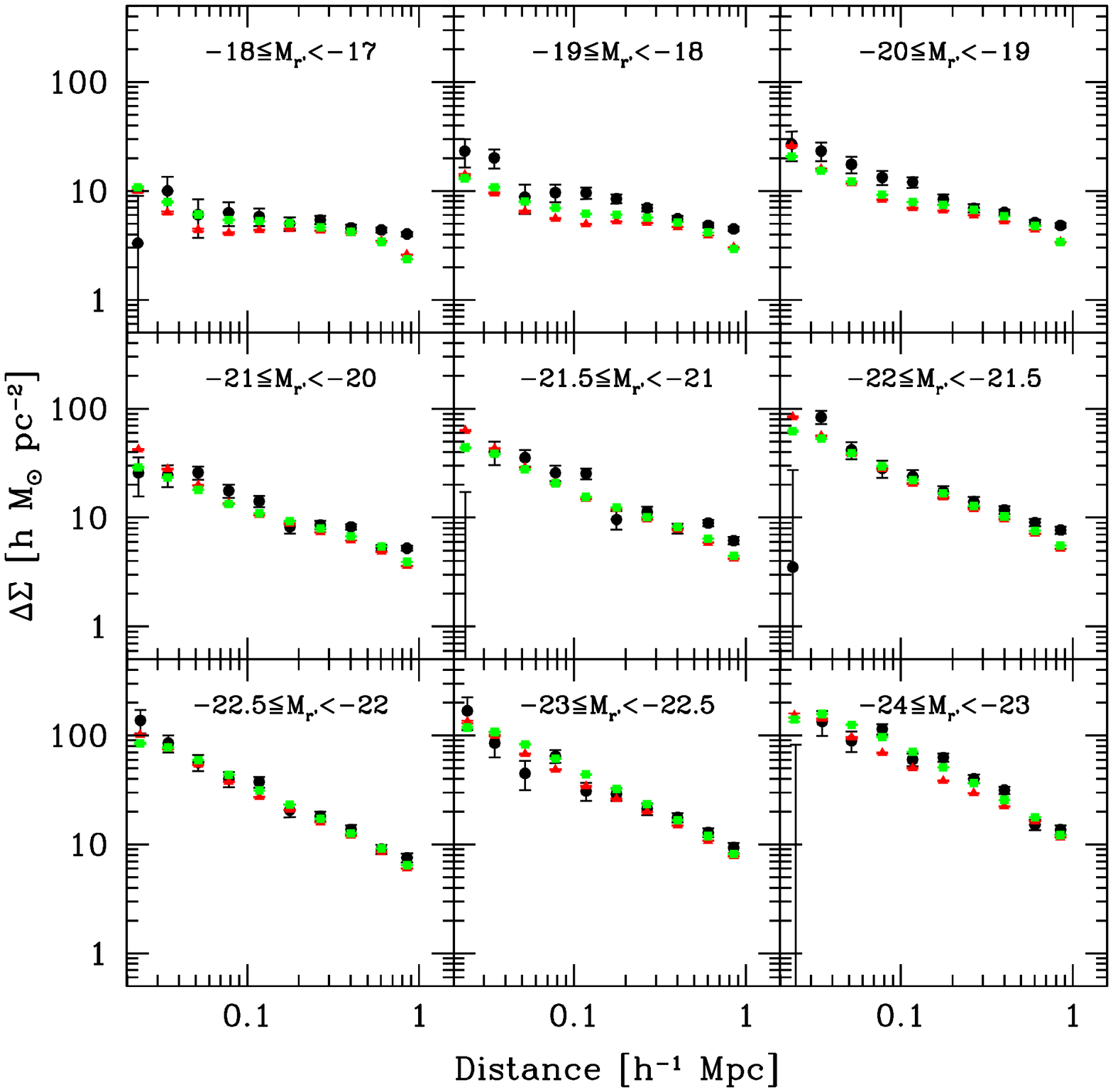}
\caption{Excess surface mass density for the lens sample in high density environments. Black circles are the observational data points, red triangles come from the BBS simulation and green squares from the NFW simulation.}
\label{fig:WIDE.total.high.ds.sim}
\end{figure*}
\begin{figure*}
\includegraphics[width=18.5cm]{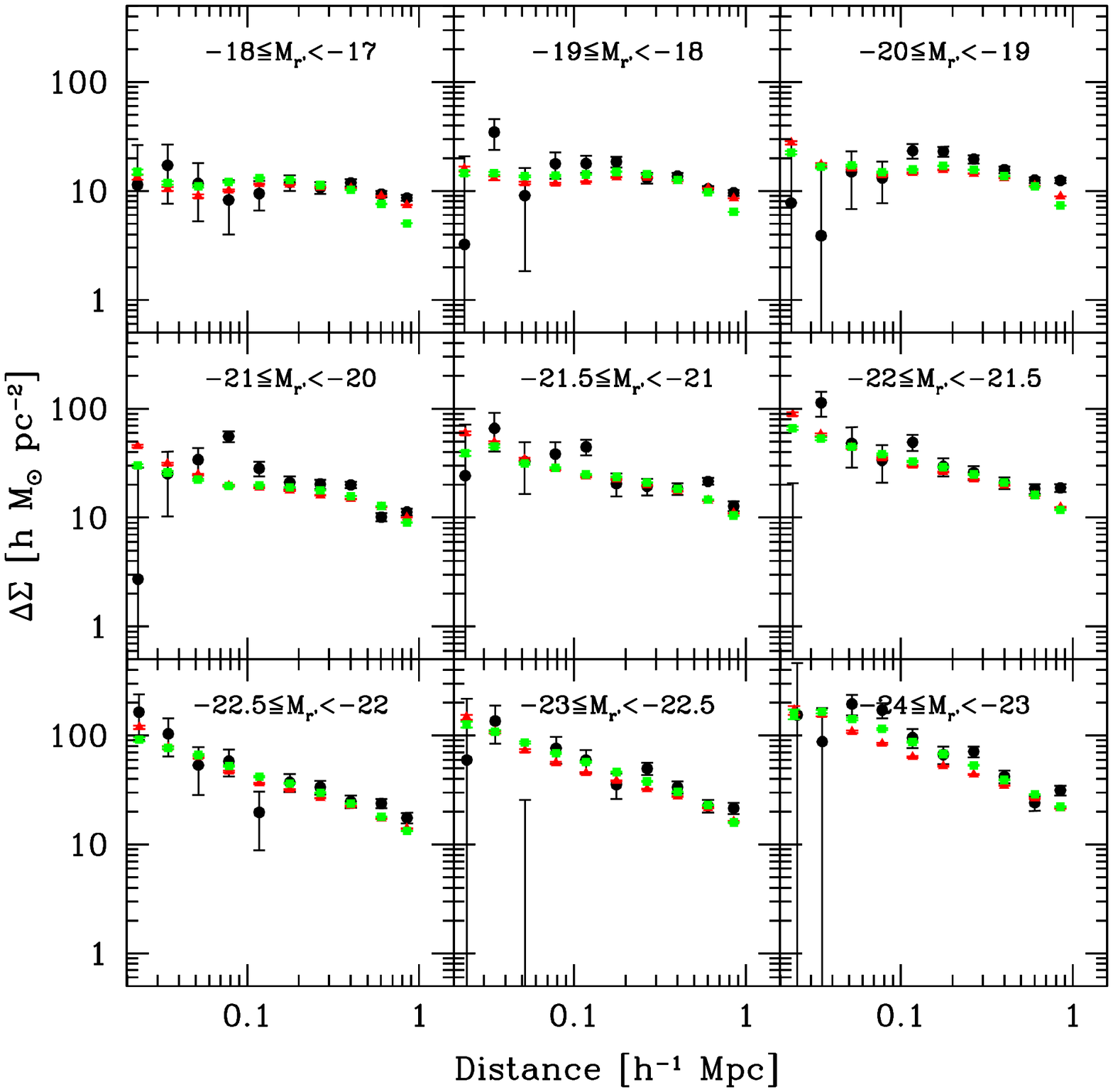}
\caption{Excess surface mass density for the lens sample in very high density environments. Black circles are the observational data points, red triangles come from the BBS simulation and green squares from the NFW simulation.}
\label{fig:WIDE.total.veryhigh.ds.sim}
\end{figure*}
\begin{figure*}
\includegraphics[width=18.5cm]{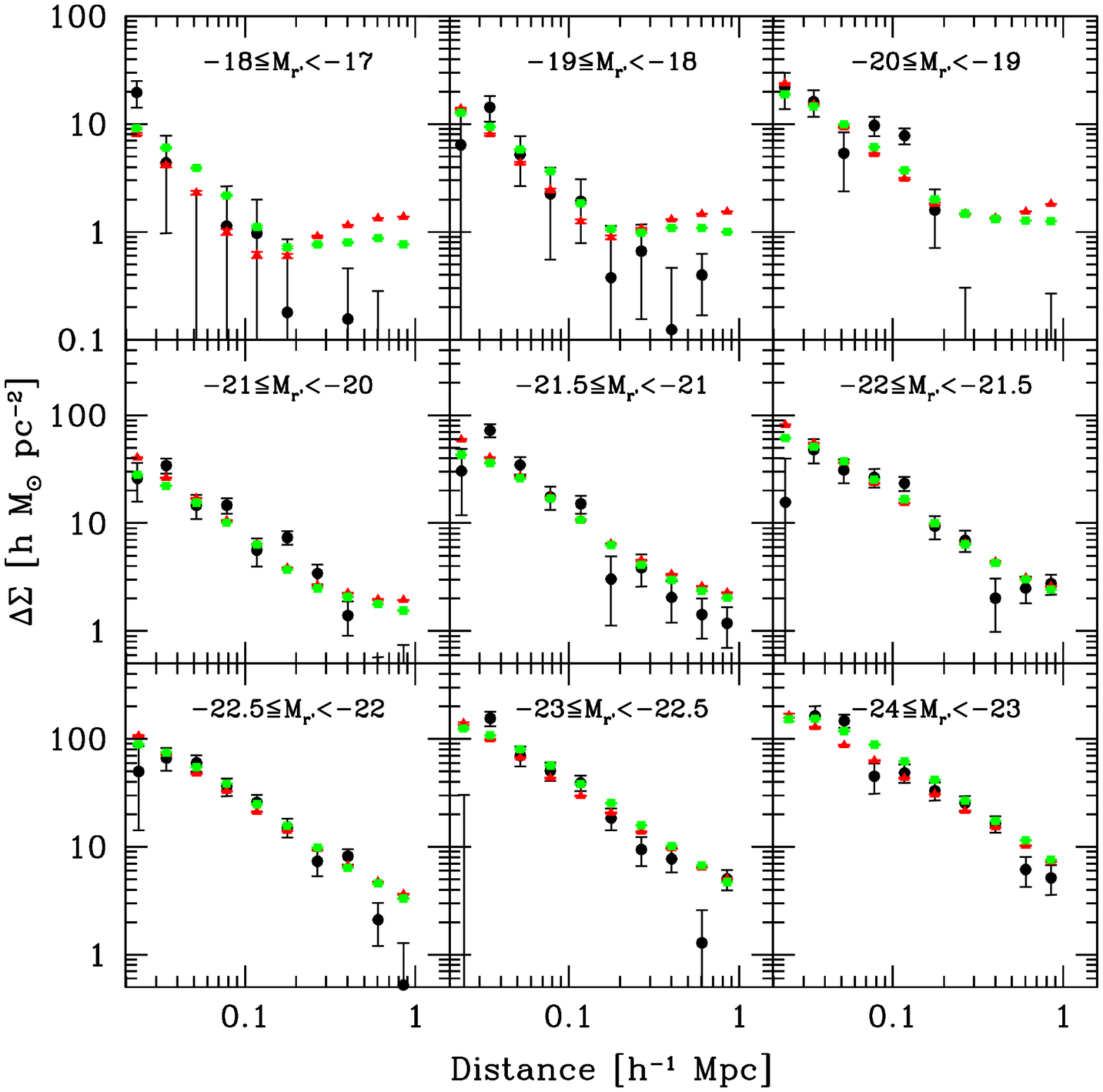}
\caption{Excess surface mass density for the lens sample in low density environments. Black circles are the observational data points, red triangles come from the BBS simulation and green squares from the NFW simulation.}
\label{fig:WIDE.total.low.ds.sim}
\end{figure*}
\begin{figure*}
\includegraphics[width=18.5cm]{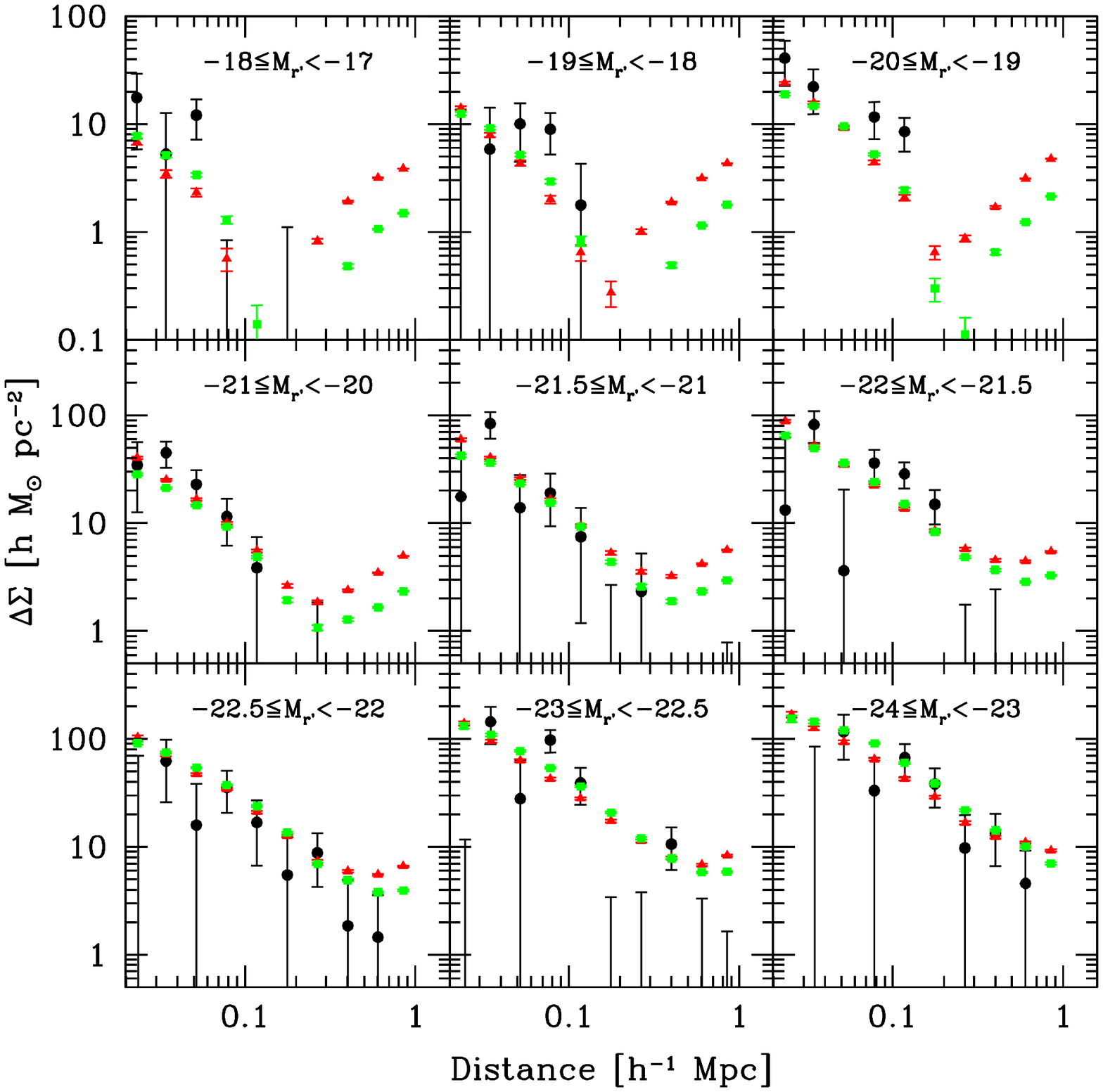}
\caption{Excess surface mass density for the lens sample in very low density environments. Black circles are the observational data points, red triangles come from the BBS simulation and green squares from the NFW simulation.}
\label{fig:WIDE.total.verylow.ds.sim}
\end{figure*}
\\
In the following we further analyse the simulated excess surface mass density as a function of the environment density. We split our lens sample as described in Section~\ref{sec:subsamples}. In the highest density environment (see Figs.~\ref{fig:WIDE.total.high.ds.sim} and \ref{fig:WIDE.total.veryhigh.ds.sim}), the simulations and the observational data match well for all luminosity bins. The profile slope for the lowest luminosity intervals is rather flat, i.e. the dominating part of the lensing signal on scales larger than $200\ h^{-1}$ kpc is not caused by galaxies from this specific luminosity bin but from brighter galaxies in the environment. The uncertainties for the faint low density environment lenses are quite large due to low number statistics. This makes it difficult to derive decisive conclusions. Nonetheless for the faintest luminosity bins the lensing signal is not inconsistent on scales shorter than $R = 200\ h^{-1}$ kpc (see Figs.~\ref{fig:WIDE.total.low.ds.sim} and \ref{fig:WIDE.total.verylow.ds.sim}). Taking into account the simple model we use to describe the mass associated to the galaxies, it is remarkable to see that the both simulations (based on the BBS and the NFW profile) agree very well with the observations over a wide scale of projected separations, galaxy luminosities and environment. Down to an absolute magnitude of $M_{r'} \sim -17$ BBS and NFW simulations mostly agree with each which makes it impossible to decide whether the observational galaxy haloes follow a BBS or an NFW profile. Only in the fainter bins the BBS and NFW signal start to differ, due to the differently modelled scaling relations for galaxies fainter than $10^{10}\ h^{-2}\ L_{\odot}$. 
\end{document}